\documentstyle[aps,preprint,tighten,floats,psfig,epsfig,rotate]{revtex}

\begin{document}

\preprint{
\vbox{
% \hbox{submitted: August 16, 2004}
\hbox{JLAB-THY-04-266}
}}

\title{Quark-Hadron Duality in Electron Scattering}

\author{W.~Melnitchouk$^1$, R.~Ent$^1$, and C.E.~Keppel$^{1,2}$}

\address{$^1$ Jefferson Lab, 12000 Jefferson Avenue,
	Newport News, VA 23606, USA}
\address{$^2$ Department of Physics, Hampton University,
	Hampton, VA 23668, USA}

\maketitle
 
\begin{abstract}

The duality between partonic and hadronic descriptions of physical
phenomena is one of the most remarkable features of strong interaction
physics.  A classic example of this is in electron-nucleon scattering, in
which low-energy cross sections, when averaged over appropriate energy
intervals, are found to exhibit the scaling behavior expected from 
perturbative QCD.  We present a comprehensive review of data on structure
functions in the resonance region, from which the global and local aspects
of duality are quantified, including its flavor, spin and nuclear medium
dependence.  To interpret the experimental findings, we discuss various
theoretical approaches which have been developed to understand the
microscopic origins of quark-hadron duality in QCD.  Examples from other
reactions are used to place duality in a broader context, and future
experimental and theoretical challenges are identified.

\end{abstract}

\vspace*{6cm}

% Keywords:
% structure functions; duality; higher twists


% \vspace*{1cm}
%
% Physics Reports {\bf xx} (2004) xxx.

%%%%%%%%%%%%%%%%%%%%%%%%%%%%%%%%%%%%%%%%%%%%%%%%%%%%%%%%%%%%%%%%%%%%%%%%%%
\cleardoublepage
\tableofcontents
\cleardoublepage

%%%%%%%%%%%%%%%%%%%%%%%%%%%%%%%%%%%%%%%%%%%%%%%%%%%%%%%%%%%%%%%%%%%%%%%%%%
\section{Introduction}
\label{sec:introduction}

Three decades after the establishment of QCD as the theory of the
strong nuclear force, understanding {\em how} QCD works remains one
of the great challenges in nuclear physics.
A major obstacle arises from the fact that the degrees of freedom
observed in nature (hadrons and nuclei) are totally different from
those appearing in the QCD Lagrangian (current quarks and gluons).
The remarkable feature of QCD at large distances --- quark confinement
--- prevents the individual quark and gluon constituents making up
hadronic bound states to be removed and examined in isolation.
Making the transition from quark and gluon (or generically, parton)
to hadron degrees of freedom is therefore the key to our ability to
describe nature from first principles.

\begin{figure}
\begin{minipage}[ht]{4.25in}
\epsfig{figure=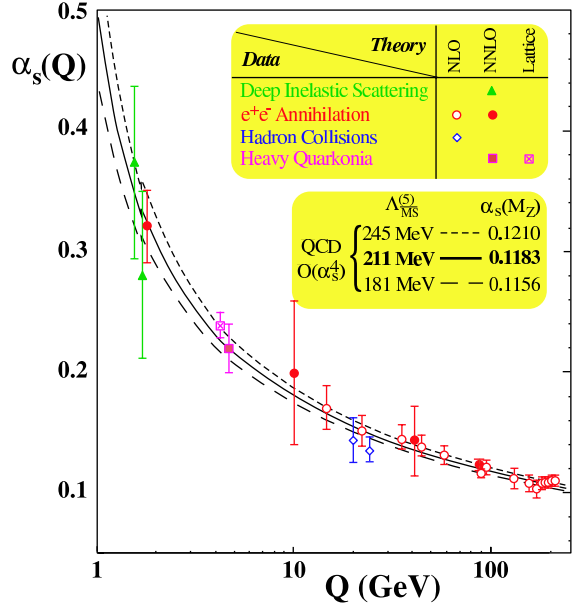, height=11cm}
\end{minipage}
\begin{centering}
\vspace*{-2cm}
\caption{\label{fig:asqplot}
	Summary of measurements of $\alpha_s(Q)$.
	The curves shown differ in their choice of the QCD scale
	parameter, $\Lambda_{\rm QCD}$.
	(From Ref.~\protect\cite{Be02}.)}
\end{centering}
\end{figure}

The property of QCD known as asymptotic freedom, in which quarks
interact weakly at short distances, allows one to calculate hadronic
observables at asymptotically high energies perturbatively, in terms
of expansions in the strong coupling constant $g_s$, or more commonly
$\alpha_s = g_s^2/4\pi$.
Figure~\ref{fig:asqplot} shows a recent summary of all measurements
of $\alpha_s$ \cite{Be02}, as a function of the momentum scale $Q$.
The small value of $\alpha_s$ at large momentum scales (or short
distances) makes possible an efficient description of phenomena in
terms of quarks and gluons.

At low momentum scales, on the other hand, where $\alpha_s$ is large,
the effects of confinement make strongly-coupled QCD highly
nonperturbative.
Here, it is more efficient to work with collective degrees of freedom,
the physical mesons and baryons.
Because of confinement, quarks and gluons must end up in color singlet
bound states of hadrons, so that exact QCD calculations at some level
must be sensitive to multihadron effects.

Despite the apparent dichotomy between the partonic and hadronic
regimes, in nature there exist instances where the behavior of
low-energy cross sections, averaged over appropriate energy intervals,
closely resembles that at asymptotically high energies, calculated in
terms of quark-gluon degrees of freedom.
This phenomenon is referred to as {\em quark-hadron duality},
and reflects the relationship between confinement and asymptotic
freedom, and the transition from perturbative to nonperturbative
regimes in QCD.
Such duality is in fact quite general, and arises in many different
physical processes, such as in $e^+ e^-$ annihilation into hadrons,
or semi-leptonic decays of heavy mesons.
In electron--nucleon scattering, quark-hadron duality links the
physics of resonance production to the physics of scaling, and is the
focus of this review.

The observation of a nontrivial relationship between inclusive
electron--nucleon scattering cross sections at low energy, in the
region dominated by the nucleon resonances, and that in the deep
inelastic scaling regime at high energy predates QCD itself.
While analyzing the data from the early deep inelastic scattering
experiments at SLAC, Bloom and Gilman observed \cite{BG1,BG2} that
the inclusive structure function at low hadronic final state mass,
$W$, generally follows a global scaling curve which describes
high-$W$ data, to which the resonance structure function averages.
Initial interpretations of this duality used the theoretical tools
available at the time, namely finite energy sum rules, or consistency
relations between hadronic amplitudes inspired by the developments
in Regge theory which occurred in the 1960s \cite{COLLINS}.

Following the advent of QCD in the early 1970s, Bloom-Gilman duality
was reformulated \cite{DGP1,DGP2} in terms of an operator product
(or ``twist'') expansion of moments of structure functions.
This allowed a systematic classification of terms responsible for
duality and its violation in terms of so-called ``higher-twist''
operators, which describe long-range interactions between quarks
and gluons.
Ultimately, however, this description fell short of adequately
explaining {\em why} particular multi-parton correlations were
suppressed, and {\em how} the physics of resonances gave way to
scaling.
From the mid-1970s the subject was largely forgotten for almost two
decades, as attention turned from the complicated resonance region to
the more tractable problem of calculating higher order perturbative
corrections to parton distributions, and accurately describing their
$Q^2$ evolution.

With the development of high luminosity beams at modern accelerator
facilities such as Jefferson Lab (JLab), a wealth of new information
on structure functions, with unprecedented accuracy and over a wide
range of kinematics, has recently become available.
One of the striking findings of the new JLab data \cite{F2JL1}
is that Bloom-Gilman duality appears to work exceedingly well, 
down to $Q^2$ values as low as 1~GeV$^2$ or even below, which is
considerably lower than previously believed.
Furthermore, the equivalence of the averaged resonance and scaling
structure functions seems to hold for each of the prominent resonance
regions individually, indicating that a resonance--scaling duality
exists to some extent locally as well.
Even though at such low $Q^2$ values $\alpha_s$ is relatively large,
{\it on average} the inclusive scattering process appears to mimic
the scattering of electrons from almost free quarks.
All of this has subsequently led to a resurgence of interest in
questions about the origin of duality in deep inelastic scattering
and related processes, and has motivated a number of theoretical
studies which have helped to elucidate important aspects of the
transition from coherent to incoherent phenomena.

In principle, at high energies the duality between quark and hadron
descriptions of phenomena can be considered as formally exact.
However, for a limited energy range, there is no reason to expect
the accuracy to which duality holds and the kinematic regime where
it applies to be similar for different physical processes.
In fact, there could be qualitative differences between the workings
of duality in spin-dependent structure functions and spin-averaged
ones, or for different hadrons --- protons compared to neutrons,
for instance.
The new data not only allow one to study in unprecedented detail the
systematics of duality in local regions of kinematics, but also for
the first time make it possible to examine the spin and target
dependence of duality.
In addition, they allow more reliable studies of the moments of
structure functions in the intermediate $Q^2$ region, where there
are sizable contributions from nucleon resonances.
%
% traditionally defined as $W < 2$~GeV.

The recent resonance structure function studies have revealed an
important application of duality: if the workings of the resonance--deep
inelastic interplay are sufficiently well understood, the region of
% high Bjorken-$x$ ($x \agt 0.5$, where $x$ is the longitudinal momentum
high Bjorken-$x$ ($x \agt 0.7$, where $x$ is the longitudinal momentum
fraction of the hadron carried by the parton in the infinite momentum
frame) would become accessible to precision studies.
As we explain later in this report, there are many reasons why accurate
knowledge of the large-$x$ region is important.
However, due to limitations of luminosity and energy, this region has
not been mapped out with the required precision in any experiments to
date.
Other applications of duality can be found in providing an efficient
average low-energy description of hadronic physics used in the
interpretation of neutrino oscillation and high energy physics
experiments, and in a more detailed understanding of how quarks
evolve into hadrons (hadronization).
The latter is the subject of duality studies in meson electroproduction
reactions, where at present only sparse data exist.

Finally, it is important to note that the moments of polarized and
unpolarized structure functions are currently the subject of some
attention in lattice QCD simulations.
Comparisons of the experimental moments with those calculated on the
lattice over a range $Q^2 \approx 1$--10~GeV$^2$ will allow one to
determine the size of higher twist corrections and the role played by
quark-gluon correlations in the nucleon.
For the experimental moments, an appreciable fraction of the strength
resides in the nucleon resonance region, so that understanding of
quark-hadron duality is vital also for the interpretation of the
results from lattice QCD.

In view of the accumulation of high precision data on structure
functions in the resonance--scaling transition region, and the recent
theoretical developments in understanding the origins of duality,
it is timely therefore to review the status of quark-hadron duality
in electron--nucleon scattering.
Following a review of definitions and kinematics relevant for
inclusive scattering in Section~II, we give an historical
perspective of duality in Section~III, recalling the understanding
and interpretation of quark-hadron duality as it existed up to the
1970s.
Section~IV is the central experimental part of this review, where we
describe the progress in the study of duality in both spin-averaged
and spin-dependent structure functions over the last decade.
Readers familiar with Regge theory and duality in hadronic reactions
may wish to omit Sec.~III and proceed to Sec.~IV directly.

The theoretical foundations of quark-hadron duality are reviewed in
Section~V.
Here we firstly outline the basic formalism of the operator product
expansion relevant for the interpretation of duality in terms of
higher twist suppression.
This is followed by a survey of duality in various dynamical models
which have been used to verify the compatibility of scaling in the
presence of confinement.
We then proceed to more phenomenological applications of local
duality, and extensions of duality to semi-inclusive and exclusive
reactions.

To shed light on the more fundamental underpinnings of quark-hadron
duality in QCD, in Section~VI the concept of duality in electron
scattering is compared to that in closely-related fields, such as
$e^+ e^-$ annihilation into hadrons, heavy meson decays, and
proton-antiproton annihilation.
Section~VII deals with applications of duality and anticipated studies
over the next decade, and some concluding remarks are given in
Section~VIII.

\clearpage
%%%%%%%%%%%%%%%%%%%%%%%%%%%%%%%%%%%%%%%%%%%%%%%%%%%%%%%%%%%%%%%%%%%%%%%%%%
\section{Lepton--Nucleon Scattering: Kinematics and Cross Sections}
\label{sec:formalism}

In this section we present the kinematics relevant for inclusive
lepton--nucleon scattering, and introduce notations and definitions
for cross sections, structure functions, and their moments, both
for unpolarized and polarized scattering.
These can be found in standard texts \cite{TW_BOOK,CLOSEBOOK},
but the most relevant formulas are provided here for completeness.

% .......................................................................
\subsection{Kinematics}

\begin{figure}[hb]
\hspace*{2.5cm}
\epsfig{figure=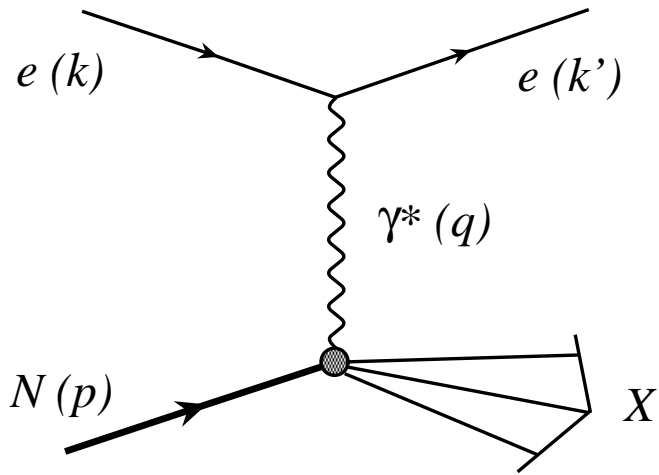, height=11cm}
\vspace*{-3.5cm}
\caption{\label{fig:fig_kin}
	Inclusive lepton--nucleon scattering in the one-photon exchange
	approximation.
	The four-momenta of the particles are given in parentheses.}
\end{figure}

The process which we focus on mainly in this report is inclusive
scattering of an electron (the case of muon or neutrino scattering
is similar) from a nucleon (or another hadronic or nuclear) target,
$e N \to e' X$, where $X$ represents the inclusive hadronic final state.
In the target rest frame, the incident electron with energy $E$ scatters
from the target through an angle $\theta$, with a recoil energy $E'$.
In the one-photon (or Born) approximation, as illustrated in
Fig.~\ref{fig:fig_kin}, the scattering takes place via the exchange of
a virtual photon (or $W^\pm$ or $Z$ boson in neutrino scattering) with
energy
\begin{equation}
\nu = E - E'\ ,
\end{equation}
and momentum $\vec q$.

Throughout we use natural units, $\hbar = c = 1$, so that momenta and
masses are expressed in units of GeV (rather than GeV/c or GeV/c$^2$).
The virtuality of the photon is then given by
$q^2 = \nu^2 - \vec q\, ^2$.
Since the photon is spacelike, it is often more convenient to work
with the positive quantity $Q^2 \equiv -q^2$, which is related to the
electron energies and scattering angle by
\begin{equation}
Q^2 = 4 E E'\sin^2{\theta \over 2}\ ,
\end{equation}
where we have also neglected the small mass of the electron.
The invariant mass squared of the final hadronic state is
\begin{equation}
W^2 = (p+q)^2 = M^2 + 2 M \nu - Q^2\ ,
\label{eq:Wsq}
\end{equation}
where $p$ and $q$ are the target nucleon and virtual photon
four-momenta, respectively, and $M$ is the nucleon mass.

The cross sections for this process in general depend on two independent
variables, which can be taken to be the scattering angle and recoil
energy, or alternatively $\nu$ and $Q^2$.
Often they are also expressed in terms of the ratio of $Q^2$ and
$\nu$, through the Bjorken $x$ variable,
\begin{equation}
x = { Q^2 \over 2M\nu }\ .
\label{eq:Bjx}
\end{equation}
In terms of $x$, the hadronic state mass $W$ can also be written as
$W^2 = M^2 + Q^2 (1-x)/x$.
For the special case of elastic scattering, one has $W = M$,
and hence $x = 1$.

% .......................................................................
\subsection{Spin-Averaged Cross Sections}

In the one-photon exchange approximation, the differential cross
section for scattering unpolarized electrons from an unpolarized
nucleon target can be written as
\begin{equation}
{ d^2\sigma \over d\Omega dE' }
= { \alpha^2 \over Q^4 } {E' \over E} L_{\mu\nu} W^{\mu\nu}\ , 
\label{eq:sigtens}
\end{equation}
where $\alpha$ is the fine structure constant, and
$\Omega = \Omega(\theta,\phi)$ is the laboratory solid angle of the
scattered electron.
The leptonic tensor $L_{\mu\nu}$ averaged over initial spins
is given by
\begin{equation}
L_{\mu\nu}
= 2 \left( k_\mu k'_\nu + k'_\mu k_\nu - g_{\mu\nu} k \cdot k' \right)\ ,
\label{eq:Lmunu}
\end{equation}
where $k$ and $k'$ are the initial and final electron momenta,
respectively.

The hadronic tensor $W^{\mu\nu}$ contains all of the information about
the structure of the nucleon target.
Using constraints from Lorentz and gauge invariance, together with
parity conservation, the hadronic tensor can be decomposed into two
independent structures,
\begin{equation}
W^{\mu\nu}
= W_1(\nu,Q^2)
  \left( { q^\mu q^\nu \over q^2} - g^{\mu\nu} \right)
+ { W_2(\nu ,Q^2) \over M^2 }
  \left( p^\mu + { p\cdot q \over q^2 } q^\mu \right)
  \left( p^\nu + { p\cdot q \over q^2 } q^\nu \right)\ ,
\label{eq:Wmunu}
\end{equation}
where $W_1$ and $W_2$ are scalar functions of $\nu$ and $Q^2$.
Using Eqs.~(\ref{eq:Lmunu}) and (\ref{eq:Wmunu}), the differential
cross section can then be written
\begin{equation}
\label{eq:w1w2}
{ d^2\sigma \over d\Omega dE' }
= \sigma_{\rm Mott}
  \left( 2 W_1(\nu,Q^2) \tan^2{\theta\over 2} + W_2(\nu,Q^2)
  \right)\ ,
\end{equation}
where $\sigma_{\rm Mott}$ is the Mott cross section for scattering
from a point particle,
\begin{equation}
\sigma_{\rm Mott}
= { 4\alpha^2 E'^2 \over Q^4 } \cos^2{\theta\over 2}\ .
\label{eq:Mott}
\end{equation}
Note that for a structureless target, $W_1$ and $W_2$ become
$\delta$-functions, and Eq.~(\ref{eq:w1w2}) reduces to the Dirac
cross section for scattering from spin-1/2 particles.

In the Bjorken limit, in which both $Q^2$ and $\nu \to \infty$, but
$x$ is fixed, the structure functions $W_1$ and $W_2$ exhibit scaling.
Namely, they become independent of $Q^2$, and are functions of the
variable $x$ only (logarithmic $Q^2$ dependence enters at finite $Q^2$
through QCD radiative effects).
It is convenient therefore to introduce the dimensionless functions
$F_1$ and $F_2$, defined by
\begin{eqnarray}
\label{eq:f1w1}
F_1(x,Q^2) &=& M W_1(\nu,Q^2)\ ,	\\
F_2(x,Q^2) &=& \nu W_2(\nu,Q^2)\ .
\label{eq:f2w2}
\end{eqnarray}
In the quark-parton model the $F_1$ and $F_2$ structure functions are
given in terms of quark and antiquark distribution functions, $q(x)$
and $\bar q(x)$,
\begin{equation}
\label{eq:f22xf1}
F_2(x) = 2 x F_1(x) = x \sum_q e_q^2 \left( q(x) + \bar q(x) \right)\ ,
\end{equation}
where $q(x)$ is interpreted as the probability to find a quark of
flavor $q$ in the nucleon with light-cone momentum fraction $x$.
The relation between the $F_1$ and $F_2$ structure functions in
Eq.~(\ref{eq:f22xf1}) is referred to as the Callan-Gross relation
\cite{CallanGross}.
Beyond the quark-parton model, the residual $Q^2$ dependence in $F_1$
and $F_2$ arises from scaling violations through perturbative QCD
corrections, as well as $1/Q^2$ power corrections which will be
discussed in the following sections.
In terms of these dimensionless functions, the differential cross
section can be written as
\begin{equation}
\label{eq:f1f2}
{ d^2\sigma \over d\Omega dE' }
= \sigma_{\rm Mott}
  \left( {2\over M} F_1(x,Q^2) \tan^2{\theta\over 2}
       + {1\over\nu} F_2(x,Q^2)
  \right)\ .
\end{equation}
Expressed in this way, the functions $F_1$ and $F_2$ reflect
the possibility of magnetic as well as electric scattering,
or alternatively, the photoabsorption of either transversely
(helicity $\pm 1$) or longitudinally (helicity 0) polarized photons.
From this perspective, the cross section can be expressed
in terms of $\sigma_T$ and $\sigma_L$, the cross sections for the
absorption of transverse and longitudinal photons,
\begin{equation}
\label{eq:gameps}
\sigma \equiv { d^2\sigma \over d\Omega dE' }
= \Gamma \left( \sigma_T(x,Q^2) + \epsilon \sigma_L(x,Q^2) \right)\ .
\end{equation}
Here $\Gamma$ is the flux of transverse virtual photons,
\begin{equation}
\label{eq:Gamma}
\Gamma = { \alpha \over 2 \pi^2 Q^2 } { E' \over E }
	 { K \over 1-\epsilon }\ ,
\end{equation}
where, in the Hand convention, the factor $K$ is given by
\begin{equation}
\label{eq:K}
K = { W^2 - M^2 \over 2M } = \nu (1-x)\ .
\end{equation}
The ratio of longitudinal to transverse virtual photon polarizations,
\begin{equation}
\label{eq:epsilon}
\epsilon
= \left[ 1 + 2 (1 + {\nu^2\over Q^2}) \tan^2{\theta\over 2}
  \right]^{-1}\ ,
\end{equation}
ranges between $\epsilon=0$ and 1.

In terms of $\sigma_T$ and $\sigma_L$, the structure
functions $F_1$ and $F_2$ can be written as
\begin{eqnarray}
\label{eq:f1sigt}
F_1(x,Q^2)
&=& { K \over 4\pi^2\alpha } M \sigma_T(x,Q^2)\ ,	\\
\label{eq:f2sigs}
F_2(x,Q^2)
&=& { K \over 4\pi^2\alpha } { \nu \over (1 + \nu^2/Q^2) }
    \left[ \sigma_T(x,Q^2) + \sigma_L(x,Q^2) \right]\ . 
\end{eqnarray}
The ratio of longitudinal to transverse cross sections can also be
expressed as
\begin{equation}
\label{eq:Rlt}
R \equiv {\sigma_L \over \sigma_T}
= { F_2 \over 2xF_1 } \left( 1 + { 4 M^2 x^2 \over Q^2} \right) - 1\ .
\end{equation}
Note that while the $F_1$ structure function is related only to the
transverse virtual photon coupling, $F_2$ is a combination of both
transverse and longitudinal couplings.
It is useful therefore to define a purely longitudinal structure
function $F_L$,
\begin{equation}
\label{eq:flsigl}
F_L = \left( 1 + { Q^2 \over \nu^2 } \right) F_2 - 2x F_1\ ,
\end{equation}
in which case the ratio $R$ can be written
\begin{equation}
\label{eq:Rfl}
R = { F_L \over 2x F_1 }\ .
\end{equation}
Using the ratio $R$, the $F_2$ structure function can be extracted
from the measured differential cross sections according to
\begin{equation}
\label{eq:F2extract}
F_2 = { \sigma \over \sigma_{\rm Mott} } \nu \epsilon
      { (1 + R) \over (1 + \epsilon R) }\ .
\end{equation}
Knowledge of $R$ is therefore a prerequisite for extracting
information on $F_2$ (or $F_1$) from inclusive electron scattering
cross sections.

To complete the discussion of unpolarized scattering, we give the
expressions for the inclusive neutrino scattering cross sections.
For the charged current reactions $\nu N \to e^- X$ or
$\bar\nu N \to e^+ X$, constraints of Lorentz and gauge invariance
allow the cross section to be expressed in terms of three functions,
\begin{eqnarray}
\label{eq:w1w2w3}
{ d^2\sigma^{\nu,\bar\nu} \over d\Omega dE' }
&=& { G_F^2 E'^2 \over 2\pi^2 }
  \left( { M_W^2 \over M_W^2 + Q^2 } \right)^2
  \left( 2 W_1^{\nu,\bar\nu}(\nu,Q^2) \sin^2{\theta\over 2}
       + W_2^{\nu,\bar\nu}(\nu,Q^2) \cos^2{\theta\over 2}
  \right.				\nonumber\\
& & \hspace*{4cm} 
  \left.
  \pm\ W_3^{\nu,\bar\nu} {E + E' \over M} \sin^2{\theta\over 2}
  \right)\ ,
\end{eqnarray}
where $G_F$ is the Fermi weak interaction coupling constant, and
$M_W$ is the $W$-boson mass (with analogous expressions for the
neutral current cross sections).
In analogy with Eqs.~(\ref{eq:f1w1}) and (\ref{eq:f2w2}), one can
define dimensionless structure functions for neutrino scattering as
\begin{eqnarray}
\label{eq:f1w1nu}
F_1^{\nu,\bar\nu}(x,Q^2) &=& M W_1^{\nu,\bar\nu}(\nu,Q^2)\ , \\
\label{eq:f2w2nu}
F_2^{\nu,\bar\nu}(x,Q^2) &=& \nu W_2^{\nu,\bar\nu}(\nu,Q^2)\ , \\
\label{eq:f3w3nu}
F_3^{\nu,\bar\nu}(x,Q^2) &=& \nu W_3^{\nu,\bar\nu}(\nu,Q^2)\ .
\end{eqnarray}
The main difference between the electromagnetic and weak scattering
cases is the presence in Eq.~(\ref{eq:w1w2w3}) of the parity-violating
term proportional to the function $W_3$.
Because of its parity transformation properties, it is also odd
under charge conjugation, so that in the parton model the $F_3$
structure function of an isoscalar nucleon ($N = (p+n)/2$)
is proportional to the {\em difference} of quark distributions rather
than their sum,
\begin{equation}
%x F_3^{\nu N}(x) = x \sum_q e_q^2 \left( q(x) - \bar q(x) \right)\ .
x F_3^{\nu N}(x) = x \sum_q \left( q(x) - \bar q(x) \right)\ .
\label{eq:xf3_q}
\end{equation}

% .......................................................................
\subsection{Spin Structure Functions}

Inclusive scattering of a polarized electron beam from a polarized
nucleon target allows one to study the internal spin structure of the
nucleon.
Recent technical improvements in polarized beams and targets have made
possible increasingly accurate measurements of two additional structure
functions, $g_1$ and $g_2$.

For spin-dependent scattering, the differential cross section
can be written as a product of leptonic and hadronic tensors,
$L^A_{\mu\nu} W_A^{\mu\nu}$, in analogy with Eq.~(\ref{eq:sigtens}),
where both tensors are now antisymmetric in the Lorentz indices $\mu$
and $\nu$.
The antisymmetric leptonic tensor is given by
\begin{equation}
L^A_{\mu\nu} 
= \mp\ 2 i \epsilon_{\mu\nu\rho\lambda} k^\rho k'^\lambda
\label{eq:LAmunu}
\end{equation}
for electron helicity $\pm 1$.
The antisymmetric hadron tensor is written in terms of the spin
dependent $g_1$ and $g_2$ structure functions as
\begin{equation}
W_A^{\mu\nu}
= i \epsilon^{\mu\nu\rho\lambda} {q_\rho \over p \cdot q}
  \left( g_1(x,Q^2) s_\lambda
       + g_2(x,Q^2)
	 \left[ s_\lambda - {s \cdot q \over p \cdot q}\ p_\lambda
	 \right]
  \right)\ ,
\end{equation}
where $s_\lambda$ is the spin four-vector of the target nucleon,
with $s^2=-1$ and $p \cdot s=0$.

The structure functions $g_1$ and $g_2$ can be extracted from
measurements where longitudinally polarized leptons are scattered
from a target that is polarized either longitudinally or transversely
relative to the electron beam.
For longitudinal beam and target polarization, the difference between
the spin-aligned and spin-antialigned cross sections is given by
\begin{eqnarray}
\label{eq:sig_long}
{ d^2\sigma^{\uparrow\Downarrow} \over d\Omega dE' }
- { d^2\sigma^{\uparrow\Uparrow} \over d\Omega dE' }
&=& \sigma_{\rm Mott} { 1 \over M \nu }
  4 \tan^2{\theta \over 2}
  \left( [E+E'\cos\theta] g_1(x,Q^2)
	- 2 M x g_2(x,Q^2)
  \right)\ ,
\end{eqnarray}
where the arrows $\uparrow$ and $\Uparrow$ denote the electron
and nucleon spin orientations, respectively.
Because of the kinematic factors associated with the $g_1$ and $g_2$
terms in Eq.~(\ref{eq:sig_long}), at high energies the $g_1$ structure
function dominates the longitudinally polarized cross section.
The $g_2$ structure function can be extracted if one in addition
measures the cross section for a nucleon polarized in a direction
transverse to the beam polarization,
\begin{eqnarray}
\label{eq:sig_transv}
{ d^2\sigma^{\uparrow\Rightarrow} \over d\Omega dE' }
- { d^2\sigma^{\uparrow\Leftarrow} \over d\Omega dE' }
&=& \sigma_{\rm Mott} { 1 \over M \nu }
  4 E' \tan^2{\theta \over 2}\ \sin\theta
  \left( g_1(x,Q^2) + { 2 E \over \nu } g_2(x,Q^2) \right)\ .
\end{eqnarray}

In practice, it is often easier to measure polarization asymmetries,
or ratios of spin-dependent to spin-averaged cross sections.
The longitudinal $(A_{\|})$ and transverse $(A_{\perp})$ polarization
asymmetries are defined by
\begin{eqnarray}
\label{eq:Apar}
A_{\|}
&=& { \sigma^{\uparrow\Downarrow} - \sigma^{\uparrow\Uparrow}
\over \sigma^{\uparrow\Downarrow} + \sigma^{\uparrow\Uparrow} }\ ,   \\
\label{eq:Aperp}
A_{\perp}
&=& { \sigma^{\uparrow\Rightarrow} - \sigma^{\uparrow\Leftarrow}
\over \sigma^{\uparrow\Rightarrow} + \sigma^{\uparrow\Leftarrow} }\ ,
\end{eqnarray}
where for shorthand we denote
$\sigma^{\uparrow\Downarrow}
\equiv d^2\sigma^{\uparrow\Downarrow}/d\Omega dE'$, {\em etc.}
The $g_1$ and $g_2$ structure functions can then be extracted from
the polarization asymmetries according to
\begin{equation}
\label{eq:g1}
g_1(x,Q^2)
= F_1(x,Q^2) { 1 \over d'}
  \left( A_{\|} + \tan{\theta \over 2} A_{\perp} \right)\ ,
\end{equation}
and
\begin{equation}
\label{eq:g2}
g_2(x,Q^2)
= F_1(x,Q^2) { y \over 2d' }
  \left( { E + E' \cos\theta \over E' \sin\theta } A_{\perp} - A_{\|} 
  \right)\ ,
\end{equation}
where $d' = (1 - \epsilon)(2 - y) / [y (1 + \epsilon R(x,Q^2))]$,
and $y = \nu/E$.

One can also define virtual photon absorption asymmetries $A_1$ and
$A_2$ in terms of the measured asymmetries,
\begin{eqnarray}
\label{eq:Apar12}
A_{\|} &=& D (A_1 + \eta A_2)\ ,	\\
\label{eq:Aperp12}
A_{\perp} &=& d (A_2 - \zeta A_1)\ ,
\end{eqnarray}
where the photon depolarization factor
is $D = (1 - E' \epsilon/E)/(1 + \epsilon R(x,Q^2))$,
and the other kinematic factors are given by
$\eta = \epsilon \sqrt{Q^2}/(E - E' \epsilon)$,
$d = D\sqrt{2 \epsilon /(1 + \epsilon)}$,
and $\zeta = \eta(1 + \epsilon) / 2\epsilon$.
%
% A_1 = { d A_\parallel - D \eta A_\perp \over
%	  d D (1 + \zeta \eta) }
%
% A_2 = { d \zeta A_\parallel + D A_\perp \over
%	  d D (1 + \zeta \eta) }
%
The $A_1$ asymmetry can also directly be expressed in terms of the
$g_1$, $g_2$ and $F_1$ structure functions,
\begin{eqnarray}
A_1(x,Q^2) =
{ 1 \over F_1(x,Q^2) }
\left( g_1(x,Q^2) - {4M^2x^2 \over Q^2} g_2(x,Q^2) \right)\ . 
\end{eqnarray}
At small values of $x^2/Q^2$, one then has $A_1 \approx g_1/F_1$.
If the $Q^2$ dependence of the polarized and unpolarized structure
functions is similar, the polarization asymmetry $A_1$ will be
weakly dependent on $Q^2$.
This may be convenient when comparing resonance region data with deep
inelastic data.
On the other hand, a presentation of the data in terms of $g_1$ is less
sensitive to the detailed knowledge of $g_2$ or $A_2$.
Note that both the spin structure functions and the polarization
asymmetries depend on the unpolarized structure function $F_1$,
and hence require knowledge of $R$ to determine $F_1$ from the
measured unpolarized cross sections.
Furthermore, positivity constrains lead to bounds on the magnitude
of the virtual photon asymmetries,
\begin{eqnarray}
|A_1| \le 1\ ,\ \ \ \ \ |A_2| \le \sqrt{R(x,Q^2)}\ . 
\end{eqnarray}

Finally, in the quark-parton model the $g_1$ structure function is
expressed in terms of differences between quark distributions with spins
aligned ($q^\uparrow$) and antialigned ($q^\downarrow$) relative to that
of the nucleon, $\Delta q(x) = q^\uparrow(x) - q^\downarrow(x)$,
\begin{equation}
\label{eq:g1quark}
g_1(x) = {1 \over 2} \sum_q e_q^2
	 \left( \Delta q(x) + \Delta \bar q(x) \right)\ .
\end{equation}
The $g_2$ structure function, on the other hand, does not have a
simple partonic interpretation.
However, its measurement provides important information on the
so-called higher twist contributions, which form a main focus
in this review.

% .......................................................................
\subsection{Moments of Structure Functions}

Having introduced the unpolarized and polarized structure
functions above, here we define their moments, or $x$-weighted
integrals.
Following standard notation, the $n$-th moments of the spin-averaged
$F_1$, $F_2$ and $F_L$ structure functions are defined as
\begin{eqnarray}
\label{eq:M1DEF}
M_1^{(n)}(Q^2)     &=& \int_0^1 dx\ x^{n-1} F_1(x,Q^2)\ ,	\\
\label{eq:MnDEF}
M_{2,L}^{(n)}(Q^2) &=& \int_0^1 dx\ x^{n-2} F_{2,L}(x,Q^2)\ ,
\end{eqnarray}
and similarly for the neutrino structure functions
$F_{1,2,3}^{\nu,\bar\nu}$.
With this definition, in which the moments are usually referred
to as the Cornwall-Norton moments \cite{CN69}, the $n=1$ moment
of the $F_1$ structure function in the parton model effectively counts
quark charges, while the $n=2$ moment of the $F_2$ structure function
corresponds to the momentum sum rule.
In the Bjorken limit, the moments of the $F_1$ and $F_2$ structure
functions are related via the Callan-Gross relation,
Eq.~(\ref{eq:f22xf1}), as $M_2^{(n)} = 2 M_1^{(n)}$.

As discussed in Sec.~\ref{sssec:ope} below, formally the operator
product expansion in QCD defines the moments for $n=2,4,6\ldots$.
To obtain moments for other values of $n$ requires an analytic
continuation to be made in $n$.
Alternatively, if the $x$ dependence of the structure functions
is known, one can define the moments operationally via
Eqs.~(\ref{eq:M1DEF}) and (\ref{eq:MnDEF}).
Note that formally the moments include also the elastic point at $x=1$,
which, while negligible at high $Q^2$, can give large contributions at
small $Q^2$.

The Cornwall-Norton moments defined in terms of the Bjorken $x$
scaling variable are appropriate in the region of kinematics where
$Q^2$ is much larger than typical hadronic mass scales, where
corrections of the type $M^2/Q^2$ can be neglected.
In this case only operators with spin $n$ contribute to the $n$-th
moments (see Sec.~\ref{ssec:qcd}).
For finite $M^2/Q^2$, however, the $n$-th moments receive
contributions from spins $n$ and higher, which can complicate the
physical interpretation of the moments.

By redefining the moments in terms of a generalized scaling variable
$\xi$ which takes target mass corrections into account, Nachtmann
\cite{EONacht} showed that the new $n$-th moments still receive
contributions from spin $n$ operators only, even at finite $M^2/Q^2$.
Specifically, for the $F_2$ structure function one has
\cite{EONacht,ROBERTS}
\begin{eqnarray}
M_2^{N(n)}(Q^2)
&=& \int_0^1 dx\ { \xi^{n+1} \over x^3 }
\left\{ { 3 + 3(n+1)r + n(n+2)r^2 \over (n+2)(n+3) } \right\}
F_2(x,Q^2)\ ,
\label{eq:NachtMom}
\end{eqnarray}
where
\begin{equation}
\xi = { 2x \over 1 + \sqrt{ 1 + 4 M^2 x^2/Q^2 } }
\label{eq:xi}
\end{equation}
is the Nachtmann scaling variable,
and $r = \sqrt{1 + 4 M^2 x^2/Q^2}$.
In the limit $Q^2 \to \infty$ one can easily verify that the
moment $M_2^{N(n)} \to M_2^{(n)}$ in Eq.~(\ref{eq:MnDEF}).
Similarly, for the longitudinal Nachtmann moments,
one has \cite{EONacht,SIMULA_FL}
\begin{eqnarray}
M_L^{N(n)}(Q^2)
&=& \int_0^1 dx\ { \xi^{n+1} \over x^3 }
\left\{
  F_L(x,Q^2)
+ {4 M^2 x^2 \over Q^2} { (n+1) \xi/x - 2 (n+2) \over (n+2)(n+3) }
  F_2(x,Q^2)
\right\}\ ,
\label{eq:NachtLMom}
\end{eqnarray}
which approaches $M_L^{(n)}$ in the $Q^2 \to \infty$ limit.
The Nachtmann $\xi$ variable and the corresponding moments can
also be generalized to include finite quark mass effects
\cite{BARBIERI,BEGR}, although in practice this is mainly relevant
for heavy quarks.

For spin-dependent scattering, the $n$-th Cornwall-Norton moments
of the $g_1$ and $g_2$ structure functions are defined analogously
to Eqs.~(\ref{eq:M1DEF}) and (\ref{eq:MnDEF}) as
\begin{eqnarray}
\Gamma_{1,2}^{(n)}(Q^2) &=& \int_0^1 dx\ x^{n-1} g_{1,2}(x,Q^2)\ ,
\label{eq:GnDEF}
\end{eqnarray}
for $n=1,3,5\ldots$ in the case of the $g_1$ structure function,
and $n=3,5\ldots$ for $g_2$.
With this definition the $n=1$ moment of $g_1$ corresponds to the
nucleon axial vector charge.
As for the unpolarized moments, for other values of $n$ one needs
to either analytically continue in $n$, or define the moments
operationally via Eq.~(\ref{eq:GnDEF}).
In the text we will sometimes refer to the lowest ($n=1$) moments
$\Gamma_{1,2}^{(1)}$ simply as $\Gamma_{1,2}$, without the superscript.

The finite-$Q^2$ generalization of the $\Gamma_1^{(n)}$ moment
of the $g_1$ structure function in terms of the Nachtmann $\xi$
variable is given by \cite{NACHT_POL}
\begin{eqnarray}
\Gamma_1^{N(n)}(Q^2)
&=& \int_0^1 dx\ {\xi^{n+1} \over x^2}
\left\{
\left[ {x \over \xi}
     - {n^2 \over (n+2)^2} {M^2 x^2 \over Q^2} {\xi \over x}
\right] g_1(x,Q^2)
- {4n \over n+2} {M^2 x^2 \over Q^2} g_2(x,Q^2) \right\}\ ,
\label{eq:G1n_N}
\end{eqnarray}
which approaches $\Gamma_1^{(n)}$ in the limit $Q^2 \to \infty$.
For the $g_2$ structure function, the most direct generalization
is actually one which contains a combination of $g_1$ and $g_2$
(corresponding to ``twist-3'' --- see Sec.~\ref{sssec:twist})
\cite{NACHT_POL},
\begin{eqnarray}
\Gamma_2^{N(n)}(Q^2)
&=& \int_0^1 dx\ {\xi^{n+1} \over x^2}
\left\{
  {x \over \xi} g_1(x,Q^2)
+ \left[ {n \over n-1} {x^2 \over \xi^2}
       - {n \over n+1} {M^2 x^2 \over Q^2}
  \right] g_2(x,Q^2)
\right\}\ ,
\label{eq:G2n_N}
\end{eqnarray}
so that in the limit $Q^2 \to \infty$, one has
$\Gamma_2^{N(n)} \to \Gamma_1^{(n)} + n/(n-1) \Gamma_2^{(n)}$.

%%%%%%%%%%%%%%%%%%%%%%%%%%%%%%%%%%%%%%%%%%%%%%%%%%%%%%%%%%%%%%%%%%%%%

\clearpage

%%%%%%%%%%%%%%%%%%%%%%%%%%%%%%%%%%%%%%%%%%%%%%%%%%%%%%%%%%%%%%%%%%%%%%%%%%
\section{Quark-Hadron Duality: An Historical Perspective}
\label{sec:hist}

Before embarking on the presentation of the recent data on structure
functions in the resonance region and assessing their impact on our
theoretical understanding of Bloom-Gilman duality, it will be
instructive to trace the origins of this phenomenon back to the late
1960s in order to appreciate the context in which the early discussions
of duality took place.
The decade or so preceding the development of QCD saw tremendous
effort devoted to describing hadronic interactions in terms of
$S$-matrix theory and self-consistency relations.
One of the profound discoveries of that era was the remarkable
relationship between low-energy hadronic cross sections and their
high-energy behavior, in which the former on average appears to
mimic certain features of the latter.
In this section we briefly review the original findings on duality
in hadronic reactions, and describe how this led to the descriptions
of duality in the early electron scattering experiments.

% ------------------------------------------------------------------------
\subsection{Duality in Hadronic Reactions}
\label{ssec:regge}

\begin{figure}[hbt]
\begin{center}
\epsfig{file=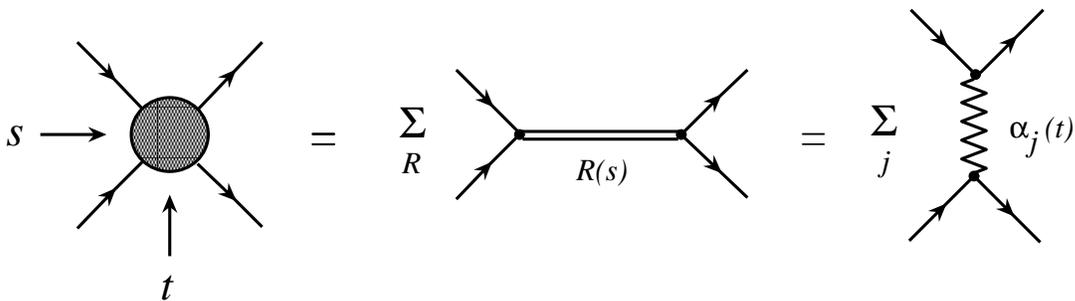,height=10cm}
\vspace*{-4.5cm}
\caption{Dual descriptions of the scattering process,
	in terms of a sum over $s$-channel resonances $R(s)$,
	and in terms of $t$-channel Reggeon exchanges $\alpha_j(t)$
	(see text).}
\label{fig:dual_st}
\end{center}
\end{figure}

Historically, duality in strong interaction physics represented
the relationship between the description of hadronic scattering
amplitudes in terms of $s$-channel resonances at low energies,
and $t$-channel Regge poles at high energies, as illustrated in
Fig.~\ref{fig:dual_st}.
The study of hadronic interactions within Regge theory is an extremely
rich subject in its own right, which preoccupied high energy physicists
for much of the decade prior to the formulation of QCD.
In this section we outline those aspects of Regge theory and
resonance--Regge duality which will help to illustrate the concept
of duality as later applied to deep inelastic scattering.
More comprehensive discussions of Regge phenomenology can be
found for example in the classic book of Collins \cite{COLLINS},
or in the more recent account of Donnachie {\em et al.} \cite{DDLN}.
A review of duality in hadronic reactions can also be found in the
report by Fukugita \& Igi \cite{FUKUGITA_IGI}.

% .......................................................................
\subsubsection{Finite Energy Sum Rules}
\label{sssec:ampl}

Consider the scattering of two spinless particles, described by the
amplitude ${\cal A}(s,t)$, where $s$ and $t$ are the standard
Mandelstam variables.
At low energies, it is convenient to write the scattering amplitude
as a partial wave series \cite{COLLINS,QUEEN},
\begin{eqnarray}
{\cal A}(s,t)
&=& 16\pi \sum_l (2l+1) {\cal A}_l(s)\ P_l(\cos\theta_s)\ ,
\label{eq:partialwave}
\end{eqnarray}
where $\theta_s$ is the $s$-channel center of mass scattering angle,
and ${\cal A}_l$ is the partial wave amplitude of angular momentum $l$.
(The generalization to non-zero intrinsic spin is straightforward,
with replacement of $l$ by the total angular momentum $J$.)
The elastic cross section is proportional to $|{\cal A}(s,t)|^2$,
and by the optical theorem the total cross section is related to
the imaginary part of the amplitude,
$\sigma \sim \Im m {\cal A}(s,t)$.

\begin{figure}[h]
\begin{center}
\vspace*{-2cm}
\hspace*{1cm}
\epsfig{file=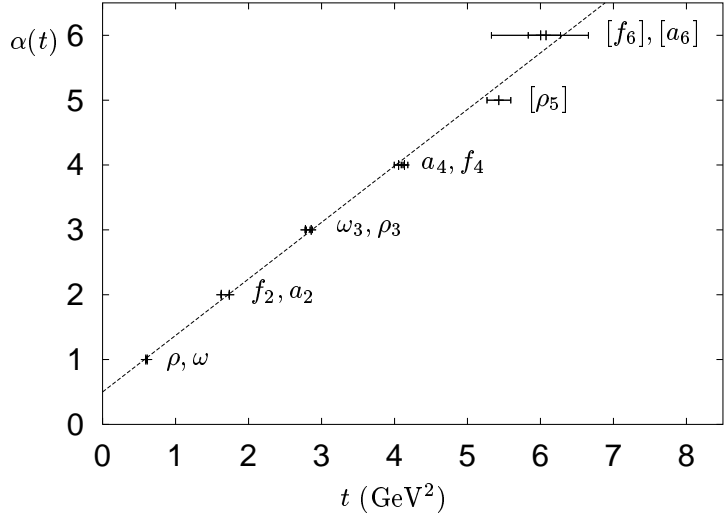,height=28cm}
\vspace*{-18cm}
\caption{Chew-Frautschi plot for several degenerate meson families
	on the Regge trajectory
	$\alpha(t) = 0.5 + 0.9\ t$.
%	(From Ref.~\protect\cite{KAIDALOV}.)}
	(From Ref.~\protect\cite{DDLN}.)}
\label{fig:regge}
\end{center}
\end{figure}

If the interaction forces are of finite range $r$, then for a given
$s$ only partial waves with $l \alt r \sqrt{s}$ will be important
in the sum.
At low energies the partial wave amplitudes ${\cal A}_l$ are then
dominated by just a few resonance poles, $R$,
\begin{eqnarray}
{\cal A}_l(s) &\approx&
\sum_R { g_R \over M_R^2 - s - i M_R \Gamma_R }\ ,
\end{eqnarray}
where $g_R$ is the coupling strength, $M_R$ is the mass of the
resonance and $\Gamma_R$ its width.
As $s$ increases, however, the density of resonances in each partial
wave, as well as the number of partial waves which must be included
in the sum (\ref{eq:partialwave}), also increases, making it harder
to identify contributions from individual resonances.
At high $s$ it becomes more useful therefore to describe the scattering
amplitude in terms of a $t$-channel partial wave series, which can be
expressed as an integral over complex $l$ via the Sommerfeld-Watson
transformation \cite{COLLINS}.
This allows the amplitude to be written as a sum of $t$-channel Regge
poles and cuts, which at high energy leads to the well-known linear
Regge trajectories,
\begin{eqnarray}
{\cal A}(s,t)\ \sim\ s^{\alpha(t)}\ ,\ \ \ \ \ s \to \infty\ ,
\label{eq:Aregge}
\end{eqnarray}
where $\alpha(t) = \alpha(0) + \alpha' t$.
This implies that at large $s$, with $t$ fixed, the total cross
section behaves as $\sigma \sim s^{\alpha(0) - 1}$.
The trajectory $\alpha(t)$, which is characterized by the slope
$\alpha'$ and intercept $\alpha(0)$, is shown in Fig.~\ref{fig:regge}
in the so-called Chew-Frautschi plot \cite{CHEW} for several 
well-established meson families.
A remarkable feature is the near degeneracy of each of the $\rho$,
$\omega$, $f_2$ and $a_2$ trajectories.
Similar linearity is observed in the baryon trajectories.

While the $s$- and $t$-channel partial wave sums describe the low-
and high-energy behaviors of scattering amplitudes, respectively,
an important question confronting hadron physicists of the 1960s was
how to merge these descriptions, especially at intermediate $s$,
where the amplitudes approach their smooth Regge asymptotic behavior,
but some resonance structures still remain.
More specifically, how do the $s$-channel resonances contribute to the
asymptotic $s$ behavior, and where do these resonances appear in the
Sommerfeld-Watson representation?

Progress towards synthesizing the two descriptions came with the
development of Finite Energy Sum Rules (FESRs), which are
generalizations of superconvergence relations in Regge theory
\cite{SUPERCVG} relating dispersion integrals over the amplitudes
at low energies to high-energy parameters.
The formulation of FESRs stemmed from the sum rule of Igi \cite{IGI},
which used dispersion relations to express the crossing symmetric
$\pi N$ forward scattering amplitude in terms of its high-energy
behavior.
An implicit assumption here is that beyond a sufficiently large energy
$\nu > \bar\nu$, the scattering amplitude can be represented by its
asymptotic form, ${\cal A}_{\rm {I\!R}}$, calculated within Regge
theory \cite{SERTORIO}.
The resulting sum rules relate functions of the high-energy parameters
to dispersion integrals which depend on the amplitude over a finite
range of energies.

Formally, the FESRs can be written as relations between (moments of)
the imaginary part of the scattering amplitude at finite energies
and the asymptotic high energy amplitude \cite{COLLINS,DDLN},
\begin{eqnarray}
\int_0^{\bar\nu} d\nu\ \nu^n\ \Im m\ {\cal A}(\nu,t)
&=& \int_0^{\bar\nu} d\nu\ \nu^n\
    \Im m\ {\cal A}_{\rm {I\!R}}(\nu,t)\ \ \ \ \ \ \ [\rm FESR]
\label{eq:fesr}
\end{eqnarray}
where here $\nu$ is defined in terms of the Mandelstam variables as
$\nu \equiv (s-u)/4$, and the integration includes the Born term.
Assuming analyticity and Regge pole dominance for
$\nu \geq \bar\nu$, the integral over the Regge amplitude in
Eq.~(\ref{eq:fesr}) can be written in terms of the Regge
trajectories $\alpha_j(t)$ and functions $\beta_j(t)$ characterizing
the residues of the poles in the complex-$l$ plane,
\begin{eqnarray}
\int_0^{\bar\nu} d\nu\ \nu^n\
  \Im m\ {\cal A}_{\rm {I\!R}}(\nu,t)
&=& \sum_j { \beta_j(t)\ \bar\nu^{\alpha_j(t) + n + 1}
		\over (\alpha_j(t) + n + 1) \Gamma(\alpha_j(t)+1) }\ ,
\label{eq:fesr2}
\end{eqnarray}
where $\Gamma$ is the Euler gamma function.
The FESRs (\ref{eq:fesr}) therefore encapsulate a duality between
the $s$-channel resonance and $t$-channel Regge descriptions of the
scattering amplitude, as illustrated in Fig.~\ref{fig:dual_st}.
For the lowest moment, $n=0$, Eq.~(\ref{eq:fesr}) reduces to the
dispersion sum rule originally derived by Logunov {\em et al.}
\cite{LST} and Igi \& Matsuda \cite{IGI_MATSUDA}.

\begin{figure}[t]
\begin{minipage}{3.0in}
\hspace*{-3cm}
\epsfig{file=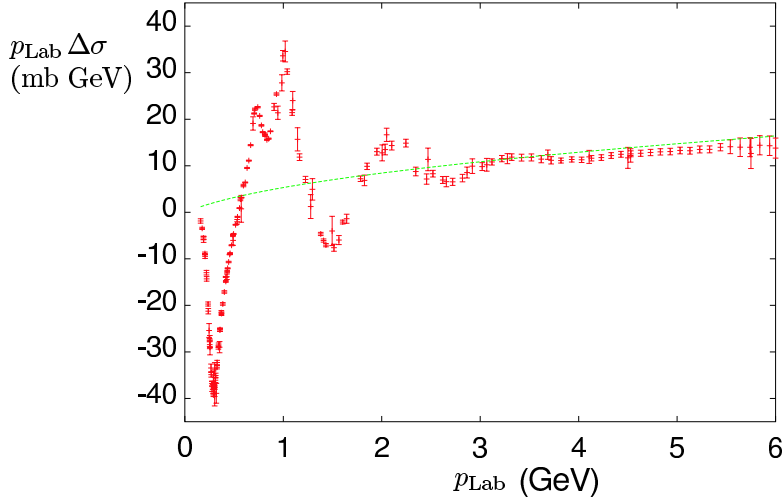,height=25cm}
\vspace*{-18cm}
\end{minipage}
\begin{centering}
\caption{\label{fig:piNdif}
	Isovector $\pi p$ cross section,
	$p_{\rm Lab} \Delta\sigma
	= p_{\rm Lab} (\sigma^{\pi^+ p} - \sigma^{\pi^- p})$,
	as a function of laboratory momentum, $p_{\rm Lab}$,
	compared with the Regge fit to high energy data (dotted line).
%	(Adapted from Ref.~\protect\cite{IGI}.)}
	(Adapted from Refs.~\protect\cite{DDLN,IGI}.)}
\end{centering}
\end{figure}

For higher moments, the sum rules require a more local duality,
${\cal A}(\nu,t)\ \approx\ {\cal A}_{\rm {I\!R}}(\nu,t)$,
and are therefore less likely to work at low energies.
Such local duality could only be expected at very high $s$,
where the density of resonances is large, and the bumps have
been smoothed out.
Note that the equality of all the moments would require the
amplitude at low $\nu$ to be identical to ${\cal A}_{\rm {I\!R}}$.
Given that the former contains poles in $s$, whereas the latter
does not, this places some restrictions on how local the duality
between the low and high energy behaviors can be.
Nevertheless, the sum rules (\ref{eq:fesr}) represent a powerful tool
which allows one to use experimental information on the low energy
cross sections for the analysis of high energy scattering, and to
connect low energy parameters (such as resonance widths and coupling
strengths) to parameters describing the behavior of cross sections
at high energies.

An important early application of FESRs was made for the case of
$\pi N$ scattering amplitudes.
In their seminal analysis, Dolen, Horn \& Schmid \cite{DHSLET,DHS}
observed that summing over contributions of $s$-channel resonances
yields a result which is approximately equal to the leading ($\rho$)
pole contribution obtained from fits to high energy data, extrapolated
down to low energies.
This equivalence (or ``bootstrap'', as it was referred to in the
early literature) is illustrated in Fig.~\ref{fig:piNdif} for the
total isovector $\pi p$ scattering cross section.
The data at small laboratory momenta show pronounced resonant
structure for $p_{\rm Lab} \alt 2$--3~GeV, which oscillates around
the Regge fit to high energy data, with the amplitude of the
oscillations decreasing with increasing momenta.
Averaging the resonance data over small energy ranges thus exposes
a semi-local duality between the $s$-channel resonances and the
Regge fit.

% .......................................................................
\subsubsection{Veneziano Model and Two-Component Duality}
\label{sssec:twocomp}

With the phenomenological confirmation of duality in $\pi N$
scattering, the quest was on to find theoretical representations
of the scattering amplitude which would satisfy the FESR relations
(\ref{eq:fesr}) and unify the low and high $s$ behaviors.
Such a representation was found to be embodied in the Veneziano
model \cite{VEN,VEN_REP}.
Observing that the simplest function with an infinite set of poles
in the $s$-channel on a trajectory $\alpha(s)=$~integer ($> 0$) is
$\Gamma(1-\alpha(s))$, Veneziano proposed a solution to (\ref{eq:fesr})
of the form ${\cal A}(s,t) + {\cal A}(s,u) + {\cal A}(t,u)$, where
%
% \begin{eqnarray}
% T(s,t,u) &=& A(s,t) + A(t,u) + A(s,u)\ ,
% \end{eqnarray}
%
\begin{eqnarray}
{\cal A}(s,t)
&=& g\ { \Gamma\left(1-\alpha(s)\right) \Gamma\left(1-\alpha(t)\right)
	 \over \Gamma\left(2-\alpha(s)-\alpha(t)\right) }\
 =\ g\ \int_0^1 dz\ z^{\alpha(s)} (1-z)^{-\alpha(t)}\ ,
\label{eq:ven}
\end{eqnarray}
with $g$ the constant strength.
The amplitude (\ref{eq:ven}) is manifestly analytic and crossing
symmetric, having the same pole structure and Regge behavior in both
the $s$ and $t$ channels.
It explicitly satisfies the FESRs and duality, and reproduces linear
Regge trajectories.
The latter can be verified by using Stirling's formula,
\begin{eqnarray}
\Gamma(z) &\to& \sqrt{2\pi}\ e^{-z} z^{z-1/2}\ ,\ \ \ z \to \infty\ ,
\end{eqnarray}
which yields, for fixed $t$,
\begin{eqnarray}
{\cal A}(s,t) &\to&
g\ { \pi (\alpha' s)^{\alpha(t)} \over
     \Gamma(\alpha(t))\ \sin(\pi\alpha(t)) }
   e^{-i \pi \alpha(t) }\
\sim\ (\alpha' s)^{\alpha(t)}\ ,
\label{eq:venHE}
\end{eqnarray}
where $\alpha(s) \to \alpha(0) + \alpha' s$ at large $s$.

\begin{figure}
\begin{minipage}[ht]{2.8in}
\hspace*{-6.5cm}
\epsfig{file=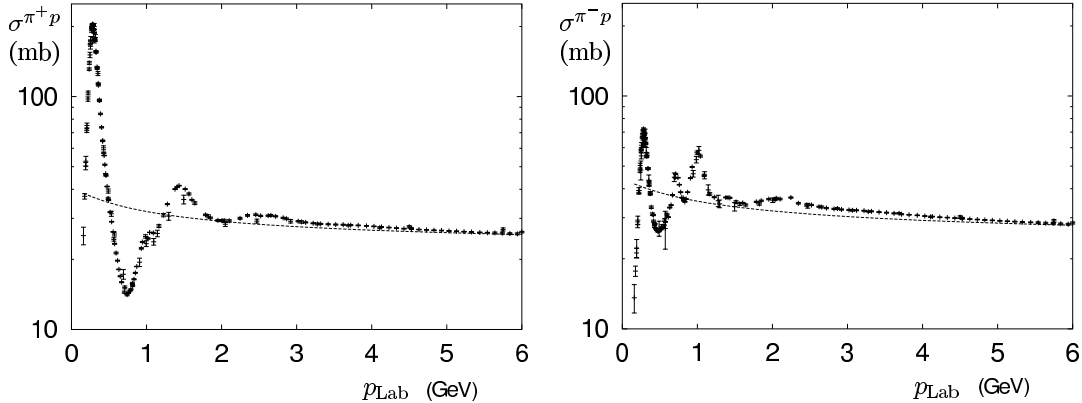,height=24cm}
\vspace*{-18cm}
\end{minipage}
\begin{centering}
\caption{\label{fig:piNdual}
	Total $\pi^+ p$ (left plot) and $\pi^- p$ (right plot)
	cross sections as a function of laboratory momentum,
	$p_{\rm Lab}$, compared with Regge fits to high energy data.
	(Adapted from Ref.~\protect\cite{DDLN}.)}
\end{centering}
\end{figure}

Much of the progress in applying the concept of duality in hadronic
reactions was due to the success of the Veneziano model, even though
the model is now regarded more as a toy model.
One of the shortcomings of the Veneziano formula (\ref{eq:ven}) is the
presence of an infinite set of zero-width resonances on the positive
real $s$ axis, which destroys the Regge behavior on the real axis.
Moreover, the solution (\ref{eq:ven}) is not unique: the functions
$\alpha(s)$, $\alpha(t)$ can be replaced by $m-\alpha(s)$, $n-\alpha(t)$
for any integer $m, n$, while still satisfying the FESRs.
This means that there are effectively no constraints on the resonance
parameters without making additional assumptions \cite{DDLN}.
A number of attempts to alleviate some of these problems have been
made in the literature --- see for instance
Refs.\cite{DAMA_COHEN,BUGRIJ,DAMA_GASKELL}.
Nevertheless, the Veneziano amplitude does provide an explicit
realization of duality, and in fact indirectly led to the development
of modern string theory (see Sec.~\ref{sec:outlook}).

The duality hypothesis embodied in the FESR (\ref{eq:fesr}) is of
course incomplete: it does not include Pomeron (${\rm {I\!P}}$)
exchange.
Pomeron exchange (exchange of vacuum quantum numbers) was introduced in
Regge theory to describe the behavior of cross sections at large $s$
\cite{COLLINS}.
Since the known mesons lie on Regge trajectories with intercepts
$\alpha_{\rm {I\!R}}(0) < 1$, from Eq.~(\ref{eq:Aregge}) the resulting
cross sections will obviously decrease with $s$.
To obtain approximately constant cross sections at large $s$ requires
an intercept $\alpha_{\rm {I\!P}}(0) \approx 1$.
While there are no known mesons on such a trajectory, the exchange of
a Pomeron with vacuum quantum numbers (which can be modeled in QCD
through the exchange of two or more gluons) is introduced as an
effective description of the high-energy behavior of cross sections.
The leading Reggeon-exchange contributions (for instance due to
$\rho$ exchange) have intercept $\alpha_{\rm {I\!R}}(0) \approx 0.5$,
and are more important at smaller $s$.

Since it is even under charge conjugation, the ${\rm {I\!P}}$-exchange
contribution to the isovector $\pi p$ cross section in
Fig.~\ref{fig:piNdif} cancels, thus exposing the duality between
$s$-channel resonances and the nondiffractive Reggeon $t$-channel
exchanges.
On the other hand, a comparison of the individual $\pi^+ p$ and
$\pi^- p$ cross sections in Fig.~\ref{fig:piNdual} suggests that
at low energies the cross sections themselves on average display
some degree of duality with the high-energy behavior.
In both cases the prominent resonances at $p_{\rm Lab} \alt 1$~GeV
oscillate around the high energy fit extrapolated to these energies.

A generalization of the $s$- and $t$-channel duality to include
contributions from both resonances and the nonresonant background
upon which the resonances are superimposed was suggested by Harari
\cite{HARARI} and Freund \cite{FREUND}.
In this ``two-component duality'', resonances are dual to the
nondiffractive Regge pole exchanges, while the nonresonant background
is dual to Pomeron exchange \cite{QUEEN},
\begin{eqnarray}
{\cal A}(s,t)
&=& \sum_{\rm res} {\cal A}_{\rm res}(s,t)\
 +\ {\cal A}_{\rm bkgd}(s,t)\		\\
&=& \sum_{\rm {I\!R}} {\cal A}_{\rm {I\!R}}(s,t)
 +\ {\cal A}_{\rm {I\!P}}(s,t)\ .
\end{eqnarray}
The data on $\pi^\pm p$ scattering in Figs.~\ref{fig:piNdif} and
\ref{fig:piNdual} demonstrate as much: since both the nondiffractive
and total cross sections satisfy duality, then so must the
diffractive, ${\rm {I\!P}}$-exchange component.

\begin{figure}[th]
\begin{center}
\epsfig{file=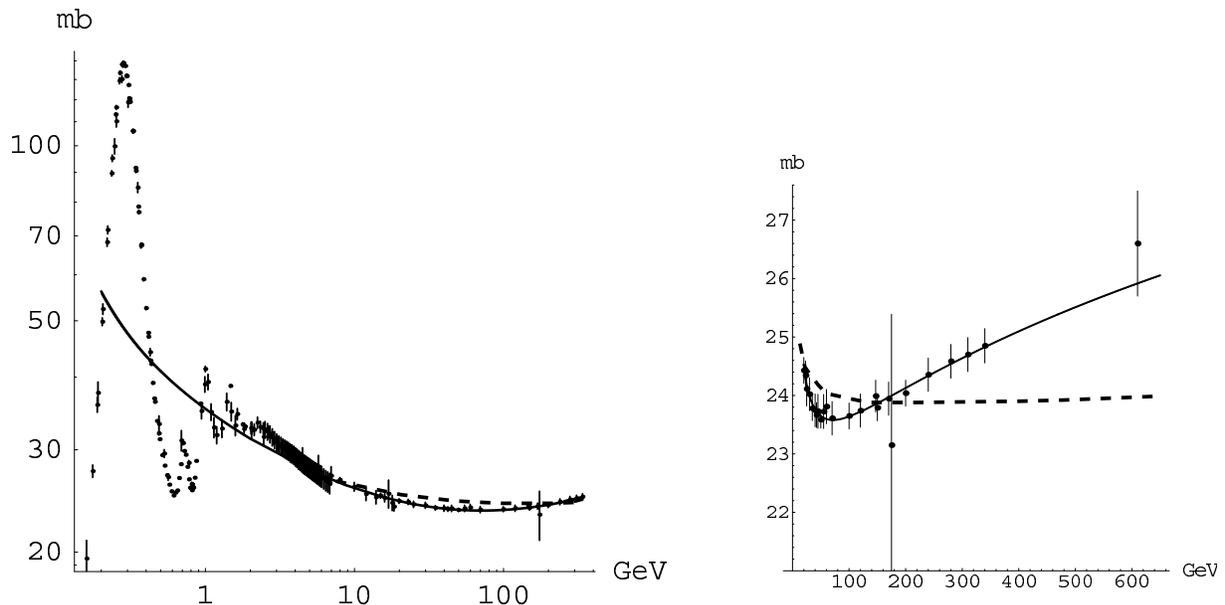,height=8cm}
\vspace*{1cm}
\caption{Total $\pi^+ p$ + $\pi^- p$ cross section (in mb) as a function of the center of mass energy $\protect\sqrt s$ (in GeV). The left panel ($\protect\sqrt s$ on a logarithmic scale) emphasizes the resonance region, while the right panel (linear scale) emphasizes the high energy region. The solid (dashed) lines represent fits with a $\protect\log^2 s$ ($\protect\log s$) asymptotic behavior.	(Adapted from Ref.~\protect\cite{IGIISHIDA}.)}
\label{fig:piNfesr}
\end{center}
\end{figure}

The practical utilization of duality and the FESRs was demonstrated
recently by Igi and Ishida \cite{IGIISHIDA} in a combined fit to both
low- and high-energy total $\pi p$ cross sections.
While it has been known for some time that the increase of total
cross sections at high energy cannot exceed the Froissart unitarity
bound, $\sigma \sim \log^2 s$ \cite{FROISSART}, experimentally it
has not been possible to distinguish a $\log s$ behavior from a
$\log^2 s$ using high energy data alone \cite{CUDELL}.
Constraining the fit by the averaged cross section data in the
resonance region at low $s$, on the other hand, as implied by
the FESR (\ref{eq:fesr}), clearly favors the $\log^2 s$ asymptotic
behavior, as seen in Fig.~\ref{fig:piNfesr} (solid curves).
The $\log s$ fit (dashed curves) overestimates the data at
$\sqrt{s} \sim 50$--100~GeV, and cannot reproduce the observed
rise in the cross section at $\sqrt{s} \agt 300$~GeV,
especially the new data point at $\sim 600$~GeV from the SELEX
Collaboration at Fermilab \cite{SELEX}.

Similar constraints have also been used by Block and Halzen
\cite{BLOCK_HALZEN} to fit the total photoproduction cross sections
at high energy.
By matching the high-$s$ fit smoothly to the average of the resonance
data at $\sqrt s \sim 4$~GeV, the results strongly favor a $\log^2 s$
behavior at large $s$.
Furthermore, the evidence for the saturation of the Froissart bound
in the $\gamma p$ cross section is confirmed by applying the same
analysis to $\pi p$ data using vector meson dominance
\cite{BLOCK_HALZEN}.

\begin{figure}[ht]
\begin{center}
\epsfig{file=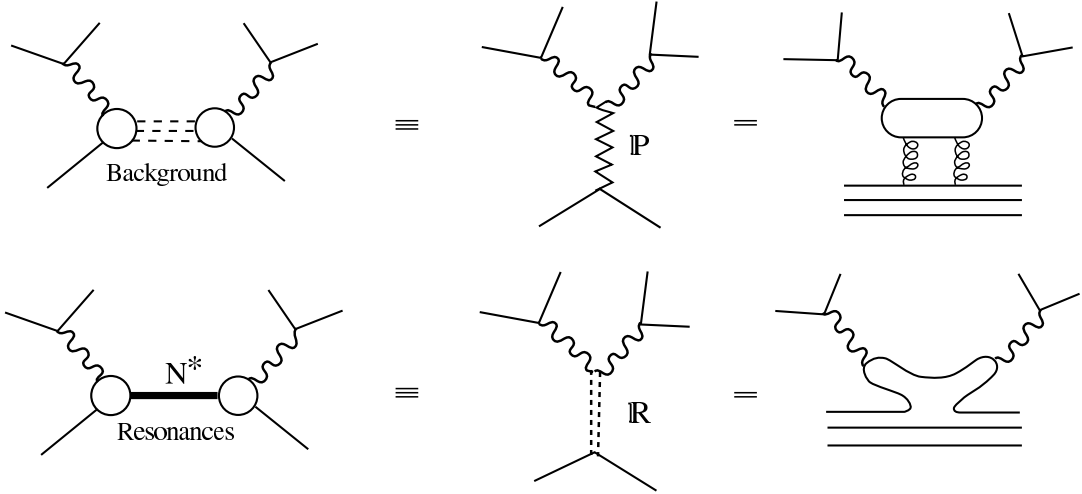,height=14cm}
\vspace*{-7cm}
\caption{Illustration of two-component duality in $eN \to eX$:
	(top) duality between the nonresonant background and
	Pomeron (${\rm {I\!P}}$) exchange, thought to be
	associated with gluon exchange in QCD;
	(bottom) duality between resonances and Reggeon
	(${\rm {I\!R}}$) exchange, corresponding to quark 
	exchange diagrams.
	(Adapted from Ref.~\protect\cite{COLLINS}.)}
\label{fig:reggedual}
\end{center}
\end{figure}

For the case of electroproduction, the two-component duality model
has immediate application in deep inelastic scattering, which we discuss
in more detail in the next section.
In inclusive electroproduction from the nucleon the behavior of the
cross sections at large $s \equiv W^2 = M^2 + Q^2 (\omega-1)$,
where $\omega = 2M\nu/Q^2$, corresponds to the $\omega \to \infty$
behavior of structure functions.
Two-component duality therefore suggests a correspondence between
resonances and valence quarks, whose behavior at large $\omega \sim s$
at fixed $Q^2$ is given by poles on the $\rho$ meson Regge trajectory,
\begin{eqnarray}
% q_{val}(x) &\sim& x^{\alpha_{\rm {I\!R}}}\ ,
F_2^{\rm val}(\omega) &\sim&
\omega^{\alpha_{\rm {I\!R}} - 1}\ ,
\end{eqnarray}
with the background dual to sea quarks or gluons, for which the
large-$\omega$ behavior is determined by Pomeron exchange,
\begin{eqnarray}
% q_{sea}(x) &\sim& x^{\alpha_{\rm {I\!P}}}\ .
F_2^{\rm sea}(\omega) &\sim& \omega^{\alpha_{\rm {I\!P}} - 1}\ .
\end{eqnarray}
This is illustrated schematically in Fig.~\ref{fig:reggedual}.

This duality imposes rather strong constraints on the production of
resonances and on the $Q^2$ dependence of the $\gamma^* N \to N^*$
transition form factors, as will be discussed below.
In fact, a dual model of deep inelastic scattering based on Regge
calculus was developed by Landshoff and Polkinghorne \cite{LPDUAL}
to describe the early deep inelastic scattering data, in which duality
was introduced by identifying the contribution of exotic states to
scattering amplitudes with diffractive processes.
More recently, dual models based on generalizations of the Veneziano
amplitude \cite{GEN_VEN} to include Mandelstam analyticity
\cite{BUGRIJ} and nonlinear trajectories \cite{SCHIERHOLZ} have been
constructed \cite{JENK,FIORE} to relate structure functions at small
and large $\omega$.

\clearpage
% ------------------------------------------------------------------------
\subsection{Duality In Inclusive Electron Scattering}
\label{ssec:dualhist}

The unique feature of inclusive electroproduction is that one can
measure points at the same $\omega = 1 + (W^2-M^2)/Q^2 \equiv 1/x$
at different values of $Q^2$ and $W^2$, both within and outside the
resonance region.
Unlike in hadronic reactions, the fact that one can vary the mass of
the probe, $Q^2$, means that duality here can be studied by directly
measuring the scaling function describing the high energy cross
section which averages the resonances.

% .......................................................................
\subsubsection{Bloom-Gilman Duality}
\label{sssec:f2hist}

By examining the early inclusive electron--proton scattering data
from SLAC, Bloom and Gilman observed \cite{BG1,BG2} a remarkable
connection between the structure function $\nu W_2(\nu,Q^2)$ in the
nucleon resonance region and that in the deep inelastic continuum.
The resonance structure function was found to be equivalent on average
to the deep inelastic one, with the averages obtained over the same
range in the scaling variable
\begin{eqnarray}
\omega^\prime &=& { 2 M \nu + M^2 \over Q^2 }\
	       =\ 1 + { W^2 \over Q^2 }\
	       =\ \omega + { M^2 \over Q^2 }\ .
\label{eq:omegapr}
\end{eqnarray}
More generally, one can define $\omega' = \omega + m^2/Q^2$, for some
arbitrary mass $m \sim 1$~GeV$^2$, although in practice the choice
$m = M$ was usually made in the early analyses.
The range of $W$ over which the structure function exhibits scaling
was found \cite{SLAC_DIS} to increase (from down to
$W^2 \approx 7$~GeV$^2$ to down to $W^2 \approx 3$~GeV$^2$)
if $\nu W_2$ were plotted as a function of $\omega'$ instead of
$\omega$.
While the physical interpretation of this modified scaling variable
was not clear at the time, it did naturally allow for the direct
comparison of data at higher $W^2$ to data at lower $W^2$, over a
range of $Q^2$.
Using the variable $\omega'$, Bloom and Gilman were able to make the
first quantitative observations of quark-hadron duality in electron
scattering.

\begin{figure} 
\hspace*{-0.5cm}
\begin{minipage}{3.1in}
\epsfig{figure=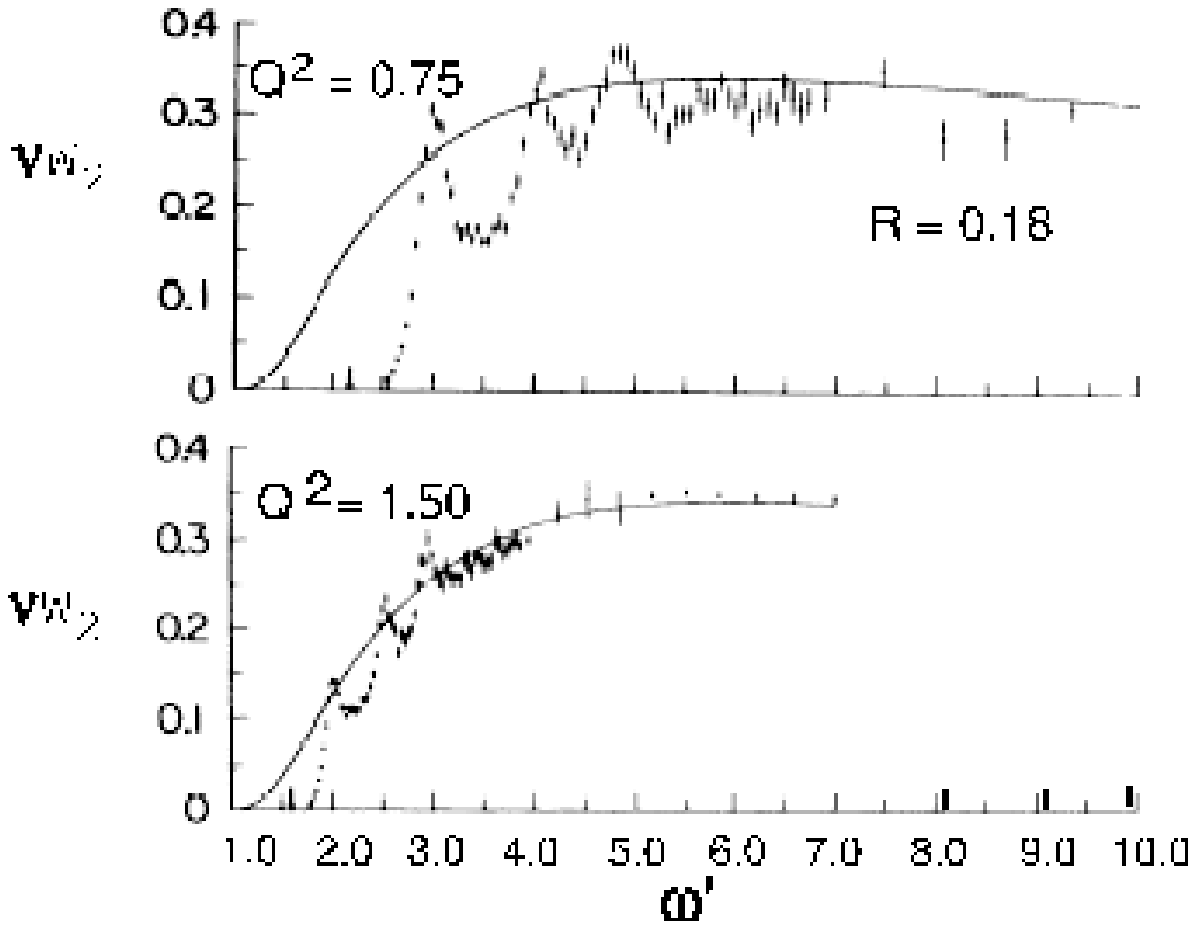, height=6.5cm} % width=7cm}
\end{minipage}
\begin{minipage}{3.1in}
\epsfig{figure=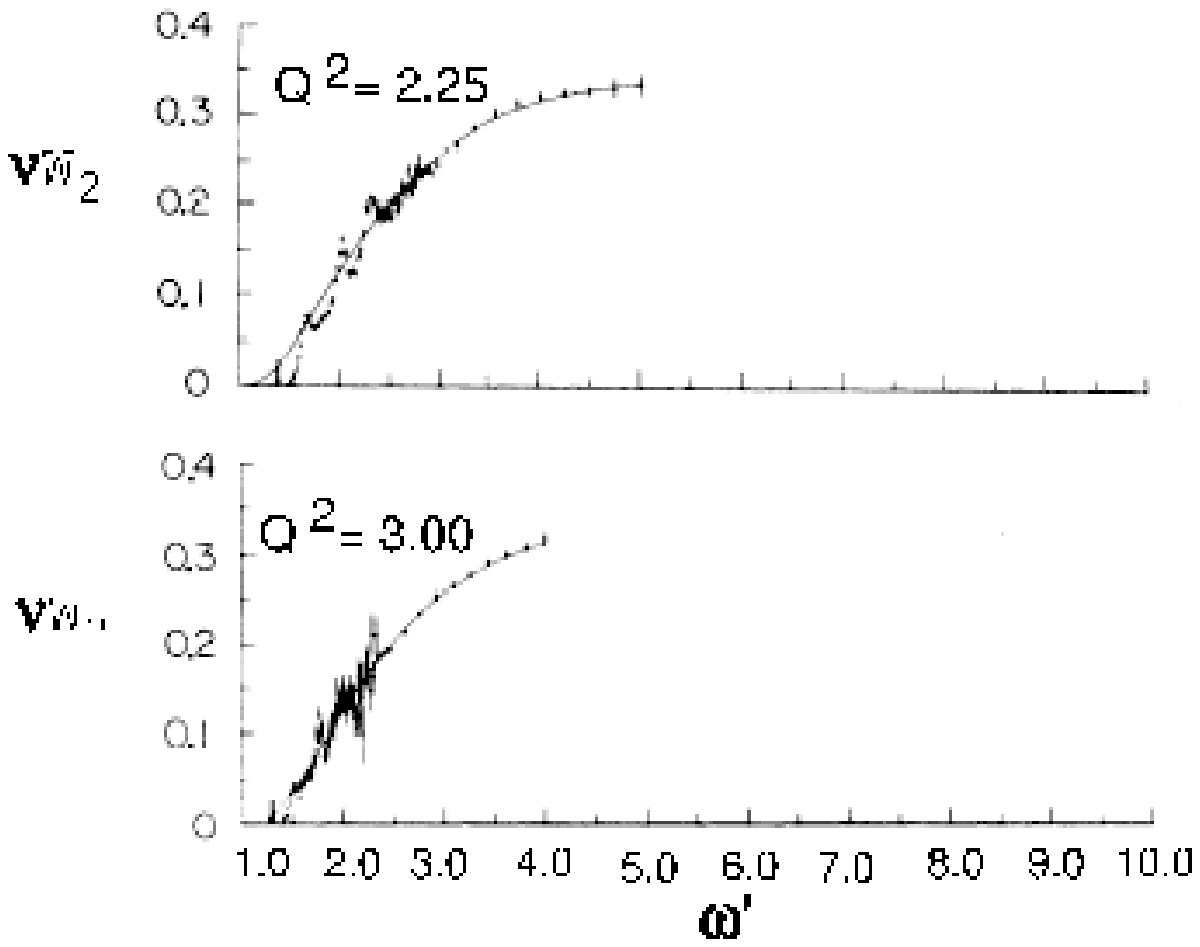, height=6.5cm} % width=7cm}
\end{minipage}
\vspace*{0.5cm}
\begin{centering}
\caption{\label{fig:BG}
	Early proton $\nu W_2$ structure function data in the
	resonance region, as a function of $\omega^\prime$,
	compared to a smooth fit to the data in the scaling region
	at larger $Q^2$. The resonance data were obtained at the
	indicated kinematics, with $Q^2$ in GeV$^2$, for the
	longitudinal to transverse ratio $R=0.18$.
	(Adapted from Ref.~\protect\cite{BG2}.)}
\end{centering}
\end{figure}

The original data on the proton $\nu W_2(\nu, Q^2)$ structure
function in the resonance region are illustrated in Fig.~\ref{fig:BG}
for several values of $Q^2$ from 0.75~GeV$^2$ to 3~GeV$^2$.
This is a characteristic inclusive electron--proton spectrum in the
resonance region, where the almost twenty 
well-established nucleon resonances with masses
below 2~GeV give rise to three distinct enhancements in the measured
inclusive cross section.
Of these, only the first is due to a single resonance, the
$P_{33}(1232)$ $\Delta$ isobar, while the others are a composite
of overlapping states.
The second resonance region, which comprises primarily the
$S_{11}(1535)$ and $D_{13}(1520)$ resonances, is generally
referred to as the ``$S_{11}$'' region due to the dominance
of this resonance at higher $Q^2$.
Since the data shown here are from inclusive measurements, they may
contain tails of higher mass resonances as well as some nonresonant
components.
The $\nu W_2(\nu, Q^2)$ structure function data were extracted from
the measured cross sections using a fixed value of the longitudinal
to transverse cross section ratio, $R = \sigma_L/\sigma_T = 0.18$.

The scaling curve shown in Fig.~\ref{fig:BG} is a parameterization
of the high-$W$ (high-$Q^2$) data available in the early 1970s
\cite{SLAC_DIS}, when deep inelastic scattering was new and data
comparatively scarce.
Presented in this fashion, the resonance data are clearly seen to
oscillate about, and average to, the scaling curve. 
A more modern comparison would include in addition the $Q^2$
evolution of the structure function from perturbative QCD
(as will be discussed in Sec.~\ref{sec:bgstatus}).
Nevertheless, the astute observations made by Bloom and Gilman
are still valid, and may be summarized as follows:
\begin{description}
\item[\ \ I]
The resonance region data oscillate around the scaling curve. 

\item[\ II]
The resonance data are on average equivalent to the scaling curve. 

\item[III]
The resonance region data ``slide'' along the deep inelastic
curve with increasing $Q^2$. 

\end{description}
These observations led Bloom and Gilman to make the far-reaching
conclusion that {\em ``the resonances are not a separate entity but
are an intrinsic part of the scaling behavior of $\nu W_2$''}
\cite{BG1}.
% and that a substantial part of the observed
% scaling behavior of inelastic electron--proton scattering is
% nondiffractive in nature''}

In order to quantify these observations, Bloom \& Gilman drew on the
work on duality in hadronic reactions to determine a FESR equating
the integral over $\nu$ of $\nu W_2$ in the resonance region, to the
integral over $\omega'$ of the scaling function \cite{BG1},
\begin{eqnarray}
{2 M \over Q^2} \int_0^{\nu_m} d\nu\ \nu W_2 (\nu,Q^2)
&=& \int_1^{1 + W_m^2/Q^2} d\omega'\ \nu W_2 (\omega')\ .
\label{eq:BGFESR}
\end{eqnarray}
Here the upper limit on the $\nu$ integration,
$\nu_m = (W_m^2 - M^2 + Q^2)/2M$, corresponds to the maximum value
of $\omega' = 1 + W_m^2/Q^2$, where $W_m \sim 2$~GeV, so that the
integral of the scaling function covers the same range in $\omega'$
as the resonance region data.
The FESR (\ref{eq:BGFESR}) allows the area under the resonances in
Fig.~\ref{fig:BG} to be compared to the area under the smooth
curve in the same $\omega'$ region to determine the degree to which
the resonance and scaling data are equivalent.
A comparison of both sides in Eq.~(\ref{eq:BGFESR}) for $W_m = 2$~GeV
showed that the relative differences ranged from $\sim 10\%$ at
$Q^2 = 1$~GeV$^2$, to $\alt 2\%$ beyond $Q^2 = 2$~GeV$^2$ \cite{BG2},
thus demonstrating the near equivalence on average of the resonance
and deep inelastic regimes (point II above).
Using this approach, Bloom and Gilman's quark-hadron duality
was able to qualitatively describe the data in the range
$1 \alt Q^2 \alt 10$~GeV$^2$.

Moreover, observation III implies a deep connection between the 
$Q^2$ dependence of the structure functions in the resonance
and deep inelastic scattering regimes.
The prominent resonances in inclusive inelastic electron--proton
scattering do not disappear with increasing $Q^2$ relative to the
``background'' underneath them (which scales), but instead fall at
roughly the same rate with increasing $Q^2$.
The prominent nucleon resonances are therefore strongly correlated
with the scaling behavior of $\nu W_2$.

% .......................................................................
\subsubsection{Duality in the Context of QCD}
\label{sssec:momhist}

Following the initial SLAC experiments, inclusive deep inelastic
scattering quickly became the standard tool for investigating the
quark substructure of nucleons and nuclei.
The development of QCD shortly after the discovery of Bloom-Gilman
duality enabled a rigorous description of structure function scaling
and scaling violations at high $Q^2$ and $W$.
In the Bjorken limit ($Q^2, \nu \to \infty$, $x = Q^2/2M\nu$ fixed),
the asymptotic freedom property of QCD reduces the structure function
$\nu W_2$ to a function of a single variable,
$\nu W_2(\nu,Q^2) \to F_2(x)$, which is related to the parton
distribution functions in the quark-parton model 
(see Sec.~\ref{sec:formalism}).

At large but finite $Q^2$, perturbative QCD (pQCD) predicts
logarithmic $Q^2$ scaling violations in $F_2$, arising from gluon
radiation and subsequent $q\bar q$ pair creation.
The observation of scaling violations in $F_2$ in fact played a
crucial role in establishing QCD as the accepted theory of the
strong interactions.
At low $Q^2$, however, perturbative QCD breaks down, and the
description of structure functions in terms of single parton
densities is no longer applicable.
Corrections which at high $Q^2$ are suppressed as powers in
$1/Q^2$ (such as those arising from multi-parton correlations
-- see Sec.~\ref{ssec:qcd}) can no longer be neglected.

A reanalysis of the resonance region and quark-hadron duality
within QCD was performed by De~R\'ujula, Georgi and Politzer
\cite{DGP1,DGP2,GP}, who reinterpreted Bloom-Gilman duality in
terms of moments $M_2^{(n)}(Q^2)$ of the $F_2$ structure function,
defined in Eq.~(\ref{eq:MnDEF})
(or $M_2^{N(n)}$ in Eq.~(\ref{eq:NachtMom})).
For $n=2$ one recovers the analog of Eq.~(\ref{eq:BGFESR}) by
replacing the $\nu W_2$ structure function on the right hand side
by the asymptotic structure function, $F_2^{\rm asy}(x)$,
so that the FESR can be written in terms of the moments as
\begin{eqnarray}
M_2^{(2)}(Q^2) &=& \int_0^1 dx\ F_2^{\rm asy}(x)\ .
\label{eq:CNn2MOM}
\end{eqnarray}
Since the moments are integrals over all $x$, at fixed $Q^2$, they
contain contributions from both the deep inelastic continuum and
resonance regions.
At large $Q^2$ the moments are saturated by the former; at low $Q^2$,
however, they are dominated by the resonance contributions.
One may expect therefore a strong $Q^2$ dependence in the low-$Q^2$
moments arising from the $1/Q^2$ power behavior associated with the
exclusive resonance channels.
A comparison of the high-$Q^2$ moments with those at low $Q^2$ then
allows one to test the duality between the resonance and scaling
regimes.

Empirically, one observed only a slight difference, consistent with
logarithmic scaling behavior in $Q^2$, between moments obtained at
$Q^2 = 10$ GeV$^2$, and those at lower $Q^2$, $Q^2 \sim 2$~GeV$^2$,
that were dominated by resonances.
This suggested that changes in the moments of the $F_2$ structure
function due to power corrections were small, and that averages of
$F_2$ over a sufficiently large range in $x$ were approximately the
same at high and low $Q^2$.
Duality would be expected to hold so long as the $1/Q^2$ scaling
violations were small \cite{DGP2}.

Note that at the energies where duality was observed the ratio
$M^2/Q^2$ is not negligible.
Application of perturbative QCD requires not only that $Q^2$ be
large enough to make expansions in $\alpha_s(Q^2)$ meaningful,
but also that $Q^2$ be large compared to all relevant masses.
Some of the $M^2/Q^2$ effects are purely kinematical in origin,
not associated with the dynamical multi-parton effects that give
rise to the $1/Q^2$ power behavior.
The reason why the variable $\omega'$ is a better scaling variable
than $\omega$ is that it partially compensates for the effects of the
target mass $M^2/Q^2$, allowing approximate scaling to be manifest
down to lower $Q^2$ values (sometimes referred to as ``precocious''
scaling).

\begin{figure}
\begin{minipage}[ht]{3.0in}
\hspace*{-3cm}
\epsfig{figure=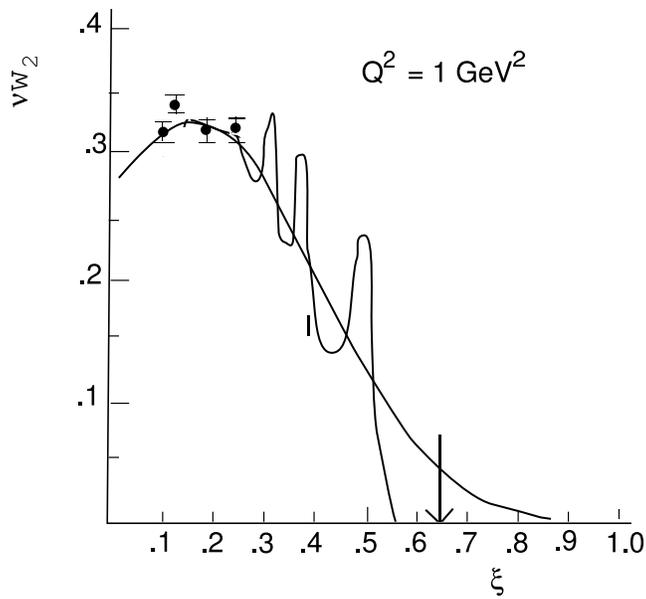, height=15cm}
\end{minipage}
\begin{centering}
\vspace*{-4cm}
\caption{\label{fig:dgpxi}
	Proton $\nu W_2$ structure function data at $Q^2 = 1$~GeV$^2$
	in the resonance (curve with oscillations) and deep inelastic
	(data points) regions as a function of the Nachtmann variable
	$\xi$.  The data are compared to a smooth curve at the same
	$\xi$ values, representing the scaling function from higher
	$Q^2$ and $x$.  The vertical arrow indicates the elastic
	point, $x=1$.  (Adapted from Ref.~\protect\cite{DGP1}.)}
\end{centering}
\end{figure}

In QCD the target mass corrections may be included via the Nachtmann
\cite{EONacht} scaling variable (or generalizations including non-zero
quark masses \cite{BARBIERI}), which are discussed along with others
in the Appendix.
Georgi and Politzer \cite{GP} suggested that the use of the Nachtmann
scaling variable $\xi$ (as in Eq.~(\ref{eq:xi})), rather than $\omega'$
or $x$, would systematically absorb all target mass corrections,
and permit duality to remain valid to lower $Q^2$.
This was indeed borne out by the proton $\nu W_2$ structure function
data, as displayed in Fig.~\ref{fig:dgpxi} as a function of $\xi$ at
$Q^2 = 1$~GeV$^2$.
The Nachtmann variable is in fact the minimal variable which includes
target mass effects, and has been used widely in studies of structure
functions at intermediate $Q^2$ \cite{NACHT_POL,Baluni}.
Further discussions on the use of the Nachtmann variable in moment
analyses can be found in Refs.~\cite{BEGR,GTW,dgptm,EPR,BJT,FG}.

The equivalence of the moments of structure functions at high $Q^2$
with those in the resonance-dominated region at low $Q^2$ is usually
referred to as ``global duality''.
If the equivalence of the averaged resonance and scaling structure
functions holds over restricted regions in $W$, or even for individual
resonances, a ``local duality'' is said to exist.
Once the inclusive--exclusive connection via local duality is taken
seriously, one can in principle use the measured inclusive structure
functions at large $Q^2$ and $\xi$, together with $Q^2$ evolution,
to directly extract resonance transition form factors at lower values
of $Q^2$ over the same range in $\xi$.
As an extreme example, it is even possible to extract elastic form
factors from the inclusive inelastic data below the pion production
threshold \cite{DGP1} to within $\sim 20\%$ \cite{DGP2}.

Bloom and Gilman's observation that the $\nu W_2$ structure function
in the resonance region tracks, with changing $Q^2$, a curve whose
shape is the same as the scaling limit curve is expressly a
manifestation of local duality, in that it occurs resonance by
resonance.
The scaling $F_2$ function becomes smaller at the larger values
of the scaling variable, associated with higher values of $Q^2$.
Therefore, the resonance transition form factors must decrease
correspondingly with $Q^2$.

Carlson and Mukhopadhyay \cite{CM90} quantified the pQCD
expectations for the exclusive resonance transition form factors,
finding the leading behavior to be $1/Q^4$.
They note that pQCD further constrains the $x \to 1$ behavior of
the inclusive nucleon structure function, $\nu W_2^p \sim (1-x)^3$
\cite{FJ},
as predicted also by dimensional scaling laws \cite{BF1,COMPTON90}.
This is yet another manifestation of the inclusive--exclusive relation
arising from local Bloom-Gilman duality.
We shall discuss this and other phenomenological applications of local
Bloom-Gilman duality in Sec.~\ref{ssec:local}.

Following this historic prelude, where we set in context the original
observations of duality in electron--nucleon scattering, we are now
ready to explore in detail the modern phenomenology of Bloom-Gilman
duality.
In the next section we discuss the current experimental status of
duality in electron--nucleon scattering, and present an in-depth
account of available data for both spin-averaged and spin-dependent
processes.

\clearpage

%%%%%%%%%%%%%%%%%%%%%%%%%%%%%%%%%%%%%%%%%%%%%%%%%%%%%%%%%%%%%%%%%%%%%%%%%
\section{Bloom-Gilman Duality: Experimental Status}
\label{sec:bgstatus}

Bloom and Gilman's initial discovery of the resonance--scaling
relations in inclusive electron--nucleon scattering was indeed
quite remarkable, particularly given the relatively poor statistics
and limited coverage of the early data.
As higher energy accelerated beams became increasingly available
in the 1970s and 1980s, focus naturally shifted to higher $Q^2$ with
experimental efforts geared towards investigating the predictions of
perturbative QCD.
This was of course a necessary step in order to establish whether QCD
itself was capable of describing hadronic substructure in regions
where the applicability of perturbative treatments was not in doubt. 
More recently, however, there has been a growing realization that
understanding of the resonance region in inelastic scattering,
and the interplay between resonances and scaling in particular, 
represents a critical gap which must be filled if one is to fully
fathom the nature of the quark--hadron transition in QCD.

The availability of high luminosity (polarized) beams, together with
polarized targets, has allowed one to revisit Bloom-Gilman duality
at a much more quantitative level than previously possible, and an
impressive amount of data, of unprecedented quantity and quality,
has now been compiled in the resonance region and beyond.
In this section we review the recent data on various spin-averaged
and spin-dependent structure functions, together with their moments,
which have been instrumental in deepening our understanding of the
resonance--scaling transition.

% -----------------------------------------------------------------------
\subsection{Duality in the $F_2$ Structure Function}
\label{ssec:f2p}

Much of the new data have been collected in inclusive electron
scattering on the proton.
At high $Q^2$, the differential cross section given in
Eq.~(\ref{eq:f1f2}) is usually expressed in terms of the $F_2$
structure function, because of the elegant interpretation which $F_2$
has in the parton model (in terms of quark momentum distributions),
and the crucial role it played in understanding scaling violations
in QCD.
Since the original observations of Bloom-Gilman duality in inclusive
structure functions, $F_2(x,Q^2)$ has become one of the best measured
quantities in lepton scattering, with measurements from laboratories
around the world contributing to a global data set spanning over five
orders of magnitude in $x$ and $Q^2$.

Here we first present $F_2$ data of particular interest to duality
studies, both on the proton and on nuclear targets, and then turn to
the extraction of the purely longitudinal and transverse structure
functions, $F_L$ and $F_1$, respectively, in Sec.~\ref{ssec:lt}.
While it is clear that longitudinal--transverse separated data are
necessary to accurately extract $F_2$ from measured cross sections,
we have chosen here to present $F_2$ results because of the
historical significance of this structure function both in Bloom \&
Gilman's original work, and also as the most widely measured
quantity in deep inelastic scattering over the past three decades.

% .......................................................................
\subsubsection{Local Duality for the Proton}
\label{sssec:f2local}

\begin{figure}[t]
\begin{minipage}{4.0in}
\hspace*{-2.5cm}
\epsfig{figure=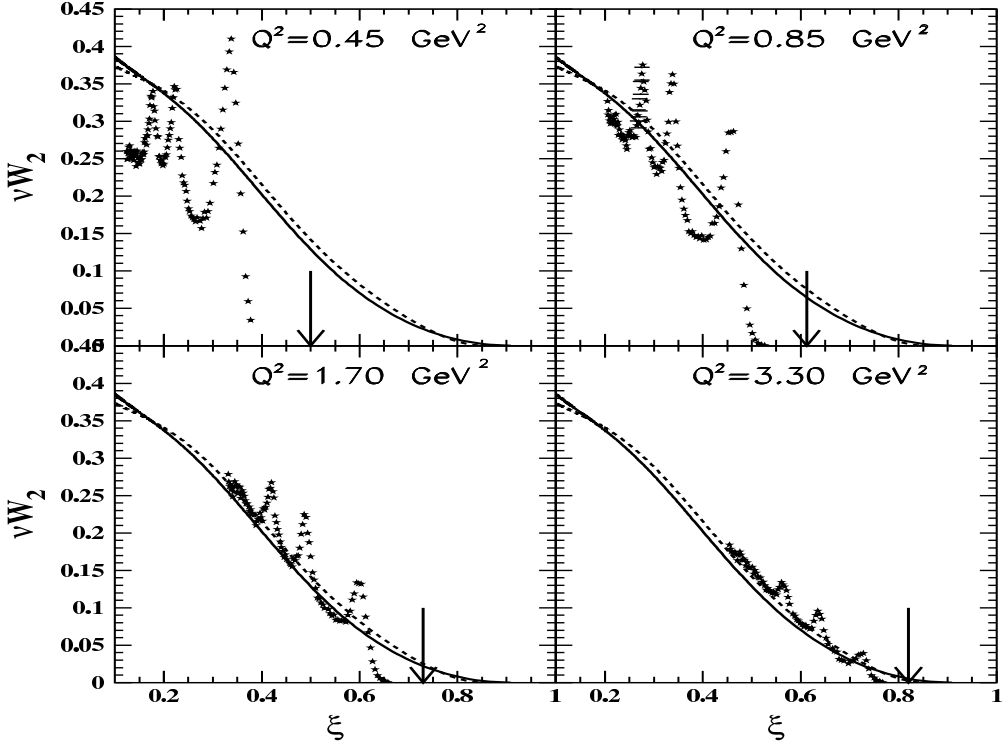, height=12cm, width=14cm}
\end{minipage}
\begin{centering}
\vspace*{-1.5cm}
\caption{\label{fig:dualspectra}
	Proton $\nu W_2^p = F_2^p$ structure function data in the
	resonance region as a function of $\xi$,
	at $Q^2 = 0.45$, 0.85, 1.70, and 3.30~GeV$^2$ from Hall~C
	at Jefferson Lab \protect\cite{F2JL1,F2JL2}.
	The arrows indicate the elastic point, $\xi = \xi(x=1)$.
	The curves represent fits to deep inelastic structure
	function data at the same $\xi$ but higher ($W^2,Q^2$) from 
	NMC \protect\cite{NMC} at $Q^2 = 5$~GeV$^2$ (dashed) and
	$Q^2 = 10$~GeV$^2$ (solid).}
\end{centering}
\end{figure}

A sample of proton $F_2^p$ structure function data from Jefferson Lab
\cite{F2JL1,F2JL2} in the resonance region is depicted in
Fig.~\ref{fig:dualspectra}, where it is compared with fits to a large
data set of higher-$W$ and $Q^2$ data from the New Muon Collaboration
\cite{NMC}.
Figure~\ref{fig:dualspectra} is in direct analogy to
Fig.~\ref{fig:dgpxi} above, where the Nachtmann variable $\xi$ has
replaced the more {\it ad hoc} variable $\omega'$ as a means to relate
high-$(W^2, Q^2)$ deep inelastic data to data at the lower $(W^2, Q^2)$
values of the resonance region, as well as to include proton target
mass corrections.
Both the $\xi$ and $\omega^\prime$ variables depend on ratios of $x$
to $Q^2$ (or, correspondingly, $W$ to $Q^2$), thereby allowing direct
comparison of structure functions in the resonance and scaling
regimes by plotting the scaling and resonance data at the same
ordinate point.
For example, $\xi = 0.6$ can correspond to a point in the $\Delta$
resonance region around $Q^2 = 1.5$ GeV$^2$, or a point in the deep
inelastic region of $W^2 = 14$ GeV$^2$ at $Q^2 = 20$ GeV$^2$.

The kinematics for the resonance data in Fig.~\ref{fig:dualspectra}
range from the single pion production threshold to $W^2 = 5$~GeV$^2$.
The elastic peak position at $\xi = \xi(x=1)$ is indicated by the
vertical arrows, and the lower $\xi$ values correspond to the
higher-$W^2$ kinematics.
Of the three prominent enhancements, the lowest mass $\Delta$
resonance falls at the highest $\xi$ values.
The statistical uncertainties are included in the error bars on
the data points, and the total systematic uncertainty was estimated
to be less than $4\%$ \cite{F2JL1}.
The latter includes some uncertainty associated with the choice of
$R$ used to extract $F_2$ from the measured cross sections
(see Eq.~(\ref{eq:Rlt})).

The resonance data are compared to a global fit curve to deep
inelastic scattering (DIS) data from Ref.~\cite{NMC}, here shown
for two fixed values of $Q^2 = 5$ and 10~GeV$^2$.
The curves are plotted at these fixed $Q^2$ (somewhat higher than the
resonance region data) and the $\xi$ values corresponding to those
of the resonance region data.
This ($\xi$, $Q^2$) choice kinematically determines the $x$ and $W^2$
values in the DIS regime, and therefore establishes the $(x, Q^2)$
values at which to utilize the DIS parameterization.
It is important to note that this causes an effective target mass
correction to the scaling curve, which can increase the structure 
function strength by tens of percent.

Several important features are worth noting in
Fig.~\ref{fig:dualspectra}.
Firstly, the data clearly display the signature oscillations around
the DIS curve, qualitatively averaging to it.
Quantitatively, scaling curves were found to describe the average
of the resonance region $F_2^p$ spectra in Refs.~\cite{F2JL1,F2JL2}
to better than 10\%.
Next, the resonance data closely follow the scaling curves to higher
$\xi$ as $Q^2$ increases, such that the $\xi$ shape of the DIS curve
determines the $Q^2$ dependence of the resonance region structure
function.
Put figuratively, the resonances slide down the scaling curve with
increasing $Q^2$.
In all, the current precision resonance and DIS data conclusively
verify the original observations of Bloom \& Gilman.

The $Q^2$ dependence of the scaling structure function is not 
drastic, as the $Q^2 = 5$ and 10~GeV$^2$ values of the structure
function are quite similar.
However, the $Q^2$ dependence of $F_2^p$ in the resonance region is
significant, as can be seen in the difference between the $Q^2 = 0.45$
and 3.3~GeV$^2$ spectra.
Knowledge of the $Q^2$ dependence of the scaling structure function
is an important improvement over the original data sets available to
Bloom \& Gilman \cite{BG1,BG2}.

\begin{figure}
\begin{minipage}[ht]{4.0in}
\epsfig{figure=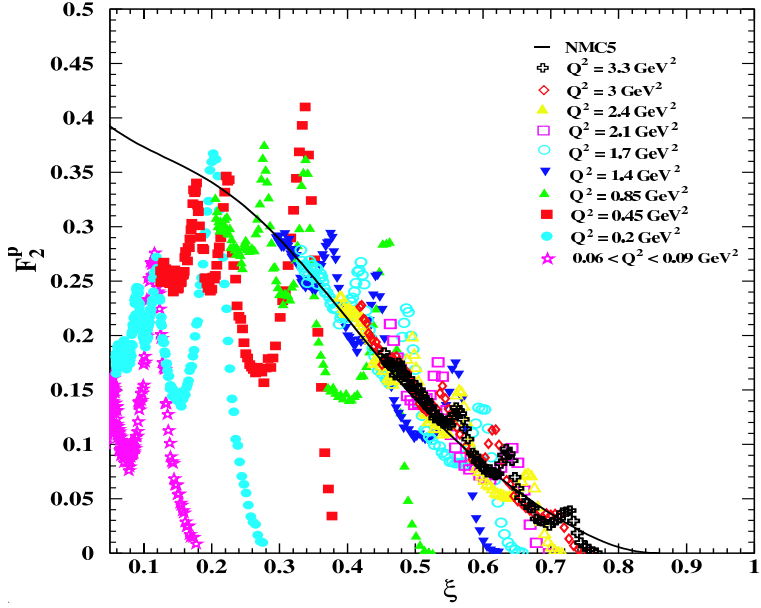, height=8cm, width=10cm}
\end{minipage}
\begin{centering}
\vspace*{0.5cm}
\caption{\label{fig:dualbanner}
	Proton $\nu W_2^p = F_2^p$ structure function data from SLAC
	and  Jefferson Lab in the resonance region in the range 
	$0.06 < Q^2 < 3.30$ GeV$^2$, as indicated.  
	The solid curve is a fit to deep inelastic data at the same
	$\xi$ but higher ($W^2,Q^2$) from Ref.~\protect\cite{NMC},
	shown here at $Q^2 = 5$~GeV$^2$.}
\end{centering}
\end{figure}

The same data set, combined with some lower $Q^2$ data from SLAC,
is depicted in Fig.~\ref{fig:dualbanner} in a single plot.
Here, the salient features of duality are even more striking:
above $\xi \sim 0.2$ the data all average to the scaling curve.
Moreover, the position of the resonance peaks relative to the scaling
curve is determined by $Q^2$, with the higher $Q^2$ values at higher
$\xi$.
Therefore, {\it both} the size {\it and} momentum dependence of the
resonance region structure function are apparently determined by the
scaling limit curve.
The lower-$Q^2$ data (below $\xi \sim 0.2$) will be discussed in more
detail in Sec.~\ref{sssec:real}, below.
We note, however, that it may not be surprising that the scaling
curve at higher $Q^2$ deviates from the resonance region data in 
this lower $\xi$ (or $x$) range, since here sea quark effects are
large and vary rapidly with $Q62$.

\begin{figure}
\begin{minipage}[ht]{4in}
\hspace*{-1cm}
\epsfig{figure=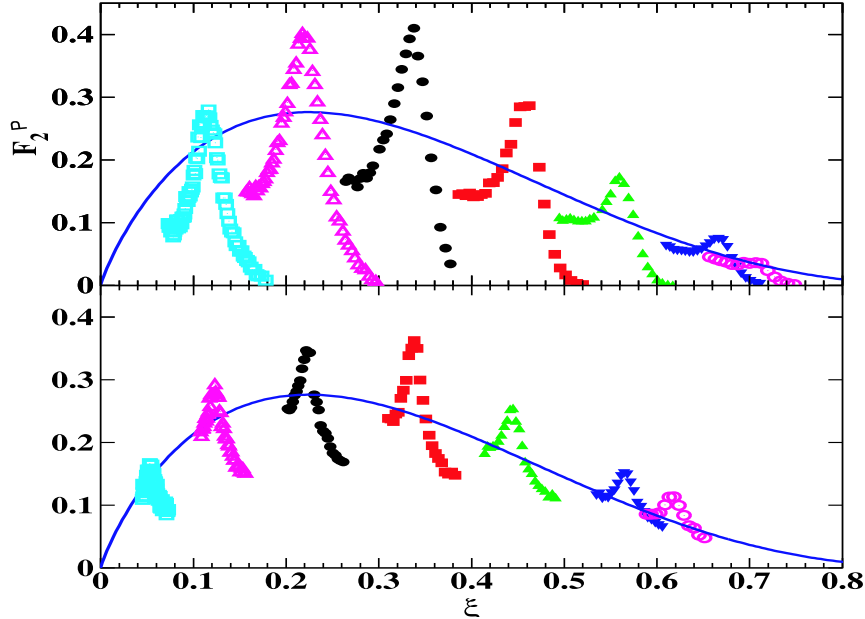, height=10cm, width=12cm}
\end{minipage}
\vspace*{-1.5cm}
\begin{centering}
\caption{\label{fig:local_res}
	Proton $F_2$ structure function in the $\Delta$ (top) and
	$S_{11}$ (bottom) resonance regions from Jefferson Lab Hall~C,
	compared with the scaling curve from Ref.~\protect\cite{F2JL1}.
	The resonances move to higher $\xi$ with increasing $Q^2$,
	which ranges from $\sim 0.5$~GeV$^2$ (smallest $\xi$ values)
	to $\sim 4.5$~GeV$^2$ (largest $\xi$ values).}
\end{centering}
\end{figure}

Analyses such as this demonstrate a global duality for the entire
resonance regime.
However, one can observe in Figs.~\ref{fig:dualspectra} and
\ref{fig:dualbanner} that the average strength of the individual
resonance structures is also consistent with that of the scaling
curve.
This ``local'' duality is more evident in Fig.~\ref{fig:local_res},
wher the $F_2^p$ structure function for the first ($P_{33}(1232)$
or $\Delta$) and second ($S_{11}(1535)$) resonance regions from 
Fig.~\ref{fig:dualspectra} are plotted versus $\xi$ for $Q^2$ values
from 0.5~GeV$^2$ to 4.5~GeV$^2$.
The sliding of the individual $\Delta$ and $S_{11}$ resonance
regions along the scaling curve is dramatically illustrated here,
where the resonances are clearly seen to move up in $\xi$ with
increasing $Q^2$.
One observes therefore that the $Q^2$ behavior of the resonances is
determined by the position on the $\xi$ scaling curve on which 
they fall.
The resonance contributions to $F_2^p$ track, with changing $Q^2$,
a curve whose shape is the same as the scaling limit curve.
Note, however, that it is always necessary to average the resonance
data over some region for local duality to hold.
For example, the data point at the maximum of the resonance peak
will stay above, and never equal, the scaling strength. 
In other words, local duality has a limit --- a point which we shall
return to again.

The classic presentation of duality in electron--proton scattering,
as depicted in Figs.~\ref{fig:dualspectra} and \ref{fig:dualbanner}, 
is somewhat ambiguous in that resonance data at low $Q^2$ values are
being compared to scaling curves at higher $Q^2$ values.
It is difficult to evaluate precisely the equivalence of the two if
$Q^2$ evolution \cite{EVOLVE} is not taken into account.
Furthermore, the resonance data and scaling curves, although at the
same $\xi$ or $\omega^\prime$, are at different $x$ and sensitive
therefore to different parton distributions.
A more stringent test of the scaling behavior of the resonances would
compare the resonance data with fundamental scaling predictions for the
{\it same} low-$Q^2$, high-$x$ values as the data.

Such predictions are now commonly available from several groups around
the world, for instance, the Coordinated Theoretical-Experimental
Project on QCD (CTEQ) \cite{metapage}; Martin, Roberts, Stirling,
and Thorne (MRST) \cite{MRSTgeneral}; Gluck, Reya, and Vogt (GRV)
\cite{GRVgeneral}; and Bl\"umlein and B\"ottcher \cite{Blumlein:2002be},
to name a few.
These groups provide results from global QCD fits to a full range of
hard scattering processes --- including lepton-nucleon deep inelastic
scattering, prompt photon production, Drell-Yan measurements, jet
production, {\em etc.} --- to extract quark and gluon distribution
functions (PDFs) for the proton. 
The idea of such global fitting efforts is to adjust the fundamental
PDFs to bring theory and experiment into agreement for a wide range
of processes.
These PDF-based analyses include pQCD radiative corrections which
give rise to logarithmic $Q^2$ dependence of the structure function.
In this report, we use parameterizations from all of these groups,
choosing in each case the most straightforward implementation for our
needs.
It is not expected that this choice affects any of the results
presented here.

\begin{figure}
\begin{minipage}[ht]{3.0in}
\hspace*{-4cm}
\epsfig{figure=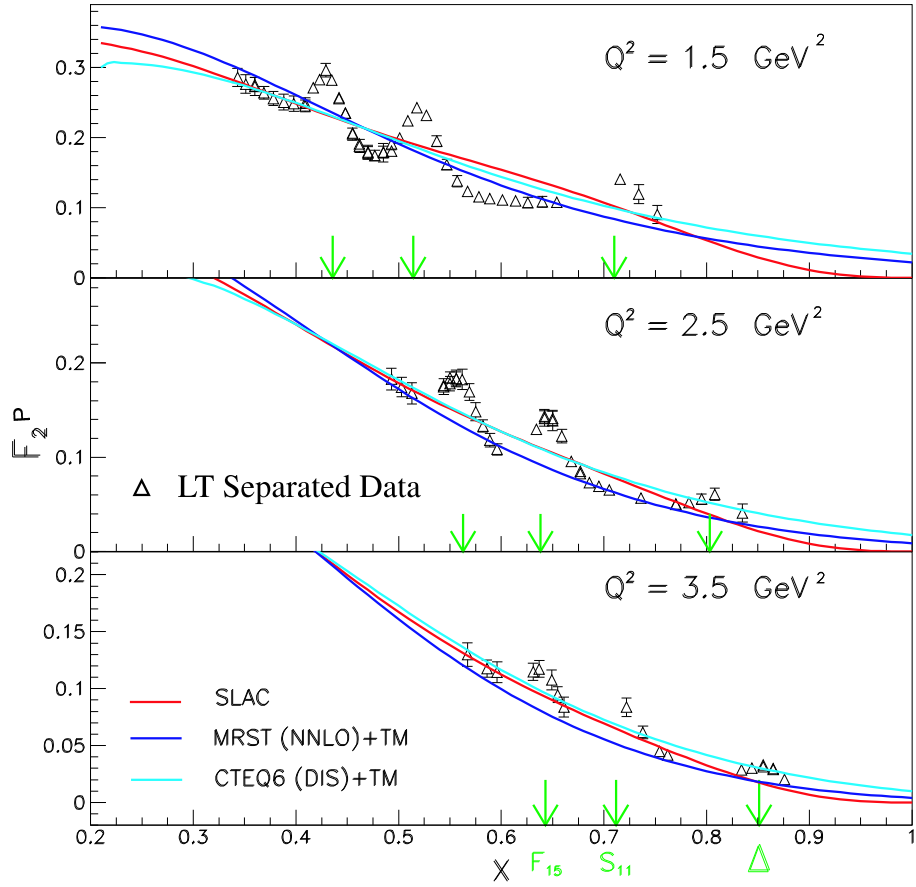, height=15cm} %, width=9cm}
\end{minipage}
\vspace*{-2cm}
\begin{centering}
\caption{\label{fig:f2}
	Proton $F_2^p$ structure function in the resonance region
	for several values of $Q^2$, as indicated.
	Data from Jefferson Lab Hall~C \protect\cite{YONGGUANG,E94110}
	are compared with some recent parameterizations of the deep
	inelastic data at the same $Q^2$ values (see text).}
\end{centering}
\end{figure}

Comparison of resonance region data with PDF-based global fits allows
the resonance--scaling comparison to be made at the same values of
($x,Q^2$), making the experimental signature of duality less
ambiguous. 
Such a comparison is presented in Fig.~\ref{fig:f2} for $F_2^p$
data from Jefferson Lab experiment E94-110 \cite{YONGGUANG,E94110},
with the data bin-centered to the values $Q^2 = 1.5$, 2.5 and
3.5~GeV$^2$ indicated.
These $F_2^p$ data are from an experiment capable of performing
longitudinal/transverse cross section separations, and so are
even more precise than those shown in
Figs.~\ref{fig:dualspectra}--\ref{fig:local_res}.

The smooth curves in Fig.~\ref{fig:f2} are the perturbative QCD
fits from the MRST \cite{MRST} and CTEQ \cite{CTEQ} collaborations,
evaluated at the same $Q^2$ values as the data.
These are shown with target mass (TM) corrections included,
which are calculated according to the prescription of Barbieri
{\it et al} \cite{BEGR}.
The SLAC curve is a fit to deep inelastic scattering data
\cite{WHITLOW}, which implicitly includes target mass effects
inherent in the actual data. 
The target mass corrected pQCD curves appear to describe, on average,
the resonance strength at each $Q^2$ value.
Moreover, this is true for all of the $Q^2$ values shown, indicating
that the resonance averages must be following the same perturbative
$Q^2$ evolution \cite{EVOLVE} which governs the pQCD parameterizations
(MRST and CTEQ).
This demonstrates even more emphatically the striking duality between
the nominally highly-nonperturbative resonance region and the
perturbative scaling behavior.

\begin{figure}
\begin{minipage}[ht]{4.0in}
\epsfig{figure=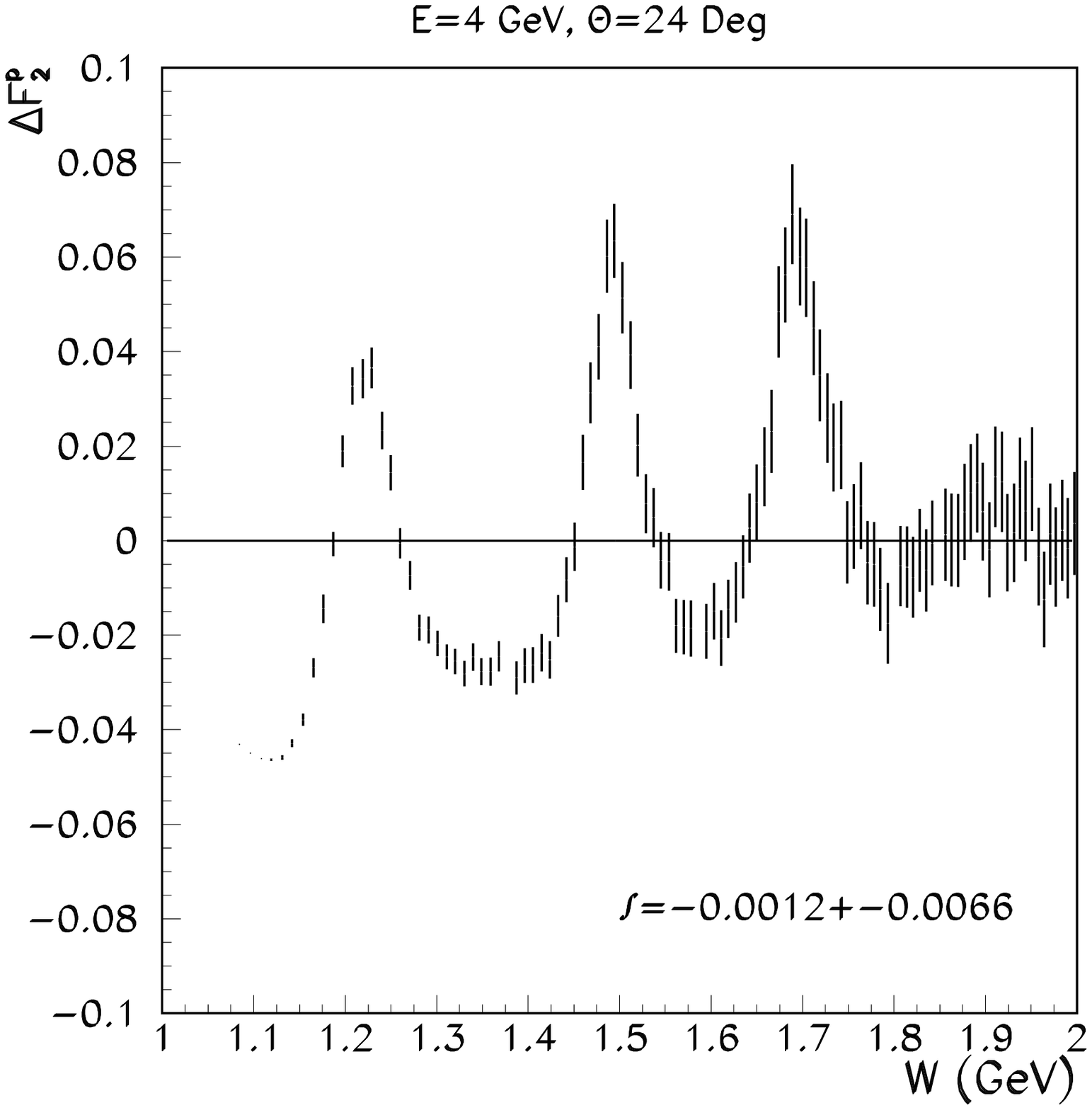, height=10cm, width=10cm}
\end{minipage}
\begin{centering}
\vspace*{-2cm}
\caption{\label{fig:sergey}
	The difference $\Delta F_2^p$ between proton $F_2^p$ 
	structure function data (at the indicated kinematics)
	from Jefferson Lab Hall C and the scaling curve of
	Ref.~\protect\cite{alekhin} as a function of missing
	mass $W$.  The integrated difference yields a value of
	--0.0012 $\pm$ 0.0066 for this particular $W$-spectrum.}
\end{centering}
\end{figure}

An alternate approach to quantifying the observation that the
resonances average to the scaling curve has been used recently
by Alekhin \cite{alekhin}.
Here the differences between the resonance structure function
values and those of the scaling curve, $\Delta F_2^p$, are used
to demonstrate duality, as shown in Fig.~\ref{fig:sergey},
where the differences are seen to oscillate around zero.
Integrating $\Delta F_2^p$ over the resonance region, the resonance
and scaling regimes are found to be within 3\% in all cases above
$Q^2 = 1$ GeV$^2$ \cite{alekhinprivate}.
One should note that in Ref.~\cite{alekhin} a different set of PDFs
was employed, extracted only from DIS scattering data.

Equivalently, quark-hadron duality can also be quantified by
computing integrals of the structure function over $x$ in the
resonance region at fixed $Q^2$ values,
\begin{equation}
\label{eq:pcnmom1}
\int_{x_{\rm th}}^{x_{\rm res}}\; dx \; F_2^p(x,Q^2)\ ,
\end{equation}
where $x_{\rm th}$ corresponds to the pion production threshold at
the fixed $Q^2$, and $x_{\rm res} = Q^2/(W_{\rm res}^2 - M^2 + Q^2)$
indicates the $x$ value at that same $Q^2$ where the traditional
delineation between the resonance and deep inelastic scattering
regions at $W = W_{\rm res} \equiv 2$~GeV falls.
These integrals may then be compared to the analogous integrals of
the ``scaling'' structure function at the same $Q^2$ and over the
same range of $x$.

\begin{figure}
\begin{minipage}[ht]{3.0in}
\hspace*{-2cm}
\epsfig{figure=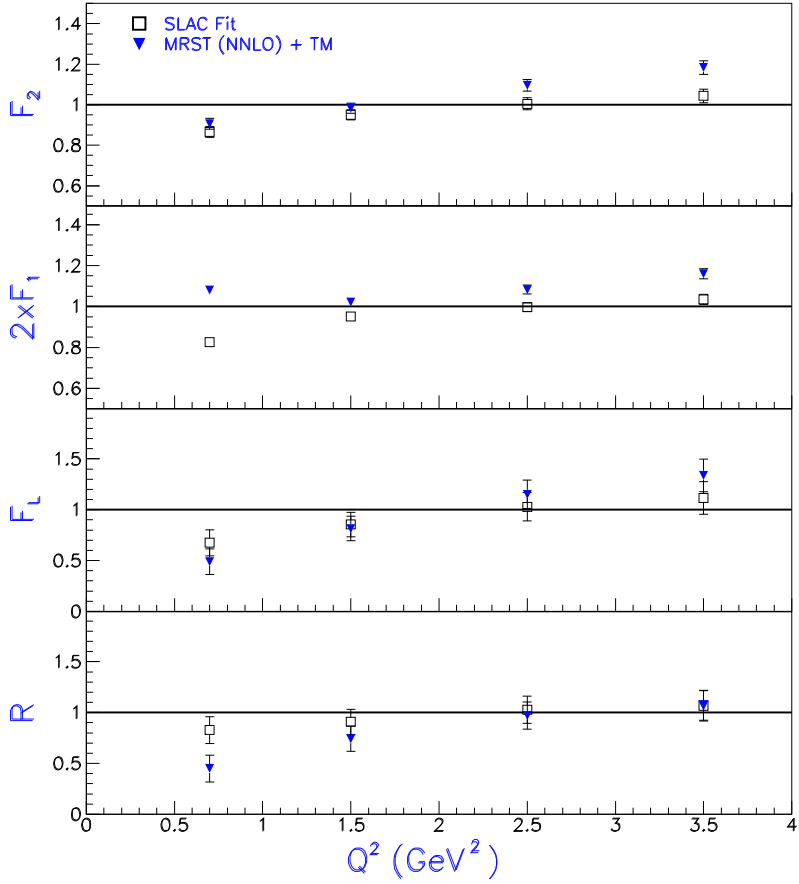, height=14cm} %, width=9cm}
\end{minipage}
\begin{centering}
\vspace*{-1.5cm}
\caption{\label{fig:dual_ratios}
	Ratios of the resonance to scaling integrals of the proton
	structure functions $F_2^p$, $F_L^p$, $2xF_1^p$, and $R^p$ 
	integrated over $x$. The integration limits are defined by
	the pion threshold at the highest $x$, and by $W = 2$~GeV at
	the lowest $x$, for the $Q^2$ values of the resonance data.
	The scaling functions in the ratios are the SLAC
	parameterization \protect\cite{WHITLOW} (squares) and the
	target mass corrected MRST fit \protect\cite{MRST} (triangles)
	at the same $(x, Q^2)$ values.}
\end{centering}
\end{figure}

The ratios of the integrals (\ref{eq:pcnmom1}) of the resonance data
to the scaling structure functions, extrapolated to the same $x$, are
shown in Fig.~\ref{fig:dual_ratios} for the proton $F_2^p$ structure
function, as well as for the $F_1^p$, $F_L^p$, and $R^p$ structure
functions discussed in Sec.~\ref{ssec:lt} below.
The perturbative strength is calculated in one case from the MRST
parameterization \cite{MRST}, with the target mass corrections applied
following Ref.~\cite{BEGR}, and in the other from a parameterization
of SLAC deep inelastic data ~\cite{WHITLOW}.
In most cases, the integrated perturbative strength is equivalent
to the resonance region strength to better than 5\% above
$Q^2 = 1$~GeV$^2$.
This shows unambiguously that duality is holding quite well on
average in all of the unpolarized structure functions; the total
resonance strength over a range in $x$ is equivalent to the
perturbative, PDF-based prediction.

Of some concern is the seeming deviation from this observation in the MRST 
ratio at the highest values of $Q^2$ in Fig.~\ref{fig:dual_ratios},
where the ratio rises above unity.
This rise is not a violation of duality, but rather is most likely
due to an underestimation of large-$x$ strength in the pQCD 
parameterizations. Higher $Q^2$ corresponds to large $x$ here 
and, for comparison with resonance region data at the larger $Q^2$
values, accurate predictions at large $x$ are crucial.
There exists uncertainty in the PDFs at large $x$, largely 
due to the ambiguity in the $d/u$ quark distribution function ratio
beyond $x \sim 0.5$, which arises from the model dependence of the
nuclear corrections when extracting neutron structure information
from deuterium data (see Refs.~\cite{NP,KUHLMANN,SLAC_HIX,alekhin_np}).
Even if nominally deep inelastic data at higher $W^2$ and $Q^2$,
rather than resonance region data, are compared to the available
pQCD parameterizations, the scaling curves do not show enough
strength at large $x$ ($x \agt 0.5$) and fall uniformly below the
data points.

If one assumes duality, it is also possible to obtain a scaling curve
by averaging the resonance region data.
Here, average values may be calculated for discrete data bins in $\xi$.
A fit to these averages has been obtained by Niculescu {\it et al.}
\cite{F2JL1}, who found that the resonances oscillate around the fit
to within $10\%$, even down to $Q^2$ values as low as 0.5~GeV$^2$.
These lower $Q^2$ values are below $\xi = 0.2$ in
Fig.~\ref{fig:dualspectra}, where the resonance data fall below the
$Q^2 = 10$~GeV$^2$ scaling curve. 

\begin{figure}
\begin{minipage}[ht]{3.0in}
\hspace*{-2cm}
\epsfig{figure=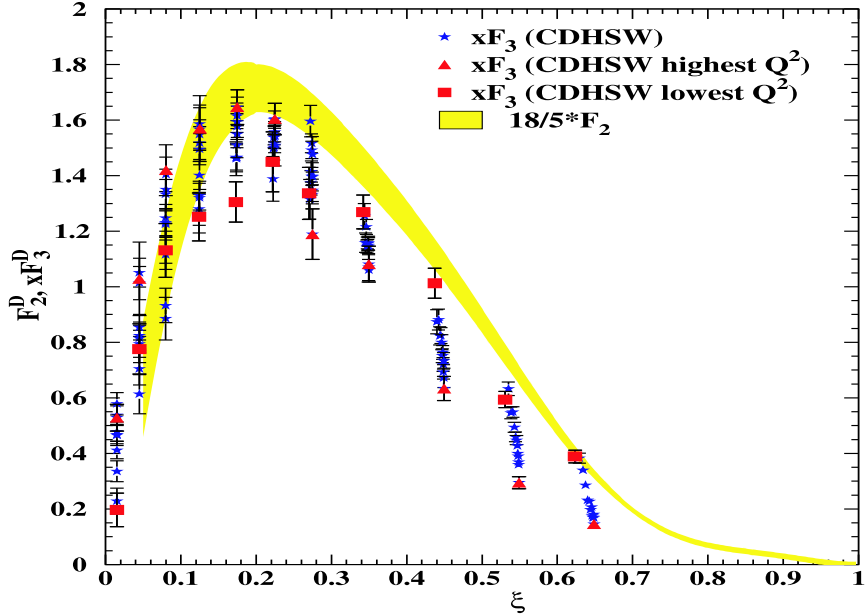, height=10cm, width=12cm}
\end{minipage}
\begin{centering}
\vspace*{-1cm}
\caption{\label{fig:xf3}
	A comparison of the duality-averaged deuteron $F_2^d$ scaling
	curve determined from the nucleon resonance region data, 
	multiplied by 18/5, with the CDHSW data~\protect\cite{BERGE}
	on the deuteron $xF_3^d$ structure function obtained from 
	deep inelastic neutrino--iron scattering data.}
\end{centering}
\end{figure}

The scaling curve obtained for the deuteron $F_2^d$ structure function
by averaging the resonance data is shown as a band in Fig.~\ref{fig:xf3},
to indicate the relevant uncertainty.
This average curve is in good agreement with extrapolations from deep
inelastic scattering above $Q^2 \sim 1.5$~GeV$^2$, and also represents
a smooth average of the resonance data even at lower $Q^2$ and $x$
values.
Note that this curve does not account for the $Q^2$ evolution of the
resonance region, having been obtained from a fit to average resonance
region data spanning a range of values in $Q^2$ within a finite-$\xi$
bin.  
However, the evolution in the range of the Jefferson Lab data
($0.5 < Q^2 < 4.5$~GeV$^2$) is not expected to be large.

When viewed over the entire range in $x$, including at low $x$ and
$Q^2$, the duality-averaged curve yields a clear valence-like shape,
which is in qualitative agreement with the neutrino/antineutrino data
on the valence $xF_3$ structure function.
To enable a direct comparison, the Jefferson Lab average scaling
curve has been multiplied by a factor 18/5 to account for the quark
charges, and a neutron excess correction has been applied to the $xF_3$
data to obtain neutrino-deuterium data \cite{EMC_NPCORR}.
The $xF_3$ structure function, which is typically accessed in deep
inelastic neutrino-iron scattering \cite{BERGE,OLTMAN,KIM}, is
associated with the parity-violating term in the hadronic current
and is odd under charge conjugation.
In the quark-parton model it is therefore expressed as a difference
between quark and antiquark distributions, as in Eq.~(\ref{eq:xf3_q}).
This suggests a unique sensitivity of the duality-averaged $F_2$ data
\cite{F2JL2} to valence quarks.

Although the agreement between the averaged $F_2$ scaling curve in
the resonance region and the deep inelastic neutrino $xF_3$ data
is not perfect, the similarity is compelling.
The observation by Bloom \& Gilman that there may be a common origin
for the electroproduction of resonances and deep inelastic scattering
seems to be true, even at the lowest values of $Q^2$, if one assumes
a sensitivity to a valence-like quark distribution only.
We shall discuss the possible origin of the valence-like behavior
of $F_2$ at low $Q^2$ in Sec.~\ref{sssec:real}.

% .......................................................................
\subsubsection{Low $Q^2$ Moments}
\label{sssec:f2moment}

The commonly accepted, QCD-based formulation of duality
\cite{DGP1,DGP2} relates the moments of structure functions at high
$Q^2$, where deep inelastic phenomena make the primary contribution,
to the low-$Q^2$ moments, which are dominated by contributions from
the resonance region.
The $Q^2$ dependence of the moments between the two regions is
expected to reflect both perturbative evolution \cite{EVOLVE},
associated with single quark scattering, and the $1/Q$ power behavior
arising from interactions between the struck quark and the remaining
``spectator'' quarks in the nucleon.
This formulation is discussed in detail in Secs.~\ref{sssec:momhist}
and \ref{ssec:qcd}, where duality is expressed in terms of the
operator product expansion in QCD.
For the purposes of this section, it is sufficient to note that the
experimental observation of duality is related to the fact that the
$1/Q$ multiparton contributions to the $F_2$ moments are small or
canceling on average, even in the low $Q^2$ region where they should
become increasingly important.
Conversely, deviations from duality would attest to the presence
of significant multiparton effects.

\begin{figure}[h]
\begin{minipage}{3.0in}
\hspace*{-2.5cm}
\epsfig{figure=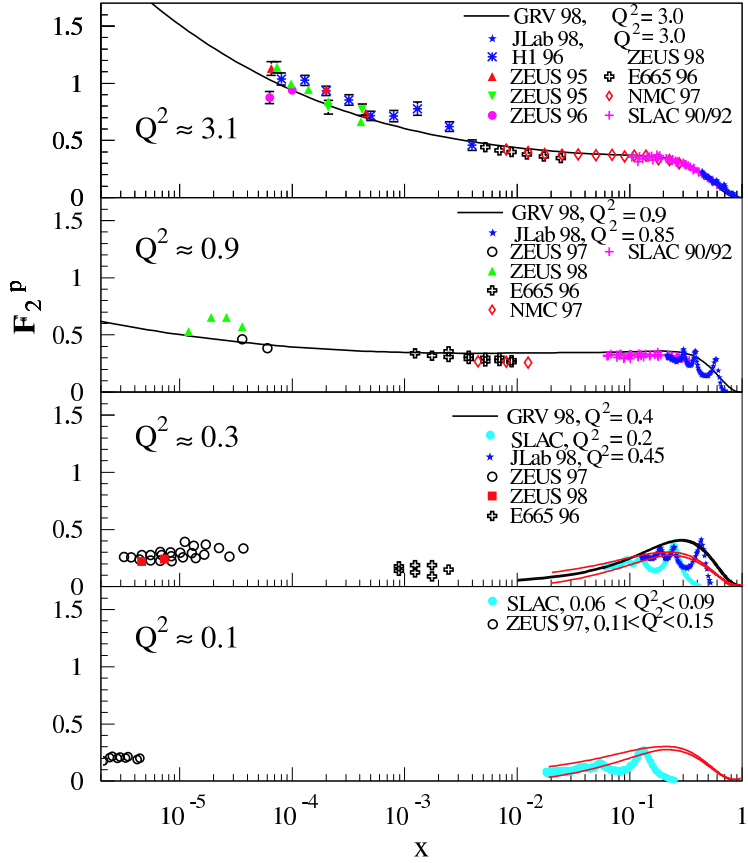, height=14cm}
\end{minipage}
\begin{centering}
\vspace*{-2cm}
\caption{\label{fig:f2lowq}
	Global data on the proton $F_2^p$ structure function versus
	$x$ at four values of $Q^2$ (note the logarithmic $x$ scale).
	The solid curves in the top two panels represent $F_2^p$
	calculated from parton distribution function parameterizations
	by the GRV collaboration \protect\cite{GRV98}, evolved from
	$Q^2 = 0.4$~GeV$^2$.
	The central (black) solid curve in the third panel represents
	the distribution at $Q^2 = 0.4$~GeV$^2$.  The two outer (red)
	curves in the bottom two panels represent the uncertainty
	range of the duality averaged curve discussed in
	Sec.~\ref{sssec:f2local}.
	(Adapted from Ref.~\protect\cite{MOMENTS}.)}
\end{centering}
\end{figure}

Duality expressed in terms of moments is demonstrated most
incontrovertibly by extending the integration limits of the duality
integrals in (\ref{eq:pcnmom1}) to include the entire $x$ range
$0 \leq x \leq 1$.
In this case, the duality integral (\ref{eq:pcnmom1}) becomes the
$n=2$ (Cornwall-Norton) moment of the $F_2$ structure function,
given in Eq.~(\ref{eq:MnDEF}).

To construct the moments accurately, data covering a large range in 
$x$ must be available at each fixed value of $Q^2$ so as to minimize
uncertainties associated with small-$x$ and other kinematic
extrapolations. 
Figure~\ref{fig:f2lowq} illustrates a compilation of global $F_2^p$
data over several orders of magnitude in $x$, for values of $Q^2$
between 0.1~GeV$^2$ and 3.1~GeV$^2$ \cite{MOMENTS}.
Resonance region data from Jefferson Lab are indicated by the stars
at large $x$.
These are the same data depicted in
Figs.~\ref{fig:dualspectra}--\ref{fig:local_res}.
Data at higher $W$ from SLAC, NMC, Fermilab and HERA are shown
at smaller $x$ for the same $Q^2$ values.
Such an extensive combined global data set facilitates the extraction
of unpolarized structure function moments with minimal uncertainties.
Also shown in the top two panels are curves representing the structure
function calculated from PDF parameterizations by the GRV group
\cite{GRV98}, evolved from $Q^2 = 0.4$~GeV$^2$ to the respective
values indicated.
The central solid curve in the third panel represents the input
parton distribution at $Q^2 = 0.4$~GeV$^2$.
The two outer curves in the bottom two panels represent the average
scaling curve from the Jefferson Lab data, encompassing its
uncertainty band, as discussed in Sec.~\ref{sssec:f2local}. 
It is interesting to note that, while there is a dramatic $Q^2$
dependence at low $x$ associated with the collapse of the nucleon sea,
there is very little $Q^2$ dependence evident in this range at
large $x$.
It has been suggested \cite{Li02} that large-$x$ evolution may require
a modification of the usual $Q^2$ evolution equations \cite{EVOLVE}
(which assume massless, on-shell quarks) to take into account the
fact that quarks at large $x$ are highly off-shell.

\begin{figure}[ht]
\begin{minipage}{4.0in}
\vspace*{0.5cm}
\hspace*{-0.5cm}
\epsfig{figure=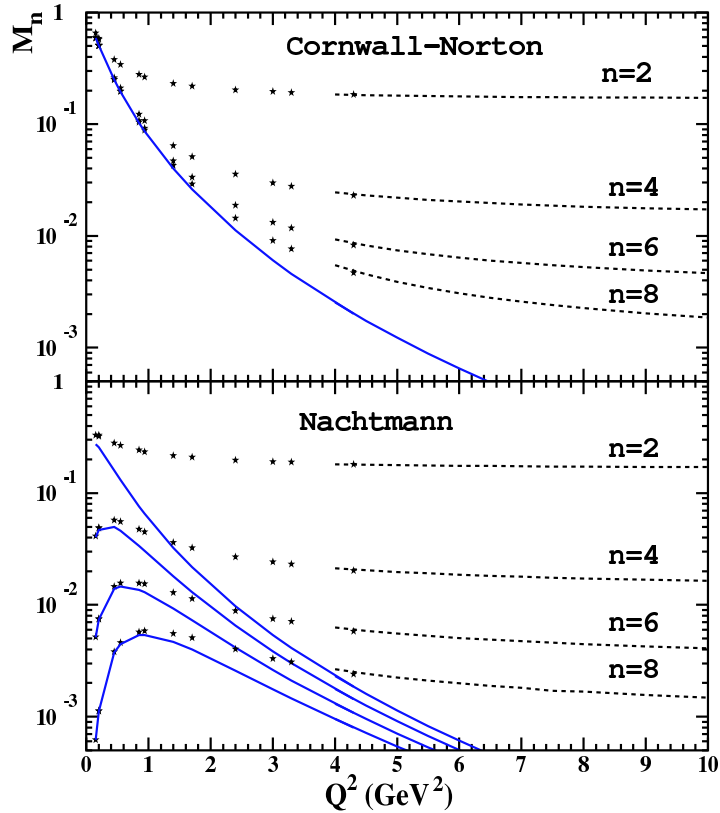, height=13cm}
\end{minipage}
\begin{centering}
\vspace*{-2cm}
\caption{\label{fig:ioanamoments}
	Moments of the proton $F_2^p$ structure function
	(upper panel: Cornwall-Norton, lower panel: Nachtmann)
	extracted from the world's electron-proton scattering data,
	for $n = 2$, 4, 6 and 8 (top to bottom on each plot).
	The elastic contributions are indicated by the solid lines.
	The low-$Q^2$ ($< 4.3$~GeV$^2$) moments (stars) are
	constructed directly from the data, while the larger-$Q^2$
	moments (dotted lines) are extracted from global fits to
	the nucleon elastic, resonance, and deep inelastic regions.}
\end{centering}
\end{figure}

The $n=2, 4, 6$ and $8$ moments of $F_2^p$, constructed from the
global data set in Fig.~\ref{fig:f2lowq}, are shown in
Fig.~\ref{fig:ioanamoments}.
The upper panel shows the Cornwall-Norton moments, while the lower
panel shows for comparison the moments calculated in terms of the
Nachtmann variable $\xi$.
The total experimental uncertainty in the constructed moments
is estimated to be less than 5\%.

Note that each of the moments necessarily includes the elastic
contribution at $x=1$, which dominates the moments at the lowest
$Q^2$ values.
To demonstrate this, the elastic contributions are shown as solid
curves in Fig.~\ref{fig:ioanamoments}.
To include the elastic contribution, we use a fit to the world's
global data set compiled in Ref.~\cite{bostedfit}.
Note that the Cornwall-Norton moments will become unity (the proton
charge squared) at $Q^2$ = 0, as expected from the Coulomb sum rule.
The Nachtmann moments, however, vanish at $Q^2$ = 0 since (in the
absence of quark mass scales) $\xi/x$ vanishes in this limit.

Although below $Q^2 \sim 1$~GeV$^2$ there is a more rapid variation
of the moments with $Q^2$, the lowest ($n=2$) moment is very weakly 
$Q^2$ dependent beyond $Q^2 \approx 1$~GeV$^2$, while the higher
moments reach a similar plateau at correspondingly larger $Q^2$. 
This observed shallow $Q^2$ dependence in Fig.~\ref{fig:ioanamoments}
is consistent with the slowly varying logarithmic behavior associated
with the perturbative, PDF-based predictions. 
In the Nachtmann moments, which take into account an additional $Q^2$
dependence due to target mass effects, even the higher moments display
a weak $Q^2$ dependence at low $Q^2$ values ($Q^2 \sim 2$~GeV$^2$).

Without the elastic contribution, which is a highly nonperturbative,
coherent effect and behaves as $\sim 1/Q^8$ at high $Q^2$, both the
Cornwall-Norton and Nachtmann moments for low $n$ are nearly constant
down to $Q^2 \sim 0.5$~GeV$^2$.
This suggests that the inelastic part of the moments may resemble
the high-$Q^2$, scaling moments and exhibit duality at lower $Q^2$.

\begin{figure}[h]
\begin{minipage}{3.0in}
\vspace*{1cm}
\hspace*{-2.5cm}
\epsfig{figure=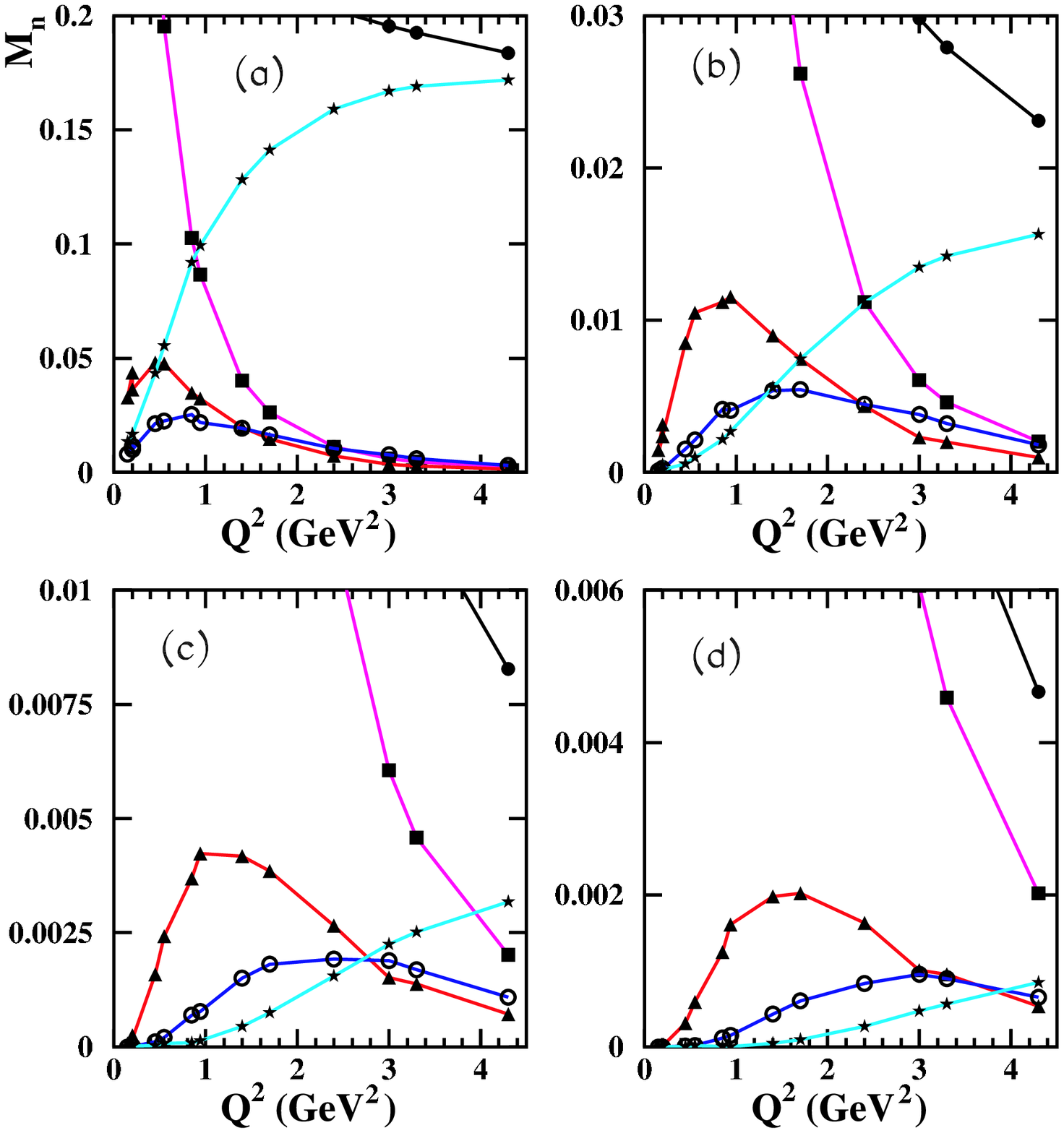, height=15cm}
\end{minipage}
\begin{centering}
\vspace*{-2cm}
\caption{\label{fig:momconts}
	Cornwall-Norton moments of the $F_2^p$ structure function,
	for (a) $n=2$, (b) $n=4$, (c) $n=6$, and (d) $n=8$.
	Contributions to the moment are shown separately for the
	elastic peak (squares),
	the regions $1.2 < W^2 < 1.9$~GeV$^2$ (triangles),
	$1.9 < W^2 < 2.5$~GeV$^2$ (open circles), and
	$W^2 > 4$~GeV$^2$ (stars),
	together with the total moment (filled circles).
	The curves connect the data points to guide the eye.}
\end{centering}
\end{figure}

The relative strength of the $W^2 < 4$ GeV$^2$ region(s) is
illustrated in Fig.~\ref{fig:momconts} for the $n=2$, 4, 6 and 8
(Cornwall-Norton) moments for $Q^2 < 5$~GeV$^2$.
The moments are separated into the elastic contribution (squares);
the contribution of the $N-\Delta$ transition region,
$1.2 < W^2 < 1.9$~GeV$^2$ (triangles);
the second resonance region, $1.9 < W^2 < 2.5$~GeV$^2$ (open circles);
and the deep inelastic contribution, $W^2 > 4$~GeV$^2$ (stars).
The total moment is indicated by the filled circles.
The vertical scale is chosen to enhance the individual region
contributions, so that the total moment is sometimes only visible
at higher $Q^2$.
The lines connecting the data points are to guide the eye.

The elastic contribution dominates the moments at low $Q^2$,
saturating the integrals near $Q^2=0$, but falls off rapidly at
larger $Q^2$.
As $Q^2$ increases from zero the contributions from the inelastic,
finite-$W^2$ regions increase and compensate some of the loss of
strength of the elastic.
At larger $Q^2$ these also begin to fall off.
On the other hand, the contribution of the $W^2 > 4$~GeV$^2$
region does not die off.
Since this contribution is not bound from above, higher-$W^2$
resonances and the inelastic nonresonant background start
becoming important with increasing $Q^2$, eventually yielding
approximately the logarithmic $Q^2$ scaling behavior of the
moments prescribed by pQCD \cite{EVOLVE}.

As evidenced by the difference between the $W^2 > 4$~GeV$^2$
data and the total moments, the contribution of the traditionally
defined resonance region ($W < 2$~GeV) is non-negligible up to
$Q^2 \approx 5$~GeV$^2$.
Considering $n=4$ in Fig.~\ref{fig:momconts} (b), for example,
the difference between the total and deep inelastic curves leaves
about a 30\% contribution to the moment at $Q^2 = 4.5$ GeV$^2$
coming from the resonance region.
In Fig.~\ref{fig:ioanamoments}, the $n=4$ moment at this $Q^2$ 
nonetheless exhibits a largely perturbative behavior. 
The significance of the resonance contributions to the moments and
their corresponding $Q^2$ behavior will be discussed in more detail
in Sec.~\ref{ssec:qcd} in the context of the twist expansion.

% .......................................................................
\subsubsection{Duality in Nuclei and the EMC Effect}
\label{sssec:emc}

While most of the recent duality studies have focused on the proton,
there have been measurements on deuterium and heavy nuclei in the
high-$x$ and low- to moderate-$Q^2$ region \cite{89008_2,89008,NE9}
which have also revealed additional information about duality.
Inclusive electron--nucleus experiments at SLAC designed to probe the
$x > 1$ region in the $F_2^A$ structure function concluded that the
data began to display scaling indicative of local duality \cite{NE9},
while citing the need for larger $\xi$ data for verification.

\begin{figure}[h]
\vspace*{1cm}
\begin{minipage}{3.0in}
\hspace*{-3cm}
\epsfig{figure=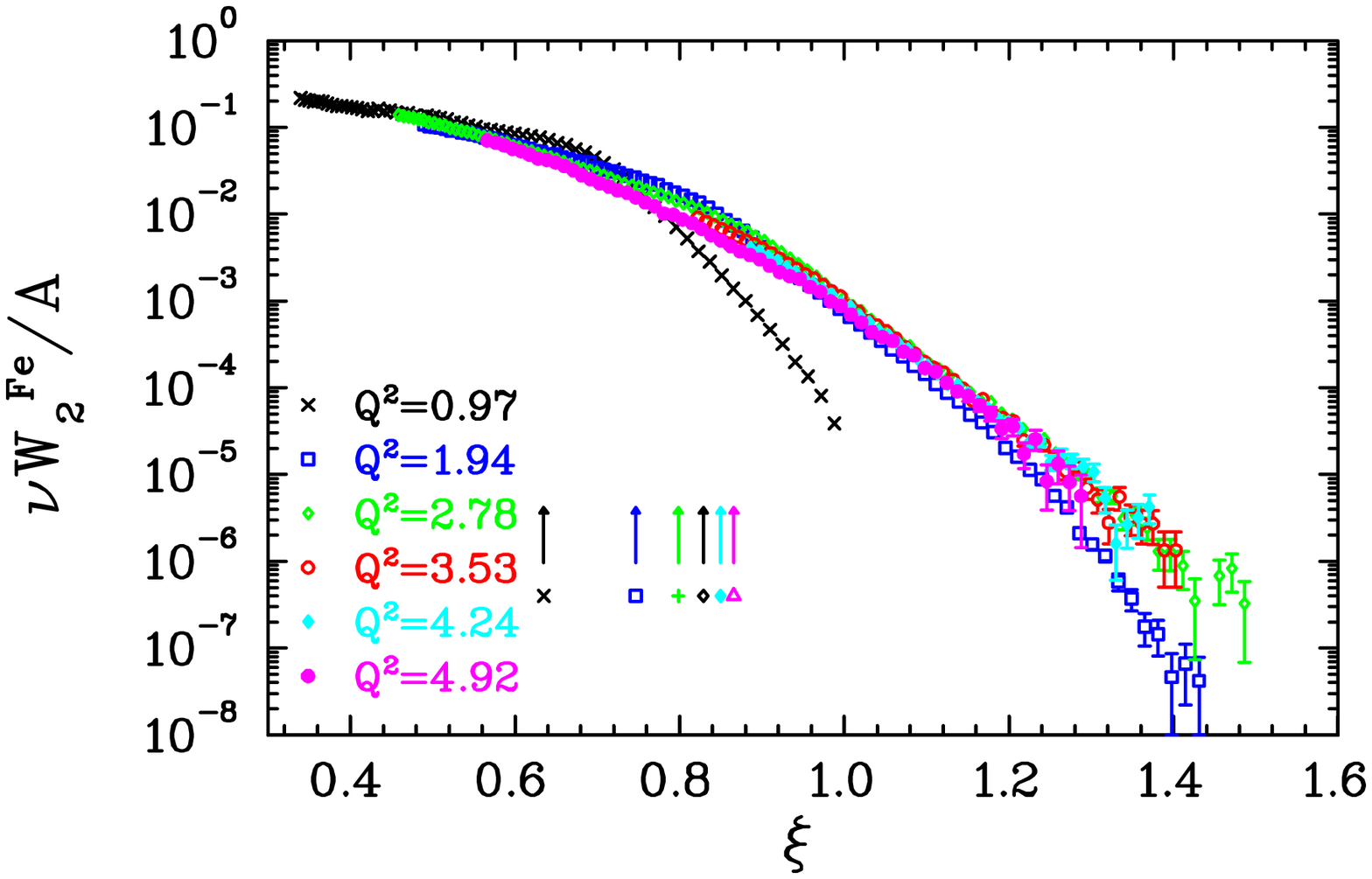, height=16cm}
\end{minipage}
\begin{centering}
\vspace*{-8cm}
\caption{\label{fig:nuc1}
	The $\nu W_2^{\rm Fe} = F_2^{\rm Fe}$ structure function
	for iron (per nucleon) as a function of $\xi$.
	The data were obtained at fixed electron scattering angle, 
	and the quoted $Q^2$ (in units of GeV$^2$) are the values
	for $x = 1$.
	The arrows indicate the values of $\xi$ corresponding
	to the quasi-elastic peak for each setting.}
\end{centering}
\end{figure}

This was studied further at Jefferson Lab, and Fig.~\ref{fig:nuc1}
is a sample plot from these newer duality studies.
Here, $F_2^A/A$ for iron is plotted as a function of $\xi$
\cite{89008_2}.
The first thing to note is that the smearing caused by nucleon Fermi
motion causes the visible resonance mass structure clearly observable
for the free nucleon, and even the quasi-elastic peak, to vanish.
Once the resonance structure is washed out, scaling is observed at all
$\xi$, and it is impossible to differentiate the DIS and resonance
regimes other than by calculating kinematics.
Other than at the lowest $Q^2$ values, the data at all $\xi$ fall on
a common, smooth scaling curve.
As in Fig.~\ref{fig:dualspectra}, any $Q^2$ dependence of the scaling 
curve should not be large here.
In this nuclear $\xi$ scaling duality can be observed even more
dramatically than for the proton: rather than appearing as a local
agreement on average between deep inelastic and resonance data,
scaling in nuclear structure functions in the resonance region is
directly observed at all values of $\xi$ {\em without} averaging.

Because nucleons in the deuteron have the smallest Fermi momentum
of all nuclei, $\xi$ scaling is not expected to work in deuterium
as well as in heavier nuclei at low $W^2$ and $Q^2$.
However, $\xi$ scaling is observed even in deuterium at extremely
low values of $W^2$ and relatively low momentum transfers.
For $Q^2 \ge 3$~GeV$^2$, the resonance structure is completely
washed out, so that even the most prominent $\Delta$ resonance
is no longer visible.

\begin{figure}[h]
\vspace*{0cm}
\begin{minipage}{3.0in}
\hspace*{-2.5cm}
\epsfig{figure=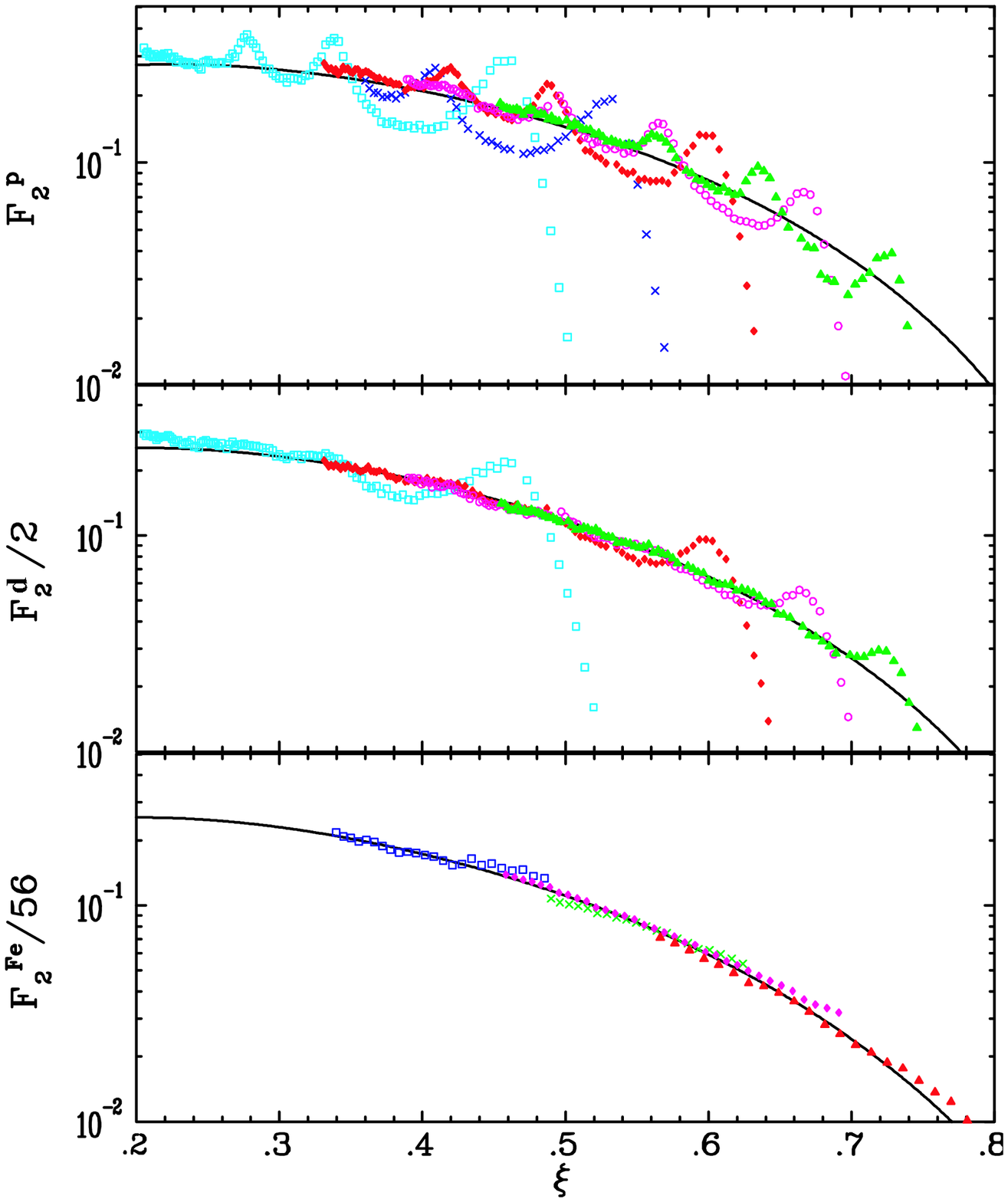, height=14cm}
\end{minipage}
\begin{centering}
\vspace*{-2cm}
\caption{\label{fig:nuc_dual}
	$F_2$ structure function per nucleon as a function of 
	$\xi$ for hydrogen, deuterium, and iron.
	The curves are the GRV parameterization \protect\cite{GRV98}
	at $Q^2 = 1$~GeV$^2$, corrected for the nuclear EMC effect.
	Errors shown are statistical only.}
\end{centering}
\end{figure}

A compilation of recent $F_2$ structure function data above 
$W^2 = 1.2$ GeV$^2$ is shown in Fig.~\ref{fig:nuc_dual} for hydrogen,
deuterium, and iron as a function of $\xi$, for a variety of momentum
transfers ranging from $Q^2 = 0.5$~GeV$^2$ at low $\xi$ to
$Q^2 = 7$~GeV$^2$ at the higher $\xi$ values.
Also shown is the $F_2$ scaling curve for the nucleon (from the GRV
parameterization \cite{GRV98}), corrected for the known nuclear
medium modifications to the structure function.
For the proton, the resonance structure is clearly visible and $F_2$
is seen to oscillate around the scaling curve.
For deuterium, and even more so for iron, the resonances become less
pronounced, being washed out by the Fermi motion of the nucleons
inside the nucleus.
The prominent peak present in the deuterium data in
Fig.~\ref{fig:nuc_dual} (center panel) corresponds to the $\Delta$
resonance.
This peak follows the scaling curve as for the proton, but the other
resonance peaks are smeared so much as to be indistinguishable from
the scaling structure function.
For heavier nuclei, even the quasi-elastic peak is washed out by the
smearing at higher $Q^2$, and scaling is seen at all values of $\xi$.
Here the resonance region is essentially indistinguishable from the
scaling regime.

\begin{figure}
\begin{minipage}[ht]{3.0in}
\hspace*{-3cm}
\epsfig{figure=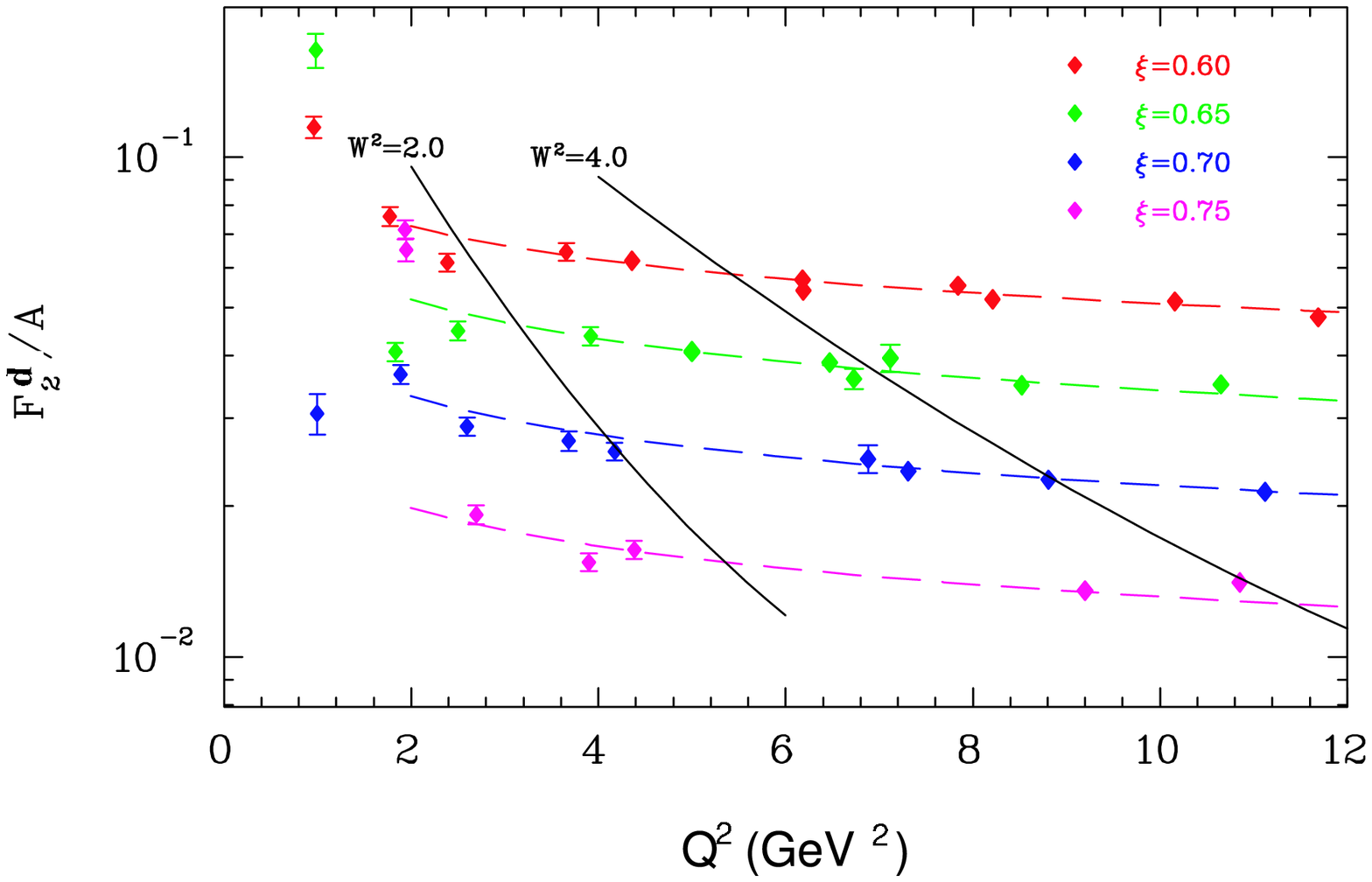, height=12cm}
\end{minipage}
\begin{centering}
\vspace*{-4cm}
\caption{\label{fig:deut}
	$F_2^d$ structure function as a function of $Q^2$
	at fixed values of $\xi$.
	The dashed lines are $d\ln F_2^2/d\ln Q^2$ fits to
	higher $Q^2$ data.
	The solid lines denote fixed values of $W^2=2$ and 4~GeV$^2$.
	Errors are statistical only, and systematic uncertainties
	vary between $\sim 3\%$ and $\sim 7\%$.
	The data above $W^2=4$~GeV$^2$ are mostly from SLAC,
	and those below $W^2=4$~GeV$^2$ from Jefferson Lab,
	as described in Ref.~\protect\cite{nucl_dual}.}
\end{centering}
\end{figure}

The same observation can also be made from Fig.~\ref{fig:deut},
which shows the deuteron $F_2^d$ structure function as a function
of $Q^2$, for several values of $\xi$.
The dashed lines are $d\ln F_2^d/d\ln Q^2$ fits to higher-$Q^2$ 
data, and the solid lines indicate the boundaries at $W^2 = 2$ and
4~GeV$^2$.
Essentially all the data, both above {\it and} below $W^2 = 4$~GeV$^2$, 
lie on the perturbative curves, making it practically impossible to
distinguish between the hadronic and partonic regimes.
Deviations appear only at very low $Q^2$, $Q^2 \sim 1$--2~GeV$^2$,
where the quasi-elastic peaks become visible.

The limited kinematic coverage of the available nuclear resonance
region data, combined with the uncertainty in modeling nuclear
effects at large $x$, does not yet permit precision duality studies
at the level of those that have been done for the proton.
However, interesting studies have been performed with the existing
data to test the practicality of using duality-averaged scaling to
access high-$x$ nucleon structure.
Rather than comparing the nuclear structure functions in the
resonance region to deep inelastic parameterizations at low $Q^2$,
as in Fig.~\ref{fig:deut}, the nuclear dependence in the resonance
region has been compared directly to measurements made in the DIS
regime.

\begin{figure}[h]
\begin{minipage}{3.0in}
\hspace*{-2cm}
\epsfig{figure=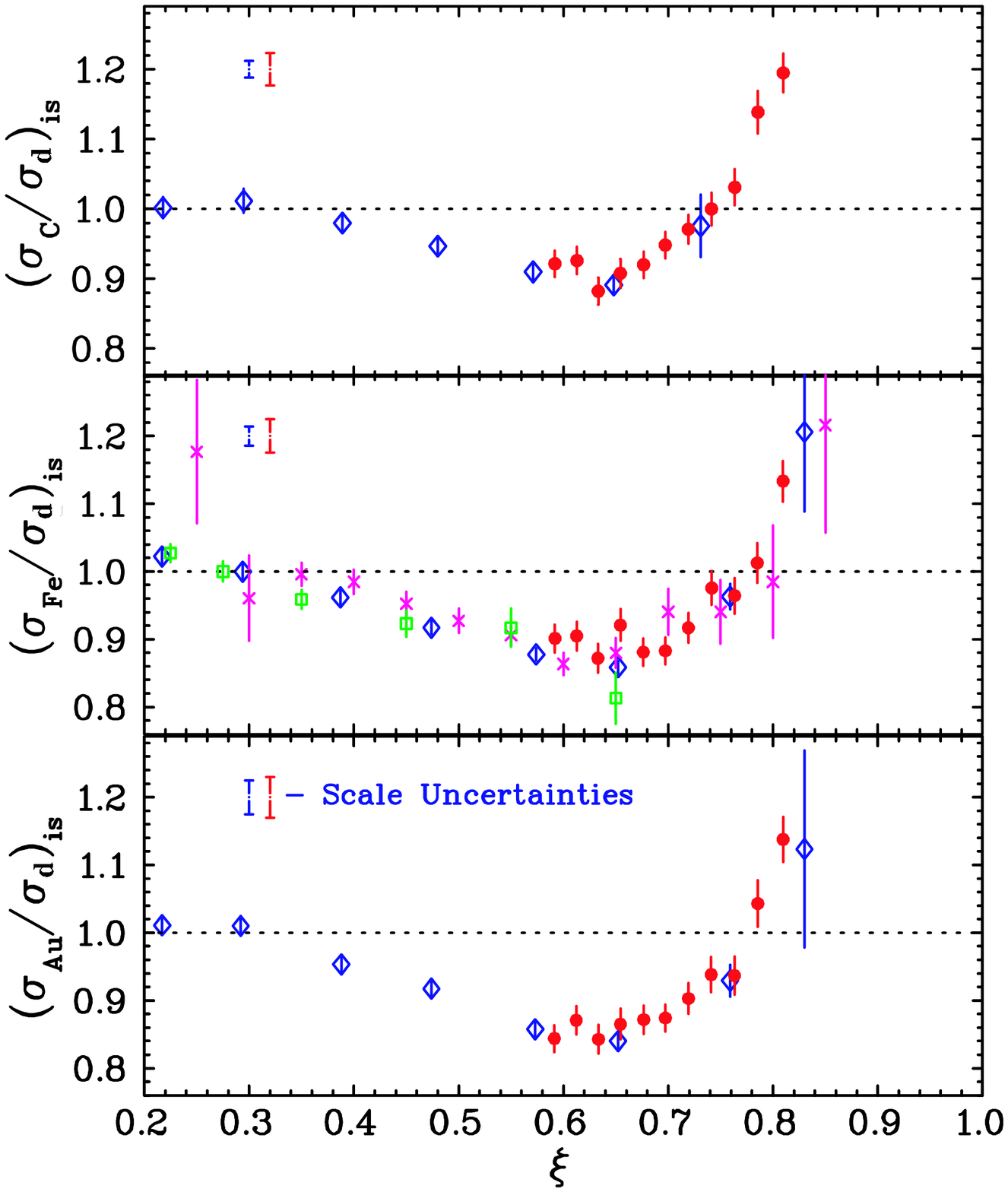, height=13cm} %, width=8cm}
\end{minipage}
\begin{centering}  
\vspace*{-1cm}
\caption{Ratio of nuclear to deuterium cross sections
	per nucleon, corrected for neutron excess, for carbon (top),
	iron (center) and gold (bottom) versus $\xi$.
	The resonance data at low $W$ and $Q^2$ from Jefferson Lab
	(circles) are compared with the deep inelastic data at
	high $W$ and $Q^2$ from SLAC E139 (diamonds), SLAC E87
	(crosses), and BCDMS (squares). The scale uncertainties
	for the SLAC (left) and JLab (right) data are shown in the
	figure.}
\label{fig:EMCplot}
\end{centering}
\end{figure}

Figure~\ref{fig:EMCplot} depicts the ratio of nuclear to deuteron
cross sections per nucleon for carbon, iron, and gold, corrected for
non-isoscalarity effects \cite{nucl_dual}.
The characteristic $\xi$ dependence of the ratio
$\sigma_{\rm Fe}/\sigma_{\rm d} \approx F_2^{\rm Fe}/F_2^{\rm d}$,
namely a dip at $\xi \sim 0.6$--0.7 and a rapid rise above unity
for $\xi \agt 0.8$ (known as the ``EMC effect''), has been
well-established from many deep inelastic measurements \cite{EMC_EFFECT}
and has been interpreted in terms of a nuclear medium modification of
the nucleon structure function.
The unique feature of the plot, however, is the additional inclusion
of resonance region data from Jefferson Lab.

Qualitatively, the nuclear effects in the resonance region appear
to be similar to those in the deep inelastic region.
This is somewhat surprising, since the nuclear dependence of the
scaling structure functions is not at all expected to be the same
as the nuclear dependence of resonance production.
While nuclear medium modifications of proton form factors has been
observed (in polarization transfer measurements of the elastic proton
$G_E^p/G_M^p$ form factor ratio, for instance \cite{HE4,STRAUCH}),
there is {\it a priori} no reason why these modifications would
be the same as those for structure functions measured in deep
inelastic scattering.
On the other hand, this may be viewed as another consequence of
quark-hadron duality.
In Sec.~\ref{sssec:elastic} we explore some consequences of local
duality in relating the nuclear medium modifications of structure
functions at large $x$ and electromagnetic form factors.

% -----------------------------------------------------------------------
\subsection{Longitudinal and Transverse Cross Sections}
\label{ssec:lt}

In the preceding section we have shown that duality has been clearly
established in the $F_2$ structure function, both locally as a
function of $x$ (or $\xi$), and globally in terms of moments.
From its definition, however, the $F_2$ structure function contains
contributions from scattering of both longitudinal and transverse
photons.
The question then arises of whether, and to what extent, duality
holds in either or both of the longitudinal and transverse channels 
separately.

The extraction of the $F_2$ structure function from cross section
data can only proceed with some input for the ratio $R$ of the
longitudinal to transverse cross sections.
At high $Q^2$ the scattering of longitudinal photons from spin-1/2
quarks is suppressed, and one expects $R \to 0$ as $Q^2 \to \infty$.
At low $Q^2$, however, $R$ is no longer suppressed, and could be
sizable, especially in the resonance region and at large $x$.
A model-independent determination of unpolarized structure functions
from inclusive cross section data requires, therefore, precision
longitudinal/transverse (LT) separations to simultaneously extract
$F_2$ and $R$, or equivalently $F_1$ and the longitudinal structure
function, $F_L$, as in Eqs.~(\ref{eq:gameps})--(\ref{eq:F2extract}).

Until recently very little data on $R$ existed in the region of
the resonances, rendering reliable LT separations impossible.
The few measurements that existed below $Q^2 = 8$ GeV$^2$ in this
region yielded $R$ in the range $-0.1 \alt R \alt 0.4$,
and had typical errors of 100\% or more.
New precision measurements of proton cross sections at Jefferson Lab
\cite{YONGGUANG} have allowed for the first time detailed duality
studies in all of the unpolarized structure functions and their
moments.

% .......................................................................
\subsubsection{Duality in the Separated Structure Functions}
\label{sssec:ltdual}

Within the framework of the naive parton model with free, massless
spin-1/2 quarks, the $F_1$ and $F_2$ structure functions are related
via the Callan-Gross relation \cite{CallanGross}, Eq.~(\ref{eq:f22xf1}),
and the longitudinal to transverse cross section ratio $R$ is zero.
By allowing quarks to have an intrinsic transverse momentum $k_T$,
and a non-zero mass $m_q$, the value of $R$ is no longer zero, and
is given by $R = 4 (m_q^2 + k_T^2)/Q^2$ \cite{FeynmanBOOK}.
Furthermore, the inclusion of hard gluon bremsstrahlung and
photon-gluon interactions also contributes to $R$ by generating
additional transverse momentum $k_T$ \cite{altar,Ellis1,Ellis2}.
%
% The gluon contributions to $p_T$ rise as $Q^2$, while the $p_T$
% contribution to $R$ falls as $Q^{-2}$.
%
In leading order pQCD, the contribution to $R$ from gluon
radiation varies as the strong coupling constant, $\alpha_s$,
$R \sim 1/\ln Q^2$.
Because the pQCD contributions to $R$ are quite small, the $1/Q^2$
power corrections, which are nonperturbative in origin, are expected
to play a significant, if not dominant, role at low $Q^2$.
Since the latter are not directly calculable, precision measurements
of $R$, or equivalently, accurate LT-separated structure functions,
are crucial to observing duality in the moderate to low-$Q^2$ regime.

It has been reported in Ref.~\cite{tao} that $R$ measured at 
intermediate $Q^2$ in the DIS region \cite{bodek,whitlow2,dasu}
is significantly higher than the next-to-leading-order pQCD
predictions, even with the inclusion of corrections due to target
mass effects.
This enhanced strength in $R$ relative to pQCD was argued to be
evidence for dynamical ``higher twist'' effects \cite{tao,dasu}
(see Sec.~\ref{ssec:qcd}).
Quark-hadron duality would suggest, on the other hand, that even
in the resonance region nonperturbative, $1/Q^2$ effects would 
be small for $Q^2$ as low as 1 GeV$^2$ when the structure function
is averaged over any of the prominent resonance regions.
The separated structure functions, therefore, are particularly 
interesting quantities for duality studies.

\begin{figure}
\begin{minipage}[ht]{4.0in}
\hspace*{-1.5cm}
\epsfig{figure=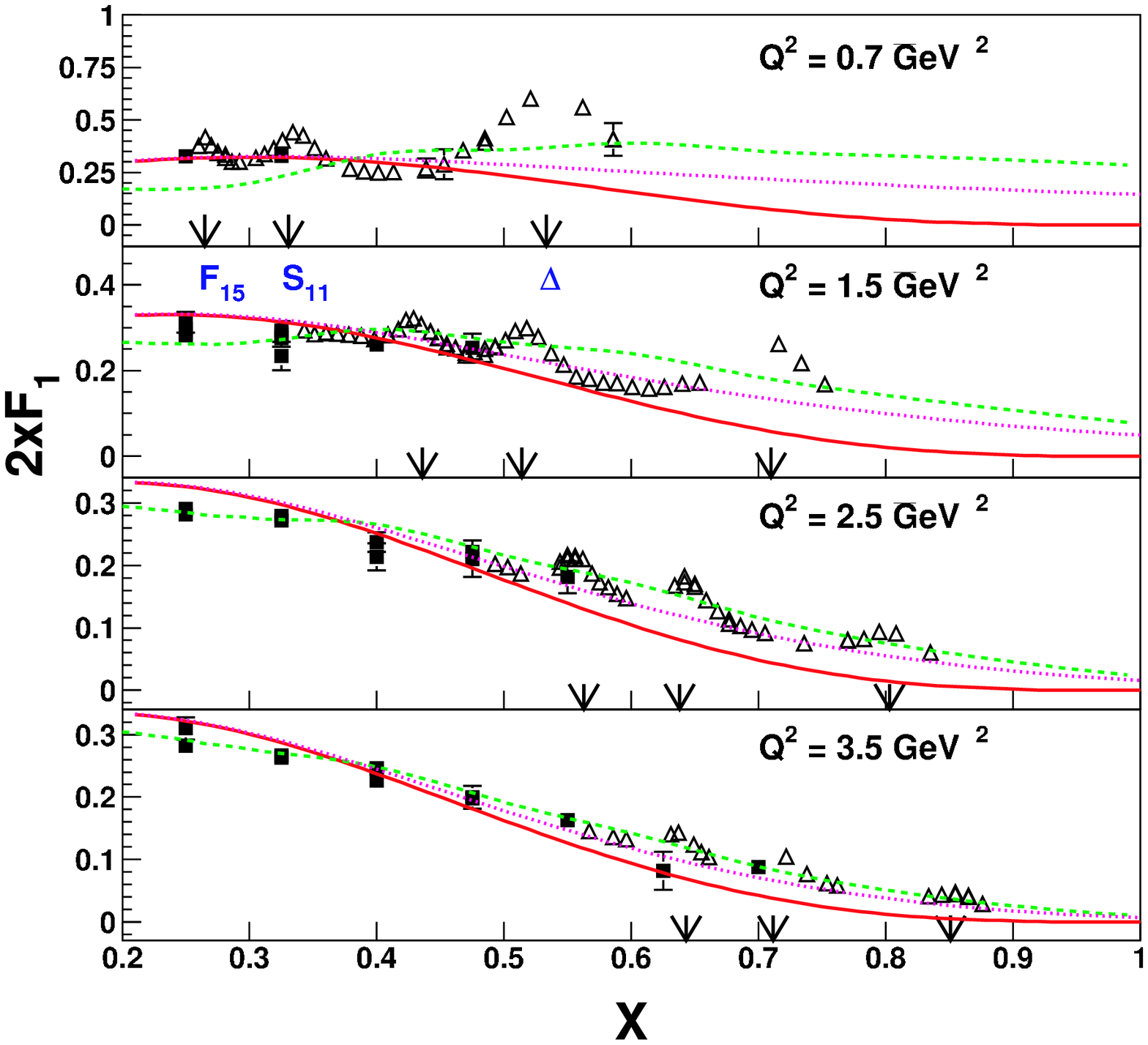, height=13.5cm}
\end{minipage}
\begin{centering}
\vspace*{-2.5cm}
\caption{\label{fig:F_1}
	The purely transverse proton structure function $2xF_1^p$,
	measured in the resonance region (triangles) as a function 
	of $x$, compared with existing high-precision DIS
	measurements from SLAC (squares).
	The curves are from Alekhin (dashed) \protect\cite{alekhin},
	and from MRST \protect\cite{MRST}, both at NNLO, with (dotted)
	and without (solid) target mass effects included, as described
	in the text.  The prominent resonance regions ($\Delta$,
	$S_{11}$, $F_{15}$) are indicated by the arrows.}
\end{centering}
\end{figure}

\begin{figure}
\begin{minipage}[ht]{4.0in}
\hspace*{-1.5cm}
\epsfig{figure=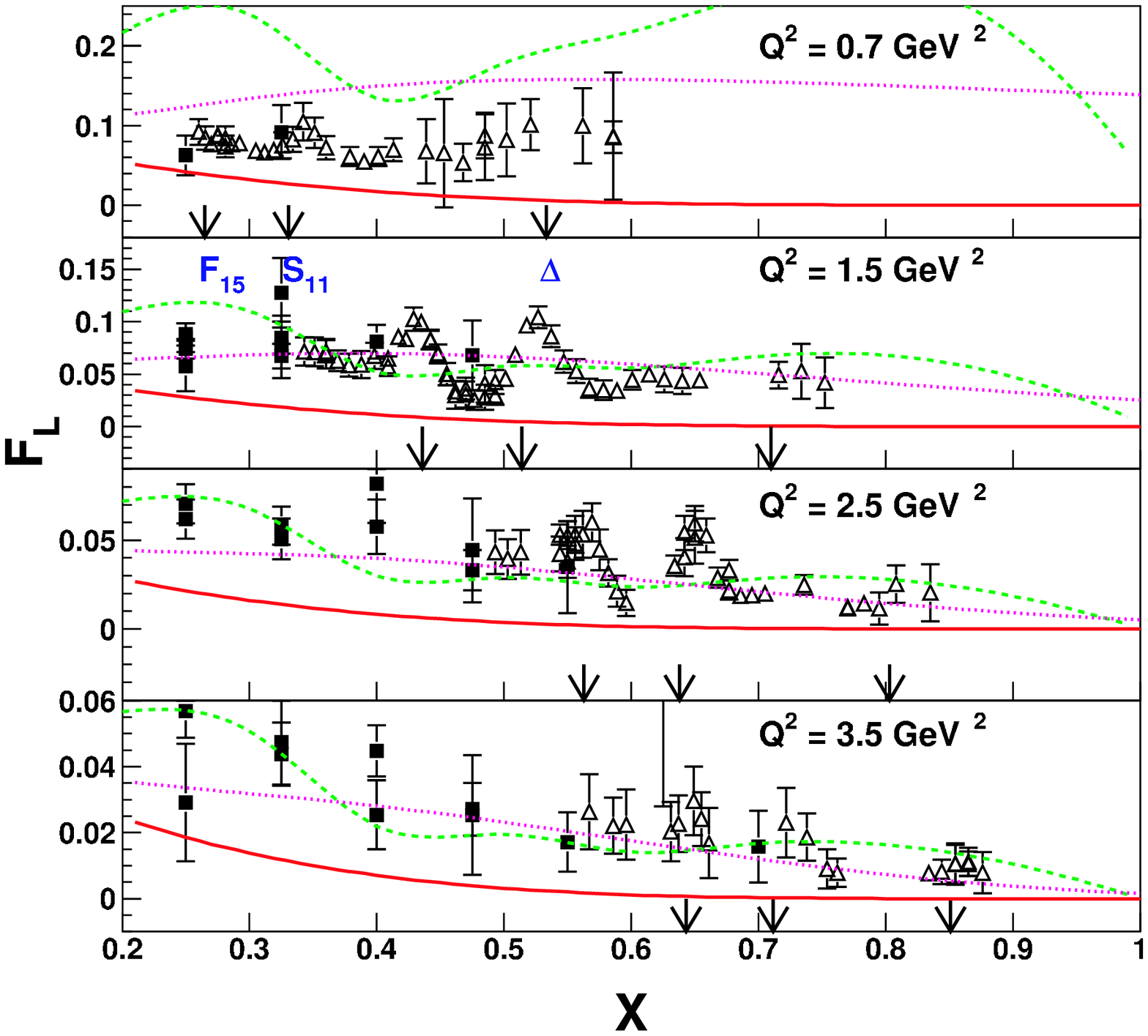, height=13.5cm}
\end{minipage}
\begin{centering}
\vspace*{-2.5cm}
\caption{\label{fig:F_L}
	As in Fig.~\protect\ref{fig:F_1}, but for the longitudinal
	structure function $F_L$.}
\end{centering}
\end{figure}

New data from Jefferson Lab experiment E94-110 on the separated
proton transverse ($F_1^p$) and longitudinal ($F_L^p$) structure
functions in the resonance region are shown in Figs.~\ref{fig:F_1}
and \ref{fig:F_L}, respectively \cite{E94110}. 
LT-separated data from SLAC, which are predominantly in the DIS
region, are also shown for comparison \cite{tao,dasu}.
Where coincident, the Jefferson Lab and SLAC data are in excellent
agreement, providing confidence in the achievement of the demanding
precision required of this type of experiment. 
In all cases, it is also interesting to note that the resonance and
DIS data smoothly move toward one another in both $x$ and $Q^2$.

The curves in Figs.~\ref{fig:F_1} and \ref{fig:F_L} are from
Alekhin's next-to-next-to leading order (NNLO) analysis \cite{alekhin},
including target mass effects as in Ref.~\cite{GP}, and 
from the MRST NNLO analysis
\cite{MRST}, with and without target mass effects according to
\cite{BEGR} included. 
It is clear that target mass effects are required to describe the
data.
However, other than the target mass corrections, no additional
nonperturbative physics seems necessary to describe the {\it average}
behavior of the resonance region for $Q^2 > 1$ GeV$^2$.
Furthermore, this is true for a range of different $Q^2$ values,
indicating that the scaling curve describes as well the average $Q^2$
dependence of the resonance region.
These results are analogous to those in Fig.~\ref{fig:f2} for the
$F_2$ structure function, and are a clear manifestation of 
quark-hadron duality in the separated transverse and longitudinal
channels.

The scaling curves in all of the unpolarized structure functions
appear dual therefore to the average of the resonance region
strength.
This statement is quantified in Fig.~\ref{fig:dual_ratios} of
Sec.~\ref{sssec:f2local} above for the ratios of integrals of the
resonance to scaling functions.
The main difficulty in the integrated ratio approach of testing
duality was the lack of knowledge of the correct perturbative
structure function at large $x$.
Nonetheless, for all of the spin-averaged structure functions of the
proton ($F_1$, $F_2$, $F_L$ and $R$), the integrated resonance region
strength for $Q^2 \agt 1$~GeV$^2$ is similar to the integrated
perturbative strength over the same range in $x$.
This strongly suggests that, at least for the unpolarized structure
functions, duality is a fundamental property of nucleon structure.

% .......................................................................
\subsubsection{Moments of $F_1$ and $F_L$}
\label{sssec:ltmoment}

In this section we present moments of new, {\em LT-separated},
spin-averaged, structure function data.
Previously, $F_2$ moments were constructed using assumed values for $R$.
Since hardly any measurements of $R$ existed in the nucleon resonance
region before the Jefferson Lab E94-110 experiment
\cite{YONGGUANG,E94110}, one may expect small changes to the
low-$Q^2$ moments of $F_2$ constructed from the earlier data.

At lower values of $Q^2$ ($<$ 5 GeV$^2$), the region of the nucleon
resonances covers larger intervals of $x$, and consequently
resonances provide increasingly dominant contributions to structure
function moments.
Since bound state resonances are associated with nonperturbative
effects in QCD, one expects deviations from perturbative behavior
to be strongest in this regime.
This is especially true in the longitudinal channel, where long-range
correlations between quarks are expected to play a greater role, as
discussed in Sec.~\ref{sssec:ltdual}, above.

\begin{figure}
\begin{minipage}[ht]{3.5in}
\epsfig{figure=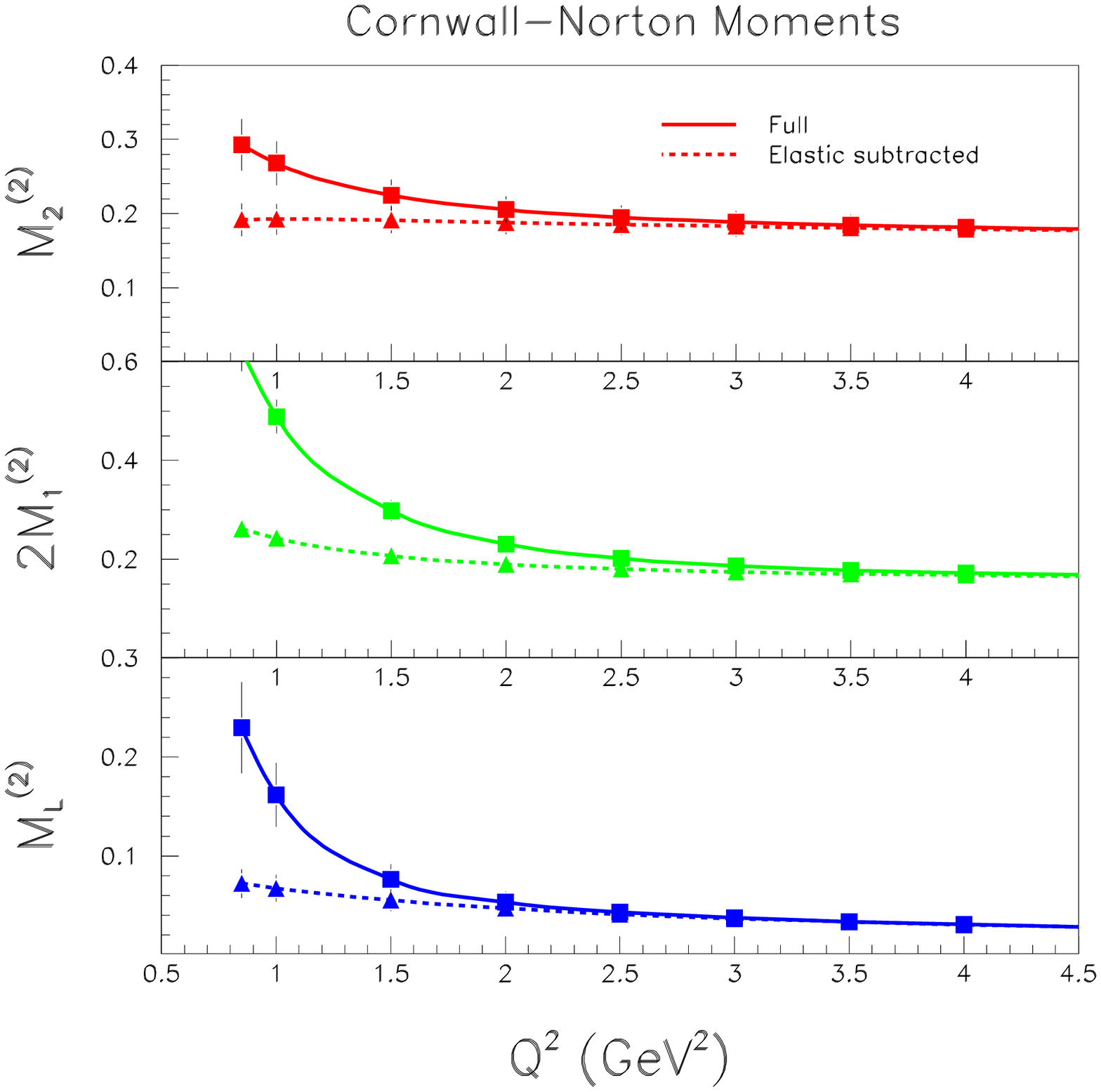, height=9.5cm} % width=9cm}
\end{minipage}
\begin{centering}
\vspace*{0.5cm}
\caption{\label{fig:F1FLmom1}
	Second ($n=2$) Cornwall-Norton moments of the $F_2$~(top),
	$2xF_1$~(center) and $F_L$~(bottom) structure functions,
	evaluated from the preliminary Jefferson Lab Hall~C data
	\protect\cite{YONGGUANG,E94110}.
	The total moments are connected by solid lines,
	and elastic-subtracted moments by dashed lines.}
\end{centering}
\end{figure}

\begin{figure}
\begin{minipage}[ht]{3.5in}
\hspace*{-0.3cm}
\epsfig{figure=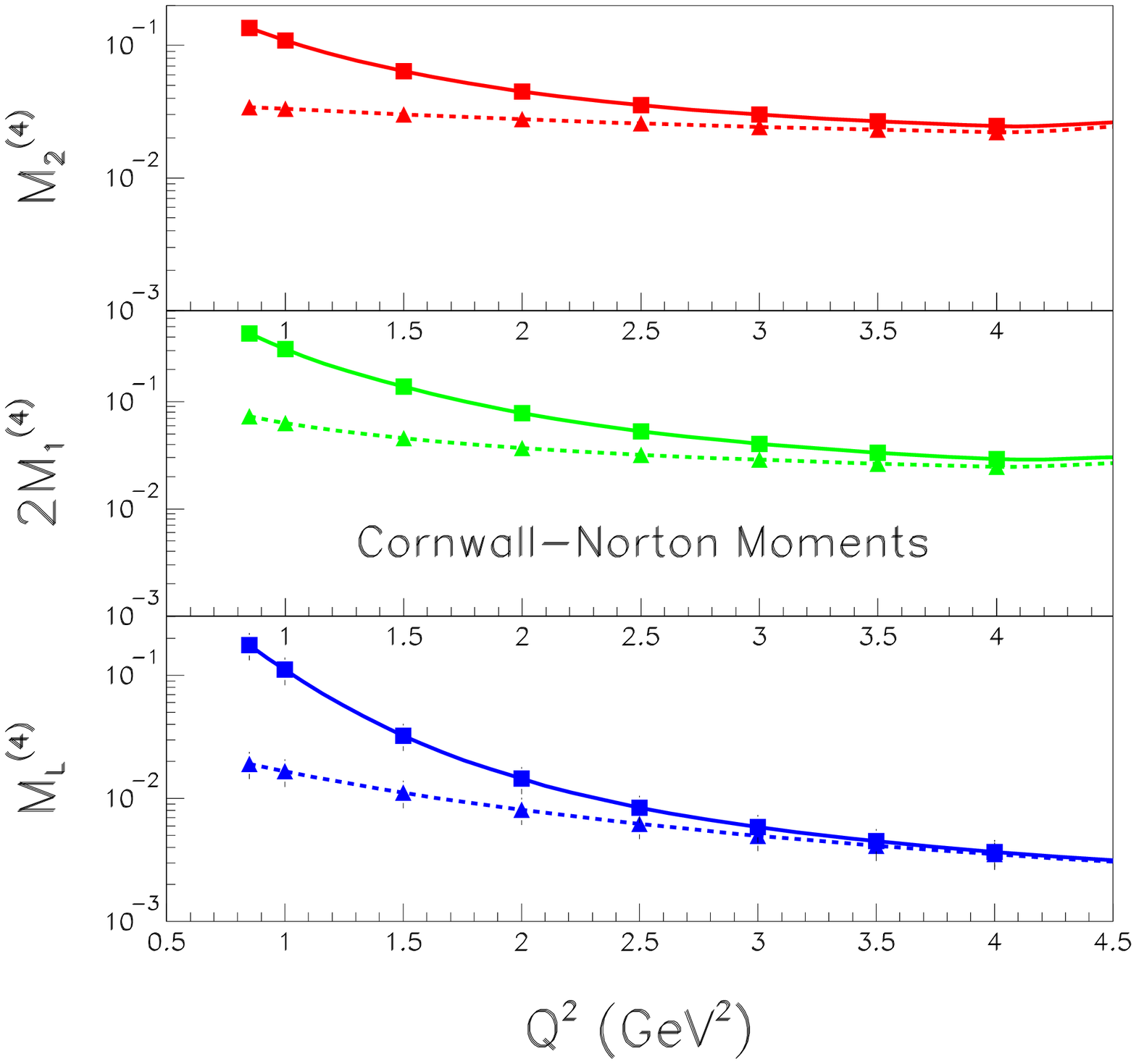, height=9.8cm} % width=9cm}
\end{minipage}
\begin{centering}
\vspace*{0.5cm}
\caption{\label{fig:F1FLmom2}
	As in Fig.~\protect\ref{fig:F1FLmom1} but for the $n=4$ moments.}
\end{centering}
\end{figure}

As can be seen in Figs.~\ref{fig:F1FLmom1} and \ref{fig:F1FLmom2},
nonperturbative effects (other than the elastic contribution)
appear to be small in the new Jefferson Lab data above
$Q^2 = 0.7$~GeV$^2$.
Here, the $n=2$ and $n=4$ moments of the $F_2^p$~(top), $2xF_1^p$~(center),
and $F_L^p$~(bottom) structure functions are extracted from fits to the
Jefferson Lab Hall~C \cite{YONGGUANG,E94110} and SLAC \cite{tao,dasu}
data.
This moment analysis is still preliminary \cite{ericpc}, and is
ultimately expected to have $\approx 5\%$ errors.
The total moments, which include the full range in $0 \le x \le 1$,
are connected by the solid lines, while the moments without the
elastic contribution are connected by the dashed lines.
%
% The elastic contribution is purely higher twist.
%
At high $Q^2$ the elastic contribution rapidly vanishes (structure
functions are identically zero at $x=1$ in the Bjorken limit),
so that the difference between the two sets of curves becomes
negligible by $Q^2 \approx 2$~GeV$^2$ for the $n=2$ moments,
and by $Q^2 \approx 2.5$~GeV$^2$ for the $n=4$ moments.

One of the most remarkable features of the results in
Figs.~\ref{fig:F1FLmom1} and \ref{fig:F1FLmom2} is that the
elastic-subtracted moments exhibit little or no $Q^2$ dependence
even for $Q^2 < 1$~GeV$^2$.
In the region where the moments are completely dominated by the
nucleon resonances, the $n=2$ and $n=4$ moments of all of the
unpolarized structure functions appear to behave just as in the
deep inelastic region at high $Q^2$.
In both cases, nonperturbative corrections to the $Q^2$ dependence must 
be quite small.

\begin{figure}[h]
\begin{minipage}{3.5in}
\epsfig{figure=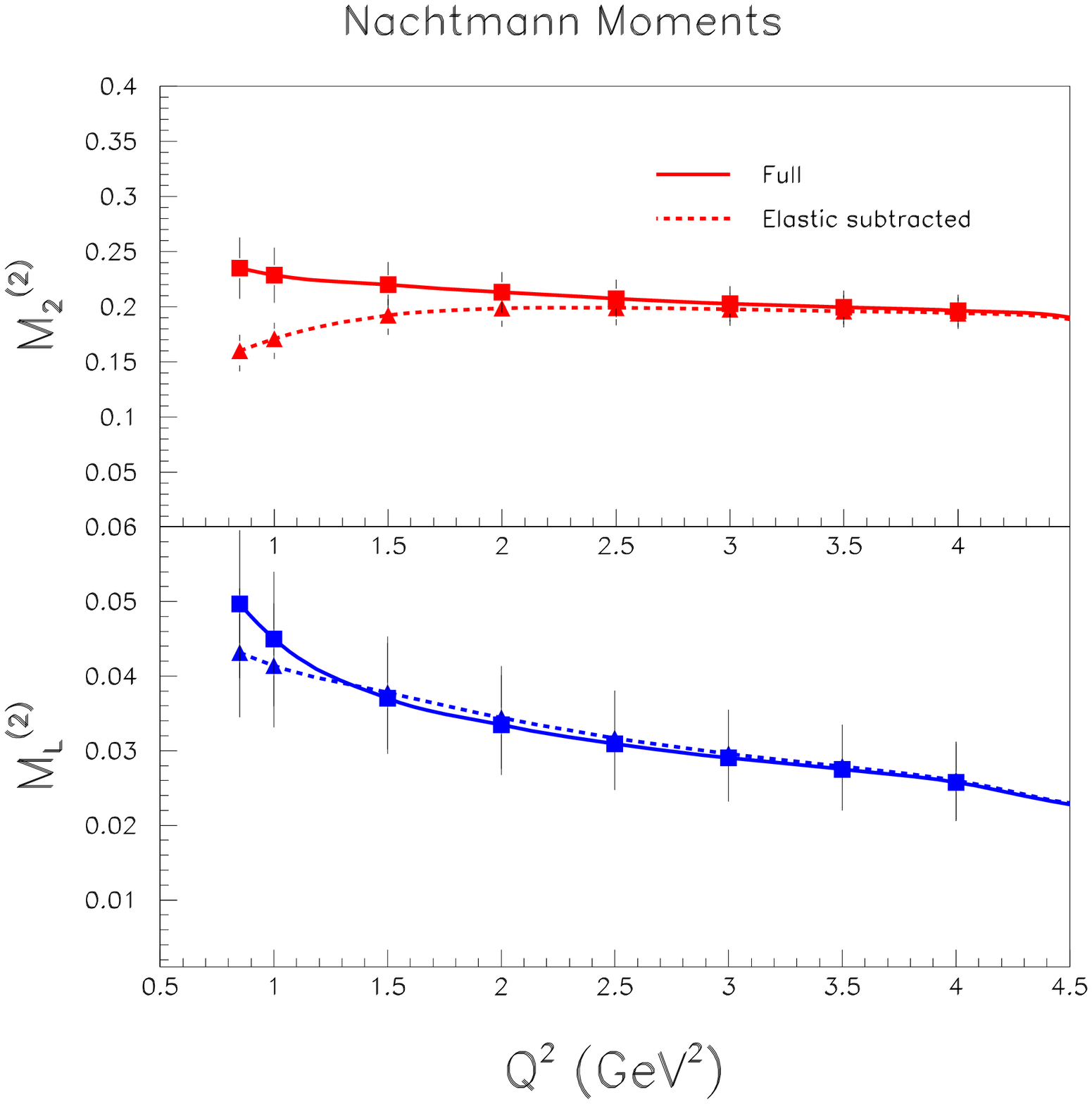, height=9.5cm} % width=9cm}
\end{minipage}
\begin{centering}
\vspace*{0.5cm}
\caption{\label{fig:F1FLmom3}
	Second ($n=2$) Nachtmann moments of the $F_2$~(top) and
	$F_L$~(bottom) structure functions, evaluated from the
	preliminary Jefferson Lab Hall~C data
	\protect\cite{YONGGUANG,E94110}.
	The total and elastic-subtracted moments are shown by
	the solid and dashed lines, respectively.}
\end{centering}
\end{figure}

Finally, in Fig.~\ref{fig:F1FLmom3} we show the ($n=2$) Nachtmann
moments, in a comparable $Q^2$ range to Figs.~\ref{fig:F1FLmom1}
and \ref{fig:F1FLmom2}.
Interestingly, the target-mass corrections to the Nachtmann moments
reduce even further the remaining $Q^2$ dependence of the structure
function moments at low $Q^2$.
The full $F_2^p$ moment with the elastic included exhibits very
limited $Q^2$ dependence, and less than the comparable
Cornwall-Norton moment.
There remains an observable $Q^2$ dependence of the full $F_L^p$ ($n=2$)
moment, on the other hand, at lower $Q^2$, but it is also reduced
compared to the Cornwall-Norton moment.
Note that the scale of Fig.~\ref{fig:F1FLmom3} is necessarily
different from that of Figs.~\ref{fig:F1FLmom1} and \ref{fig:F1FLmom2}.
The effect of neglecting the elastic contribution is reduced for
both, but more dramatically so in the latter.

We shall discuss the implications of these findings in terms of the
operator product expansion in Sec.~\ref{sssec:ope}.
In the next section, however, we examine duality for spin-dependent
structure functions.

\clearpage
% ------------------------------------------------------------------------
\subsection{Duality in Spin-Dependent Structure Functions}
\label{ssec:spin}

In the previous section we have explored the transition between the
partonic and hadronic regimes in unpolarized electron scattering,
and established the degree to which quark-hadron duality holds in
the $F_1$ and $F_2$ structure functions.
In principle, there should also exist kinematic regions in
spin-dependent electron--nucleon scattering, where descriptions
in terms of both hadron and parton degrees of freedom coexist.
Indeed, duality in spin-dependent structure functions has been
predicted from both perturbative \cite{CM98} and nonperturbative
QCD arguments \cite{CG,CI}.
%
% However, spin dependent reactions by necessity deal with a smaller
% set of nucleon states, as the spin filter applied selects certain
% production processes --- some of the spin-averaged electromagnetic
% response of the nucleon does not contribute to its spin dependence.

The feature which most distinguishes the study of duality in
spin-dependent scattering from spin-averaged is that since spin
structure functions are given by differences of cross sections,
they no longer need be positive.
A dramatic example of this is provided by the $\Delta$ resonance,
whose contribution to the $g_1$ structure function of the proton
is negative at low $Q^2$, but changes sign and becomes positive
at high $Q^2$.
In spin-dependent scattering several new questions for the
investigation of quark-hadron duality therefore arise:
\begin{enumerate}
\item{Does quark-hadron duality work better (or only) for
	positive definite quantities such as cross sections,
	in contrast to polarization asymmetries?}
\item{Is there a quantitative difference between the onset of
	quark-hadron duality for spin-averaged and spin-dependent
	scattering, and if so, what can this be attributed to?}
\item{Does quark-hadron duality also hold for local regions in $W$
	for spin-dependent structure functions, and if so, how do
	these regions differ from those in unpolarized scattering?}
\end{enumerate}

Expanding on the last question, the example above of the $\Delta$
resonance contribution to the polarization asymmetry is sometimes
used as evidence against quark-hadron duality in spin-dependent
scattering \cite{SIM02}.
However, this argument is still not complete: the $\Delta$ resonance
{\em region} consists of both a resonant {\it and} a nonresonant
contribution, and it is the interplay between these that is crucial
for the appearance of duality \cite{CM98,CM93}.
The more relevant question is at which value of $Q^2$ does the $\Delta$
resonance region turn positive (in the case of the proton $g_1^p$),
and whether quark-hadron duality holds at lower $Q^2$ if one averages
over the elastic or other nearby resonances in addition to the $\Delta$.
Clearly duality cannot be too local at low $Q^2$.

In this section we will examine the degree to which local quark-hadron
duality exists in spin-dependent electron scattering, and how this is
reflected in the moments of the $g_1$ structure function.
We begin by reviewing measurements of the proton $g_1$ structure
function, following which we discuss experiments with deuterium
and $^3$He (neutron) targets.
The latter can be combined with the proton data to resolve the
isospin dependence of duality in spin structure functions.
Several sum rules, most notably the generalized Gerasimov-Drell-Hearn
sum rule, are discussed, and we conclude by reviewing the relevance
of the $g_2$ structure function for quark-hadron duality studies.

% .......................................................................
\subsubsection{Proton $g_1$ Structure Function}
\label{sssec:g1}

A large quantity of precision spin structure function data has been
collected over the past two decades \cite{PDG02} in the deep inelastic
region ($W > 2$~GeV) over a large range of $Q^2$.
This has allowed for initial studies of the logarithmic scaling
violations in the $g_1^p$ structure function, and more recently has
enabled one to embark upon dedicated investigations of quark-hadron
duality in spin-dependent scattering.

The spin structure functions $g_1$ and $g_2$ are typically extracted
from measurements of the longitudinal ($A_\parallel$) and transverse
($A_\perp$) polarization asymmetries (see Sec.~\ref{sec:formalism}).
The early resonance region measurements of $A_\parallel$ from SLAC,
over twenty years ago \cite{BAUM}, covered the range
$Q^2 \approx 0.5$~GeV$^2$ to 1.5~GeV$^2$.
The data showed that the asymmetries in the resonance region,
apart from the $\Delta$, were indeed positive.
From comparisons of the measured asymmetries with a fit to deep
inelastic data, it was concluded \cite{BAUM} that the behavior
of the spin-dependent asymmetries was consistent with duality,
in analogy with the unpolarized case.
The noted exception was a major oscillation away from the deep
inelastic behavior in the $\Delta$ region, for $Q^2$ = 0.5~GeV$^2$.

The first modern experiment accessing the spin structure functions
in the resonance region was SLAC experiment E143 \cite{ABE97,ABE98},
which measured both $A_\parallel$ and $A_\perp$ for protons (and
deuterons) over a wide range of kinematics.
Significant structure was observed in $g_1^p$, and, within
uncertainties, agreement with the previous SLAC data \cite{BAUM}
taken at similar kinematics.
Again, a negative contribution in the region of the $N-\Delta$
transition was observed, and a large positive contribution for
$W^2 > 2$~GeV$^2$.

\begin{figure}[t]
\begin{minipage}{3.5in}
\hspace*{-0.5cm}
\epsfig{figure=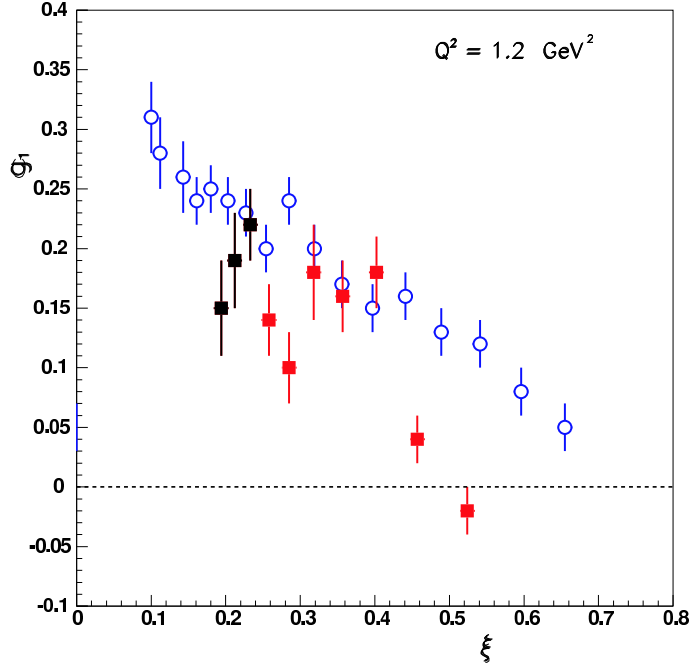, height=11cm}
\vspace*{-2cm}
\end{minipage}
\begin{centering}
\caption{\label{fig:g1p_e143}
	Proton $g_1^p$ structure function measured by SLAC experiment
	E143 \protect\cite{ABE97,ABE98}.
	The open circles denote the deep inelastic region at
	$Q^2 = 3.0$~GeV$^2$, and the solid squares represent
	the nucleon resonance region at $Q^2 \approx 1.2$~GeV$^2$.
	(The three solid squares at the lowest $\xi$ are
	beyond the nucleon resonance region, at $4 < W^2 < 5$~GeV$^2$.)
	The data are shown as a function of the Nachtmann variable
	$\xi$ to take target mass effects into account, and to
	facilitate comparison of these disparate kinematics.}
%\vspace*{1.0cm}
\end{centering}
\end{figure}

The E143 data at $Q^2 \approx 1.2$~GeV$^2$ are shown in
Fig.~\ref{fig:g1p_e143}, together with data in the deep inelastic
region at $Q^2 = 3.0$~GeV$^2$ (data at $Q^2 \approx 0.5$~GeV$^2$
were also taken).
To facilitate comparison at different $Q^2$ the data are shown
as a function of the Nachtmann scaling variable $\xi$, which
accounts for target mass corrections.
Target mass effects can also be incorporated in perturbative
QCD-based calculations, as was done for the unpolarized structure
functions.
We will show such comparisons with the more recent data below.

One can see from Fig.~\ref{fig:g1p_e143} that the resonance region
data at $Q^2 \approx 1.2$~GeV$^2$ seem to approach the deep inelastic
results, with the exception of the $N-\Delta$ transition region
(which occurs at $\xi \approx 0.5$).
When integrating over the region of $\xi$ corresponding to the nucleon
resonances at $Q^2 \approx 1.2$~GeV$^2$, one finds about 60\% of
the corresponding deep inelastic strength at $Q^2 = 3.0$~GeV$^2$.
Obviously, a large source of this missing strength lies in the
$\Delta$ region, which is still negative, and indeed
the integrated strength in the region $2 < W^2 < 4$~GeV$^2$ amounts
to about 80\% of the corresponding deep inelastic strength.

\begin{figure}
\begin{minipage}[ht]{3.5in}
\epsfig{figure=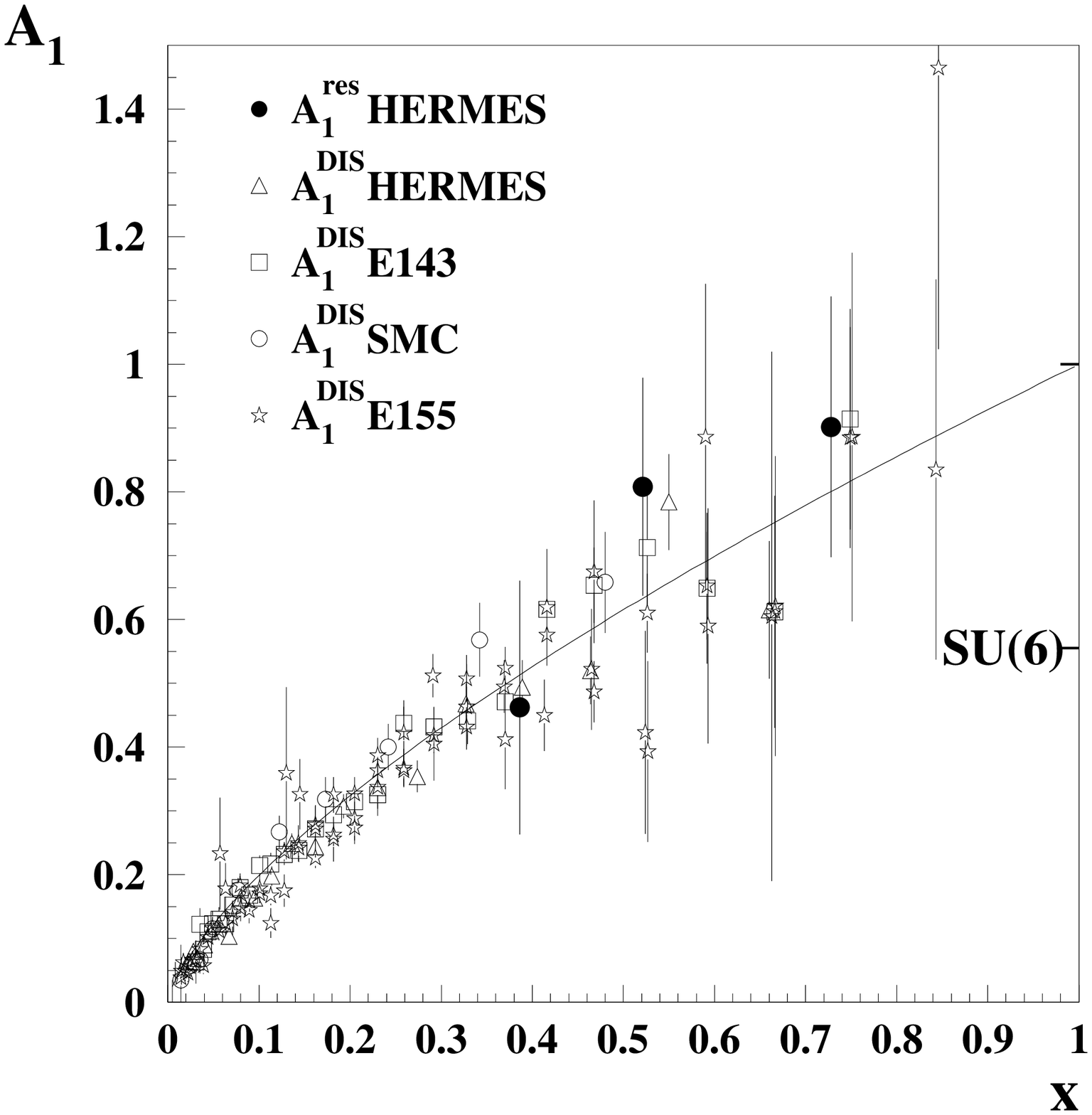, height=8cm}
\vspace*{0.5cm}
\end{minipage}
\begin{centering}
\caption{\label{fig:a1p_hermes}
	Proton spin asymmetry $A_1^p$ as a function of $x$ in the
	resonance region (solid circles) by the HERMES Collaboration
	\protect\cite{AIR02}.
	The errors are statistical only, with the systematic
	uncertainty in the resonance region about 16\%.
	Open symbols are previous results obtained in the deep
	inelastic region.
	The curve represents a power law fit to the deep inelastic
	data at $x > 0.3$, and the SU(6) prediction ($A_1^p=5/9$)
	\protect\cite{CLOSEBOOK} for the $x \to 1$ limit is
	indicated.}
%\vspace*{1.0cm}
\end{centering}
\end{figure}

Recently, the HERMES Collaboration at DESY reported $A_1^p$ spin
asymmetry data in the nucleon resonance region for
$Q^2 > 1.6$~GeV$^2$ \cite{AIR02} --- see Fig.~\ref{fig:a1p_hermes}.
The resonance region data are in agreement with those measured in
the deep inelastic region \cite{ABE98,AIR98,ADE98,ANT00}.
The curve in Fig.~\ref{fig:a1p_hermes} is a power law fit to the
world deep inelastic data at $x > 0.3$, $A_1^p = x^{0.7}$.
Such a parameterization is constrained to approach unity at $x = 1$,
which is consistent with the trend of the data shown.
The $A_1^p$ data in the resonance region exceed the prediction from
the SU(6) symmetric quark model ($A_1^p = 5/9$) \cite{CLOSEBOOK}
for $x \agt 0.5$ (see Sec.~\ref{sssec:elastic} below).
The chosen parameterization is independent of $Q^2$, as supported
by the experimental data in this range of $x$ \cite{ANT00}.

The average ratio of the measured $A_1^p$ asymmetry in the
resonance region to the deep inelastic power law fit is
$1.11~\pm~0.16$ (stat.)~$\pm$~0.18 (syst.) \cite{AIR02}.
This suggests that for $Q^2 > 1.6$~GeV$^2$, the description
of the spin asymmetry in terms of quark degrees of freedom is,
on average, also valid in the nucleon resonance region.
The implication of this result is the tantalizing possibility of
measuring the partonic content of $A_1^p$ at large values
of $x$, almost up to $x = 1$, by extending such measurements into
the nucleon resonance region.
Measurements of spin structure functions in the nucleon resonance
region at $Q^2 > 1$~GeV$^2$, with both good statistical and systematic
precision, would be very welcome to investigate this in detail.
Recently, the E01-006 experiment at Jefferson Lab \cite{E01006,YUNPR}
measured $A_\parallel$ and $A_\perp$ to high precision at
$Q^2 = 1.3$~GeV$^2$, and the data, which are currently being analyzed,
will allow a sensitive test of the assumptions made to extract $A_1^p$.
However, to investigate the mechanisms and the applications of
quark-hadron duality, precise measurements at higher values of $Q^2$
are required.

\begin{figure}[t]
\hspace*{0.5cm}
\epsfig{figure=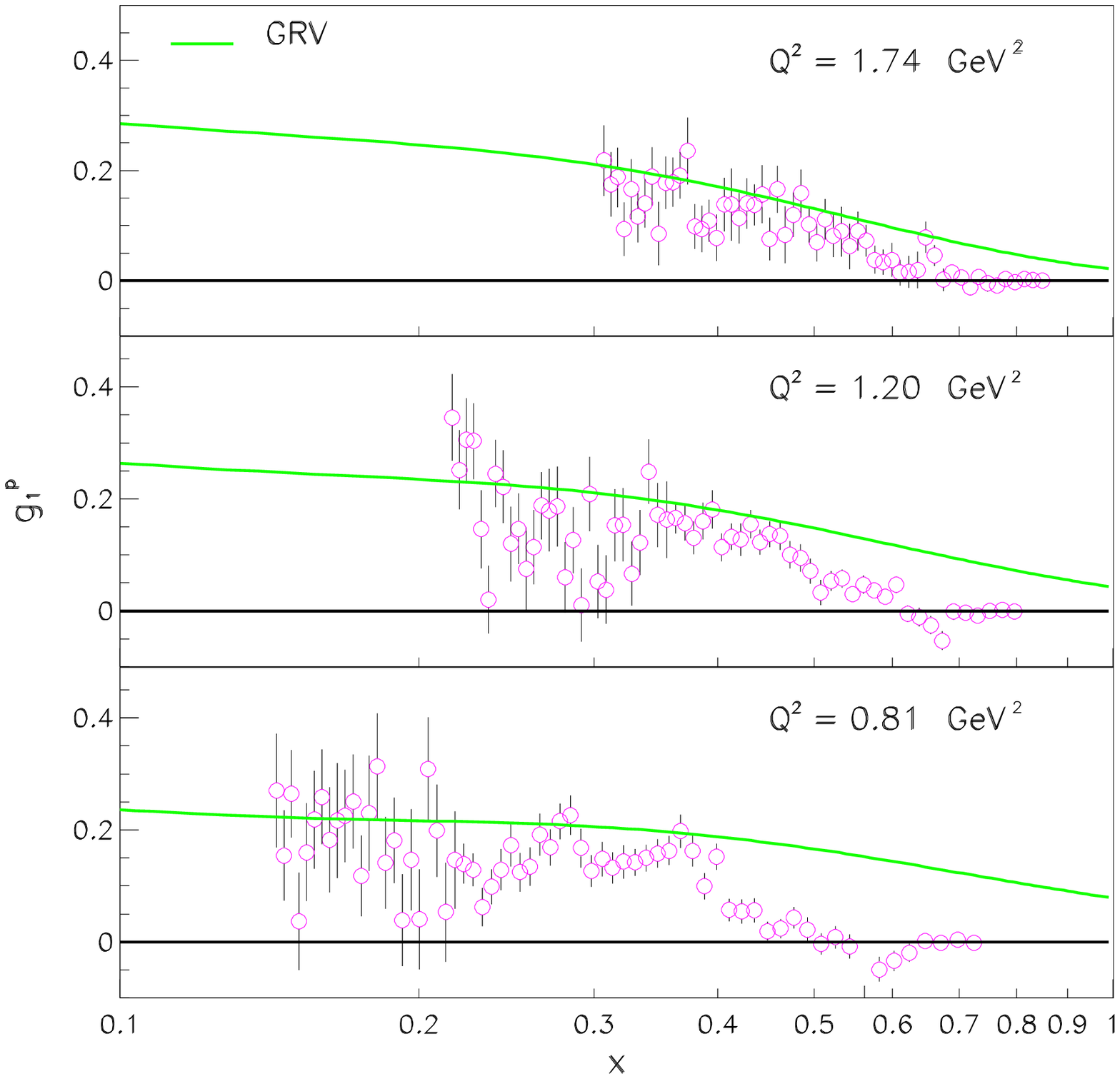, height=15cm}
\vspace*{-2.5cm}
\caption{\label{fig:clasg1}
	Proton spin structure function $g_1^p$ from CLAS
	\protect\cite{CLASg1} in the resonance region, at three
	values of $Q^2$ indicated.
	The curves are the global parameterizations of the spin
	structure functions from Ref.~\protect\cite{GRVg1}.}
\end{figure}

As mentioned in Sec.~\ref{sec:formalism}, there are advantages in
presenting spin-dependent data in terms of $g_1$ rather than $A_1$,
as the former is less sensitive to the precise knowledge of $g_2$
(or $A_2$).
The CLAS Collaboration at Jefferson Lab carried out inclusive
polarized scattering experiments at energies of 2.6 and 4.3~GeV,
using polarized ${\rm NH_3}$ as the target material \cite{CLASg1}.
Some of the results, for $Q^2 > 0.7$~GeV$^2$, are shown in
Fig.~\ref{fig:clasg1}.
In the lowest-$Q^2$ bin, the contribution of the $\Delta$ resonance
region to $g_1^p$ is negative, whereas the contributions of the
higher-mass states are positive.
The negative $\Delta$ contribution obviously prevents a naive
local duality interpretation at low $Q^2$.
However, in some models \cite{CI,NOSU6} local duality is only
expected to arise after averaging over the $\Delta$ and the
(positive) elastic contribution (see Sec.~\ref{sssec:qm} below).
Indeed, addition of the nucleon elastic and $N-\Delta$ transition
contributions would render a positive definite value for the
averaged $g_1^p$.

At higher $Q^2$ values, the role of the nonresonant background becomes
more prominent, and the magnitude of the (negative) contribution of the
$\Delta$ region rapidly decreases, becoming comparable to the (positive)
contribution from elastic scattering.
In contrast, the $g_1^p$ structure function at the higher-$W$ regions
shows less $Q^2$ variation, and in fact already closely resembles the
global structure function parameterizations \cite{GRVg1}.
As found earlier in the $F_2^p$ structure function, the nucleon
resonance region data seem to ``heal'' towards the perturbative
expectation.
This onset is slower for the $\Delta$ region, due to the still large,
but rapidly decreasing, elastic contribution.
Apart from the $\Delta$ region, which still shows no clear sign of
local duality at the $Q^2$ values of the present data, one can
conclude that some evidence for quark-hadron duality does exist
for the proton spin structure function $g_1^p$.

\begin{figure}[t]
\begin{minipage}{4.0in}
\epsfig{figure=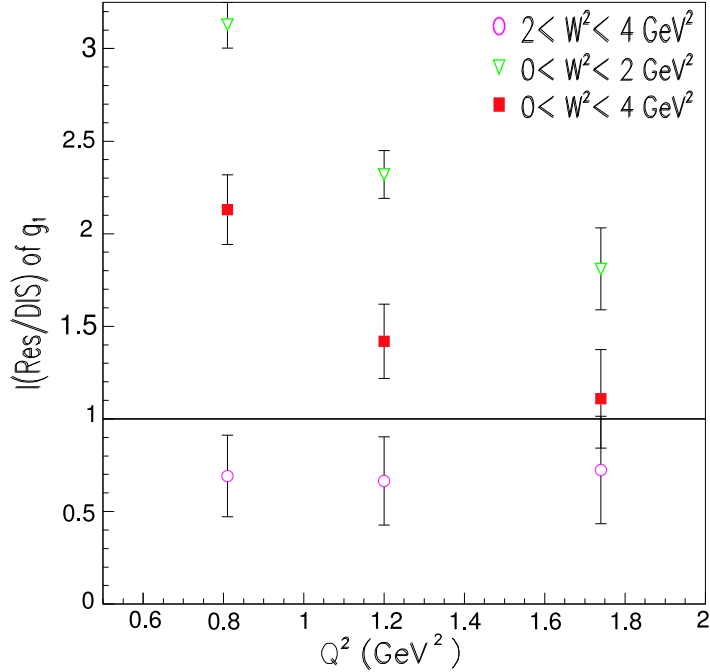, height=11cm}
\vspace*{-1.5cm}
\end{minipage}
\begin{centering}
\caption{\label{fig:clasint}
	Ratio of the integrated strength of the $g_1^p$ data in
	Fig.~\protect\ref{fig:clasg1} to that of the global
	parameterization from Ref.~\protect\cite{GRVg1}.
	Both the data and the QCD parameterization are integrated
	for each $Q^2$ over the $x$ regions corresponding to the
	indicated $W^2$ regions (with the elastic contribution
	included).}
%\vspace*{1.0cm}
\end{centering}
\end{figure}

This is further illustrated in Fig.~\ref{fig:clasint}, where we show
the integrated strength of the nucleon resonance region data in
Fig.~\ref{fig:clasg1} as compared to the integrated strength from
the global structure function parameterizations \cite{GRVg1}.
Here the data have been split into two regions --- the region
$W^2 < 2$~GeV$^2$ (with the elastic contribution included),
and $2 < W^2 < 4$~GeV$^2$ --- and then integrated for each $Q^2$
over the $x$ regions corresponding to the chosen $W^2$.
Clearly the elastic region {\em overcompensates} for the negative
$\Delta$ region contribution, and the ratio for the region
$W^2 < 2$~GeV$^2$ falls as a function of $Q^2$.
The region $2 < W^2 < 4$~GeV$^2$ has $\sim 75\%$ of the strength
of the global QCD parameterization \cite{GRVg1}, close to the 80\%
found in the SLAC-E143 data \cite{ABE97,ABE98} at $Q^2 = 1.2$~GeV$^2$.
The complete nucleon resonance region, with the elastic contribution
included, closely resembles what one expects from the QCD
parameterization at $Q^2 \approx 1.7$~GeV$^2$.
However, an even earlier onset is observed when both the elastic
and $\Delta$ regions are left out.

The special role played by the $\Delta$ resonance in spin-dependent
scattering means that quark-hadron duality sets in later (at higher
$Q^2$) here than in the corresponding spin-averaged case.
The $\Delta$ region in the $g_1^p$ structure function remains
negative at least until $Q^2 \approx 2$~GeV$^2$.
On the other hand, one could also argue that the $\Delta$ region
is not negative enough!
This is clear from Fig.~\ref{fig:clasint}, where the $\Delta$
resonance region together with the elastic contribution included
still has {\em too much} strength at low $Q^2$.
This is consistent with the fact that higher-twist analyses of the
lowest moment of $g_1^p$ do not show large higher twist effects
(see Sec.~\ref{sssec:twist}).

Summarizing the current experimental evidence, we see that some
form of local duality is clearly evident for $Q^2 > 1.6$~GeV$^2$ 
from at least two observations:
the spin asymmetry $A_1^p$ in the nucleon resonance region agrees
well on average with a deep inelastic power law fit, and the $g_1^p$
integrated strength (with the elastic contribution included) agrees
well with that from a global structure function parameterization
\cite{GRVg1} at $Q^2 > 1$~GeV$^2$.
This leads us to conclude that the onset of duality in the proton
spin structure function occurs somewhere in the region of
$1 < Q^2 < 2$~GeV$^2$.
Furthermore, the evidence for quark-hadron duality in both the
spin-averaged and the spin-dependent scattering process suggests
that the helicity-1/2 and helicity-3/2 photoabsorption cross
sections exhibit quark-hadron duality separately.

% ........................................................................
\subsubsection{Experiments with Polarized $^2$H and $^3$He Targets}
\label{sssec:2h3he}

The absence of free neutron targets means that the neutron spin
structure function $g_1^n$ is usually obtained from polarized lepton
scattering off either polarized deuterium or polarized $^3$He targets.
In the former case, since the deuteron has spin 1, the spins of the
bound proton and neutron are predominantly aligned, with a small
($\approx$~5\%) probability (due to the nuclear tensor force) of
finding the nucleons in a relative $D$-state with spins antialigned.
In the case of a spin-1/2 $^3$He nucleus, the protons pair off with
opposite spins with $\approx 90\%$ probability, leaving the neutron
to carry most of the polarization of the nucleus \cite{OSCAR}.

The extraction of the free neutron structure function $g_1^n$ from
either the $g_1^d$ or $g_1^{^3{\rm He}}$ data requires corrections
to be made for the neutron depolarization, as well as for other
nuclear effects such as nuclear binding and Fermi motion.
These have been studied extensively in Refs.~\cite{NUCLEAR_D,NUCLEAR_HE3},
and are found to be important mostly at large $x$.
They have also been calculated recently for the structure functions
in the resonance region, at low and intermediate values of $Q^2$
\cite{NUCLEAR_LOWQ2}.
For the low moments of $g_1^n$ the magnitude of the correction is
relatively small, however.

The first experiment measuring the deuteron spin structure function
$g_1^d$ in the nucleon resonance region was the SLAC experiment E143
\cite{ABE97,ABE98}, utilizing a polarized ${\rm ND_3}$ target.
As in the proton case, the $Q^2 \approx 1.2$~GeV$^2$ data showed
a clear negative contribution in the region of the $N-\Delta$
transition, and a positive contribution for $W^2 > 2$~GeV$^2$.
The measured $g_1^d$ structure function amounts to about half of the
$g_1^p$ structure function, leading to an almost null, but slightly
negative, contribution of $g_1^n$.
This is essentially the same behavior as that found in the DIS data
at higher $W$ and $Q^2$.
The overall strength (integrated over $\xi$) of $g_1^n$ in the nucleon
resonance region ({\em not} including the quasi-elastic contribution)
amounts to about 60\% of the corresponding deep inelastic strength at
$Q^2 = 3.0$~GeV$^2$ of the same experiment \cite{ABE98}, or about 80\%
for the region $2 < W^2 < 4$~GeV$^2$, similar to that found for the
proton.

\begin{figure}[t]
\begin{minipage}{4.0in}
\epsfig{figure=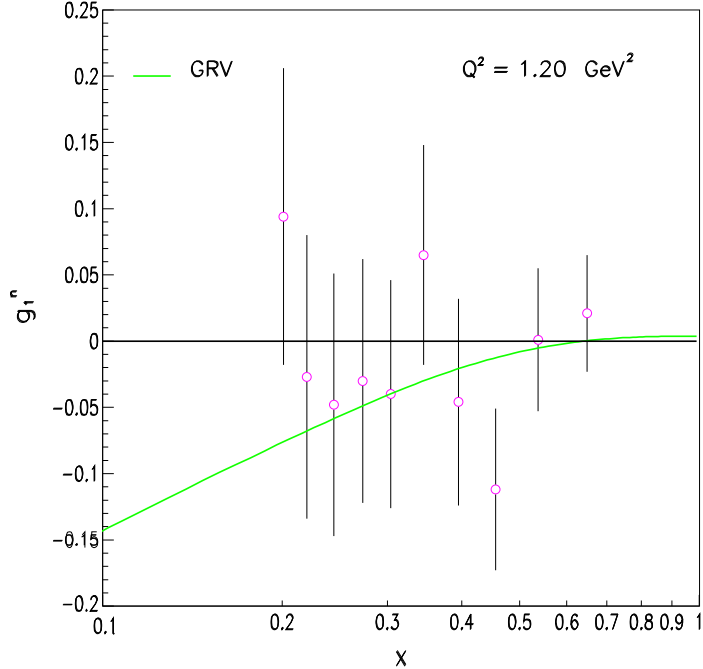, height=11cm}
\vspace*{-1.5cm}
\end{minipage}
\begin{centering}
\caption{\label{fig:g1n_e143}
	Neutron $g_1^n$ structure function data, as extracted
	from the difference of SLAC-E143 $g_1^d$ and $g_1^p$ data
	\protect\cite{ABE97,ABE98}.
	The open circles represent the nucleon resonance region
	data at $Q^2 \approx 1.2$~GeV$^2$ (the three lowest-$x$
	data points are technically beyond the nucleon resonance
	region, at $4 < W^2 < 5$~GeV$^2$).
	The curve is a global parameterization of the spin structure
	functions from Ref.~\protect\cite{GRVg1}.}
%\vspace*{1.0cm}
\end{centering}
\end{figure}

The $g_1^n$ structure function extracted from the difference of the
SLAC-E143 $g_1^d$ and $g_1^p$ data,
$g_1^n = 2g_1^d/(1-1.5\omega_D) - g_1^p$, with $\omega_D \approx 5\%$
the $D$-state probability, is shown in Fig.~\ref{fig:g1n_e143}.
Nuclear corrections other than the $D$-state probability are not
included, as these are small compared to the statistical
uncertainties of the experiment \cite{NUCLEAR_D}.
The $g_1^n$ data in Fig.~\ref{fig:g1n_e143} are compared with
the results from a global structure function parameterization
at similar $Q^2$ \cite{GRVg1}. 
Although the statistics in the $g_1^n$ data are rather limited,
some evidence for duality is visible, at a similar level as for
the $g_1^p$ data (at identical $Q^2$) in Fig~\ref{fig:g1p_e143}.
The $g_1^n$ nucleon resonance region data are {\em negative} on
average, so that quark-hadron duality appears to work both for
positive-definite and negative-definite quantities.

The CLAS Collaboration at Jefferson Lab collected $g_1^d$ data with
significantly smaller statistical uncertainties than the SLAC-E143
experiment, and better resolution in $W$ \cite{YUN02}.
Unfortunately, the maximum average $Q^2$ in their data is currently
limited to $\approx 1.0$~GeV$^2$, which precludes any conclusions
on the onset of duality beyond what can be inferred from the SLAC
data.
The higher-precision CLAS data does show an unambiguously positive
$g_1^d$ for $W^2 > 2$~GeV$^2$, indicating that the helicity-1/2
transition amplitudes dominate even at rather low values of $Q^2$
($Q^2 \approx 0.5$~GeV$^2$).
They conclude that the onset of local duality is slower for
polarized structure functions than for unpolarized, as only the
highest $Q^2 = 1.0$~GeV$^2$ data, beyond the $\Delta$ region,
show fairly good agreement with a fit to DIS data at
$Q^2 = 5$~GeV$^2$ \cite{YUN02}.
For the unpolarized $F_2^d$ structure function, local duality was
observed to hold well already for $Q^2 = 0.5$~GeV$^2$, from a
similar comparison.
Recently, the CLAS Collaboration extended the momentum transfer
region of their data to $Q^2 \approx 5$~GeV$^2$.
The results of this analysis will be very valuable in determining
the value of $Q^2$ for the onset of duality in $g_1^d$.

Similarly, an experiment in Hall A at Jefferson Lab has accumulated
data to investigate the onset of local duality in polarized electron
scattering from a polarized $^3$He target \cite{E01012}.
The experiment measured the $g_1^{^3{\rm He}}$ structure function over
the full nucleon resonance region, up to $Q^2 \approx 5$~GeV$^2$.
As mentioned above, the polarized $^3$He target acts to good
approximation as a source of polarized neutrons, although nuclear
corrections will become more important in the large-$x$ region
($x \agt 0.5$) covered by these data than for the case of the
deuteron.

% .......................................................................
\subsubsection{Sum Rules at Low and High $Q^2$}
\label{sssec:gdh}

Sum rules involving the spin structure of the nucleon offer an
important opportunity to study fundamental properties of QCD.
A classic example is the Bjorken sum rule, which at high $Q^2$
relates the lowest moment of the isovector nucleon $g_1$ structure
function to the nucleon axial charge $g_A$ \cite{BJSR},
\begin{eqnarray}
\Gamma_1^{p-n}(Q^2)
&=& {1 \over 6} g_A
    \left( 1 + { \alpha_s(Q^2) \over \pi }\ + \cdots \right)\ .
\end{eqnarray}
Sum rules for the individual proton and neutron moments, $\Gamma_1^p$
and $\Gamma_1^n$, can also be derived \cite{EJ74}, assuming knowledge
of the octet and singlet axial charges
(see also Ref.~\cite{EJ_RIGOR}.)

At the other extreme of real photon scattering, $Q^2=0$, there is
another fundamental sum rule, derived independently by Gerasimov,
and Drell and Hearn (GDH) \cite{GDH}.
The GDH sum rule relates the helicity-dependent total absorption cross
section for circularly polarized photons on linearly polarized nucleons
to the nucleon anomalous magnetic moment $\kappa$,
\begin{equation}
\label{eq:gdh}
I_{\rm GDH}\
\equiv\ \int_{\nu_0}^\infty {d\nu \over \nu} 
\left( \sigma_{1/2}(\nu) - \sigma_{3/2}(\nu) \right)\
=\ - 2 \pi^2 \alpha {\kappa^2 \over M^2}\ ,
\end{equation}
where $\sigma_{1/2}$ and $\sigma_{3/2}$ are the total helicity-1/2
and 3/2 photoabsorption cross sections, $\alpha$ is the
electromagnetic fine structure constant,
and $\nu_0 = m_\pi (1 + m_\pi/2M)$ is the inelastic pion production
threshold energy.
This sum rule thus provides a fascinating link between the
helicity-dependent dynamics at low and high energies, and static
ground state properties of the nucleon.
In terms of the $g_1$ structure function, the GDH sum rule can be
equivalently written as
\begin{equation}
\label{eq:gdh_g1}
\int_{\nu_0}^\infty {d\nu \over \nu^2}\ g_1(\nu,Q^2=0)\
=\ -{\kappa^2 \over 2 M}\ .
\end{equation}

The derivation of the GDH sum rule follows from a general dispersion
relation applied to forward Compton scattering, and the applicability
of the low energy theorem (LET) and the no-subtraction hypothesis for
the spin-flip part of the Compton scattering amplitude.
The use of unsubtracted dispersion relations follows from causality,
while the LET originates from gauge invariance and relativity.
Because of the 1/$\nu$ weight in the integral in Eq.~(\ref{eq:gdh}),
the GDH sum rule is mostly sensitive to the low-energy part of the
photoabsorption cross section, in the region where baryon resonances
dominate and single pion production is the main contribution.
The generality of the assumptions in deriving the GDH sum rule
have prompted a concerted experimental effort to test its validity
directly.

From Eqs.~(\ref{eq:gdh}) or (\ref{eq:gdh_g1}) one observes that
the right hand side of the GDH sum rule at $Q^2=0$ is negative.
On the other hand, the corresponding integral at non-zero $Q^2$,
\begin{equation}
\label{eq:gdh_hiQ}
\int_{\nu_0}^\infty {d\nu \over \nu^2}\ g_1(x,Q^2)\
=\ { 2 M \over Q^2 }\ \Gamma_1^{\rm inel}(Q^2)\ ,
\end{equation}
is determined by the inelastic contribution to the moment
$\Gamma_1(Q^2)$, which for the case of the proton is known to be
positive.
This illustrates a striking example of the workings of quark-hadron
duality in spin structure functions: as one moves from the real
photon point where duality is clearly violated, the integral
(\ref{eq:gdh_g1}) is forced to change sign and approach a positive
value at large $Q^2$.
This is in contrast to the unpolarized proton $F_2$ structure
function, for instance, where only the magnitude of the $n=2$
moment changes, from $\approx 0.2$ at intermediate $Q^2$ to unity
(the proton charge) at $Q^2 = 0$.

The GDH sum rule can be formally generalized to virtual photons
at finite $Q^2$ by defining \cite{AIL89,JIOS01,DKT01}
\begin{eqnarray}
\label{eq:gdh_q}
I_{\rm GDH}(Q^2)\
&\equiv& \int_{\nu_0}^\infty {d\nu \over \nu} 
   \left( \sigma_{1/2}(\nu,Q^2) - \sigma_{3/2}(\nu,Q^2) \right)  \\
&=& { 8\pi^2 \alpha \over M^2} \int_0^{x_0} {dx \over K\ x}
   \left( g_1(x,Q^2) - \gamma^2 g_2(x,Q^2) \right)\ ,
\end{eqnarray}
where $x_0 = Q^2/2M\nu_0$ is the value of $x$ at the pion production
threshold, $\gamma = Q^2/\nu^2$, and $K$ is the virtual photon flux
\cite{DKT01} (see Eq.~(\ref{eq:K}) in Sec.~\ref{sec:formalism}).
In the limit $Q^2 \to 0$ the integral (\ref{eq:gdh_q}) reduces to
the GDH sum rule,
$I_{\rm GDH}(Q^2) \to I_{\rm GDH}$, while in the Bjorken limit
it is given by the moment of $g_1(x)$,
\begin{equation}
I_{\rm GDH}(Q^2)\
\to { 16\pi^2\alpha \over Q^2 }\ \Gamma_1(Q^2)\ \equiv\ I(Q^2).
\end{equation}
At finite but non-zero $Q^2$ the integral $I_{\rm GDH}(Q^2)$
therefore interpolates between the two limits, allowing one to
study the evolution of the sum rule from large distances, where
effects of confinement are dominant, towards short distances,
where a partonic description is possible through asymptotic freedom.
The generalized GDH sum rule is hence ideal for the study of
quark-hadron duality.

A similar phenomenon also occurs in the lowest moment of the
unpolarized $F_2$ structure function, which interpolates between
the nucleon's electric charge at $Q^2=0$ (Coulomb sum rule)
and the momentum sum rule at asymptotic values of $Q^2$.
The only difference is that the spin-dependent sum rules result
from interference effects, and as such may lead to a deeper
understanding of the transition from confinement to asymptotic
freedom.

Before proceeding with the experimental results on the generalized
GDH integral $I_{\rm GDH}(Q^2)$, we note that while the $Q^2 \to 0$
and $Q^2 \to \infty$ limits are well defined, there are two avenues
for exploring the transition at intermediate $Q^2$.
From its definition in Eq.~(\ref{eq:gdh}), the integral
$I_{\rm GDH}$ for real photons includes only inelastic contributions,
whereas the deep inelastic integral is formally defined as a sum over
all possible final states, including the elastic.
The latter is zero at asymptotically large $Q^2$, but can be
significant at $Q^2 \alt 1$--2~GeV$^2$.
In constructing the generalized GDH integral, one can therefore either
add the elastic to the GDH sum at $Q^2=0$ and match to the total DIS
moment at high $Q^2$, or subtract the elastic component from the DIS
integral and evolve the inelastic integral to low $Q^2$.
For higher-twist analyses (see Sec.~\ref{sssec:twist} below), which
rely on the formal operator product expansion, the former choice
must be made.
On the other hand, the evolution of the integral is more dramatically
illustrated by considering the elastic-subtracted sum rule.
The choice is in principle arbitrary, but it is important to ensure
that like quantities are being compared.

The GDH sum rule for real photons has been studied for photon
energies up to 2.5~GeV \cite{Ah01}.
The current experimental result deviates from the theoretical
prediction by about 10\%, although higher photon energy data are
required for a more definitive conclusion.
The Bjorken sum rule has been verified at the 5\% level for
$Q^2 \agt 2$~GeV$^2$.
In the remainder of this section we will focus on experimental
results on the integral $I_{\rm GDH}(Q^2)$ at low and intermediate
values of $Q^2$ ($Q^2 < 2$~GeV$^2$), which is most relevant for
the study of quark-hadron duality.

\begin{figure}[t]
\begin{minipage}{4.0in}
\hspace*{-1.5cm}
\epsfig{figure=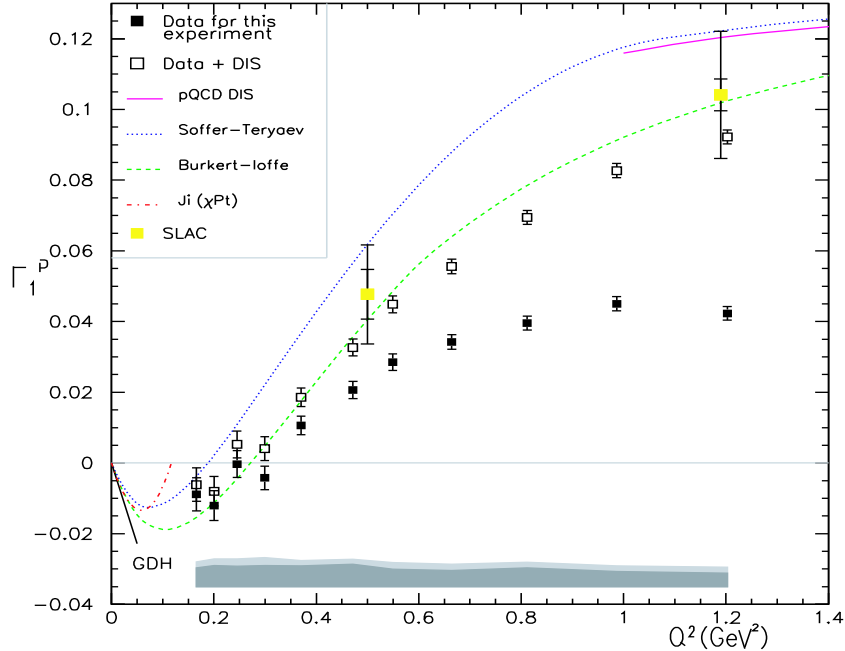, height=11cm, width=12cm}
\vspace*{-2cm}
\end{minipage}
\begin{centering}
\caption{\label{fig:intg1p}
	Inelastic contribution to the moment $\Gamma_1^p$ of the
	proton $g_1$ structure function as a function of $Q^2$.
	The filled squares correspond to the measured values from
	CLAS \protect\cite{CLASg1}, while the open squares include in
	addition contributions from the unmeasured low-$x$ region.
	The light shaded squares are from the SLAC E143 experiment
	\protect\cite{ABE98}.
	The various curves are explained in the text.}
\end{centering}
\end{figure}

As discussed above, measurements on polarized proton targets have
been performed at SLAC by the E143 Collaboration \cite{ABE97,ABE98},
at DESY by the HERMES Collaboration \cite{Ai03}, and at Jefferson Lab
by the CLAS Collaboration \cite{CLASg1}.
To construct the integral $\Gamma_1^p$, parameterizations were used
to extrapolate beyond the experimentally accessible regions of $x$,
to $x=0$ and $x=1$.
The results for the elastic-subtracted $\Gamma_1^p$ integral from the
CLAS and SLAC E143 experiments are shown in Fig.~\ref{fig:intg1p}
for $Q^2 < 1.4$~GeV$^2$.

The most characteristic feature of $\Gamma_1^p(Q^2)$ is the strong
$Q^2$ dependence for $Q^2 < 1$~GeV$^2$, with a zero crossing near
$Q^2 \approx 0.2$--0.25~GeV$^2$.
The zero crossing is due largely to an interplay between the
excitation strengths of the $\Delta$ and $S_{11}$(1535) resonances,
and the rapid change in the helicity structure of the $D_{13}$(1520)
from helicity-3/2 dominance at the real photon point to helicity-1/2
dominance at $Q^2 > 0.5$~GeV$^2$ \cite{CLASg1}.
The dramatic evolution of $\Gamma_1^p(Q^2)$ is therefore due to the
intrinsic sensitivity of the spin-dependent structure functions to
the interference between various resonant and nonresonant transition
states, whereas the spin-averaged structure functions are sensitive
to the square of their sum.
In addition, in the limit $Q^2 \to 0$ one enhances the effect of the
spin-3/2 ground state, the $\Delta$(1232).
In the nonrelativistic SU(6) quark model this effect would be even
more spectacular, as will be discussed in Sec.~\ref{sssec:qm} below.

The data in Fig.~\ref{fig:intg1p} slightly underestimate the
perturbative QCD curve evolved down to $Q^2 \approx 1$~GeV$^2$.
This deviation can be mostly attributed to the negative contribution
of the $\Delta$ resonance, which is still sizable even at
$Q^2 \approx 1$~GeV$^2$.
The data are well described by the model of Burkert and Ioffe
\cite{BI}, which includes resonance excitations and connects to
the deep inelastic region assuming vector meson dominance.
The description in the model of Soffer and Teryaev \cite{SO97},
without explicit nucleon resonance contributions, is not as good.
In this model, the low-$Q^2$ behavior of $g_1$ is governed by the
$Q^2$ dependence of a linear combination of the electric and magnetic
form factors.
Heavy baryon chiral perturbation theory has been proposed as a tool
to describe the evolution of the GDH sum rule to small non-zero
values of $Q^2$ ($Q^2 \alt 0.1$~GeV$^2$), although the existing
calculations \cite{JI01} (labeled ``$\chi$Pt'' in
Fig.~\ref{fig:intg1p}) are still at too low $Q^2$ to compare with
the data shown.

\begin{figure}[t]
\begin{minipage}{3.5in}
\hspace*{-2cm}
\epsfig{figure=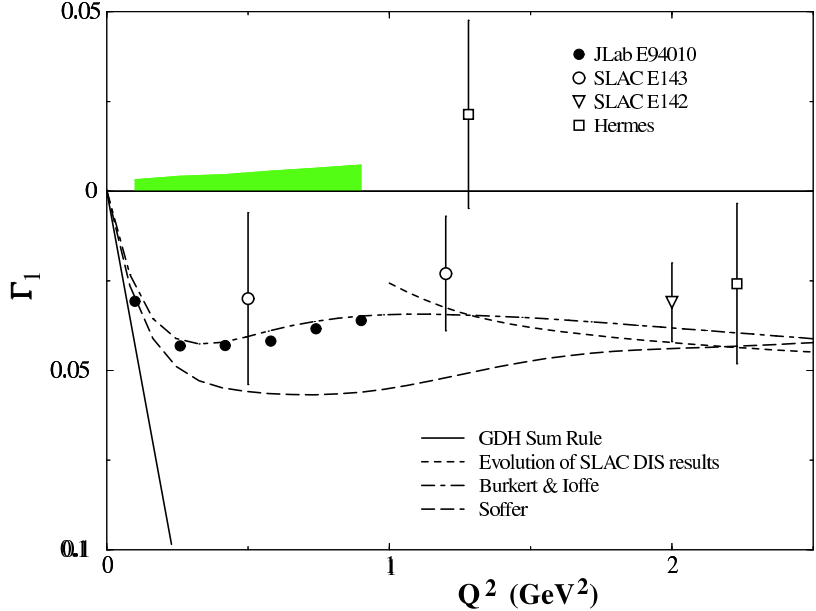, height=12cm}
\vspace*{-3.5cm}
\end{minipage}
\begin{centering}
\caption{\label{fig:intg1n}
	Inelastic contribution to the moment $\Gamma_1^n$ of the
	neutron $g_1$ structure function as a function of $Q^2$.
	The solid circles are from the Jefferson Lab Hall~A
	experiment E94-010 \protect\cite{E94010}, with the band
	indicating the size of the systematic uncertainties.
	The open symbols are from SLAC and HERMES experiments.
	The curves are as in Fig.~\protect\ref{fig:intg1p} and
	described in the text.}
\end{centering}
\end{figure}

To extract information on the neutron's first moment, $\Gamma_1^n$,
experiments have been performed using both polarized deuterium and
$^3$He targets at SLAC \cite{ABE97,ABE98}, at DESY \cite{Ai03}, and
at Jefferson Lab \cite{YUN02,E94010}, as discussed in the previous
section.
We will focus only on the results of the $^3$He experiments,
as they have the largest overlap with our region of interest.
After correcting for nuclear effects and accounting for the
unmeasured low-$x$ part, the elastic-subtracted moment
$\Gamma_1^n(Q^2)$ is shown in Fig.~\ref{fig:intg1n}.
Again, the model of Ref.~\cite{BI} including resonance
contributions and assuming a vector meson dominance inspired
connection with the perturbative region describes the data fairly
well.
Also, as mentioned in Sec.~\ref{sssec:2h3he} above, $\Gamma_1^n$
remains negative from high $Q^2$ down to low $Q^2$, highlighting
the fact that quark-hadron duality works well even for quantities
which are not positive-definite.

\begin{figure}
\begin{minipage}[ht]{4.0in}
\epsfig{figure=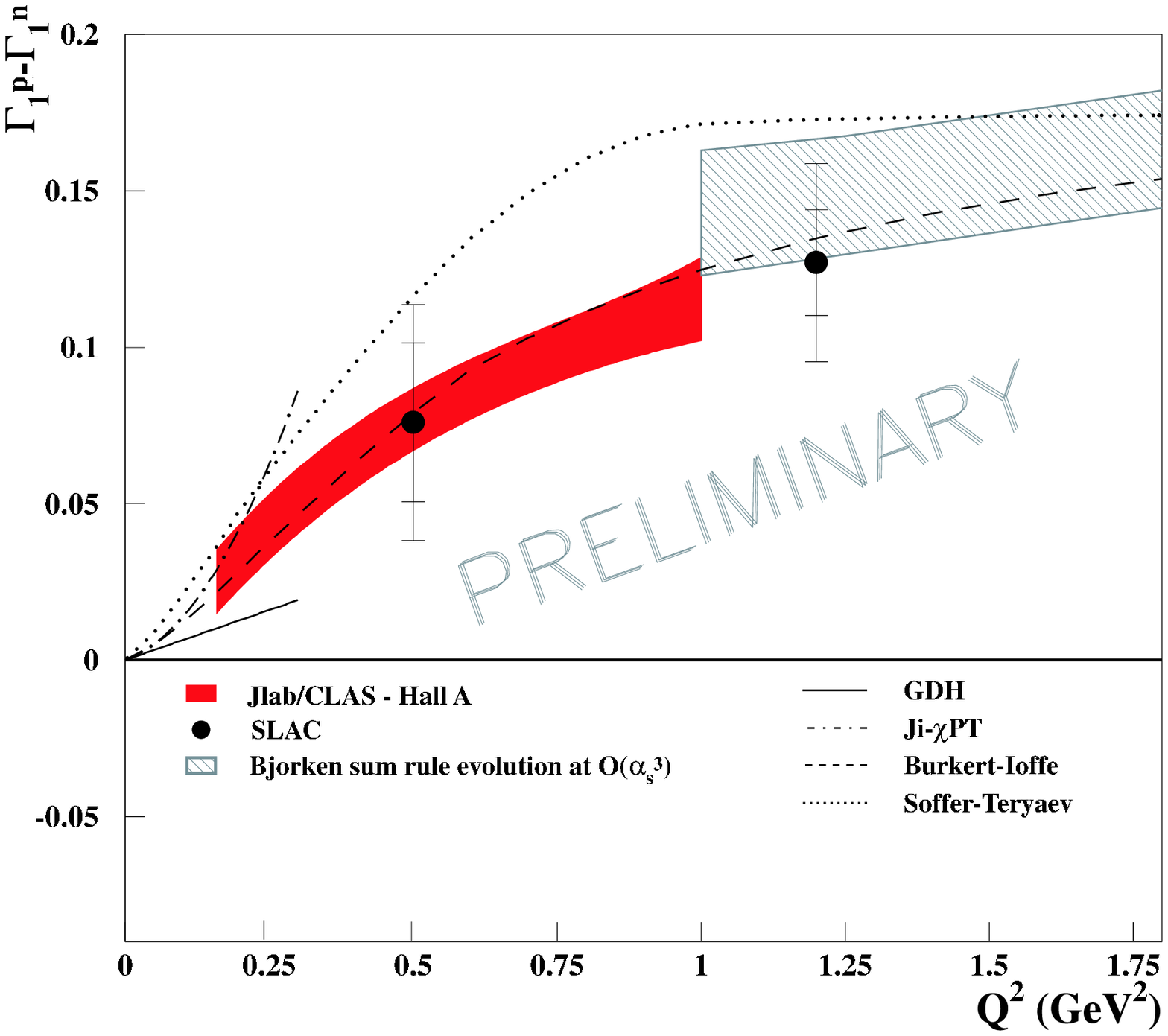, height=12cm}
\vspace*{-3cm}
\end{minipage}
\begin{centering}
\caption{\label{fig:intp-n}
	Difference of the proton and neutron moments of the $g_1$
	structure function, including only inelastic contributions.
	The shaded band below $Q^2 = 1$~GeV$^2$ parameterizes
	the results derived from Refs.~\protect\cite{E94110} and
	\protect\cite{BUR02}.
	The light-shaded band above $Q^2 = 1$~GeV$^2$ corresponds
	to the perturbative evolution of the Bjorken integral from
	large $Q^2$ including ${\it O}(\alpha_s^3)$ corrections.
	The curves are as in Fig.~\ref{fig:intg1p} and described
	in the text.}
\end{centering}
\end{figure}

Finally, using the results for the proton and neutron, the $Q^2$
dependence of the (elastic-subtracted) Bjorken integral is displayed
in Fig.~\ref{fig:intp-n}.
Here, the contributions from isospin-3/2 resonances, such as the
$\Delta$(1232), cancel out exactly, thereby removing the zero
crossing from the isovector integral.
Since the results on the proton and on the neutron (from $^3$He)
were obtained at somewhat different $Q^2$ values, a smooth
interpolation was used to evolve to common $Q^2$ values.
The results for the proton--neutron difference are at the
centroid of the shaded band in Fig.~\ref{fig:intp-n}.
As expected from the comparison with the separate $\Gamma_1^p$ and
$\Gamma_1^n$ moments, the model of Ref.~\cite{BI} also provides
a good description of the difference $\Gamma_1^p - \Gamma_1^n$.

It is perhaps not very surprising that this sum rule smoothly joins
to the perturbative expectation already at $Q^2 \approx 1$~GeV$^2$.
We have seen before that duality violations appear strong for the
region where the well-isolated ground states of the spin-1/2 (elastic)
and spin-3/2 ($\Delta$) are prominent (see {\em e.g.},
Figs.~\ref{fig:ioanamoments},~\ref{fig:momconts},
and ~\ref{fig:clasint}).
With the removal of the elastic contribution, the cancellation of the
isospin-3/2 resonances, and the partial cancellation of contributions
from other resonances at low $Q^2$, the transition from a
confinement-based hadronic world to an asymptotically free quark-gluon
world may appear fairly smooth down to low $Q^2$.

% ........................................................................
\subsubsection{The $g_2$ Structure Function}
\label{sssec:g2}

The structure function $g_1$ can be understood within the
quark-parton model in terms of spin-dependent quark distributions
(see Eq.~(\ref{eq:g1quark}) of Sec.~\ref{sec:formalism}).
The interpretation of the structure function $g_2$, on the other
hand, is less straightforward.
In the language of the operator product (or twist) expansion in QCD
(see Sec.~\ref{ssec:qcd} below), the $g_2$ structure function receives
contributions from a scaling (or ``leading twist'') part, derived by
Wandzura \& Wilczek \cite{WW77} and denoted by $g_2^{WW}$, a component
which arises from transverse quark polarization (which is proportional
to the quark mass $m_q$, and usually neglected), and a ``higher-twist''
contribution associated with nonperturbative quark-gluon interactions.
Since the parton model includes neither transverse momentum nor
quark-gluon interactions, there is no direct interpretation of
$g_2$ within this framework.

For studies of quark-hadron duality, $g_2$ is of particular interest
specifically because, unlike for the other structure functions, the
effects of quark-gluon correlations are not suppressed by powers of
$1/Q^2$, but enter at the same order as the leading-twist terms.
One could argue, therefore, that measurement of $g_2$ provides one
of the most direct windows on duality and its violation.
We will present in this section data on $g_2^p$ and $g_2^n$,
in anticipation of an analysis in terms of higher-twist matrix
elements in Sec.~\ref{sssec:twist} below.

An important sum rule involving the $g_2$ structure function is the
the Burkhardt-Cottingham (BC) sum rule \cite{BC70},
\begin{equation}
\Gamma_{\rm 2}(Q^2) = \int_0^1 dx\ g_2(x,Q^2) = 0\ ,
\label{eq:BC}
\end{equation}
which follows from a dispersion relation for the forward spin-flip
Compton amplitude, and is expected to be valid at {\em all} $Q^2$.
Its validity assumes the absence of singularities at low $x$,
similar to the assumption made in the derivation of the GDH sum rule.
A comprehensive discussion of the BC sum rule and what it tests can
be found in Ref.~\cite{BGJ73}.
The BC sum rule is of interest from the point of view of quark-hadron
duality, as various elastic, resonance, and deep inelastic
contributions {\it must} cancel for the sum rule to hold.

\begin{figure}
\begin{minipage}[ht]{4.0in}
\hspace*{-4.5cm}
\epsfig{figure=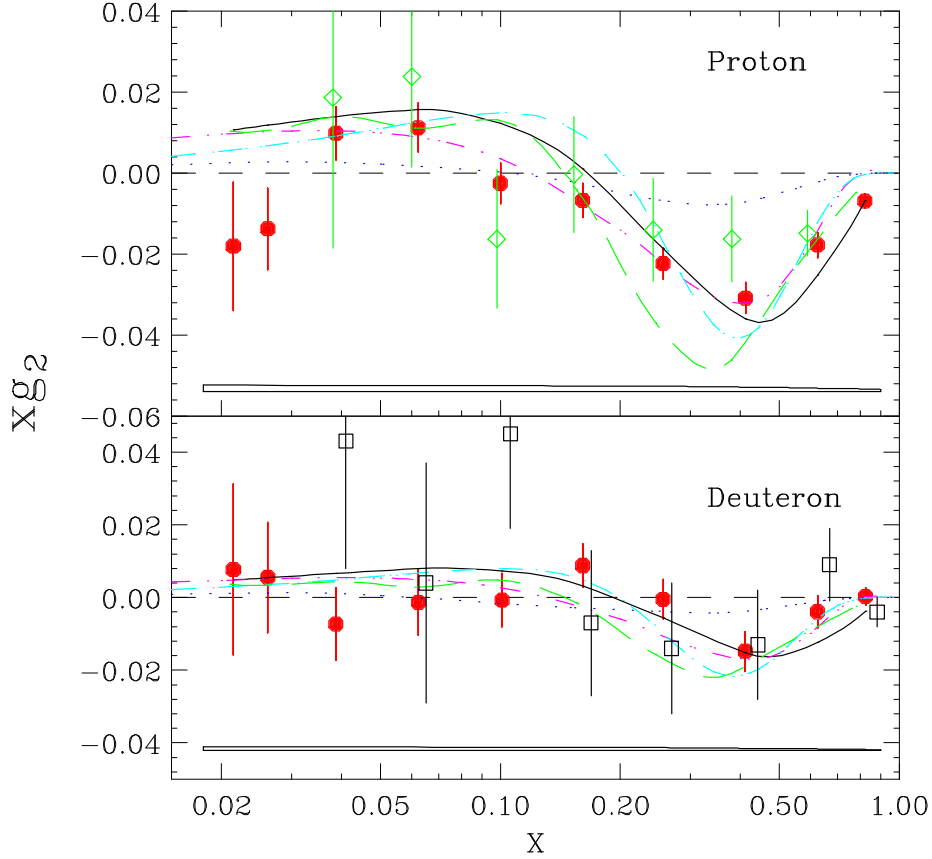, height=18cm}
\vspace*{-6cm}
\end{minipage}
\begin{centering}
\caption{\label{fig:e155g2}
	The structure function $xg_2$ from SLAC experiments E155
	(filled circles) \protect\cite{E155}, E143 (open diamonds)
	\protect\cite{ABE96}, and E155 (open squares)
	\protect\cite{ANT99}.
	The errors are statistical; the systematic errors are
	shown as bands at the bottom of each panel.
	The curves are the leading-twist Wandzura-Wilczek
	contribution \protect\cite{WW77} (solid), the bag model
	calculations of Stratmann \protect\cite{STR93} (dot-dashed)
	and Song \protect\cite{SON96} (dotted), and the chiral soliton
	models of Weigel \& Gamberg \protect\cite{WG00} (short-dashed)
	and Wakamatsu \protect\cite{WAK00} (long-dashed).
	(From Ref.~\protect\cite{E155}.)}
\end{centering}
\end{figure}

The SLAC E155 Collaboration has made the most precise measurements
of the proton and deuteron $g_2$ structure functions, over a large
range in $x$ and $Q^2$ in the deep inelastic ($W^2 \geq 3$~GeV$^2$)
region \cite{E155}.
Figure~\ref{fig:e155g2} shows the $Q^2$-averaged proton and deuteron
$x g_2$ structure functions, with $Q^2$ ranging from 0.8~GeV$^2$
(at low $x$) to 8.4~GeV$^2$ (at high $x$).
The combined data for the proton disagree with the leading-twist
$g_2^{WW}$ prediction, whereas the data for the deuteron agree.
The latter indicates that the nonperturbative quark-gluon
correlations are small for the deuteron. 
The derived BC sum rule is found to be violated at the level of
three standard deviations for the proton, and found to hold within
uncertainties for the deuteron.
This can be most readily explained by assuming more $g_2$ strength
for the proton from the unmeasured $x < 0.02$ region than for the
deuteron.

\begin{figure}
\begin{minipage}[ht]{4.0in}
\hspace*{-3.5cm}
\epsfig{figure=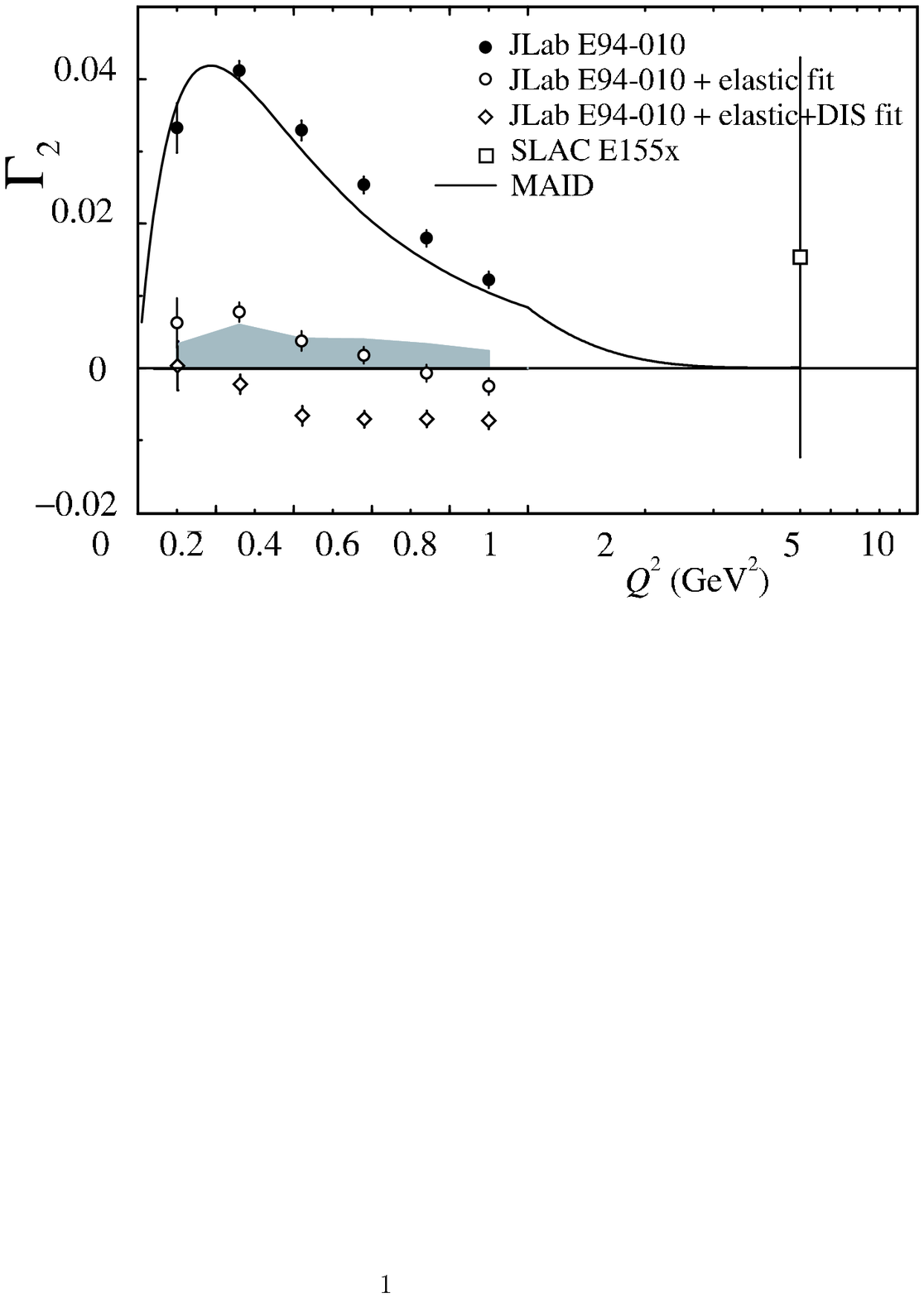, height=16cm}
\vspace*{-9cm}
\end{minipage}
\begin{centering}
\caption{\label{fig:mezianig2}
	Contributions to the moment $\Gamma_2^n$ of the neutron
	$g_2^n$ structure function from the resonance region from
	JLab experiment E94-010 \protect\cite{Ama03} (filled circles)
	along with the world data from deep inelastic scattering
	\protect\cite{E155} (open square), and the resonance
	contribution evaluated using MAID \protect\cite{DKT01}
	(solid line).
	The open circles include the elastic contribution, whereas
	the open diamonds include both the elastic and an estimate
	of the unmeasured deep inelastic contributions.
	The light grey band corresponds to the total systematic error,
	including uncertainties in the nuclear corrections.
	(Adapted from Ref.~\protect\cite{Ama03}.)}
\end{centering}
\end{figure}

The E94-010 Collaboration in Hall~A at Jefferson Lab recently
measured \cite{Ama03} the $g_2$ structure function of the neutron
using a polarized $^3$He target.
Excitation energies covered the resonance region and part of the
deep inelastic region, for $0.1 < Q^2 < 0.9$~GeV$^2$.
Figure~\ref{fig:mezianig2} shows the extracted $\Gamma_2$ for the
neutron, in the measured region (filled circles), after adding
also the elastic component (open circles), and after adding an
estimated contribution from the unmeasured deep inelastic region
assuming $g_2$ is given there by $g_2^{WW}$ (open diamonds).
To extract neutron information from the $^3$He data, nuclear
corrections were performed as described in
Ref.~\cite{NUCLEAR_HE3}.
The resonance contribution calculated in the MAID model \cite{DKT01}
(solid line) agrees well with the measured resonance data.

The interplay between strength in the resonance region and the
elastic contribution is striking.
The two contributions nearly cancel, such that the BC sum rule is
verified, within uncertainties, over the $Q^2$ range measured,
for the limited $x$ range of this experiment.
This result appears at odds, however, with the violation of the
BC sum rule on the proton reported at high $Q^2$ in Ref.~\cite{E155}.
On the other hand, the BC sum rule result extracted for the
neutron at high $Q^2$ ($\approx 5$~GeV$^2$) \cite{E155} is
consistent within the large error bar.
The difference between the behavior of the proton and neutron $g_2$
data is indeed intriguing.
In passing, we note that quark-hadron duality in the $g_1$ structure
function, as we saw in Secs.~\ref{sssec:g1} and \ref{sssec:2h3he},
also seemed to be more readily obeyed for the neutron than for the
proton.
We can only look forward to future high-precision data providing
a definitive resolution of the BC sum rule's validity.

% ------------------------------------------------------------------------
\subsection{Scaling in Electro-Pion Production}
\label{ssec:pion}

Scaling is a well established phenomenon in inclusive deep inelastic
scattering.
The cross section is proportional to structure functions that depend
only on the scaling variable $x$, up to calculable logarithmic $Q^2$
corrections \cite{EVOLVE}.
Both the observation of scaling and subsequently the (logarithmic)
scaling violations in the measured structure functions played a
crucial role in establishing QCD as the accepted theory of strong
interactions, and in mapping out the distributions of the constituents
of protons and neutrons.

The observation of duality between the various inclusive structure
functions measured in the resonance region and those in the deep
inelastic limit further indicates that the single quark scattering
process is decisive in setting the scale of the reaction,
even in the region dominated by nucleon resonances.
Apparently, the role of final state interactions required to form
the resonances becomes almost insignificant when averaged over the
resonances.

\begin{figure}
\begin{minipage}[ht]{2.5in}
\epsfig{figure=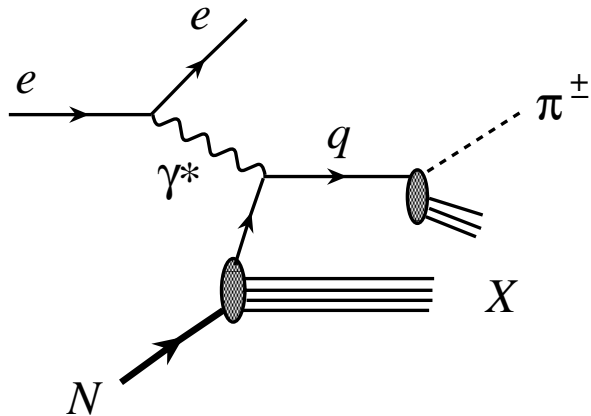, height=6cm}
\end{minipage}
\begin{centering}
\vspace*{0.5cm}
\caption{\label{fig:eepiplot}
	A representation of semi-inclusive electroproduction of
	$\pi^\pm$ mesons from nucleons.  The produced $\pi^\pm$
	is detected in coincidence with the scattered electron,
	and $X$ denotes the remaining inclusive hadronic final
	state.}
\end{centering}
\end{figure}

Given this situation it seems worthwhile to examine other electron
scattering processes that are closely related to deep inelastic
scattering, but where scaling and scaling violations are not as well
established. 
The prime example of such a process is semi-inclusive deep inelastic
electroproduction of mesons $m$ from nucleons,
\begin{equation}
e\ N \to e^\prime\ m~X\ ,
\label{eq:semiexcl}
\end{equation}
where the meson is detected in coincidence with the scattered
electron.
In this section we examine inclusive pion ($m$ = $\pi^\pm$)
electroproduction, as illustrated in Fig.~\ref{fig:eepiplot},
paying special attention to both the onset of scaling and the
appearance of quark-hadron duality.
While the phenomenon of duality in inclusive scattering is now well
established, only preliminary experimental studies of duality exist
in semi-inclusive scattering and quantitative tests are only just
beginning.

The outgoing pion is characterized by the elasticity, $z$,
defined in terms of the target nucleon ($p$), virtual photon ($q$),
and pion ($p_\pi$) momentum four vectors,
$z = p \cdot p_\pi/p\cdot q$.
In the target rest (or laboratory) frame this becomes the
fraction of the virtual photon's energy taken away by the pion,
$z = E_\pi/\nu$.
In the elastic limit the pion carries away all of the photon's
energy, in which case $z$ = 1.
Here we will consider processes where the electroproduced pion
carries away a large fraction, but not all, of the exchanged
virtual photon's energy.

The invariant mass $W^\prime$ of the undetected hadronic system is
reconstructed from the momenta of the target nucleon, photon and
produced pion, $W^{\prime 2} = (p + q - p_\pi)^2$.
Neglecting the mass of the pion with respect to $Q^2$
\cite{ACW}, one finds
\begin{equation}
W^{\prime 2}
= W^2 - 2z\nu(M + \nu - \left| \vec{q} \right| \cos \theta_{q\pi})\ ,
\end{equation}
where $\theta_{q\pi}$ is the angle
between the virtual photon momentum $\vec q$ and the outgoing pion
momentum $\vec p_\pi$ in the laboratory frame.
As in the inclusive scattering case, the square of the total
invariant hadronic mass is given by $W^2 = M^2 + Q^2(1/x - 1)$.
If the outgoing pion is further limited to be collinear with the
virtual photon ({\em i.e.}, parallel kinematics, $\theta_{q\pi}$ = 0),
and if $Q^2/\nu^2 \ll 1$, the invariant mass $W^{\prime}$ can be
expressed in terms of $z$, $x$ and $Q^2$ as
\begin{equation}
W^{\prime 2} \approx M^2 + Q^2 (1-z) \left( \frac{1}{x}-1 \right)\ .
\end{equation}
The mass $W'$ can play a role analogous to $W$ for duality in
inclusive scattering \cite{ACW}.
In the limit of large $z$, $W^{\prime}$ will span masses in the
nucleon resonance region, which we define to be the same as that
in the inclusive scattering case, $W^{\prime 2} < 4$~GeV$^2$.
Before proceeding with the discussion of the results of early
investigations of quark-hadron duality in pion electroproduction,
we shall first define what one means by the scaling region for
such a reaction.

As implied by Fig.~\ref{fig:eepiplot}, at high energies one expects
from perturbative QCD that there will be factorization between
the virtual photon--quark interaction and the subsequent quark
hadronization into pions.
At lowest order in $\alpha_s$, the detected pion yield
${\cal N}^{\pi^{\pm}}$ then factorizes into quark distribution
functions $q(x,Q^2)$ and fragmentation functions
$D_{q \to \pi^{\pm }}(z,Q^2)$,
\begin{equation}
\label{eq:factorization}
{\cal N}^{\pi^{\pm}}(x,z,Q^2)\
\propto\ \sum_q e_q^2
	 \left[ q(x,Q^2) D_{q \to \pi^{\pm}}(z,Q^2)
	      + \bar{q}(x,Q^2) D_{\bar q \to \pi^{\pm}}(z,Q^2)
	 \right]\ ,
\end{equation}
where $D_{q \to \pi^{\pm }}(z,Q^2)$ gives the probability that
a quark of flavor $q$ hadronizes to a pion carrying a fraction $z$
of the quark (or photon) energy.
(At higher orders one also has gluon fragmentation functions,
but we shall neglect these for the purposes of this discussion.)
A consequence of this factorization is that the fragmentation
function is independent of $x$, and the quark distribution function
is independent of $z$.
Both the quark distribution and fragmentation functions, however,
depend on $Q^2$ through perturbative $Q^2$ evolution
\cite{FRAG_EVOLVE}.

The fragmentation functions parameterize how a quark involved in a
high-energy scattering reaction evolves into the detected pion.
Initial investigations of the hadronization process were made in
electron--positron annihilation and in deep inelastic scattering.
In the latter case, high energies were used to separate the pions
produced by the struck quark (termed ``current fragmentation'')
from pions originating from the spectator quark system (``target
fragmentation'') using large intervals in rapidity, $\eta$.
Rapidity is defined in terms of the produced pion energy and the
longitudinal component of the momentum (along the $\vec q$ direction),
\begin{eqnarray}
\label{eq:rapidity}
\eta &=& {1 \over 2}
	 \ln\left( { E_\pi - p^z_\pi \over E_\pi + p^z_\pi }
	    \right)\ .
\end{eqnarray}
Earlier data from CERN \cite{BERG1,BERG2} suggest that a difference
in rapidities, $\Delta\eta$, between pions produced in the current
and target fragmentation regions (``rapidity gap'') of
$\Delta\eta \approx 2$ is needed to kinematically separate the two
regions.

At lower energies, it is not obvious that the pion electroproduction
process factorizes in the same manner as in Eq.~(\ref{eq:factorization}).
We shall return to this later in Sec.~\ref{ssec:semi}, but for the
moment will simply assume that factorization holds {\it if} one can
reach a region where kinematical separation between current and
target fragmentation is possible, {\it and} one is in the DIS region,
$W^{\prime 2} > 4$~GeV$^2$.

\begin{figure}[t]
\begin{minipage}{3.1in}
\epsfig{figure=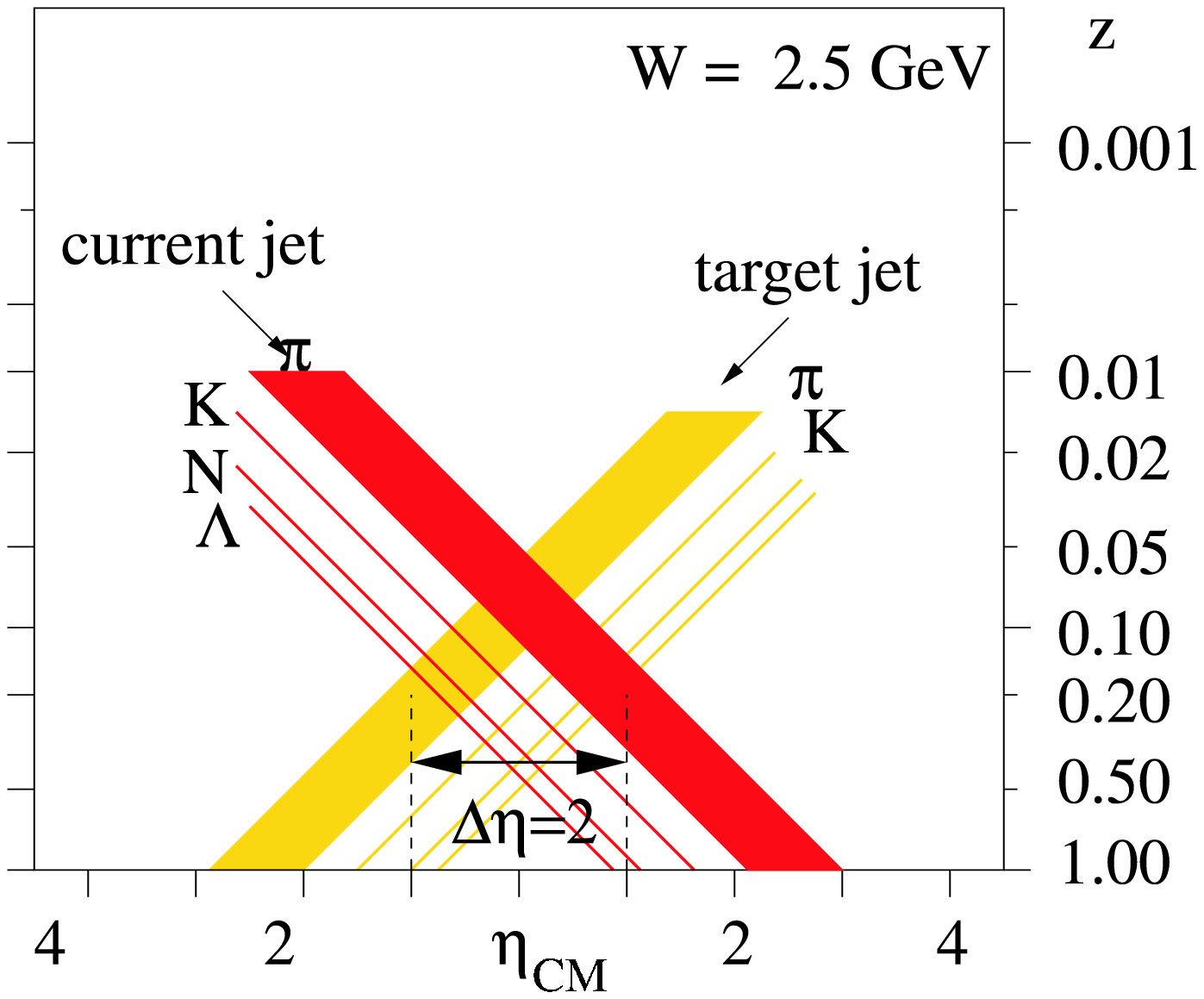, height=12cm, width=8cm}
\end{minipage}
\begin{minipage}{3.1in}
\vspace*{-0.5cm}
\hspace*{-1.5cm}
\epsfig{figure=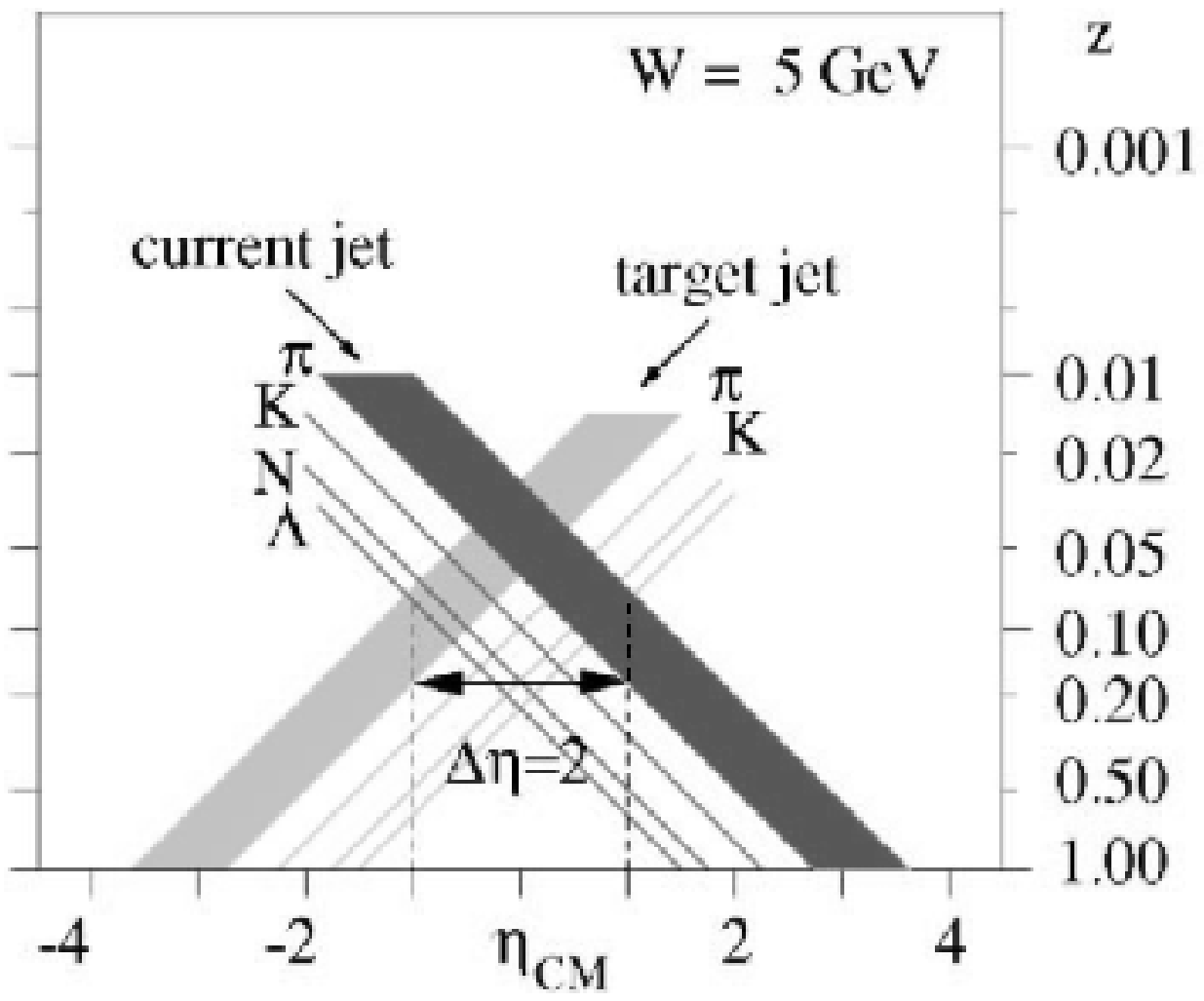, height=12.3cm, width=9cm}
\end{minipage}
\vspace*{-5cm}
\begin{centering}\caption{\label{fig:semi}
	Relation between elasticity $z$ and center of mass rapidity
	$\eta_{\rm CM}$ in semi-inclusive electroproduction of
	various hadrons for $W = 2.5$~GeV (left panel) and $W = 5$~GeV
	(right panel).
	(Adapted from Ref.~\protect\cite{MULD}.)}
\end{centering}
\end{figure}

It has been argued that such kinematic separation is possible, 
even at low $W^2$, if one considers only electroproduced
pions with large elasticity $z$, {\em i.e.}, with energies
close to the maximum energy transfer \cite{BERG2,MULD}.
Figure~\ref{fig:semi} shows a plot of rapidity versus $z$
for $W = 2.5$~GeV (left panel) and 5~GeV (right panel).
At $W = 2.5$~GeV, a rapidity gap of $\Delta \eta \geq 2$
would be obtained with $z > 0.4$ for pion electroproduction.
For larger $W$, such a rapidity gap could already be attained
at a lower value of $z$. Hence, one would anticipate a reasonable
separation between the current and target fragmentation processes
for $z > 0.4$ and $z > 0.2$ at $W = 2.5$ and 5~GeV, respectively.

In the annihilation process $e^+e^- \rightarrow h X$ \cite{SIEG,HANS}
one finds that the data beyond $z \approx 0.5$ at $W = 3$~GeV
($W^\prime = 1.94$~GeV) could be described in terms of fragmentation
functions.
The region extends to $z \geq 0.2$ for $W = 4.8$~GeV
($W^\prime = 2.84$~GeV) and to $z \geq 0.1$ for $W = 7.4$~GeV
($W^\prime = 4.14$~GeV).
For $z > 0.3$, fragmentation functions have also been obtained from
data \cite{DREWS} on $ep \rightarrow e^\prime \pi^\pm X$ at an
incident energy $E = 11.5$~GeV, with $3 < W < 4$~GeV.
All of these data are beyond the nucleon resonance region as
defined above.

At lower energies, a series of measurements of semi-inclusive pion
electroproduction was carried out at Cornell, with both hydrogen
and deuterium targets \cite{BEB75,BEB76,BEB77}, covering a region in
$Q^2$ between 1 and 4~GeV$^2$, and in $\nu$ between 2.5 and 6~GeV.
The data from these experiments were analyzed in terms of an invariant
structure function (analogous to ${\cal N}^{\pi^{\pm}}(x,z)$ in
Eq.~(\ref{eq:factorization})), written in terms of the sum of
products of parton distribution and fragmentation functions.
The authors conclude that within their region of kinematics this
invariant structure function shows no $Q^2$ dependence, and a weak
$W^2$ dependence.
This is particularly striking if one realizes that the kinematics in
these experiments covered a region in $W^2$ between 4 and 10~GeV$^2$,
and in $z$ between 0.1 and 1.
In fact, for a portion of the kinematics one is in the region
$M^2 < W'^2 < 4$~GeV$^2$, right within the nucleon resonance region.
Nonetheless, the data were surprisingly found to exhibit scaling
\cite{CALOG}.

Up to now we have neglected the dependence of measured pion yields,
as in Eq.~(\ref{eq:factorization}), on the pion transverse momentum,
$p_T$.
At high energies the dependence on $p_T$ is approximately given by
$\exp(-b p_T^2$), where $b$ reflects the average transverse momentum
of the struck quark.
At lower energies, the measured $p_T$ dependence will reflect the
decay angular distributions of the electroproduced resonances in
regions where these resonances dominate.
One would expect therefore the $p_T$ dependence to vary at low
$W^\prime$. We will come back to this in Sec.~\ref{ssec:semi}.
In the Cornell data at relatively low $W^\prime$, however,
the dependence of the measured cross sections on $p_T$
(which was only low, $< 0.5$~GeV, in these data) was found to be,
within the experimental uncertainties, independent of kinematics
\cite{CALOG}.

Empirical evidence of factorization (independence of the $x$ and $z$
distributions) in pion electroproduction at even lower energies is
apparent in the results of several test runs made in Hall~C
at Jefferson Lab \cite{E00108}. The data included measurements of
semi-inclusive pion electroproduction on $^1$H and $^2$H targets
at relatively low energy, $\nu = 3.9$~GeV, with $W^2 = 5.9$~GeV$^2$,
and $Q^2 = 2.4$~GeV$^2$.
Under the assumptions of factorization, as in
Eq.~(\ref{eq:factorization}), charge conjugation invariance,
isospin symmetry, and neglecting nuclear corrections,
the use of charged pion yields on both targets allows for the
extraction of the ratio of valence $d$ to $u$ quark
distributions in the proton, $d_v/u_v$.

\begin{figure}
\hspace*{-0.5cm}
\epsfig{figure=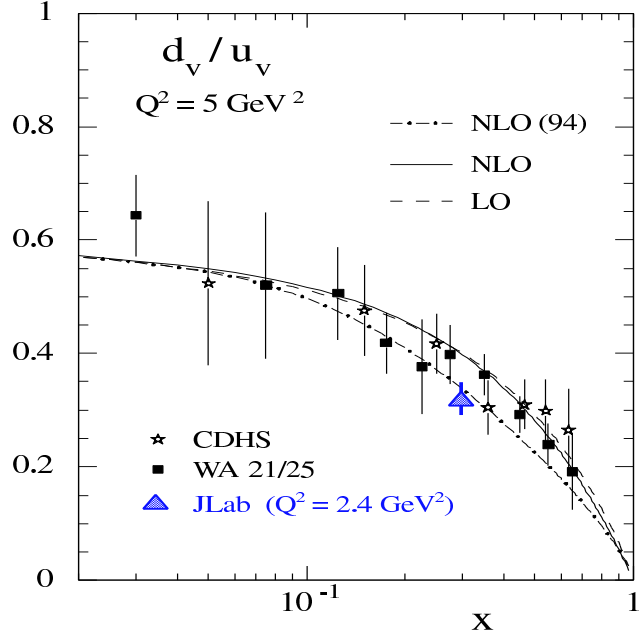, height=13cm, width=13cm}
\vspace*{-4cm}
\caption{\label{fig:grvduplot}
	The ratio $d_v/u_v$ from several high-energy experiments
	(CDHS at Fermilab and WA~21/25 at CERN),
	together with a single point (triangle) extracted from a
	low-energy measurement at Jefferson Lab \protect\cite{E00108}.
	The curves represent various global fits to the data.
	(Adapted from Ref.~\protect\cite{GRV98}.)}
\end{figure}

The single Jefferson Lab point is plotted in Fig.~\ref{fig:grvduplot}
together with a collection of data from deep inelastic neutrino
measurements at CERN and Fermilab at energies of several hundred
GeV \cite{GRV98}.
As the data were obtained at an elasticity $z > 0.5$, it may not be
surprising that reasonable agreement is found at these vastly
different kinematics, even though the $W^\prime$ of the Jefferson Lab
data is in the nucleon resonance region,
$2.0 < W^{\prime 2} < 3.3$~GeV$^2$.
In the kinematics plot of Fig.~\ref{fig:semi} (left panel), one
would anticipate factorization to work reasonably well for $z > 0.5$,
whereas the experimental data show hardly any resonance structure
at $W'^2 > 2$~GeV$^2$, as for the Cornell data
\cite{BEB75,BEB76,BEB77} described above.
Duality may follow simply from the fact that one cannot clearly
distinguish the resonance and scaling regions, and from the existence
of such low-energy factorization \cite{CI,EHK}.

It is important to stress that the existence of quark-hadron duality
does not imply that the reaction can be described by perturbative QCD
alone.
As in the inclusive DIS case, where parton distribution functions
parameterize the soft, nonperturbative nucleon structure, so too in
semi-inclusive meson electroproduction one parameterizes the soft
hadronization process in terms of fragmentation functions.
In the exclusive limit, if the total center of mass energy $W$
is much larger than $W^\prime$ (ensuring the large rapidity gap),
and the momentum transfer is sufficiently large so that the
electroproduced pion does not reinteract with the target,
then there is a one-to-one correspondence \cite{EHK,HOY02}
between the semi-inclusive process under investigation and
ordinary deep inelastic scattering.

The close analogy with DIS suggests that semi-inclusive processes
may also exhibit quark-hadron duality \cite{ACW,BLA74,SCO74,SCO75}.
It has been argued \cite{EHK,HOY02} that for $W^\prime = M$, the fully
exclusive limit, quark-hadron duality predicts the energy dependence
observed in the $\gamma p \to \pi^+ n$ and $\gamma p \to \gamma p$
(Compton scattering) data, but that the absolute normalization is off
by one to two orders of magnitude (see Sec.~\ref{sssec:excl_pi} below).
This may indicate that although the elementary single quark scattering
process also dominates the low-energy semi-inclusive reaction, even in
the nucleon resonance region, the assumption of no final state
interactions with the target is not yet valid.
Hence, both scaling and factorization should rather be viewed as
{\em precocious}.

\begin{figure}
\begin{minipage}[ht]{4.0in}
\hspace*{1.0cm}
\epsfig{figure=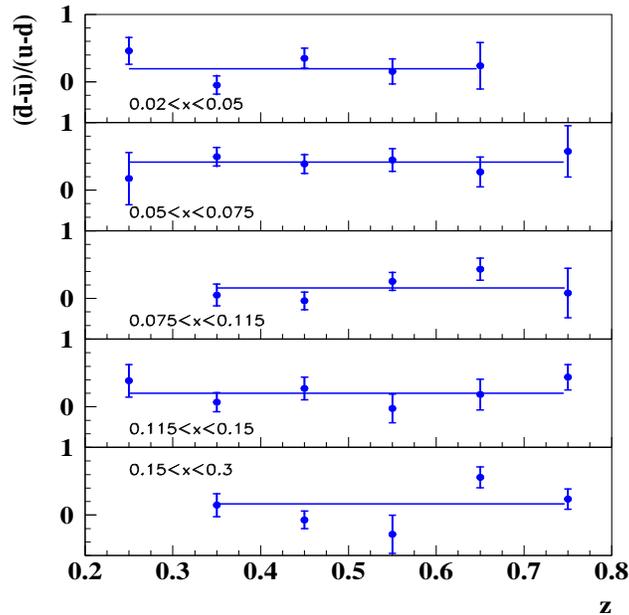, height=9cm, width=9cm}
\end{minipage}
\begin{centering}
\vspace*{0.5cm}
\caption{\label{fig:hermfact}
	The distribution ($\bar d - \bar u$)/($u - d$) as a function
	of $z$ in five $x$ bins, from the HERMES Collaboration
	\protect\cite{ACK98}.
	The points are fitted to a constant for each $x$ bin.
	The error bars represent statistical and systematic
	uncertainties added in quadrature.}
\end{centering}
\end{figure}

Recently there have been several experimental investigations of
factorization in semi-inclusive pion electroproduction in the
DIS region.
The HERMES Collaboration tested factorization for the light sea
quark distributions $\bar d$ and $\bar u$, extracted from
semi-inclusive pion electroproduction data on proton and deuteron
targets, over the range $13 < \nu < 19$~GeV and
$21 < W^2 < 35$~GeV$^2$, with an average $Q^2 = 2.3$~GeV$^2$.
The results were obtained in a region where the requirement of a
rapidity gap of $\Delta \eta \geq$ 2 was only valid for $z > 0.2$.
Within limited statistics, the data on the ratio
($\bar d - \bar u$)/($u - d$) were found to be consistent with the
factorization assumption.
The results are shown in Fig.~\ref{fig:hermfact}, as a function of $z$,
for five bins in $x$.
The values of ($\bar d - \bar u$)/($u - d$), averaged over $z$,
can be recast in the form of the absolute difference $\bar d - \bar u$
assuming a particular parametrization of the $u$ and $d$
distribution functions.
%The results shown in Fig.~\ref{fig:hermsea} are obtained from the
%ratio in Fig.~\ref{fig:hermfact} using the leading order parton
%distribution functions from Ref.~\cite{GLU95}.

An alternate method of measuring the flavor asymmetry
$\bar d - \bar u$ is via the
Drell-Yan reaction on protons and deuterons, $pp(d) \to \mu^+\mu^- X$,
which is directly sensitive to the ratio $\bar u/\bar d$.
The Fermilab E866 Collaboration measured $\bar u/\bar d$ at an
average $Q^2 = 54$~GeV$^2$ \cite{E866}, and using the CTEQ4M
\cite{CTEQ97} parameterization of $\bar u + \bar d$, converted the
ratio into the difference $\bar d - \bar u$.
%, shown in Fig.~\ref{fig:hermsea}.
Despite a factor of $\sim 20$ difference in $Q^2$ between
the two experiments, and the different experimental technique,
the data were found to be in remarkable agreement.
This suggests that the factorization assumption may be valid
at energies accessible at HERMES.

Finally, some preliminary results have just become available from
the E00-108 experiment, recently performed in Hall C at Jefferson Lab
to explicitly study duality in pion electroproduction \cite{E00108}.
A 5.5~GeV electron beam was used to study pion electroproduction off
proton and deuteron targets for $Q^2$ between 1.8 and 6.0 GeV$^2$,
for $0.3 \le x \le 0.55$, and with $z$ in the range 0.35--1.

\begin{figure}[htb]
\begin{center}
\epsfig{figure=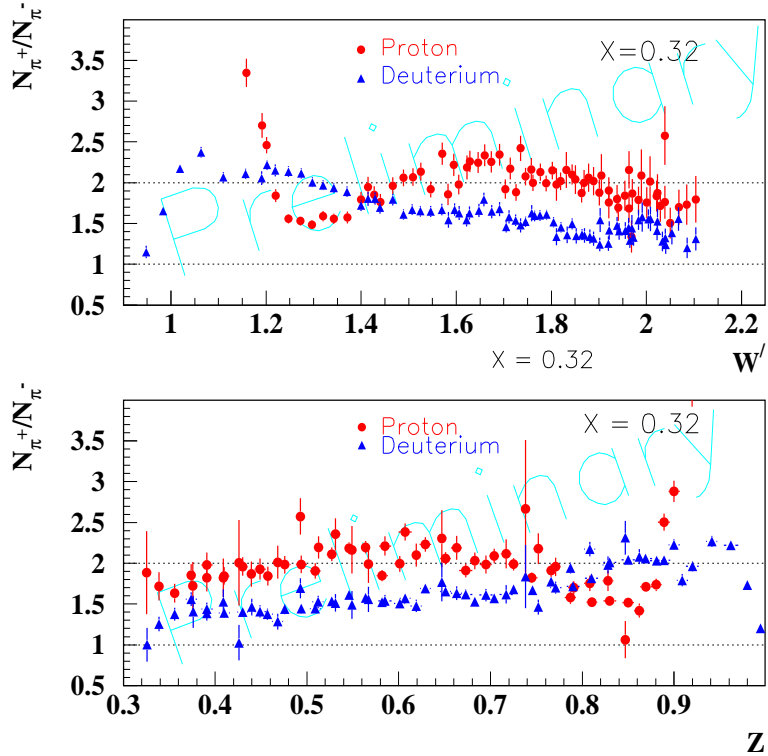, height=10cm, width=10cm}
\vspace*{0.5cm}
\caption{\label{fig:piratio}
	The ratio of $\pi^+$ to $\pi^-$ yields off proton and
	deuteron targets as a function of $W^\prime$ (upper panel)
	and $z$ (lower panel), at $x = 0.32$.}
\end{center}
\end{figure}

The preliminary results for the ratio of $\pi^+$ to $\pi^-$ cross
sections for both proton and deuteron targets are shown in
Fig.~\ref{fig:piratio} as a function of $W^\prime$ (upper panel)
and $z$ (lower panel) at $x = 0.32$.
At the present stage of analysis the $\pi^\pm$ cross sections are
known to 20\%, although the final analysis will render results
to better than 5\% accuracy.
Due to the choice of kinematics in this experiment, $z > 0.7$
directly corresponds to $W^\prime <$ 1.6 GeV since $Q^2$ and $x$
are kept nearly constant.

The behavior in the $\pi^+/\pi^-$ ratio for the proton spectrum
around $z = 0.85$ simply corresponds to the behavior of this ratio
in the $\Delta$ region (see Sec.~\ref{ssec:semi}).
Indeed, if one considers the $^1{\rm H} (e,e^\prime \pi^-) X$
spectrum as a function of missing mass of the residual system $X$,
one only sees one prominent resonance around the mass of the $\Delta$.
Apparently, above $W^\prime$ = 1.6 GeV there are already sufficient
resonances in the missing mass spectra of $^1$H(e,e$^\prime \pi^\pm$)$X$
that the $\pi^+/\pi^-$ ratio does not show obvious resonance structure
and is nearly flat as a function of $W^\prime$ (or, equivalently, $z$).
The $\pi^+/\pi^-$ ratio off deuterium shows a slightly steeper rise
with decreasing $W^\prime$ in this region.

\begin{figure}[htb]
\begin{center}
\hspace*{-2cm}
\epsfig{figure=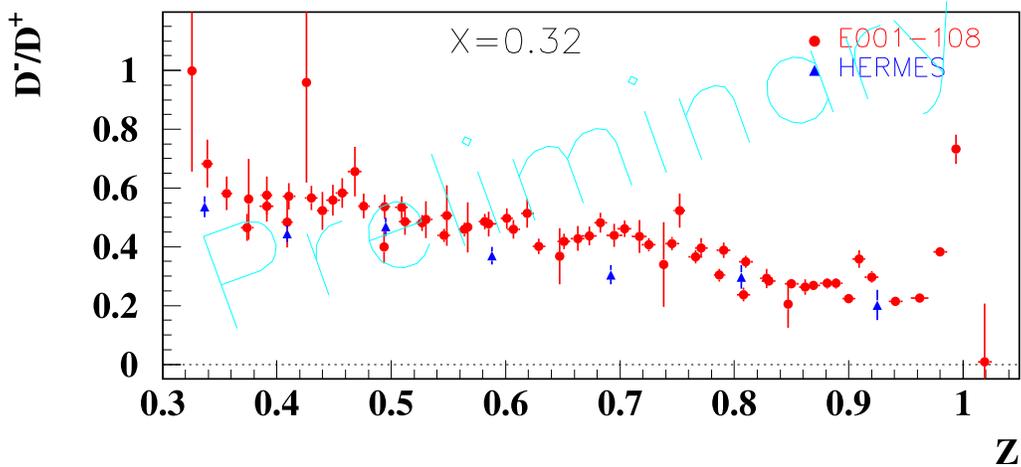, height=19cm}
\vspace*{-12cm}
\caption{\label{fig:dminusdplus}
	The ratio of unfavored to favored fragmentation function
	$D^-/D^+$ (at $z = 0.55$) as a function of $x$, using only
	deuterium data.}
\end{center}
\end{figure}

Using the deuterium data only, we can extract the ratio of unfavored
to favored fragmentation functions $D^-/D^+$.
Here the favored fragmentation function ($D^+$) corresponds to a pion
which contains the struck quark ({\em e.g.}, a $\pi^+$ after a $u$ or
$\bar d$ quark is struck), while the unfavored fragmentation function
($D^-$) describes the fragmentation of a quark not contained in the
valence structure of the pion ({\em e.g.} a $d$ quark for the $\pi^+$).
This ratio is, to a good approximation, simply given by 
$D^-/D^+ = (4 - {\cal N^{\pi^+}}/{\cal N^{\pi^-}})/
           (4{\cal N^{\pi^+}}/{\cal N^{\pi^-}} - 1)$.
The preliminary results are shown in Fig.~\ref{fig:dminusdplus} in
comparison with data from the HERMES experiment \cite{GEIGERPHD}.

The first observation to draw is that the $D^-/D^+$ ratio extracted
from the JLab data shows a smooth slope as a function of $z$.
This simply reflects the smooth rise in the $\pi^+/\pi^-$ ratio off
deuterium seen in Fig.~\ref{fig:piratio}.
This is quite remarkable given that the data cover the full
resonance region, $0.88 < W^{\prime 2} < 4.2$~GeV$^2$.
Apparently, there is some mechanism at work that removes the
resonance excitations in the $\pi^+/\pi^-$ ratio, and hence the
$D^-/D^+$ ratio.
We will discuss a possible mechanism in Sec.~\ref{ssec:semi},
but here will simply mention that this behavior is consistent
with the expectations of an onset of duality.
The second observation is that the behavior as a function of $z$
of $D^-/D^+$ measured by E00-108 in the nucleon resonance region 
closely resembles the behavior seen in the HERMES experiment
\cite{GEIGERPHD}.
The $D^-/D^+$ ratio measured by E00-108 seems slightly larger than
the HERMES ratio, however, it is premature to draw a final conclusion
from this in view of the preliminary state of the E00-108 analysis.

It would be interesting to experimentally verify the $p_T$ dependence
of the constructed $D^-/D^+$ ratio.
We argued earlier that the $p_T$ dependence of pion yields at low
$W^\prime$ is expected to reflect the decay angular distributions
of the electroproduced resonances. 
If duality is valid, however, the $p_T$ dependence of the $D^-/D^+$
ratio shown in Fig.~\ref{fig:piratio} should be similar to the
dependence found at higher energies, even though the ratio was
constructed from pion yields solely in the resonance region.

In summary, there exist strong hints in electro-pion production data
that quark-hadron duality extends to semi-inclusive scattering.
To convert these hints into conclusive evidence requires a next series
of precision semi-inclusive experiments encompassing both
the nucleon resonance and deep inelastic regions.
Among the first of them is the E00-108 experiment at JLab which is
currently being analyzed. It has been argued that the existence of
quark-hadron duality in electro-pion production may be linked to
low-energy factorization: the empirical observation of the
independence of $x$ and $z$ distributions when selecting a pion
that carries most of the energy transfer.
The data shown are in agreement with this postulate.

If the existence of quark-hadron duality in electro-pion production
and the applicability of factorization at lower energy are indeed
established experimentally \cite{E00108}, the spin and flavor
dependence of duality can be investigated in detail.
More pragmatically, confirmation of duality in inclusive hadron
production would open the way to an enormously rich semi-inclusive
program in the preasymptotic regime, allowing unprecedented access
to the spin and flavor distributions of the nucleon, especially at
large $x$.
We will discuss semi-inclusive scattering in some more detail from a
theoretical perspective in Sec.~\ref{ssec:semi}, and outline future
challenges for semi-inclusive duality in Sec.~\ref{ssec:pion_outlook}.

\clearpage

%%%%%%%%%%%%%%%%%%%%%%%%%%%%%%%%%%%%%%%%%%%%%%%%%%%%%%%%%%%%%%%%%%%%%%%%%
\newpage
\section{Theoretical Foundations}
\label{sec:thy}

After reviewing the experimental status of duality in electron--nucleon
scattering, in this section we discuss various theoretical approaches
which have been developed in an effort to understand its origin.
Because the discovery of Bloom-Gilman duality preceded QCD, much of
the early work was phenomenological or based on effective hadronic
descriptions.
We shall discuss some of these in the descriptions of models of
duality in Sec.~\ref{ssec:models} below.
To begin with, however, we present what has become the standard
framework for discussing duality in the context of QCD --- namely,
the operator product, or twist, expansion.
Following this, we will demonstrate the power of duality in
phenomenological applications in inclusive scattering,
in semi-inclusive processes, and in exclusive reactions.

% -----------------------------------------------------------------------
\subsection{QCD and the Twist Expansion}
\label{ssec:qcd}

The theoretical basis for describing Bloom-Gilman duality in QCD is
the operator product expansion (OPE) of Wilson \cite{OPE} and others
\cite{EONacht,BRANDT}, the main elements of which we shall briefly
review here.
The quantities most directly amenable to a QCD analysis are the
moments of structure functions.
An important point to realize is that the OPE analysis is
intrinsically perturbative, in the sense that a systematic expansion
of structure function moments is performed in terms of inverse powers
of a hard scale, $Q^2$.
The essence of the OPE is that it enables one to isolate the ``soft'',
nonperturbative physics contained in parton correlation functions,
from the ``hard'' scattering of the probe from the partons.

% .......................................................................
\subsubsection{The OPE, Resonances and Duality}
\label{sssec:ope}

According to the OPE, at large $Q^2 \gg \Lambda_{\rm QCD}^2$ the
moments of the structure functions can be expanded in powers of
$1/Q^2$ \cite{OPE}.
The coefficients in the expansion are matrix elements of quark \&
gluon operators corresponding to a certain {\em twist}, $\tau$,
defined as the mass dimension minus the spin, $n$, of the operator.
For the $n$-th moment of the $F_2$ structure function, $M_2^{(n)}$,
for example (see Eq.~(\ref{eq:MnDEF})), one has the expansion
\begin{eqnarray}
M_2^{(n)}(Q^2) &=&
\sum_{\tau=2,4\ldots}^\infty
{ A_\tau^{(n)}(\alpha_s(Q^2)) \over Q^{\tau-2} }\ ,\ \ \ \
n = 2, 4, 6 \ldots
\label{eq:MnOPE}
\end{eqnarray}
where $A_\tau^{(n)}$ are the matrix elements with twist $\leq \tau$.
Note that because of the crossing symmetry properties of the $F_2$
structure function under $\nu \to -\nu$ (or equivalently $x \to -x$),
under which $F_2$ is even, the OPE expansion for $M_2^{(n)}$
is defined for positive, even integers $n$.
As the argument suggests, the $Q^2$ dependence of the matrix elements
can be calculated perturbatively, with $A_\tau^{(n)}$ expressed as a
power series in $\alpha_s(Q^2)$.

Similarly, for the spin-dependent $g_1$ structure function,
the twist expansion for the $n$-th moment $\Gamma_1^{(n)}$
(see Eq.~(\ref{eq:GnDEF})) becomes
\begin{eqnarray}
\Gamma_1^{(n)}(Q^2) &=&
\sum_{\tau=2,4\ldots}^\infty
{ \mu_\tau^{(n)}(\alpha_s(Q^2)) \over Q^{\tau-2} }\ ,\ \ \ \ 
n = 1, 3, 5 \ldots
\label{eq:GnOPE}
\end{eqnarray}
Here once again the coefficients $\mu_\tau^{(n)}$ represent matrix
elements of operators with twist $\leq \tau$, and since the $g_1$
structure function is odd under the interchange $x \to -x$, its
moments are related to matrix elements of local operators for
positive, odd integers $n$.

Asymptotically, as $Q^2 \to \infty$ the leading-twist ($\tau=2$) terms
in the expansions (\ref{eq:MnOPE}) and (\ref{eq:GnOPE}) dominate the
moments.
In the absence of perturbatively generated corrections, these give rise
to the $Q^2$ independence of the structure function moments, and hence
are responsible for scaling.
For the spin-averaged moments, the coefficients $A_2^{(n)}$ of the
twist-2 terms in Eq.~(\ref{eq:MnOPE}) are given in terms of matrix
elements of spin-$n$ operators,
\begin{eqnarray}
A_2^{(n)}\ P^{\mu_1} \cdots P^{\mu_n}
&=& \langle P |
    \bar\psi\ \gamma^{\{\mu_1}\ iD^{\mu_2} \cdots\, iD^{\mu_n\}} \psi\
    | P \rangle\ ,
\label{eq:F2LTop}
\end{eqnarray}
where $P$ is the nucleon momentum, $\psi$ is the quark field,
$D^\mu$ is the covariant derivative, and the braces $\{ \cdots \}$
denote symmetrization of indices and subtraction of traces.
Similarly, for the moments of the spin-dependent $g_1$ structure
the leading-twist contributions are given by
\begin{eqnarray}
\mu_2^{(n)}\ S^{\mu_1} P^{\mu_2} \cdots P^{\mu_n}
&=& \langle P, S |
    \bar\psi\ \gamma^{\mu_1}\,\gamma_5\, iD^{\{\mu_2}
	\cdots\, iD^{\mu_n\}} \psi\
    | P, S \rangle\ ,
\label{eq:g1LTop}
\end{eqnarray}
where $S^\mu$ is the spin vector of the nucleon.

\begin{figure}[ht]
\begin{center}
\epsfig{file=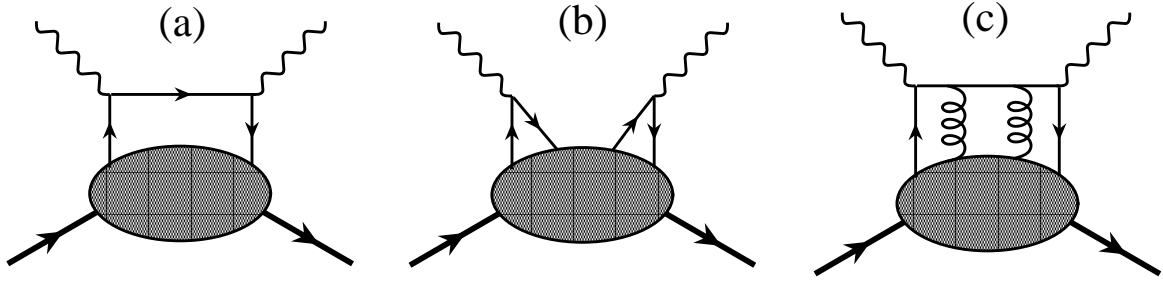,height=12cm}
\vspace*{-7.5cm}
\caption{\label{fig:diag}
	(a)~Leading-twist (``handbag diagram'') contribution to the
	structure function.
	(b)~Higher-twist (``cat's ears'') four-quark contributions.
	(c)~Higher-twist quark-gluon interactions.}
\end{center}
\end{figure}

The leading-twist terms correspond to diagrams such as in
Fig.~\ref{fig:diag}~(a), in which the virtual photon scatters
incoherently from a single parton.
The higher-twist terms in Eqs.~(\ref{eq:MnOPE}) and (\ref{eq:GnOPE})
are proportional to higher powers of $1/Q^2$ with coefficients
which are matrix elements of local operators involving multi-quark
or quark-gluon fields (see Sec.~\ref{sssec:twist} below).
Diagrammatically, these correspond to processes such as those
depicted in Fig.~\ref{fig:diag}~(b) and (c).

The relation between the higher-twist matrix elements and duality
in electron scattering was elucidated in the classic work of
De~R\'ujula, Georgi and Politzer \cite{DGP2,GP}, who reformulated
the empirical observations of Bloom and Gilman in terms of the twist
expansion of the structure function moments in QCD.
The connection follows almost immediately from the definition of the
moment expansions in Eqs.~(\ref{eq:MnOPE}) and (\ref{eq:GnOPE}).
For the $F_2$ structure function, the lowest moment, $M_2^{(2)}$,
corresponds precisely to the Bloom-Gilman integral in
Eq.~(\ref{eq:BGFESR}).
At low $Q^2$ the moments display strong $Q^2$ dependence,
violating both scaling and duality.
In the OPE language this violation is associated with large
corrections from the subleading, $1/Q^{\tau-2}$ higher-twist
terms in Eqs.~(\ref{eq:MnOPE}) and (\ref{eq:GnOPE}).

At larger $Q^2$ the moments become independent of $Q^2$, as they
would if they were given entirely by the scaling contribution.
According to Eqs.~(\ref{eq:MnOPE}) and (\ref{eq:GnOPE}), this duality
can only occur if the higher-twist contributions are either small or
cancel.
Duality is synonymous, therefore, with the suppression of higher
twists, which in partonic language corresponds to the suppression
of interactions between the scattered quark and the spectator system,
as illustrated in Fig.~\ref{fig:diag}~(b) and (c).
In other words, suppression of final state interactions is a
prerequisite for the existence of duality.

Taking the $F_2$ structure function for illustration, the appearance
of duality implies that the moment $M_2^{(n)}$ is dominated, even at
low $Q^2$, by the leading term, $A_2^{(n)}$.
The higher-twist contributions can be defined as the difference between
the total and leading-twist moments,
\begin{eqnarray}
\Delta M_2^{(n)}(Q^2)
&\equiv& M_2^{(n)}(Q^2) - A_2^{(n)}(Q^2)\
 =\ { A_4^{(n)}(Q^2) \over Q^2 }\
 +\ {\cal O}(1/Q^4)\ .
\label{eq:delM}
\end{eqnarray}
Of course, the suppression or cancellation of the higher-twist terms
cannot occur for all moments $n$ --- since higher moments weight
the large-$x$ region more than lower moments, this would require
identical $x$ distributions for the resonance and scaling functions,
which clearly cannot occur at any finite $Q^2$.
For higher $n$ one expects duality and the onset of scaling to occur
at relatively higher $Q^2$, with $n M_0^2 \sim Q^2$, where $M_0$ is
some mass scale, of the order of the transverse momentum of quarks
in the nucleon ($\sim 500$~MeV) \cite{DGP2}.

As elaborated by De~R\'ujula {\em et al.} \cite{DGP2}, and
subsequently by Ji \& Unrau \cite{JI_F2}, there should exist a region
of $n$ and $Q^2$ where the higher-twist contributions become important
but remain perturbative, in the sense that a twist expansion exists in
which they can be isolated.
In this region the structure function still exhibits the prominent
resonances, which organize themselves to approximately follow, on
average, the deep inelastic scaling function.
Here the physics can be described in terms of either resonance
production, or in terms of scattering from partons \cite{JI_F2}.
Note that since the resonances are bound states of quarks and gluons,
they necessarily involve (an infinite number of) higher twists.
The low-lying resonances contribute significantly to the low moments,
but the overall size of the higher twists is not overwhelming.
The extent to which Bloom-Gilman duality holds then reflects the size
of this region.
For higher moments $n M_0^2 \agt Q^2$, the higher-twist terms become
more important than the leading-twist, so that the twist expansion
diverges, and the OPE ceases to be reliable as a means of organizing
the different structure function contributions.

\begin{figure}[ht]
\begin{center}  
\epsfig{file=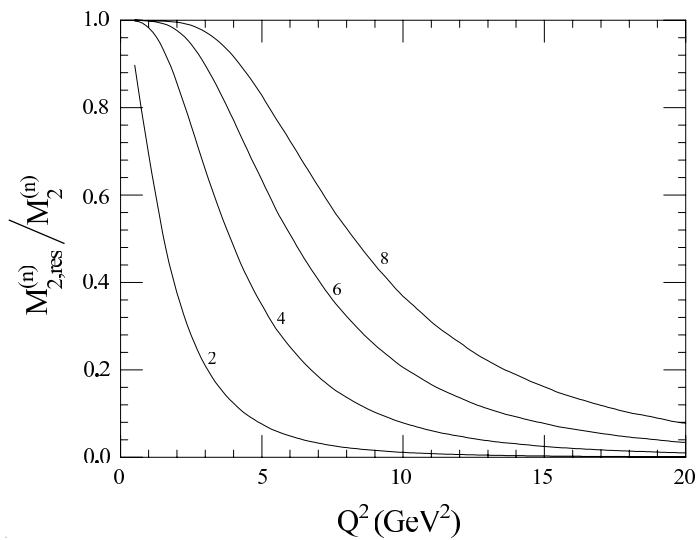,height=7cm}
\vspace*{0.5cm}
\caption{\label{fig:f2mom}
	Ratio of the $n=2, 4, 6$ and 8 moments of the proton $F_2^p$
	structure function from the resonance region, $W < 2$~GeV,
	to the total.
	(Adapted from Ref.~\protect\cite{JI_F2}.)}
\end{center}
\end{figure}

\begin{figure}[h]
\begin{center}
\hspace{-0.5cm}
\epsfig{file=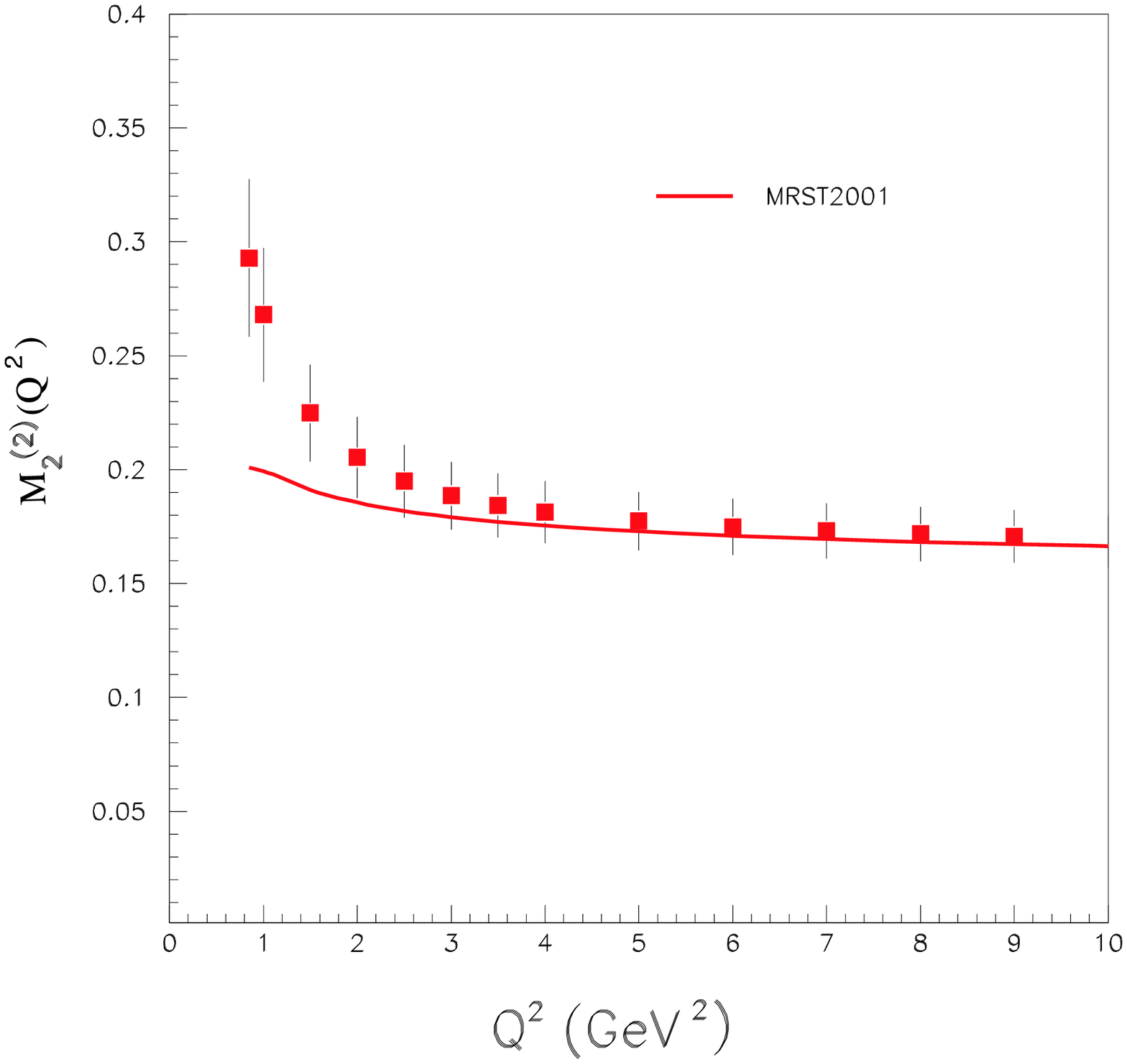,height=10.0cm}
\vspace*{-1cm}
\caption{\label{fig:f2ht}
	Total $n=2$ moment of the proton $F_2^p$ structure function
	(squares) and the leading-twist contribution (solid line)
	\protect\cite{ericpc}.}
\end{center}
\end{figure}

The interplay between resonances and higher twists can be dramatically
illustrated by considering the contribution from the resonance region,
traditionally defined (as in Sec.~\ref{sec:bgstatus}) as that restricted
to $W < W_{\rm res} = 2$~GeV, to moments of the structure function as a
function of $Q^2$.
For the proton $F_2^p$ structure function the resonance contribution to
the moment $M_2^{(n)}$ is given by
\begin{eqnarray}
M_{2, {\rm res}}^{(n)}(Q^2)
&=& \int_{x_{\rm res}}^1 dx\ x^{n-2}\ F_2(x,Q^2)\ ,
\end{eqnarray}
where $x_{\rm res} = Q^2 / (W_{\rm res}^2 - M^2 + Q^2)$.
The ratios of the resonance contributions $M_{2, {\rm res}}^{(n)}$
to the total moments is illustrated in Fig.~\ref{fig:f2mom} for the
$n=2, 4, 6$ and 8 moments.
At $Q^2 = 1$~GeV$^2$ approximately 70\% of the cross section integral
(or the $n=2$ moment) comes from the resonance region,
$W < W_{\rm res}$.
Despite this large resonant contribution, the resonances and the
deep inelastic continuum conspire to produce only about a 10--15\%
higher-twist correction at the same $Q^2$, as Fig.~\ref{fig:f2ht}
demonstrates.
Here the total $M_2^{(2)}$ moment from recent proton measurements
in Hall~C at Jefferson Lab \cite{ericpc} is plotted as a function
of $Q^2$, together with the leading-twist contribution calculated
from the PDF parameterization of Ref.~\cite{MRST}.
Remarkably, even though each bound state resonance must be built up
from a multitude of twists, when combined the resonances interfere in
such a way that they closely resemble the leading-twist component!

\begin{figure}[ht]
\begin{center}  
\hspace*{-0.5cm}
\epsfig{file=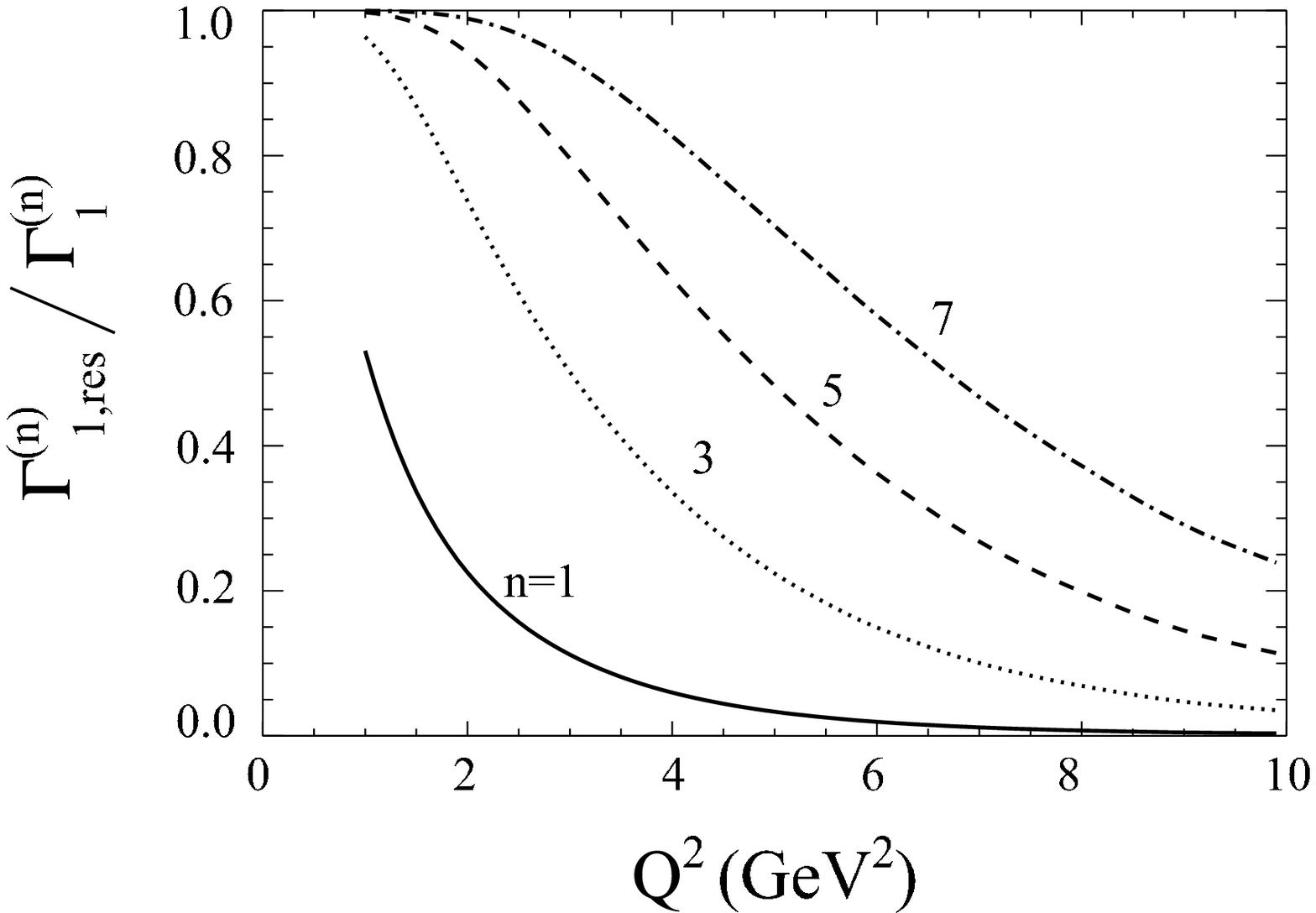,height=6.5cm}
\vspace*{0.5cm}
\caption{\label{fig:g1mom}
	Contribution to the moments $\Gamma_1^{(n)}$ from low mass
	excitations with $W < W_{\rm res}$ relative to the full
	moments, for $n=1, 3, 5$ and 7.
	From Ref.~\protect\cite{EDELMANN}.}
\end{center}
\end{figure}

A similar duality can be studied in polarized structure functions.
Defining the resonance contribution to the $n$-th moment of the $g_1$
structure function as
\begin{eqnarray}
\Gamma_{1, {\rm res}}^{(n)}(Q^2)
&=& \int_{x_{\rm res}}^1 dx\ x^{n-1}\ g_1(x,Q^2), \
\end{eqnarray}
Edelmann {\em et al.} \cite{EDELMANN} considered the ratio of
$\Gamma_{1, {\rm res}}^{(n)}$ to the total $n$-th moment for the
proton, displayed in Fig.~\ref{fig:g1mom} for $n=1$, 3, 5 and 7.
The resonance contributions to the moments $\Gamma_1^{(n)}$ are quite
sizable, especially for higher moments.
At $Q^2 = 1$ GeV$^2$ they are responsible for about $50\%$ of the first
moment, $\Gamma_1^{(1)}$, and essentially saturate the higher moments.
At large $Q^2$ the continuum contributions with $W > W_{\rm res}$ take
over.

On the other hand, the interference between the resonances and the deep
inelastic continuum leads to relatively small higher-twist corrections,
which, as for the $F_2$ structure function, can be defined as the
difference between the total and leading-twist contribution,
\begin{eqnarray}
\Delta\Gamma_1^{(n)}(Q^2)
&\equiv& \Gamma_1^{(n)}(Q^2) - \mu_2^{(n)}(Q^2)\
 =\ { \mu_4^{(n)}(Q^2) \over Q^2 }\
 +\ {\cal O}(1/Q^4)\ .
\label{eq:delGam}
\end{eqnarray}
Because the spin-dependent structure functions do not need to be
positive, however, the cancellations here will not be as dramatic
as in the unpolarized case.
For the $n=1$ moment of $g_1^p$, for instance, the higher-twist
contribution is now about a 20--25\% of the total
(see Fig.~\ref{fig:gam} below).
Similarly for the higher moments, the rather large resonant
contributions nonetheless integrate to produce a relatively
small higher-twist correction.
At $Q^2=2$~GeV$^2$ one finds that the $n=3$ moment is composed of
$\sim 75\%$ resonances, whereas only $\sim 40\%$ of the moment comes
from higher twists \cite{EDELMANN}.

These simple comparisons illustrate the important point that the
often quoted distinction between the resonance and deep inelastic
regions is, in fact, artificial.
In reality, resonances are an integral part of the nucleon structure
function, and can never be avoided in a moment analysis.
In other words, if scaling is defined as the $Q^2$ independence of
structure function moments, then the resonance region qualifies as
a scaling regime!
This will be discussed in greater detail in Sec.~\ref{sssec:conf}
within specific resonant models of structure functions.

% .......................................................................
\subsubsection{Physics of Higher Twists}
\label{sssec:twist}

In the context of global analyses of parton distributions,
higher-twist effects are often seen as unwelcome complications.
On the other hand, higher twists contain valuable information on
nucleon structure -- no less fundamental than that contained in
leading twists -- and are therefore of tremendous interest in their
own right.

The previous discussion points to an important practical consequence
of the duality between resonances and scaling in deep inelastic
scattering.
Namely, if one knows the size of the higher-twist contributions,
either experimentally or theoretically ({\em e.g.}, from lattice QCD
calculations), one can use this to extract the properties of the
resonances.
Conversely, from data on the structure function in the resonance
region one can extract matrix elements of higher-twist operators.
This use of duality is well known in QCD sum rule calculations,
as well as in applications to other physical processes
(see Sec.~\ref{sec:related}).
In this section we examine the physics content of the higher-twist
matrix elements, and illustrate how duality can be used to obtain
information about higher-twist effects in the nucleon from structure
function data in the intermediate-$Q^2$ region.

Before proceeding, we should point out that since the perturbative
expansion in $\alpha_s$ is expected to be divergent (the coefficients
of the higher-order terms grow like $n!$), and the separation of the
radiative (logarithmic) and twist (power) expansions in $Q^2$ may be
ambiguous \cite{MUELLER}.
The uncertainty in regularizing the divergent series is closely
related to the precise definition of the higher-twist contributions.
The practical solution adopted in many higher-twist analyses
\cite{JI_F2,EDELMANN,JI_G1,JM} is to utilize the available
perturbative calculations up to a given order, and define the
leading-twist contribution to that order.
As long as one works in a region of $Q^2$ where the first higher-twist
term is much larger than the smallest term in the perturbative
expansion (so that adding higher-order terms in $\alpha_s$ will not
improve the accuracy of the expansion), the ambiguity in defining the
higher-twist corrections can be neglected.

Taking the experimental moments of the proton $F_2^p$ structure
function, Ji and Unrau \cite{JI_F2} showed how information on the
coefficients of the twist-4 operators can be extracted from data at
intermediate $Q^2$.
More recent higher-twist analyses of $F_2^p$ moments measured at JLab
were made by Armstrong {\em et al.} \cite{MOMENTS} and Osipenko
{\em et al.} \cite{OSIPENKO}.
Any such extraction relies on knowledge of the twist-2 contributions,
including higher-order corrections in $\alpha_s$, which must be
subtracted from the total.
For the $n=2$ moment of $F_2$, the leading-twist contribution
corresponds to the momentum carried by quarks,
\begin{eqnarray}
A_2^{(2)}(Q^2)
&=& \sum_q e_q^2
    \int_0^1 dx\ x \left( q(x,Q^2) + \bar q(x,Q^2) \right)\ ,
\end{eqnarray}
with the sub-leading $1/Q^2$ correction given by
\begin{eqnarray}
A_4^{(2)}(Q^2)
&=& M^2
    \left( {1\over 2} A_{4,{\rm TMC}}^{(2)}(Q^2)
	 + \widetilde A_4^{(2)}(Q^2)
    \right)\ .
\end{eqnarray}
Here $A_{4,{\rm TMC}}^{(2)}$ is a target mass correction,
which arises from the kinematical $Q^2/\nu^2$ corrections
in the OPE analysis, and which would vanish for a massless
target.
Formally this is a matrix element of the $\tau=2$ operator
in Eq.~(\ref{eq:F2LTop}) with $n=4$.
At $Q^2=1-2$~GeV$^2$, the target mass term $A_{4,{\rm TMC}}^{(2)}$
amounts to around 4\% of the total moment.
The coefficient $\widetilde A_4^{(2)}$ represents matrix elements of
genuine $\tau=4$ operators \cite{Ellis2,JI_F2,SV_HT,SOLDATE}, and is
given
by
\begin{eqnarray}
2 \widetilde A_4^{(2)} \left( P^\mu P^\nu - g^{\mu\nu} M^2 \right)
&=& \sum_{q,q'} e_q e_{q'}
    \langle P |
    {1\over 2} g^2 \bar\psi_q \gamma^{\{ \mu} \gamma_5 t^a \psi_q
       \bar\psi_{q'} \gamma^{\nu\}} \gamma_5 t^a \psi_{q'}\
							\nonumber\\
& & \hspace*{1.5cm}
 +\ {5\over 16} g^2 \bar\psi_q \gamma^{\{ \mu} t^a \psi_q
	\bar\psi_{q'} \gamma^{\nu\}} t^a \psi_{q'}	\nonumber\\
& & \hspace*{1.5cm}
 +\ {1\over 16} g \bar\psi_q iD^{\{ \mu}
	\widetilde{G}^{\nu\}\lambda}
	\gamma^\lambda \gamma_5 \psi_{q'}\ \delta_{qq'}
    | P \rangle\ ,
\label{eq:A4}
\end{eqnarray}
where
$\widetilde G^{\mu\nu}
= {1\over 2} \epsilon^{\mu\nu\alpha\beta} G_{\alpha\beta}$ is the dual
gluon field strength tensor,
$t^a$ are color SU(3) matrices, and $g$ is the strong coupling constant.
The first and second terms in Eq.~(\ref{eq:A4}) represent four-quark
operators (in general nondiagonal in quark flavors $q$, $q'$),
corresponding to the ``cat's ears'' diagram in Fig.~\ref{fig:diag}~(b),
while the third term represents a quark-gluon mixed operator such as
that represented in Fig.~\ref{fig:diag}~(c).

The higher-twist contribution is $\sim 10-15\%$ of the total at
$Q^2 \sim 1$--2~GeV$^2$, as seen in Fig.~\ref{fig:f2ht}.
The $Q^2$ dependence of the higher-twist operators is governed by the
appropriate anomalous dimensions of the twist-4 operators \cite{SV_HT},
which results in a (logarithmic) $Q^2$ dependence of the matrix element
$\widetilde A_4^{(2)}(Q^2)$.
For all $n \alt 10$, $\widetilde A_4^{(n)}$ remains approximately
constant, in contrast to the rapidly decreasing size of the
leading-twist moment, $A_2^{(n)}$, as shown in Fig.~\ref{fig:f2mom}.
Phenomenologically, this reflects the fact that at fixed $Q^2$
the higher-twist corrections become more significant for higher
moments, which stems from the increased number of twist-4 operators
\cite{JI_F2}.
For $n=4$, the twist-4 term is as large as the leading twist at
$Q^2 \approx 2$~GeV$^2$, while for larger $n$ the higher twists
dominate at this $Q^2$.
It has been suggested \cite{JI_F2} that the twist expansion breaks
down when the higher-twist corrections exceed $\sim 50\%$ of the
leading twist.

Of course, neglecting $1/Q^6$ and higher corrections in such an
analysis is only justified as long as $Q^2$ is not too small.
Since the twist expansion is believed to be controlled by a scale
related to the average transverse momentum of quarks in the nucleon
\cite{GP}, typically of the order 0.4--0.5~GeV \cite{JI_F2,GP},
one may expect that the role of twist-6 effects should not be
significant for $Q^2 > 1$~GeV$^2$, and not overwhelming for
$Q^2 \agt 0.5$~GeV$^2$ \cite{JI_G1}.

Information on the spin dependence of higher twists can be extracted
from moments of the polarized $g_1$ and $g_2$ structure functions
\cite{EDELMANN,JI_G1,HT_N,HT_P}.
At leading twist, the $n=1$ moment of the $g_1$ structure function can
be written in terms of the helicity-dependent quark distributions,
$\Delta q = q^\uparrow - q^\downarrow$,
\begin{eqnarray}
\mu_2^{(1)}(Q^2)
&=& {1\over 2} \sum_q e_q^2\ \int_0^1 dx\
    \left( \Delta q(x,Q^2) + \Delta \bar q(x,Q^2) \right)\ ,
\label{mu2}
\end{eqnarray}
which can in turn be related to the axial vector charges of the
nucleon.
The triplet and octet charges are known from neutron and hyperon
$\beta$-decay, while the singlet axial charge is interpreted in
the quark-parton model as the total spin of the nucleon carried
by quarks.
The higher-twist contributions to the lowest moment $\Gamma_1^{(1)}$
contain information on spin-dependent quark-gluon correlations,
and in particular on the color polarizabilities of the nucleon
\cite{MANK_CHI,JI_CHI}, which describe how the color electric
and magnetic gluon fields respond to the spin of the nucleon.

\begin{figure}[ht]
\begin{center}
\epsfig{file=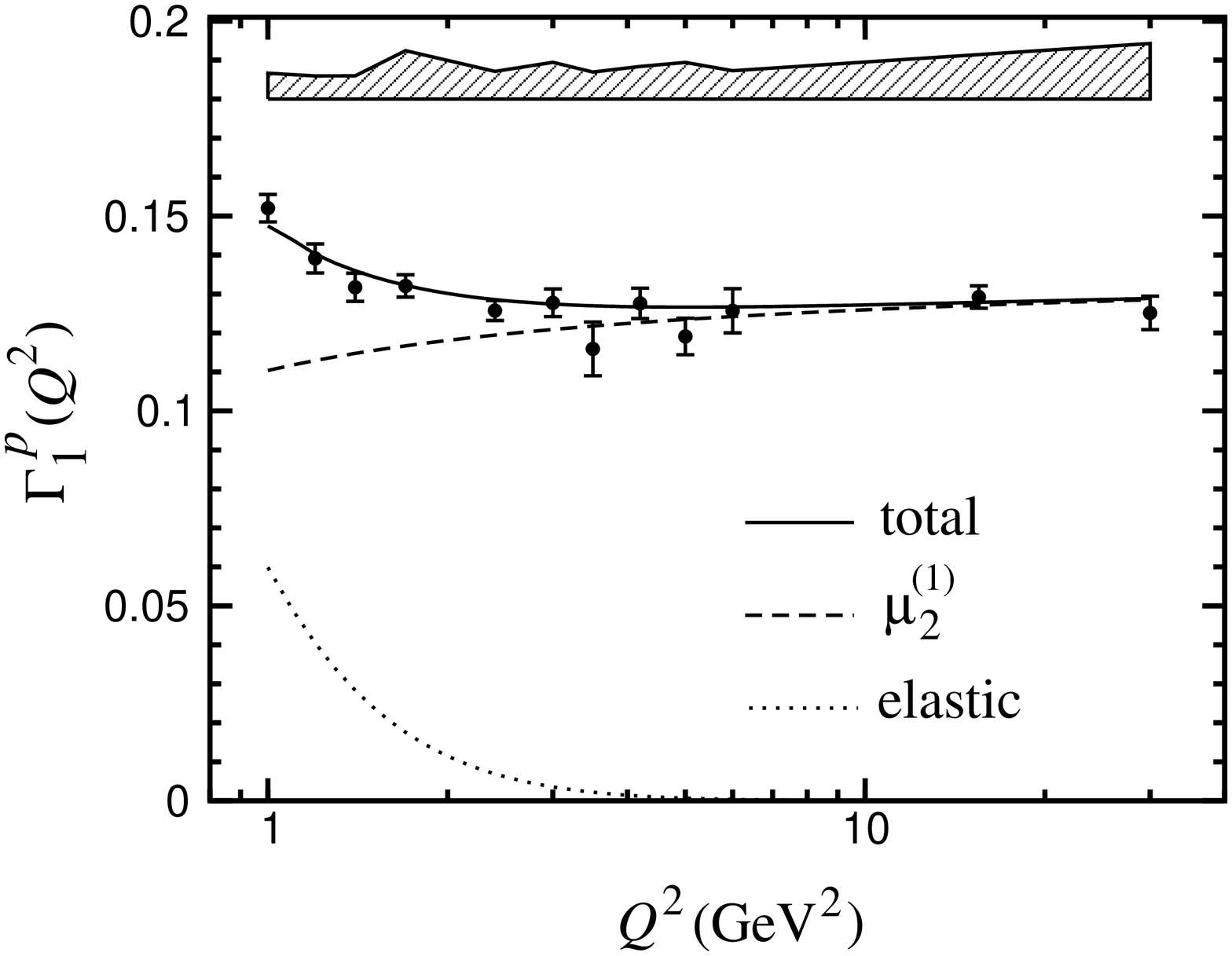,height=6.7cm,angle=270}
\epsfig{file=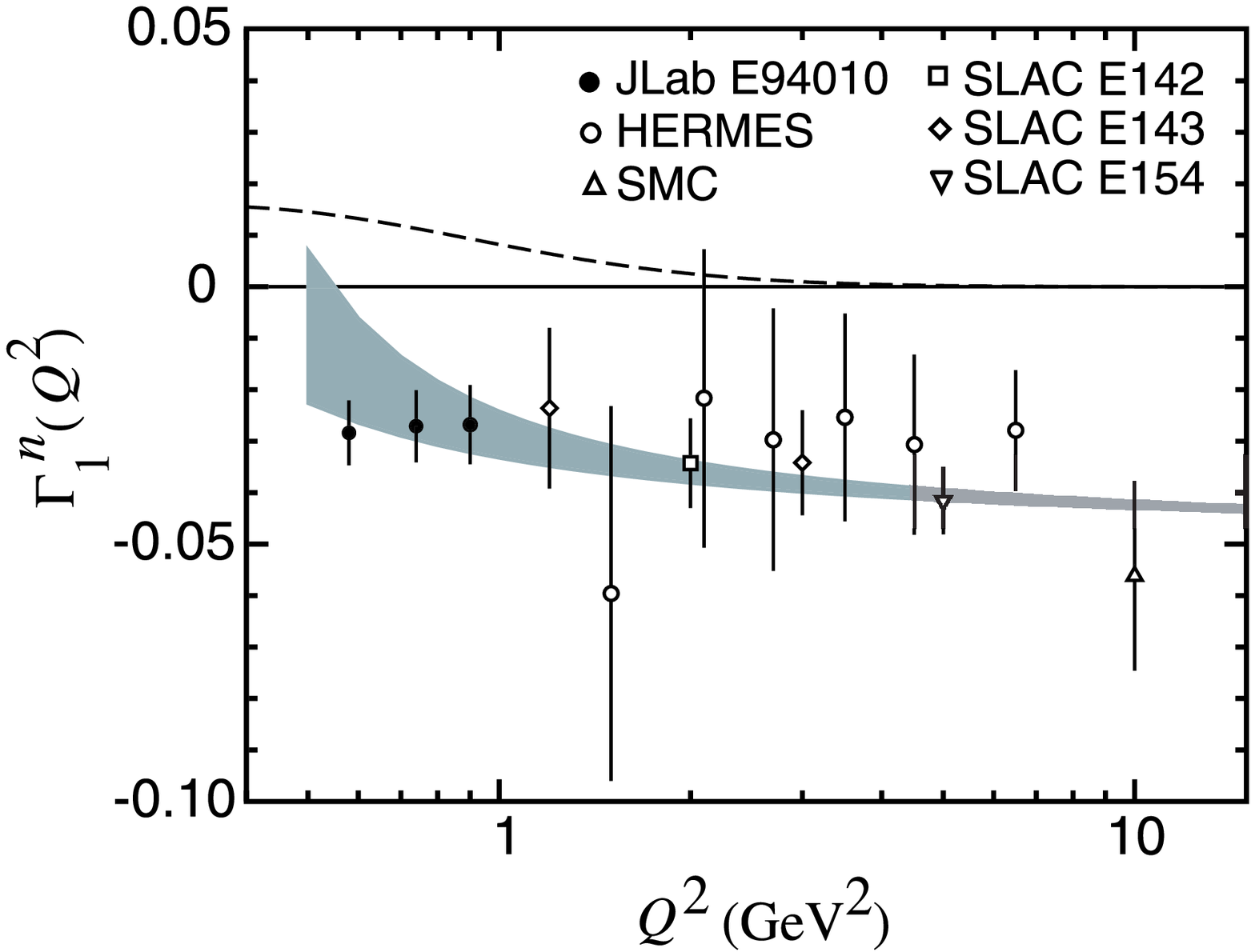,height=5.8cm}
\vspace*{0.5cm}
\caption{\label{fig:gam}
	(Left panel)
	Lowest moment of the proton $g_1^p$ structure function.
	The points are from a reanalysis of world data by
	Osipenko {\em et al.}\protect\cite{HT_P}; the error bars give
	statistical uncertainties only, while the systematic and
	low-$x$ extrapolation errors are given by the shaded band.
	(Right panel)
	Lowest moment of the neutron $g_1^n$ structure function
	\protect\cite{HT_N}.  The error bars are a quadratic sum of
	statistical and systematic errors.  The shaded band represents
	the uncertainty on the leading-twist contribution due to
	$\alpha_s$, and the dashed curve indicates the elastic
	contribution.}
\end{center}
\end{figure}

The lowest moment of the proton $g_1$ structure function is shown
in Fig.~\ref{fig:gam}~(left panel).
The data points are from a reanalysis \cite{HT_P} of world data in
which different experiments were combined using a consistent set of
assumptions for the low-$x$ extrapolations, beyond the measured region.
The elastic component of $\Gamma_1^p$ is seen to make a sizable
contribution at low $Q^2$, and in fact dominates the moment below
$Q^2 \approx 0.5$~GeV$^2$.
For $Q^2 \agt 2$--3~GeV$^2$ the moment is effectively saturated by
the leading-twist contribution, $\mu_2^{(1)}$.
At $Q^2 \alt 2$~GeV$^2$ a noticeable difference between the total
and leading-twist moments starts to appear, indicating the
presence of higher twist effects.
The corresponding neutron moment is shown in
Fig.~\ref{fig:gam}~(right panel).
The data are from various experiments at SLAC, CERN, DESY and JLab,
and have also been reanalyzed \cite{HT_N} to ensure the same low-$x$
extrapolations are applied to all the data sets.
The difference between the leading-twist contribution (shaded band)
and the data is smaller in this case, within the errors, suggesting
that the higher twists may be relatively small even for
$Q^2 \alt 1$~GeV$^2$.

The size of the higher twists can be quantified by subtracting
$\mu_2^{(1)}$ from the total moment.
The leading term in the difference is the $1/Q^2$ correction,
$\mu_4^{(1)}$, which can written as \cite{JI_G1,JM,SV_HT}
\begin{eqnarray}
\mu_4^{(1)}(Q^2)
&=& {1 \over 9} M^2
\left( a_2(Q^2) + 4 d_2(Q^2) + 4 f_2(Q^2) \right)\ .
\label{eq:mu4}
\end{eqnarray}
Here $a_2$ is related to the target mass corrections (which are
formally twist-2), whose role has also been emphasized in
Refs.~\cite{Li02,SIM02,LIUTI} in discussions of the $x$ dependence
of the higher-twist contributions, especially at large $x$.
The term $d_2$ is a twist-3 matrix element given by the second moment
of the (leading-twist parts of the) $g_1$ and $g_2$ structure
functions,
\begin{eqnarray}
d_2(Q^2)
&=& \int_0^1 dx\ x^2\ \left( 2 g_1(x,Q^2)\ +\ 3 g_2(x,Q^2) \right)\ .
\label{eq:d2}
\end{eqnarray}
Note that the $x^2$ weighting in the integrand places greater emphasis
on the large-$x$ region, and implies a greater role for the resonance
contributions which populate this region.
The $f_2$ term corresponds to the matrix element of a twist-4 operator
involving both quark and gluon fields \cite{SV_HT,MANK_CHI,JI_CHI},
\begin{eqnarray}
f_2(Q^2)\ M^2 S^\mu
&=& {1 \over 2} \sum_q e_q^2\
    \langle P,S |\
        g\ \bar\psi_q\ \widetilde{G}^{\mu\nu} \gamma_\nu\ \psi_q\
    | P, S \rangle\ .
\label{eq:f2op}
\end{eqnarray}
Note that the sign convention here follows that in
Refs.~\cite{ABE98,HT_N,HT_P,MANKIEWICZ}, and is opposite to that
adopted in Ref.~\cite{JI_G1}.
The $Q^2$ dependence of the higher-twist matrix elements is given by
the respective anomalous dimensions, which have been calculated by
Shuryak \& Vainshtein \cite{SV_HT}.

Fitting the proton data for $Q^2 > 1$~GeV$^2$, the analysis in
Ref.~\cite{HT_P} extracted a value for the $f_2^p$ matrix element
(normalized at a scale $Q^2=1$~GeV$^2$) of
\begin{eqnarray}
f_2^p
&=& 0.039\
\pm\ 0.022\ {\rm (stat.)}\
\pm\ {}^{0.000}_{0.018}\ {\rm (sys.)}
\pm\ 0.030\ ({\rm low}\ x)\ \pm\ {}^{0.007}_{0.011}\ (\alpha_s)\ ,
\label{eq:f2p}
\end{eqnarray}
where the first and second errors are statistical and systematic,
respectively, the third is due to the low-$x$ extrapolation,
and the last arises from the uncertainty in the value of $\alpha_s$
at low $Q^2$.
For the neutron data in Fig.~\ref{fig:gam}, a best fit for the
$Q^2 > 0.5$~GeV$^2$ data gives
\begin{eqnarray}
f_2^n = 0.034 \pm 0.043\ ,
\label{eq:f2n}
\end{eqnarray}
where the error includes statistical and (the more dominant) systematic
uncertainties, as well as from the $x \to 0$ extrapolation.

One should note, however, that the extracted higher-twist contribution
depends somewhat on the size of the $\tau=2$ contribution.
In particular, the best fits to the individual proton and neutron data
sets give rise to different central values of the singlet axial charge,
$\Delta\Sigma$: $\Delta\Sigma^{(p)} = 0.15 \pm 0.11$ from the proton
data, and $\Delta\Sigma^{(n)} = 0.35 \pm 0.08$ from the neutron data.
While consistent within errors, this does suggest a need for
higher-precision data on $g_1^{p,n}$ at higher $Q^2$ to test the
assumptions which go into polarized structure function analyses.

The extracted values of $f_2$ can be combined with the previously
measured $d_2$ matrix element to provide information on the color
electric ($\chi_E$) and magnetic ($\chi_B$) polarizabilities of the
nucleon \cite{JI_CHI},
\begin{eqnarray}
\chi_E\ 2 M^2 \vec S
&=& \langle P,S |\ \vec j_a \times \vec E_a\ | P,S \rangle\
 =\ 2 M^2 \vec S\ {1 \over 3} \left( 4 d_2\ +\ 2 f_2 \right)\ ,	\\
\chi_B\ 2 M^2 \vec S
&=& \langle P,S |\ j_a^0\ \vec B_a\ | P,S \rangle\
 =\ 2 M^2 \vec S\ {1 \over 3} \left( 4 d_2\ -\ f_2 \right)\ ,
\label{eq:chi}
\end{eqnarray}
where $\vec E_a$ and $\vec B_a$ are the color electric and magnetic
fields, and $j_a^\mu$ is the quark current \cite{MANK_CHI}.
The color polarizabilities reflect the response of the color
electric and magnetic fields in the nucleon to the nucleon spin,
$\vec S$ \cite{MANK_CHI,JI_CHI}; the sign on $\chi_B$,
for instance, reflects the direction of the color magnetic field
with respect to the polarization of the proton.
With the values for $f_2$ in Eqs.~(\ref{eq:f2p}) and (\ref{eq:f2n}),
and the results for $d_2^p$ from the global analysis in
Ref.~\cite{HT_P} and $d_2^n$ from the SLAC E155 measurement
\cite{E155}, one finds
\begin{eqnarray}
\chi_E^p &=&  0.026\ \pm\ 0.028\ ,\ \ \
\chi_B^p\ =\ -0.013\ \mp\ 0.014		\\
\chi_E^n &=& 0.033 \pm 0.029\ , \ \ \
\chi_B^n\ =\ -0.001 \pm 0.016
\end{eqnarray}
where the errors have been added in quadrature.
These results indicate that both the color electric and magnetic
polarizabilities in the proton and neutron are relatively small,
with the central values of the color electric polarizabilities
being positive, and the color magnetic zero or slightly negative.

The small values of the higher-twist corrections, in both polarized
and unpolarized structure functions, suggest that the long-range,
nonperturbative interactions between quarks and gluons in the nucleon
are not as dominant at $Q^2 \agt 1$~GeV$^2$ as one may have expected.
For the polarized neutron moment, they may even play a minor role
down to $Q^2 \approx 0.5$~GeV$^2$.
This would imply strong cancellations between neutron resonances
resulting in the dominance of the leading-twist contribution to
$\Gamma_1^n$.
If so, it would be a spectacular confirmation of quark-hadron
duality in spin-averaged and spin-dependent structure functions.

At lower values of $Q^2$ ($Q^2 \alt 0.5$~GeV$^2$) there will be
significant contributions from even higher twists ($\tau=6$
and higher) which will eventually render the perturbative twist
expansion unreliable.
On the other hand, structure functions and their moments have been
measured to very low $Q^2$, and a theoretical understanding of the
transition to the real photon limit is necessary for a complete
description of the quark structure of the nucleon.
In the next section we discuss the behavior of structure functions
in the limit $Q^2 \to 0$, and how duality may be realized even in
this extreme case.

% ........................................................................
\subsubsection{The Transition to $Q^2 = 0$}
\label{sssec:real}

As $Q^2$ decreases from the perturbative regime, eventually a
description of the structure function in terms of scattering from
a few parton constituents becomes unreliable.
For $Q^2 \sim \Lambda_{\rm QCD}^2$ a perturbative expansion in
$\alpha_s$ ceases to be meaningful, as obviously does an expansion
in terms of powers of $1/Q^2$.
When $Q^2$ is close to the real photon limit,
$Q^2 \ll \Lambda_{\rm QCD}^2$, an expansion in powers of $Q^2$ may
instead become more relevant.
Here one may resort to effective field theory techniques, such as chiral
perturbation theory, to describe structure functions, or their moments,
in terms of hadronic degrees of freedom.
Nevertheless, as one traverses through the intermediate-$Q^2$ region,
where neither the perturbative high-energy nor effective low-energy
expansion schemes are applicable, physical quantities should still
remain smooth functions of $Q^2$, and some sort of duality may still
hold in terms of variables other than $Q^2$.

There are indeed some empirical indications which suggest the existence
of a duality between resonances and continuum cross sections even at
the real photon point, $Q^2=0$.
This may at first sight be surprising since the physics at $Q^2=0$ is
in some ways qualitatively different from that applicable for deep
inelastic scattering at large $Q^2$.
The former limit is dominated by purely coherent processes, with cross
sections generically proportional to the squares of sums of charges
of the interacting quarks, $(\sum_q e_q)^2$.
Processes at large $Q^2$, on the other hand, are dominated by
incoherent scattering from individual quarks in the nucleon,
with a strength proportional to sums of squares of the individual
quark charges, $\sum_q e_q^2$.
The difference between these is then a sum over the nondiagonal
contributions, $\sum_{q\not= q'} e_q e_{q'}$, which in general
can be as large as the terms diagonal in $e_q$ \cite{BRODSKY_HT}.
Despite this, for real photons the nondiffractive contribution to
the ratio of the total neutron to proton cross sections is empirically
found to be
\begin{eqnarray}
{ \sigma^{\gamma n} \over \sigma^{\gamma p} }
&\sim& { 2 \over 3 }\
 \equiv\ { 2 e^2_d + e^2_u \over 2 e^2_u + e^2_d }\ .
\end{eqnarray}
The ratio thus behaves as if it were given by the squares of constituent
quark charges, even though there is no reason for the dominance of the
incoherent terms when $Q^2 = 0$.

\begin{figure}[ht]
\begin{center}
\epsfig{file=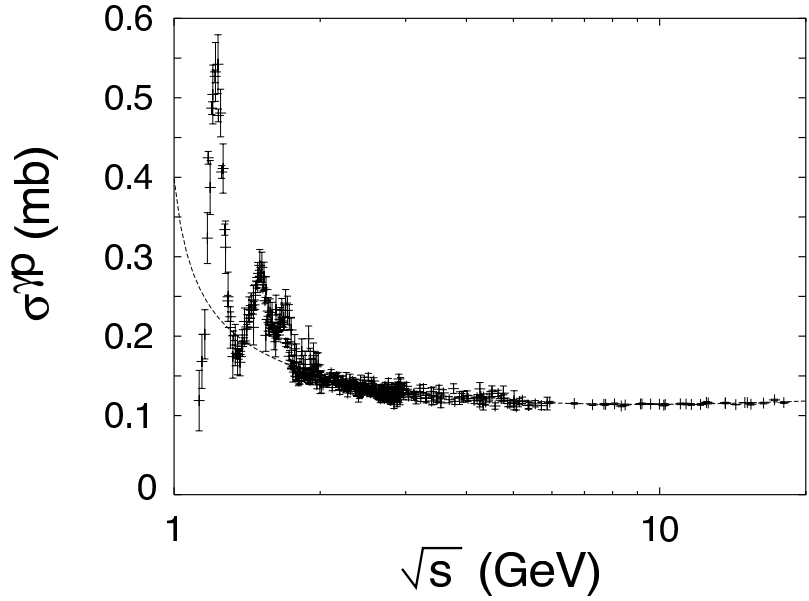,height=8cm}
\vspace*{0.5cm}
\caption{\label{fig:photon}
	Total inclusive photoproduction cross section data for the
	proton as a function of the center of mass energy, $\sqrt{s}$,
	compared with a parameterization (dashed curve) of the high
	energy data.
	From Ref.~\protect\cite{PVL}.}
\end{center}
\end{figure}

In fact, some oscillations around the high-energy behavior can be seen
in the total photon--proton cross section, $\sigma^{\gamma p}$, at low
$\nu$, as illustrated in Fig.~\ref{fig:photon}.
The high-energy ``scaling'' curve here is a fit to the large-$s$
data by Donnachie \& Landshoff \cite{DL} using a Regge-inspired
model in which the total $\gamma p$ cross section is parameterized
by the sum of diffractive and nondiffractive components,
\begin{eqnarray}
\sigma^{\gamma p}
&=& X\ (2M\nu)^{\alpha_{\rm I\!P}-1}\
 +\ Y\ (2M\nu)^{\alpha_{\rm I\!R}-1}\ ,
\label{eq:sigDL}
\end{eqnarray}
where for real photons one has $2 M \nu = s - M^2$, with $s$ the total
$\gamma p$ center of mass energy squared.
The exponents $\alpha_{\rm I\!P} = 1.0808$ and
$\alpha_{\rm I\!R} = 0.5475$ are fitted to $pp$ and $p\bar p$ total
cross section data, and the coefficients $X$ and $Y$ are given by
$X = 0.0677$ and $Y = 0.129$.
Numerically the exponents $\alpha_{\rm I\!P}$ and $\alpha_{\rm I\!R}$
are found to be very similar to the intercepts of the Pomeron and
Reggeon ($\rho$ meson) trajectories, respectively.
Although the parameters were fitted to the $\sqrt{s} \agt 6$~GeV data,
the fit appears on average to go through the resonance data at low
$\sqrt{s}$ (even at $Q^2=0$!).

The behavior of the structure functions in the $Q^2 \to 0$ limit
is to some extent constrained by global symmetries.
In particular, electromagnetic current conservation requires that
$q^\mu W_{\mu\nu} = q^\nu W_{\mu\nu} = 0$, where $W_{\mu\nu}$ is
the electromagnetic hadronic tensor (see Eq.~(\ref{eq:Wmunu})).
The requirement that there be no kinematical singularities at $Q^2=0$
implies that for fixed $\nu$, the $F_1$ and $F_2$ structure functions
at small $Q^2$ must behave as
\begin{eqnarray}
& & F_2\ =\ {\cal O}(Q^2)\ ,	\\
& & F_1 + {p \cdot q \over q^2} F_2\ =\ {\cal O}(Q^2)\ .
\end{eqnarray}
Since at fixed $\nu$ one has $x \sim Q^2$, this implies that the
longitudinal structure function $F_L$ is suppressed by an additional
power of $Q^2$ compared with $F_2$ in the $Q^2 \to 0$ limit.
In terms of the cross sections for scattering of transversely and
longitudinally polarized photons, $\sigma_T$ and $\sigma_L$,
respectively, the structure functions in the $Q^2 \to 0$ limit can be
written as ({\em cf.} Eqs.~(\ref{eq:f2sigs}) \& (\ref{eq:flsigl}))
\begin{eqnarray}
F_2 &\to& { Q^2 \over 4 \pi^2 \alpha }\ 
	\left( \sigma_T + \sigma_L \right)\ ,	\\
F_L &\to& { Q^2 \over 4 \pi^2 \alpha }\ \sigma_L\ .
\end{eqnarray}
Because only transversely polarized photons contribute to the total
cross section $\sigma^{\gamma p}$ in the $Q^2 \to 0$ limit, $\sigma_L$
must vanish for real photons, in which case
\begin{eqnarray}
\sigma^{\gamma p}
&=& \sigma_T\ +\ \epsilon\ \sigma_L\
\to\ \sigma_T\ \ \ {\rm as}\ \ \ Q^2 \to 0\ ,
\end{eqnarray}
where $\epsilon$ is the virtual photon polarization factor
(see Eq.~(\ref{eq:epsilon})).
At $Q^2=0$ the total (transverse) cross section remains finite, which
implies that the $F_2$ structure function must vanish linearly with
$Q^2$,
\begin{eqnarray}
F_2(x,Q^2)
&\to& Q^2\ \ \ {\rm as}\ \ \ Q^2 \to 0\ ,
\label{eq:F2realim}
\end{eqnarray}
while the $F_L$ structure function vanishes as
\begin{eqnarray}
F_L(x,Q^2)
&\to& Q^4\ \ \ {\rm as}\ \ \ Q^2 \to 0\ ,
\label{eq:FLlowQ}
\end{eqnarray}
and therefore the longitudinal to transverse ratio
\begin{eqnarray}
R(x,Q^2) &=& { \sigma_L \over \sigma_T }\
 \to\ Q^2\ \ \ {\rm as}\ \ \ Q^2 \to 0\ .
\end{eqnarray}
Note that each of these kinematical constraints is a direct
consequences of electromagnetic gauge invariance.
In the case of neutrino scattering, where there are no such constraints
on the axial vector current, the neutrino structure function $F_2^\nu$
does not vanish in the $Q^2 \to 0$ limit, but by PCAC is proportional
to the total $\pi p$ cross section,
$F_2^\nu \sim \sigma^{\pi p} \to$~constant.

Donnachie \& Landshoff \cite{DLF2} have also fitted the proton $F_2$
structure function at low $Q^2$ using a Regge-inspired model for the
parton distributions, modified to incorporate the kinematical
$Q^2 \to 0$ constraint in Eq.~(\ref{eq:F2realim}).
Using two simple powers of $x$, in analogy with the total cross section
fit in Eq.~(\ref{eq:sigDL}), each multiplied by a simple function of
$Q^2$, they parameterize the $Q^2 \to 0$ behavior as \cite{DLF2}
\begin{eqnarray}
F_2(x,Q^2)
&\sim& A\ x^{1-\alpha_{\rm I\!P}}
       \left( { Q^2 \over Q^2 + a } \right)^{\alpha_{\rm I\!P}}\
    +\ B\ x^{1-\alpha_{\rm I\!R}}
       \left( { Q^2 \over Q^2 + b } \right)^{\alpha_{\rm I\!R}}\ ,
\label{eq:F2DL}
\end{eqnarray}
with $A$ and $B$ constrained to reproduce the photoproduction limit
in Eq.~(\ref{eq:sigDL}).

A more intuitive interpretation of the $Q^2$ dependence of the
low-$Q^2$ structure function was provided by Badelek \& Kwiecinski
\cite{BKF2} through a generalization of the vector meson dominance
(VMD) model.
The VMD model is a quantitative realization of the fact that photon
interactions with the nucleon proceed via the hadronic components of
the photon wave function, which at low $Q^2$ are largely saturated by
the $\rho$, $\omega$ and $\phi$ meson contributions \cite{VMD}.
In the generalized VMD model the $F_2$ structure function is given
by two terms,
\begin{eqnarray}
F_2(x,Q^2)
&=& F_2^{\rm light}(x,Q^2)\
 +\ F_2^{\rm heavy}(x,Q^2)\ ,
\end{eqnarray}
representing contributions from light and heavy vector meson states.
The former is given by \cite{BKF2}
\begin{eqnarray}
F_2^{\rm light}(x,Q^2)
&=& { Q^2 \over 4 \pi }
    \sum_{v=\rho,\omega,\phi}
    { m_v^2\ \sigma_v(s) \over \gamma_v^2 (Q^2 + m_v^2)^2 }\ ,
\end{eqnarray}
where $\sigma_v$ is the vector meson--nucleon total cross section,
and the coupling $\gamma_v$ is related to the leptonic width of the
vector meson,
\begin{eqnarray}
\gamma_v^2
&=& { \alpha^2 \pi m_v \over 3\ \Gamma_{v \to e^+e^-} }\ .
\end{eqnarray}
At large $Q^2$ the low-mass contributions vanish since
$F_2^{\rm light}(x,Q^2) \sim 1/Q^2$.
The scaling of the structure function is modeled by including
contributions from an infinite number of vector meson states heavier
than some mass $M_0$,
\begin{eqnarray}
F_2^{\rm heavy}(x,Q^2)
&=& Q^2 \int_{M_0^2}^\infty dQ'^2
	{ \Phi(Q'^2,s) \over (Q^2 + Q'^2)^2 }\ ,
\label{eq:F2vmdH}
\end{eqnarray}
where
\begin{eqnarray}
\Phi(Q^2,s)
&=& -{1 \over \pi}
    \Im m \int^{-Q^2} { dQ'^2 \over Q'^2 }
    F_2^{\rm asym}(x'={Q'^2\over s-M^2+Q'^2},Q'^2)\ .
\end{eqnarray}
As $Q^2 \to \infty$ the function $F_2^{\rm heavy}$ approaches the
asymptotic (scaling) structure function $F_2^{\rm asym}$.
Badelek \& Kwiecinski \cite{BKF2} further simplify the heavy vector
meson contribution by taking
\begin{eqnarray}
F_2^{\rm heavy}(x,Q^2)
&=& { Q^2 \over Q^2 + M_0^2 }
    F_2^{\rm asym}(\bar x,Q^2+M_0^2)\ ,
\end{eqnarray}
where $\bar x = (Q^2 + M_0^2)/(s - M^2 + Q^2 + M_0^2)$,
which preserves the analytic properties of the representation in
Eq.~(\ref{eq:F2vmdH}).
Although the low-mass vector meson component dominates $F_2$ at low
$Q^2$, the large-mass, partonic contribution still gives
$\sim 20$--30\% of the total at $Q^2 \sim 1$~GeV$^2$.

\begin{figure}[h]
\begin{center}
\epsfig{file=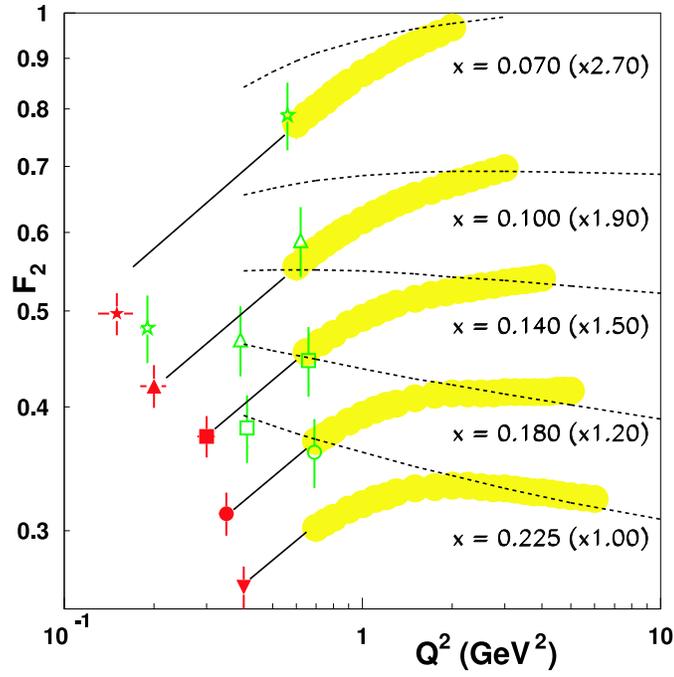,height=11cm}
\vspace*{-1cm}
\caption{\label{fig:lowF2Q}
	$Q^2$ dependence of JLab $F_2^p$ data at low $x$ and $Q^2$.
	The solid lines connect the resonance (JLab) and DIS data
	(light shaded circles), and consistently represent a
	$F_2^p \sim Q^{0.5}$ behavior.
	The dotted lines denote perturbative QCD predictions from
	Ref.~\protect\cite{Li02}.
	(Adapted from Ref.~\protect\cite{F2JL2}.)}
\end{center}
\end{figure}

Other models and parameterizations of structure functions at low $Q^2$
are reviewed in Ref.~\cite{BKREV}.
While these parameterizations provide satisfactory fits to the
low-$Q^2$ data, they do not in themselves provide answers to the
question of whether gauge invariance is the only physics which
underlies the transition to $Q^2=0$, or whether there is additional
dynamics which drives the transition.
Some clues to this question may be provided by the low-$Q^2$ data
on $F_2^p(x,Q^2)$ measured recently at Jefferson Lab \cite{F2JL2}.
The experiments there found that the $F_2^p$ structure function at low
$Q^2$ does not follow the linear $\sim Q^2$ behavior that would be
expected from gauge invariance constraints alone, but instead behaves
as $F_2 \sim Q^{0.5}$ down to $Q^2 \sim 0.3$~GeV$^2$, as displayed in
Fig.~\ref{fig:lowF2Q}.
This indicates that these values of $Q^2$ are still transitional,
and that lower $Q^2$ is needed before the behavior becomes driven
by gauge invariance.

\begin{figure}[h]
\begin{center}
\epsfig{file=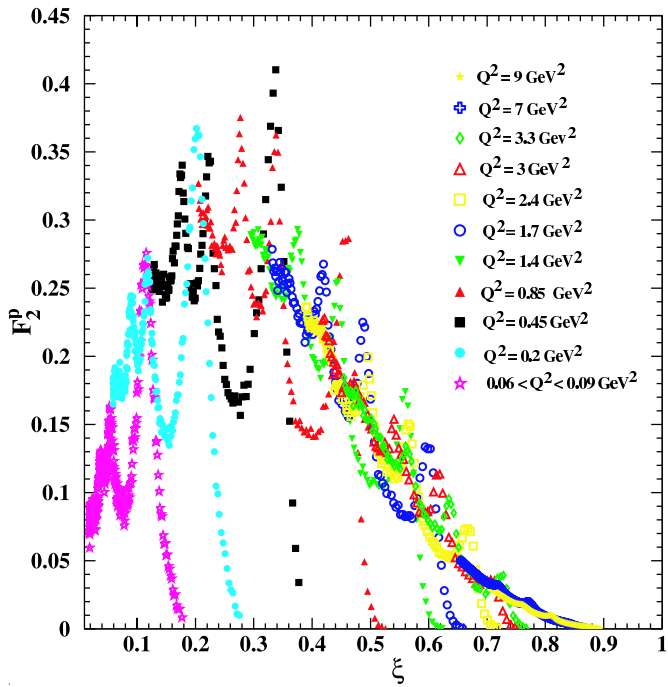,height=9cm}
\vspace*{0.5cm}
\caption{\label{fig:F2val}
	$F_2^p$ structure function versus the Nachtmann scaling
	variable $\xi$, illustrating valence-like behavior at
	low $\xi$ and $Q^2$.
	(Adapted from Ref.~\protect\cite{F2JL2}.)}
\end{center}
\end{figure}

\begin{figure}[h]
\begin{center}
\hspace*{0.5cm}
\epsfig{file=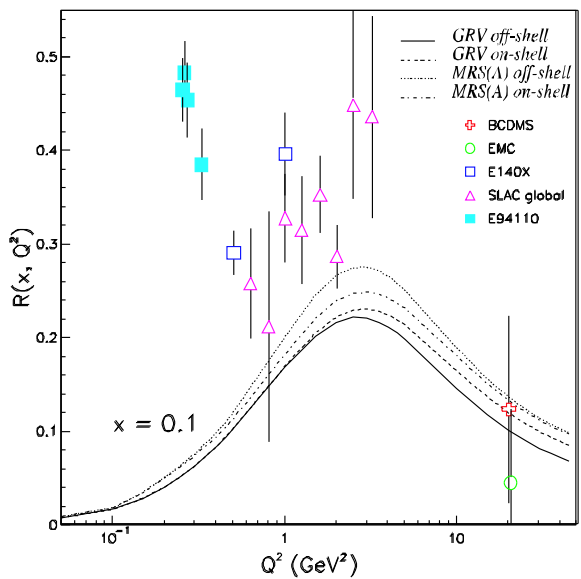,height=11cm}
\vspace*{-3cm}
\caption{\label{fig:lowRQ}
	Comparison of the calculated $R$ ratio as a function of
	$Q^2$ at $x=0.1$ for various input gluon distributions.
	Adapted from Ref.~\protect\cite{BKS}, with new data from
	JLab Experiment E94-110 \protect\cite{ERIC}. These new data
        have an additional point-to-point systematic uncertainty
        of 0.1 (not shown).}
\end{center}
\end{figure}

Furthermore, Niculescu {\em et al.} \cite{F2JL2} found that the
$F_2^p$ structure function at low $Q^2$, when averaged over
resonances, behaves much like the $x F_3$ structure function
measured in neutrino scattering which is determined by valence
quarks only --- see Fig.~\ref{fig:xf3}.
The valence-like behavior of $F_2^p$ at low $Q^2$ is shown in
Fig.~\ref{fig:F2val} for various spectra ranging from
$Q^2 = 0.06$~GeV$^2$ to 9~GeV$^2$.
This suggests that at low $Q^2$ the $F_2$ structure function is
dominated by valence quarks, with contributions from the sea
suppressed.
Such an interpretation would support a two-component duality picture
in which the valence quark (nondiffractive) contributions are dual to
resonances, while sea quark (diffractive) contributions are dual to
the nonresonant background \cite{QUEEN,HARARI,FREUND}
(see Sec.~\ref{sssec:twocomp} above).

Additional insights into low-$Q^2$ dynamics may be obtained by
comparing the expected behavior of the longitudinal cross section,
or the $R$ ratio, with low-$Q^2$ data.
For instance, Badelek {\em et al.} \cite{BKS} parameterize $F_L$
using a model based on the photon--gluon fusion mechanism, suitably
extended to low $Q^2$, and calculate $R$ utilizing a parameterization
of $F_2$ from Ref.~\cite{BKF2}.
The predictions for the $Q^2$ dependence of $R(x,Q^2)$ at $x=0.1$
are displayed in Fig.~\ref{fig:lowRQ}, using several different gluon
distribution functions.
Included in the longitudinal structure function is a higher-twist
contribution associated with the exchange of a soft Pomeron with
intercept equal to unity \cite{BKS}.
The overall low-$Q^2$ behavior of $F_L$ is $\sim Q^4$, as required
from Eq.~(\ref{eq:FLlowQ}), leading to an approximately linear $Q^2$
dependence of $R$ as $Q^2 \to 0$.
While slightly below the earlier SLAC data at $Q^2 \sim 1$~GeV$^2$,
the parameterization significantly underestimates the new Jefferson Lab
data at smaller $Q^2$ values.
As for the $F_2^p$ case, the low $Q^2$ data on $R$ indicate that at
values as low as $Q^2 \sim 0.2$~GeV$^2$ there appear to be additional
dynamics responsible for the $Q^2$ dependence, beyond that expected
from gauge invariance constraints alone.

Further progress in revealing the dynamics of the $Q^2 \to 0$
transition can be made with additional data on spin-dependent
structure functions at low $Q^2$.
For the $g_1$ and $g_2$ structure functions there are no gauge
invariance constraints as for $F_2$ and $F_L$, although sum rules
such as the Gerasimov-Drell-Hearn (Eq.~(\ref{eq:gdh})) and
Burkardt-Cottingham (Eq.~(\ref{eq:BC})) sum rules provide some
constraints on their moments at $Q^2=0$
(see Secs.~\ref{sssec:gdh} and \ref{sssec:g2}).
A parameterization similar to that for the $F_2$ structure function
was recently constructed by Badelek {\em et al.} \cite{BKG1} for
$g_1$ at low $Q^2$ within the generalized VMD model.

Beyond phenomenological studies, the theoretical tools currently
available are unable to provide us with quantitative understanding
from first principles in QCD of the nature of the low-$Q^2$ region,
where neither the twist expansion nor effective field theories are
applicable.
In the absence of a rigorous theoretical framework, one can instead
resort to models of QCD to obtain clues about the physics governing
the transition.
In the next section we discuss the insight into the origin of scaling
and low-$Q^2$ duality which can be garnered from dynamical models of
structure functions.

% -----------------------------------------------------------------------
\subsection{Scaling and Duality in Dynamical Models}
\label{ssec:models}

\begin{figure}[ht]
\begin{center}
\hspace*{0.5cm}
\epsfig{file=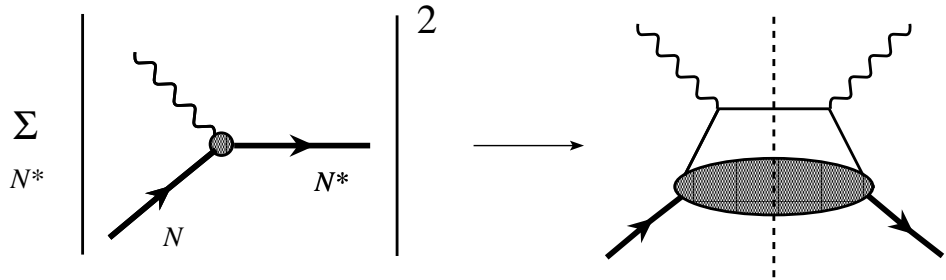,height=12cm}
\vspace*{-7.5cm}
\caption{\label{fig:dual}
	Schematic illustration of quark-hadron duality in
	inclusive electron--nucleon scattering: at high energy
	summation over excited states $N^*$ (left) is equivalent
	to the leading-twist, parton model result, given by the
	imaginary part of the virtual Compton scattering amplitude
	(right).}
\end{center}
\end{figure}

Although Bloom-Gilman duality for structure function {\em moments} at
intermediate and high $Q^2$ can be analyzed systematically within a
perturbative operator product expansion, an elementary understanding
of the origins of duality for structure functions as a function of $x$
and $Q^2$ is more elusive.
This problem is closely related to the question of how to build up
a scaling ($Q^2$-independent) structure function entirely out of
resonances \cite{IJMV}, each of which is described by a form factor
% $G_R(Q^2)$ off as some power of $1/Q^2$.
that falls rapidly with increasing $Q^2$.
The description of Bjorken scaling in DIS structure functions is of
course most elegantly formulated within the QCD quark-parton model,
which is justified on the basis of asymptotic freedom.
On the other hand, the physical final state is comprised entirely
of hadrons, so it must also be possible, in the general sense of
quark-hadron duality, to describe the process in terms of hadronic
degrees of freedom (resonances and their decays) alone.
Figure~\ref{fig:dual} is a schematic illustration of this duality.
How to obtain a scaling function from a sum over resonances,
and how to reconcile this with the perturbative description, is the
key to understanding the origins of duality, and we focus on these
aspects in this section.

We saw in Sec.~\ref{sssec:twist} above that the appearance of duality
results directly from the absence of, or cancellations between,
higher twists in structure functions.
This means that final state interactions between the scattered quark
and the remnants of the hadronic target are suppressed, and the
process can be described in terms of incoherent scattering from
individual, {\em free} quarks.
The nonexistence of final state interactions in the presence of
confinement is not at all obvious, however.
While factorization of the hard and soft parts of the Compton
scattering amplitude has been demonstrated within the operator
product expansion in perturbation theory, no derivation exists
nonperturbatively.
In the absence of a first principles proof, various methods and models
have been developed to examine the intriguing question of how confined
constituents of hadrons respond asymptotically as if they were free.
We shall see that a critical element in the appearance of scaling in
terms of hadronic degrees of freedom is access to a complete set of
hadronic states.

In this section we review several studies which illustrate how scaling
can be compatible with the presence of confining interactions.
Following this, we examine in more detail how the region dominated by
resonances merges into the scaling region, and illustrate the onset of
duality between the resonance-averaged and asymptotic structure
functions within simple models.
Finally, having demonstrated the mechanisms by which summation over
hadronic states can dual the free quark cross sections,
we apply these ideas to phenomenological examples in which resonances
at high $Q^2$ can be used to constrain the large-$x$ behavior of
structure functions, and vice versa.

% .......................................................................
\subsubsection{Confinement and Scaling}
\label{sssec:conf}

One of the central mysteries in strong interaction physics, and a key
to the question of the origin of duality, is how scattering from bound
(confined) states of quarks and gluons in QCD can be consistent with
scaling --- a property synonymous with scattering from free quarks.
To illustrate this dual nature of the strong interactions, we consider
several pedagogical examples where this transition is demonstrated
explicitly.
Starting from a discussion of scaling in the simplified case of QCD in
1+1 dimensions and in the limit of a large number of colors, $N_c$,
we then consider models with increasing levels of sophistication,
before finally progressing to more phenomenological descriptions.
A common feature in each of the model discussions will be closure,
or access to a complete set of hadronic states which dual the partonic
scattering process.
These examples also demonstrate that the distinction between scattering
to hadronic final states (resonances) and to the continuum (scaling) is
somewhat spurious, and that resonances are in fact an integral part of
the scaling structure function.

\vspace*{0.5cm}
% . . . . . . . . . . . . . . . . . . . . . . . . . . . . . . . . . . . .
\paragraph{Large $N_c$ Limit} %'t~Hooft Model}

Perhaps the simplest, and most graphic, demonstration of the interplay
between resonances and scaling is in QCD in the large-$N_c$ limit.
In the case of $q\bar q$ bound states, in this limit the hadronic
spectrum consists entirely of infinitely narrow, noninteracting
resonances of increasing mass.
On the other hand, since no element of the perturbative QCD results
for deep inelastic scattering depends on $N_c$, at the quark level
one still obtains a smooth scaling structure function.
Therefore in the large-$N_c$ world duality must be invoked even in
the scaling limit!

The derivation of a scaling function from large $N_c$ resonances
was demonstrated explicitly for the case of one space and one time
dimension \cite{EINHORN}.
QCD in 1+1 dimensions in the $N_c \to \infty$ limit, known as the
``'t~Hooft model'' \cite{THOOFT,THOOFT2}, is an exactly soluble
field theory, in which all hadronic Green's functions are calculable
in terms of quark degrees of freedom.
In the large-$N_c$ limit, even in lowest order, the exchange of a
massless gluon between quarks corresponds to an attractive $q\bar q$
potential which rises linearly with $r$ (compared with the 3+1
dimensional case which gives rise to a Coulombic $1/r$ potential).
Therefore confinement is an almost trivial consequence in the
't~Hooft model.
Furthermore, simply by power counting one can show that the theory
is asymptotically free, which automatically leads to Bjorken
scaling in structure functions.

The essential simplification which allows one to solve the 1+1
dimensional theory nonperturbatively is the freedom to choose gauges
in which the gluon self-coupling vanishes.
Only ladder diagrams thus need to be computed for $q\bar q$
interactions, in addition to rainbow diagrams for wave function
and mass renormalization \cite{THOOFT}.
The infrared behavior of the $q\bar q$ bound state wave functions
and the bound state spectrum can then be determined by solving a
Bethe-Salpeter equation (also known as the 't~Hooft equation in this
application) \cite{THOOFT},
\begin{eqnarray}
\mu_n^2\ \phi_n(x)
&=& \left( {\gamma_q-1 \over x} + {\gamma_{\bar q}-1 \over 1-x} \right)
    \phi_n(x)\
 -\ {\rm Pr} \int_0^1 {dy\ \phi_n(y) \over (y-x)^2}\ ,
\label{eq:thooft}
\end{eqnarray}
where $\phi_n(x)$ can be interpreted (in the infinite momentum frame)
as the probability amplitude for finding the quark $q$ in the $n$-th
$q\bar q$ bound state, with light-cone momentum fraction $x$.
Here $\mu_n$ is the bound state mass,
$\gamma_{q(\bar q} = (\pi/g^2 N_c)\ m_{q(\bar q)}^2$,
and the coupling constant $g$ has dimensions of mass.

In this model confinement follows from the fact that
\begin{eqnarray}
{\rm Pr} \int_0^1 {dy\ \phi_n(y) \over (y-x)^2}
&=& 0\
\end{eqnarray}
at the values of $x$ where $\phi_n(x)$ would develop poles.
The color singlet spectrum then contains an infinity of stable
$q\bar q$ bound states, with no $q\bar q$ continuum.
Finite widths of resonances only appear as $1/N_c$ corrections to
the theory.
The functions $\phi_n(x)$ can be solved numerically, for instance using
the Multhopp technique discussed in Refs.~\cite{JAF_MENDE,LEBED}.
For the lowest energy state, $n=0$, one has $\phi_0 = 1$.
As $x \to 0$ or 1, the distribution functions vanish as a power of $x$.
For large $n$, the functions $\phi_n$ behave as \cite{THOOFT2,CCG}
\begin{eqnarray}
\phi_n(x)
&\to& \sqrt{2} \sin(\mu_n^2 x/\pi)\ ,
\end{eqnarray}
with $\mu_n^2 \sim n \pi^2$.

If all the states in the 1+1 dimensional theory are stable bound
states, how is it that one can obtain scaling from a sum over
resonances?
To address this question, Einhorn \cite{EINHORN} calculated the deep
inelastic structure function of the $n$-th $q\bar q$ bound state.
In 1+1 dimensions there is only one structure function, denoted by
$W_n(Q^2,\nu)$, where $\nu = p\cdot q/\mu_n$.
Since the only final states which contribute at leading order in
$1/N_c$ are single-meson states, one can calculate the imaginary part
of the Compton scattering amplitude by squaring the $n \to m$ transition
form factors,
\begin{eqnarray}
W_n &\propto&
\sum_m \left| F^q_{nm}(Q^2) + F^{\bar q}_{nm}(Q^2) \right|^2\
	\delta((p+q)^2 - \mu_n^2)\ .
\end{eqnarray}
Note that on dimensional grounds, in 1+1 dimensions the scaling function
would correspond to $\nu^2 W_n$.
As $Q^2 \to \infty$, one can smooth the $\delta$-functions by replacing
the sum over $m$ with an integral, with $\mu_m^2 \sim m \pi^2 \to \infty$
(this is where the implicit use of duality alluded to above enters).
In this limit the transition form factors are simply given in terms of
the wave functions $\phi_n$ \cite{EINHORN},
\begin{eqnarray}
F^q_{nm}(Q^2)
&\sim& {(-1)^m \over Q^2}\ e_q\ m_q\ x \phi_n(x)\ ,
\end{eqnarray}
and similarly for the $\bar q$ contribution.
Squaring and performing the integral over $m$ then gives the structure
function in the $Q^2, \nu \to \infty$ limit \cite{EINHORN},
\begin{eqnarray}
\nu^2\ W_n(Q^2,\nu)
&=& 2\pi^2 \left( e_q^2\ m_q^2\ \phi^2_n(x)
		+ e_{\bar q}^2\ m_{\bar q}^2\ \phi^2_n(1-x)
	   \right)\ .
\label{eq:thooftW}
\end{eqnarray}
%
% \begin{eqnarray}
% W_n(Q^2,\nu)
% &\sim& { x^2\ \mu_n^2 \over Q^4 }\
%       \left| e_q\ m_q\ \phi_n(x)
%	   + (-1)^m\ e_{\bar q}\ m_{\bar q}\ \phi_n(1-x)
%       \right|^2\ .
% \label{eq:thooftW}
% \end{eqnarray}
%
In addition to the diagonal terms, Eq.~(\ref{eq:thooftW}) also
contains an interference term proportional to
$(-1)^m\ e_q\ e_{\bar q}\ \phi_n(x)\ \phi_n(1-x)$, which is analogous
to the higher-twist terms (associated with four-quark operators)
discussed in Sec.~\ref{sssec:twist}.
The phase $(-1)^m = \exp(-i (p+q)^2/\pi)$ oscillates infinitely rapidly
as $(p+q)^2 \to \infty$, and naively one may expect the interference
term to survive since the individual final states have either positive
($m$ even) or negative ($m$ odd) parity.
However, the smoothing of the $\delta$-function discontinuity means
that the interference term contributes only to the real part of the
Compton amplitude, and does not scale.

The physical picture which this discussion paints is that for
$(p+q)^2>0$ the perturbation expansion diverges and the structure
function is built up entirely from mesonic final states.
Nevertheless, the asymptotic behavior of the resonance sum replicates
exactly that which would result from the handbag diagram with
scattering from free quarks.
Furthermore, as $x \to 1$, the structure function
$\nu^2 W \sim (1-x)^{2\beta-1}$, where the exponent $\beta$ gives
the characteristic $(1/Q^2)^\beta$ fall-off of the meson form factor,
which satisfies the Drell-Yan--West relation \cite{DY,WEST}
(see Sec.~\ref{sssec:elastic} below).
Since the resonances not only contribute to but saturate the scaling
function, the model provides a graphic and quantitative illustration of
the duality between bound state resonances and the scaling function.

What conclusions can be extrapolated from the duality in 1+1
dimensions to the more realistic case of QCD in 3+1 dimensions?
A partial step in this direction was made recently by Batiz and Gross
\cite{NC_BG}, who generalized the 't~Hooft model by extending spinor
degrees of freedom to 3+1 dimensions.
One of the complications of three spatial dimensions lies in
demonstrating that the transverse degrees of freedom, such as
massless gluons, are damped and that only massive hadrons arise
\cite{EINHORN}.
Nevertheless in the large-$N_c$ limit one expects that $q\bar q$ bound
states will still be narrow, so that local duality must still be
invoked.
Beyond the $N_c \to \infty$ limit, however, resonances will acquire
finite widths, and one can expect complications with mixing of
resonant and nonresonant background contributions.
In addition, confinement has of course not been proved in 3+1
dimensions, rendering the discussion suggestive but not rigorous.
Instead, in the literature one usually resorts to quark models to
learn how duality may arise in Nature.
In the following we examine several model studies which may shed
light on how scaling can coexist with confinement in QCD.

\vspace*{0.5cm}
% . . . . . . . . . . . . . . . . . . . . . . . . . . . . . . . . . . . .
\paragraph{Nonrelativistic Models}

To obtain clues about how the disparate regimes of confinement and
asymptotic freedom could coexist in QCD, Greenberg \cite{GREENBERG}
studied a nonrelativistic model of two scalar quarks each with
mass $m$ bound by a harmonic oscillator potential,
$V(r) \sim m \omega^2 r^2$,
where $\omega$ is the harmonic oscillator
eigenfrequency.\footnote{In fact, Greenberg considered the case of
unequal quark masses, however, for clarity we shall simplify the
discussion to the equal mass case.}
The choice of potential was motivated partly by simplicity, and partly
by the expectation that quarks interacting via a harmonic oscillator
would be more free at short distances than for a Coulombic potential,
leading to a more rapid approach to scaling.
A similar model was subsequently discussed by Gurvitz \& Rinat
\cite{GURVITZ_RINAT}, in which other potentials, such as an infinite
square well, were considered.

Solving a two-body Schr\"odinger equation yields solutions for the
wave functions $\psi_n(r)$ for the $n$-th energy level of the bound
state system in terms of Hermite polynomials.
The structure function for scattering from the $n=0$ ground state can
then be written \cite{GREENBERG}
\begin{eqnarray}
{\cal W}
&=& {1 \over \pi^2}
    \sum_n {1 \over n!} \left| \phi_n \right|^2
    \delta\left( q_0 - n\omega - {\vec q \, ^2 \over 2 M} \right)\ ,
\end{eqnarray}
where $\phi_n$ is defined in terms of the wave function $\psi_n(r)$ as
\begin{eqnarray}
\phi_n^a
&=& { \sqrt{n!} \over (-i)^n }
    \int dz\ \psi_0(z)\ \psi_n(z)\ 
    \exp\left( -{i q z \over 2} \right)			\nonumber\\
&=& \left( {\vec q \, ^2 \over 2 M \omega} \right)^{n/2}
    \exp\left( -{\vec q \, ^2 \over 4 M \omega}
	\right)\ ,
\end{eqnarray}
where $M = 2 m$ is the sum of the quark masses, and $\vec q$ is chosen
to be in the $+z$ direction, with $q = |\vec q|$.
Introducing a nonrelativistic scaling variable,
\begin{equation}
x_{\rm nr} = { \vec q \, ^2 \over 2 M q_0 }\ ,
\end{equation}
and using Stirling's formula to approximate $n!$ at large $n$,
the structure function of the system becomes
\begin{eqnarray}
{\cal W}
&\approx& {1 \over \pi^2}
\exp(-2 \eta f(x_{\rm nr}))\
\sum_n \delta
  \left( {\eta (1-x_{\rm nr}) \over x_{\rm nr} - n} \omega
  \right)\ ,
\label{eq:greenbergW}
\end{eqnarray}
where $\eta = \vec q \, ^2 / 2 M \omega$, and
\begin{eqnarray}
f(x_{\rm nr})
&=& 1\ -\ {1 \over 2 x_{\rm nr}}\
 +\ {1-x_{\rm nr}\over x_{\rm nr}}
    \ln\left( {m_a \over m_b} {1-x_{\rm nr} \over x_{\rm nr}}
       \right)\ .
\end{eqnarray}
Each term in the sum in (\ref{eq:greenbergW}) comes from a different
excitation of the $ab$ bound state.
Note that the relevant variable here is $\vec q \, ^2$ rather than the
four-momentum transfer squared, $Q^2 = \vec q \, ^2 - q_0^2$.
Replacing the sum over $n$ in Eq.~(\ref{eq:greenbergW}) with an integral
over $n$ (or averaging over $\vec q \, ^2$ at fixed $x_{\rm nr}$) then
gives \cite{GREENBERG}
\begin{eqnarray}
{\cal W}
&\approx& {1 \over \pi^2 \omega}
\left[ \exp(-2 \eta f(x_{\rm nr})) \right]^2\ .
\end{eqnarray}
The function $f(x_{\rm nr})$ is positive for $0 \leq x_{\rm nr} \leq 1$,
except for a quadratic zero at $x_{\rm nr} = 1/2$, and at large
$\vec q \, ^2$ is approximately equal to
\begin{eqnarray}
f(x_{\rm nr})
&\approx& 4 \left( x - {1 \over 2} \right)^2\ .
\end{eqnarray}
The structure function therefore vanishes for large $\vec q$ ($\eta$),
except at the value of $x_{\rm nr}$ corresponding to the fraction of
the bound state momentum carried by the quark, as expected in the
parton model.
This demonstrates that the deep inelastic limit of the structure
function approaches the limit of incoherent elastic scattering off
its constituents as though the constituents were free, and illustrates
how the scaling limit can coexist with confinement.

\vspace*{0.5cm}
% . . . . . . . . . . . . . . . . . . . . . . . . . . . . . . . . . . . .
\paragraph{Relativistic Models}

The above nonrelativistic model example demonstrates how the effects
of final state interactions, which would spoil the interpretation of
the structure function in terms of incoherent scattering from quark
constituents, are suppressed at large momenta, even in the case
of confining inter-quark forces.
For deep inelastic scattering in the Bjorken limit, on the other hand,
the energy transfer $q_0$ in the target rest frame is much greater than
the mass of the hadron, $q_0 \gg M$, while the nonrelativistic approach
holds only if $q_0 \ll M$.
It is pertinent, therefore, to ask whether the effects of final state
interactions are still suppressed even as one probes the region of
relativistic momenta.

An attempt to address this question was made by Gurvitz
\cite{GURVITZ_BSE} within a relativistic Bethe-Salpeter framework.
As with the above nonrelativistic model, the constituents and the
virtual photon were all taken to be scalars.
The structure function $W$ here can be expressed in terms of the
relativistic bound state wave function $\Phi$, which represents the
solution of the Bethe-Salpeter equation in the ladder approximation,
\begin{eqnarray}
W &=& {1 \over \pi}
      \Im m \int {d^4p \over (2\pi)^4} {d^4p' \over (2\pi)^4}
\Phi(P,p)\ \langle p | G(P+q) | p' \rangle\ \Phi(P,p')\ ,
\end{eqnarray}
where $P$ and $p$ are the bound state and struck quark momenta,
with corresponding masses $M$ and $m$, respectively, and $G$ is the
full Green's function.
Expanding the structure function in powers of $1/q$,
\begin{eqnarray}
\nu W &=& {\cal F}_0 + { {\cal F}_1 \over Q^2 } + \cdots\ ,
\end{eqnarray}
the leading (scaling) term was found \cite{GURVITZ_BSE} to be
\begin{eqnarray}
{\cal F}_0
&=& {1 \over (4\pi)^2}
    {\nu \over q} \int_{|\tilde y|}^\infty dp\ { p \over E_p }
    |\Phi(P,p)|^2\ ,
\end{eqnarray}
where $E_p = \sqrt{\vec p \, ^2 + m_s^2}$ is the energy of the spectator
system (antiquark for a mesonic $q\bar q$ state or diquark for a
three-quark bound state), with $m_s$ the spectator mass.
The variable $\tilde y$ is the minimal momentum of the struck quark,
\begin{equation}
\tilde y(x,Q^2)
= { M (1-x)^2 - m_s^2/M \over
    \sqrt{(1-x)^2 + 4 m_s^2 x^2/Q^2}
  + \sqrt{(1-x)^2 + 4 M^2 x^2 (1-x)^2/Q^2} }\ ,
\end{equation}
which in the nonrelativistic limit reduces to the West scaling
variable $y$ \cite{WEST_Y},
\begin{equation}
\tilde y\ \to\ y \equiv -{q \over 2} + {m \omega \over q}\ ,
\label{eq:westy}
\end{equation}
for the case of zero binding, $m+m_s=M$.
%
% (Note that $(m-y)/M$ is the nonrelativistic analog of the Bjorken
% scaling variable $x$.)
%
After integrating over momenta, the structure function ${\cal F}_0$
was found to depend only on the scaling variable $\tilde x$,
\begin{equation}
\tilde x
= { x + \sqrt{1 + 4 M^2 x^2/Q^2} - \sqrt{(1-x)^2+4m_s^2x^2/Q^2}
   \over 1 + \sqrt{1+ 4 M^2 x^2/Q^2} }\ ,
\label{eq:xbardef}
\end{equation}
which corresponds to the light-cone fraction of the bound state
carried by off-shell struck quark.
In fact, the variables $\tilde x$ and $\tilde y$ are related by
\begin{equation}
\tilde x = 1 - {\sqrt{m_s^2+\tilde y^2}+\tilde y \over M}\ .
\end{equation}
As well as accounting for target mass effects as in the Nachtmann
variable $\xi$, Eq.~(\ref{eq:xi}), the variable $\tilde x$
includes in addition dynamical corrections to the $x$-scaling through
the $m_s$ dependent term in Eq.~(\ref{eq:xbardef}), which are not
accounted for in $\xi$.
The use of the modified scaling variables explicitly removes
kinematical $1/Q^2$ corrections, and allows a more effective
separation of the leading-twist and higher-twist effects.

Going beyond the scalar approximation, Pace, Salme \& Lev \cite{PSL}
explicitly incorporated spin degrees of freedom of the hadronic
constituents in their study of the compatibility of confinement
with scaling in DIS.
Using light-cone Hamiltonian dynamics, the authors consider a system
of two relativistic spin-1/2 particles with mass $m$ interacting via
a relativistic harmonic oscillator potential, $V(r) = (a^4/m) r^2$,
where $a$ is a constant with dimensions of mass.
The light-cone, or front-form, dynamics allows one to determine the
energy spectrum and wave functions exactly from the correspondence
between the relativistic wave equation for the mass operator and the
nonrelativistic Schr\"odinger equation.
The energy of the $n$-th excited state of this system is then
given by
\begin{equation}
E_n = (M_n^2 - 4 m^2)/4m\ ,
\end{equation}
where $M_n$ is the mass of the $n$-th excited state,
\begin{equation}
M_n = 2 \sqrt{m^2 + a^2 (2n+3)}\ ,
\end{equation}
and $n = n_x + n_y + n_z$.
In the weak binding limit, $a \ll m$, the structure function $F_1$
(or $F_2$) is calculated by summing over the discrete states $n$,
\begin{eqnarray}
F_1(x,Q^2)
&=& {m^2 x^4 \over 8 \pi^2 (1-x) Q^2}
    \sum_n \delta \left( x - {Q^2 \over Q^2 + 8 a^2 n} \right)
    \left( f^2(n,x) \right)\ ,
\label{eq:psl}
\end{eqnarray}
where $f(n,x)$ depends explicitly on the wave functions, and $x$ is
the usual Bjorken scaling variable.
In the limit $n \to \infty$, $f(n,x)$ reduces to \cite{PSL}
\begin{eqnarray}
f(n,x)
&\to& { \sqrt{8\pi}\ a\ \chi(x) \over m\ x^{3/2} }\ ,
\end{eqnarray}
where $\chi(x)$ is proportional to the ground state wave function,
\begin{eqnarray}
\chi(x)
&=& \left( { 2 \sqrt{\pi} m \over a } \right)^{1/2}
    \exp\left[ -k_z(x)/2a^2 \right]\ ,
\end{eqnarray}
with $k_z(x) = m (x-1/2)/\sqrt{x(1-x)}$.

The correspondence between the structure function in terms of a
discrete spectrum of $\delta$-functions and the continuous, smooth
scaling function versus $x$ is implemented by averaging, or smearing,
over the experimental resolution in bins of $x$ and $Q^2$.
Averaging over an interval of $\langle x \rangle \in [x,x+\delta x]$,
such that $\delta x \ll \langle x \rangle$, Pace {\em et al.} \cite{PSL}
define the smeared structure function
\begin{equation}
{\bar F}_1(\langle x \rangle,Q^2)
\equiv {1 \over \delta x} \int_x^{x+\delta x} dx\ F_1(x,Q^2)\ .
\label{eq:pslavg}
\end{equation}
At large $Q^2$ there are many states $n$ which populate the region
$Q^2/(Q^2+8a^2n) \in [x,x+\delta x]$ (see Eq.~(\ref{eq:psl})),
so that the integral in Eq.~(\ref{eq:pslavg}) becomes a smooth
function of $Q^2$.
In the limit $Q^2 \to \infty$ ($n \to \infty$) the structure function
then becomes a scaling function of $x$,
\begin{equation}
F_1(x) \longrightarrow { \chi(x)^2 \over 8 \pi x(1-x) }\ .
\end{equation}
Therefore the result for $F_1(x,Q^2)$ is indeed compatible with the
parton model once an average over bins of $x$ is performed.
The issue of averaging is a crucial one in relating structure
functions calculated in hadronic and partonic bases, and we shall
return to this in Sec.~\ref{sssec:res} below.
This example also demonstrates that the usual interpretation of the
Bjorken variable $x$ as the momentum fraction of the struck quark is
still valid in a relativistic framework, in the presence of strong
final state interactions.

\vspace*{0.5cm}
% . . . . . . . . . . . . . . . . . . . . . . . . . . . . . . . . . . . .
\paragraph{Phenomenological Models}

The above models have focused on understanding the qualitative
features of the appearance of scaling from hadronic degrees of
freedom, with only remote connections to the empirical spectrum
of resonances from which the scaling function is built up.
In a more phenomenological approach, as an early alternative to
the parton model, Domokos {\em et al.} \cite{DOM1,DOM2,DOM3} showed
that one could accommodate structure function scaling by summing over
resonances parameterized by $Q^2$-dependent form factors.

Assuming a harmonic oscillator-like spectrum of nucleon excitations,
in which the mass of the $n$-th excited state was given by
$M_n^2 = (n+1) \Lambda^2$, with $n$ an integer and $\Lambda$ some mass
scale, analytic expressions for the structure function were obtained
by including contributions from positive and negative parity states
with spin $1/2, 3/2, \cdots, n+1/2$, with $n$ even and $n$ odd
corresponding to isospin 1/2 and 3/2 excitations, respectively.
%
% The total widths are given by
% $\Gamma_n = \sqrt{s_n} \Gamma_0 (s_n-1)$,
%
The structure function $F_2$ was then given by a sum of transition
form factors weighted by kinematical factors \cite{DOM1}.
% as in Eq.~(\ref).
%
Although both electric and magnetic form factors contribute to the
resonance sum, at high $Q^2$ the structure function becomes dominated
by the magnetic coupling, in which case the transition form factors
can be parameterized by
\begin{eqnarray}
G_n(Q^2)
&=& { \mu_n \over \left( 1 + Q^2 r^2/M_n^2 \right)^2 }\ ,
\end{eqnarray}
where $\mu_n$ here is the magnetic moment for the state $n$,
and the parameter $r^2 \approx 1.41$.
In the Bjorken limit the summation over discrete states is replaced
by an integration over the variable $z \equiv M_n^2/Q^2$, 
\begin{eqnarray}
F_2
&\sim& (\omega' - 1)^{1/2} (\mu_{1/2}^2 + \mu_{3/2}^2)
       {\Gamma_0 \over \pi}
       \int_0^\infty dz
       {  z^{3/2} (1 + r^2/z)^{-4} \over
	  z + 1 - \omega' + \Gamma_0^2 z^2 }\ ,
\label{eq:dom_scale}
\end{eqnarray}
where $\omega' = \omega + M^2/Q^2$ is the scaling variable introduced
by Bloom \& Gilman \cite{BG1,BG2}, and a Breit-Wigner form has been
introduced to smear the narrow resonances,
\begin{eqnarray}
\delta(W^2 - M_n^2)
&\to& {1\over \pi}
      {\Gamma_n M_n \over (W^2 - M_n^2)^2 + \Gamma_n^2\ M_n^2}\ ,
\label{eq:bw}
\end{eqnarray}
with $\Gamma_n$ the total width for the $n$-th state.
The parameter $\Gamma_0 \approx 0.13$ in Eq.~(\ref{eq:dom_scale})
is obtained from the slope of $\Gamma_n M_n$ versus $M_n^2$ for
the existing nucleon and $\Delta$ resonances (see Ref.~[12] of
\cite{DOM1}).
The replacement of the summation over the discrete set of
$\delta$-functions by a continuous integral amounts to an averaging
over neighboring regions of $W$, which becomes a better approximation
at increasingly higher $Q^2$.

From Eq.~(\ref{eq:dom_scale}) one sees that the resonance summation
indeed yields a scaling function in the Bjorken limit.
Furthermore, in the narrow resonance approximation, $\Gamma_n \to 0$,
this simplifies even further, with the structure function depending
only on the magnetic moments and the scaling variable $\omega'$,
\begin{eqnarray}
F_2 &\sim&
(\mu_{1/2}^2 + \mu_{3/2}^2)
{ (\omega' - 1)^3 \over (\omega' - 1 + r^2)^4 }\ .
\end{eqnarray}
In particular, this form exhibits the correct $\omega' \to 1$ behavior
according to the Drell-Yan--West relation \cite{DY,WEST}, and the
empirical dependence of structure functions in the $x \to 1$ limit
(see Sec.~\ref{sssec:elastic} below).
Similar arguments have also been used to derive scaling in
spin-dependent \cite{DOM2} and neutrino structure functions
\cite{DOM3} from sums over resonant excitations.

% .......................................................................
\subsubsection{Resonances and the Transition to Scaling}
\label{sssec:res}

The above models (nonrelativistic, relativistic and phenomenological)
provide graphic illustrations of the compatibility of confinement and
asymptotic scaling in DIS, however, they do not address the question
of the origin of Bloom-Gilman duality at {\em finite} $Q^2$.
The behavior of structure functions in the region of transition from
resonance dominance to scaling, and the onset of Bloom-Gilman duality
in the {\em preasymptotic} region, was examined recently by several
authors \cite{CI,IJMV,CZ,PAND,HARRINGTON} in dynamical models.
As we shall see, the issue of averaging and smearing the
$\delta$-function spikes in the resonance sum is rather more important
here than in the Bjorken limit, since it determines to a large extent
the shape of the resonance structure function, and the speed with which
the scaling function is approached.

In this section we generalize the models introduced above to the case
of finite $Q^2$, and examine specifically how the resonance structures,
which dominate the structure function at low $Q^2$, dual the scaling
function which characterizes scattering at asymptotic $Q^2$.
While the models may give rise to scaling at high $Q^2$, it is not
{\em a priori} obvious that the resonance structure functions need
approach the scaling limit uniformly, and the origin of the empirical
oscillations about the scaling curve discussed in
Sec.~\ref{sec:bgstatus} needs to be understood.

We begin the discussion with a recent phenomenological study of duality
in which the structure function at low $Q^2$ is built up from the known
resonances below $W \approx 2$~GeV.
With increasing $Q^2$, the phenomenological approach quickly becomes
intractable, however, and a quark-level description becomes more viable.
We illustrate how low-$Q^2$ duality arises in simple quark models,
firstly considering the simplified case of scattering from a single
quark bound to an infinitely massive core, and then to the more
realistic case of several quark charges.
The latter case is important in clarifying the puzzle of how the square
of the sum of quark charges (coherent scattering) can yield the sum of
the squares of quark charges (incoherent scattering).

\vspace*{0.5cm}
% . . . . . . . . . . . . . . . . . . . . . . . . . . . . . . . . . . . .
\paragraph{Resonance Parameterizations}

A phenomenological model of the structure functions in the resonance
region was constructed recently by Davidovsky \& Struminsky
\cite{STRUM}, in the spirit of the earlier work of Domokos {\em et al.}
\cite{DOM1,DOM2,DOM3}, but with additional physical constraints for
the threshold behavior as $\vec q \to 0$, and the asymptotic behavior
as $Q^2 \to \infty$.
In terms of the helicity amplitudes $G_m^R(Q^2)$ for a given resonance
$R$ \cite{CMFF},
\begin{eqnarray}
G_m^R(Q^2)
&=& {1 \over 2M}
    \left\langle R (\lambda_R)\ \left|\
	\epsilon^\mu_{(m)} \cdot J_\mu(0)\
    \right| N (\lambda = 1/2) \right\rangle\ ,
\end{eqnarray}
where $\epsilon^\mu_{(m)}$ is the photon polarization vector for
helicity $m$ ($m = 0, \pm$), and $\lambda_R = m - 1/2$ is the helicity
of the resonance $R$, the contributions to the spin-averaged $F_1$
and $F_2$ and spin-dependent $g_1$ and $g_2$ structure functions
of the nucleon can be written \cite{STRUM,CMFF} as
\begin{eqnarray}
\label{eq:F1cm}
F_1^R
&=& M^2\ \left[ |G^R_+|^2 + |G^R_-|^2 \right]\
    \delta(W^2 - M_R^2)\ ,				\\
\label{eq:F2cm}
\left( 1 + {\nu^2 \over Q^2} \right) F_2^R
&=& M \nu\
    \left[ |G^R_+|^2 + 2|G^R_0|^2 + |G^R_-|^2 \right]\
    \delta(W^2 - M_R^2)\ ,				\\
\label{eq:g1cm}
\left( 1 + {Q^2 \over \nu^2} \right) g_1^R
&=& M^2\ 
    \left[ |G^R_+|^2 - |G^R_-|^2
	 + (-1)^{J_R-1/2} \eta_R\ {\sqrt{2 Q^2} \over \nu}\
	   G^{R *}_0\ G^R_+
    \right]\
    \delta(W^2-M_R^2)\ ,				\\
\label{eq:g2cm}
\left( 1 + {Q^2 \over \nu^2} \right) g_2^R
&=& -M^2\
    \left[ |G^R_+|^2 - |G^R_-|^2
	 - (-1)^{J_R-1/2} \eta_R\ {\nu\sqrt{2} \over \sqrt{Q^2}}\
	   G^{R *}_0\ G^R_+
    \right]\
    \delta(W^2-M_R^2)\ ,
\end{eqnarray}
with $J_R$ and $\eta_R$ the total spin and parity of the resonance $R$,
respectively.
%
% \begin{eqnarray}
% \delta(W^2 - s_n)
% &\to& {1\over \pi}
%       {\Gamma_n s_n^{1/2} \over (W^2 - s_n)^2 + \Gamma_n^2 s_n}\ .
% \end{eqnarray}
%
% In the notation in Eq.~\ref{eq:F2dom}, the form factors correspond to
% $G_{1n} \sim G_R^0$, $G_{2n} \sim G_R^-$, $G_{3n} \sim G_R^+$.
%
Apart from the nucleon elastic, and to some extent the $N \to \Delta$
transition, the detailed $Q^2$ dependence of the form factors
$G^R_m(Q^2)$ is not known.
On the other hand, there are firm predictions from perturbative QCD
for the asymptotic $Q^2 \to \infty$ behavior of the form factors, which
can be used to constrain the phenomenological parameterizations.
Using in addition the known constraints from the $|\vec q\,| \to 0$
behavior of the form factors at threshold, and the value at the photon
point, $Q^2=0$, the form factors were parameterized in
Refs.~\cite{STRUM,FIORE_REV} as
\begin{eqnarray}
\label{eq:strum1}
\left| G^R_{\pm}(Q^2) \right|^2
&=& \left| G^R_{\pm}(0) \right|^2\,
\left( {|\vec{q}| \over |\vec{q}|_0}\,
       {\Lambda^{'2} \over Q^2 + \Lambda^{'2}}
\right)^{\gamma_1}\,
\left( {\Lambda^2 \over Q^2 + \Lambda^2} \right)^{m_\pm},	\\
\left| G^R_0(Q^2) \right|^2
&=& C^2 \left( {Q^2 \over Q^2 + \Lambda^{''2}} \right)^{2a}
{q_0^2 \over |\vec{q}|^2}\,
\left( {|\vec{q}| \over |\vec{q}|_0}\,
       {\Lambda^{'2} \over Q^2 + \Lambda^{'2}}
\right)^{\gamma_2}\,
\left( {\Lambda^2 \over Q^2 + \Lambda^2} \right)^{m_0}\ ,
\label{eq:strum2}
\end{eqnarray}
where $\gamma_1 = 2 J_R - 3$ and $\gamma_2 = 2 J_R - 1$ for normal
parity $\gamma^* N \to R$ transitions
($J^P=1/2^+ \to 3/2^-, 5/2^+, 7/2^-, \cdots$), and
$\gamma_1 = 2 J_R - 1$ and $\gamma_2 = 2 J_R + 1$
for anomalous transitions
($1/2^+ \to 1/2^-, 3/2^+, 5/2^-, \cdots$),
with
\begin{eqnarray}
|\vec{q}|
&=& {\sqrt{(M_R^2-M^2-Q^2)^2+4 M_R^2 Q^2} \over 2 M_R}\ ,\ \ \ \
|\vec{q}|_0\ \equiv\ |\vec q|(Q^2=0)\ =\ {M_R^2-M^2 \over 2 M_R}\ .
\end{eqnarray}
The exponents in Eqs.~(\ref{eq:strum1}) and (\ref{eq:strum2}) are
given by $m_+=3$, $m_0=4$ and $m_-=5$, and the parameters $\Lambda^2$,
$\Lambda^{'2}$, $\Lambda^{''2}$, $a$ and $C$ are determined
empirically.
Focusing on the $F_2$ structure function, one can easily verify that
in the limit $x \to 1$ the resulting structure function behaves as
$F_2(x) \sim (1-x)^{m_+}$, as required by pQCD counting rules
(see Sec.~\ref{sssec:elastic} below).

\begin{figure}[ht]
\begin{center}  
\epsfig{file=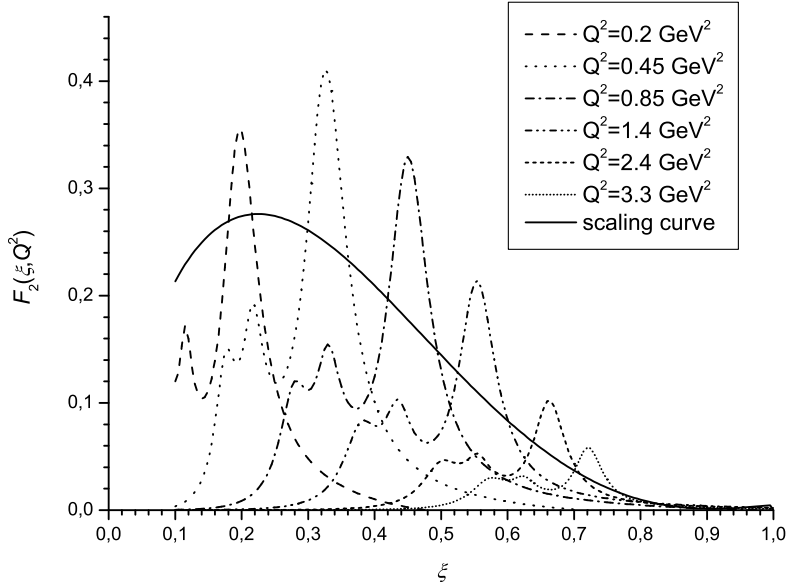,height=9cm}
\caption{\label{fig:strumF2}
	Resonance contributions to the proton $F_2^p$ structure
	function versus the Nachtmann scaling variable $\xi$ in
	the model of Ref.~\protect\cite{STRUM}.  The solid line
	is a parameterization of DIS data \protect\cite{F2JL1}.}
\end{center}
\end{figure}

Summing over a total of 21 resonance states in the isospin-1/2 and
isospin-3/2 channels with masses $M_R \alt 2$~GeV, the total $F_2$
structure function is shown in Fig.~\ref{fig:strumF2} as a function
of the Nachtmann scaling variable $\xi$.
The $\delta$-functions in Eqs.~(\ref{eq:F1cm})--(\ref{eq:g2cm}) are
smeared by a Breit-Wigner shape as in Eq.~(\ref{eq:bw}).
The $\Delta$ resonance clearly provides the largest contribution.
The nonresonant background contribution here is relatively small,
so that as $Q^2$ increases the $\Delta$ peak moves to larger $\xi$,
following the general trend of the scaling curve.
On the other hand, the higher-mass resonances lie somewhat below
(factor 2) the scaling curve at the $Q^2$ values shown, which reflects
the absence of the nonresonant backgrounds which are relatively more
important for the higher-mass resonances.
At lower $\xi$ (higher $W$), a quantitative description of the data
would require the inclusion of additional resonances beyond
$M_R \sim 2$~GeV.
This quickly becomes intractable, however, as little phenomenological
information exists on $N \to R$ transitions at high $W$, and indicates
that a quark-level description may be more feasible at these kinematics.

\vspace*{0.5cm}
% . . . . . . . . . . . . . . . . . . . . . . . . . . . . . . . . . . . .
\paragraph{Harmonic Oscillator Model}

Despite the challenges in describing the transition to scaling in terms
of phenomenological form factors, it is nevertheless vital to understand
how the dynamics of resonances gives way to scaling.
Recently Isgur {\em et al.} \cite{IJMV} addressed this problem in the
context
of a simple quark model, in which both the appearance of duality at low
$Q^2$ and the onset of scaling at high $Q^2$ was studied.
To simplify the problem Isgur {\em et al.} consider a spinless, charged
quark of mass $m$ confined to an infinitely massive core via a
harmonic oscillator potential (see also Ref.~\cite{IOFFE}).
%
% Final state interactions for a relativistic particle bound in an
% external field was considered by Ioffe {\em et al.} \cite{IOFFE}.
% However, in this model one of the particles (the source of the field)
% has an infinitely large mass.
%
For the case of scalar photons, the inclusive structure function is
given by a sum of squares of transition form factors (as in the models
discussed above) weighted by appropriate kinematic factors \cite{IJMV},
\begin{equation}
{\cal W}(\nu,\vec{q})
= \sum_{N=0}^{N_{\rm max}} {1 \over 4 E_0 E_N}
  |F_{0,N}(\vec{q})|^2
  \delta(E_N - E_0 - \nu) \ .
\label{eq:ijmv_W}
\end{equation}
The form factors $F_{0,N}$ represent transitions from the ground
state to states characterized by the principal quantum number
$N(\equiv l + 2k$, where $k$ is the radial and $l$ the orbital quantum
numbers), and the sum over states $N$ goes up to the maximum
$N_{\rm max}$ allowed at a given energy transfer $\nu$.
% Note that for fixed $Q^2 \equiv \vec q \, ^2 - \nu^2 > 0$,
% $N_{max} = \infty$.
A related discussion which focuses on the response in the time-like
region was given by Paris \& Pandharipande \cite{PAND}.

The spectrum corresponding to this system can be determined by noting
the similarity between the relativistic Klein-Gordon equation and the
Schr\"odinger
equation for a nonrelativistic harmonic oscillator with a potential
$V^2 (r) = \alpha \, r^2$, with $\alpha$ a generalized, relativistic
string constant, which yields the same solutions for the wave functions.
The energy eigenvalues in this case are given by $E = \pm E_N$,
where $E_N = \sqrt{2\beta^2(N+3/2) + m^2}$ and $\beta = \alpha^{1/4}$,
and the excitation form factors are derived using the recurrence
relations of the Hermite polynomials \cite{IJMV},
\begin{equation}
F_{0,N} (\vec q \, ^2) = {1 \over \sqrt{N!}} \, i^N \,
\left ( {|\vec q| \over \sqrt{2} \, \beta} \right ) ^N
\exp (- \vec q \, ^2 / 4 \, \beta^2)\ .
\label{eq:hoff}
\end{equation}
This form factor is in fact the sum of all form factors for excitations
from the ground state to degenerate states with the same principal
quantum number $N$.
A necessary condition for duality is that these form factors can
represent the pointlike free quark.
One can verify that
$\sum_{N=0}^{N_{\rm max}} |F_{0,N} (\vec q \,)|^2 \rightarrow 1$ as
$N_{\rm max} \rightarrow \infty$, which follows from the completeness
of the wave functions.
For any individual contribution, $F_{0,N}(\vec{q})$ reaches its
maximum value when $\vec q \, ^2 = 2 \beta^2 N$, at which point
$F^2_{0,N} = F^2_{0,N+1}$.
This coincidence is true in fact for all juxtaposed partial waves
at their peaks.
Furthermore, using $\nu_N=E_N-E_0$ and $E_N=\sqrt{2\beta^2\,N+E_0^2}$,
one finds that $\nu_N=(\vec q \, ^2_N - \nu_N^2)/2E_0$, so that the
position of the peak in the averaged structure function occurs at
$Q^2/2m\nu = m/E_0$, which is the fraction of the bound system's
light-cone momentum.

The scaling function corresponding to the structure function in
Eq.~(\ref{eq:ijmv_W}) is given by
\begin{eqnarray}
{\cal S}(u,Q^2)
&\equiv& |\vec q|\ {\cal W}\
 =\ \sqrt{\nu^2 + Q^2}\ {\cal W}\ ,
\end{eqnarray}
with dimensions [mass]$^{-2}$.
The scaling variable $u$ is defined as
\begin{eqnarray}
u &=& {1 \over 2 m}
      \left( \sqrt{\nu^2 + Q^2} - \nu \right)
      \left( 1 + \sqrt{1 + {4 m^2 \over Q^2}} \right)\ ,
\label{eq:ijmv_u}
\end{eqnarray}
and takes into account both target mass and quark mass effects
\cite{BARBIERI} ({\em cf.} the variable $\tilde x$ in
Eq.~(\ref{eq:xbardef})).
Note that the variable $u$ in Eq.~(\ref{eq:ijmv_u}) is scaled by the
quark mass, $m$, rather than the bound state mass, so that the range of
$u$ is between 0 and $\infty$.

\begin{figure}[ht]
\begin{center}  
\hspace*{-3cm}
\epsfig{file=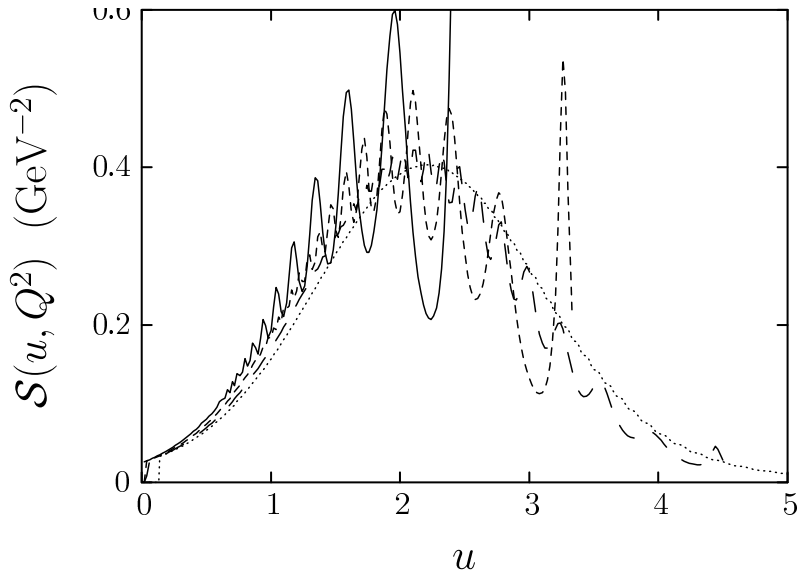,height=22cm}
\vspace*{-14cm}
\caption{\label{fig:ijmv}
	Onset of scaling for the structure function ${\cal S}(u,Q^2)$
	as a function of $u$ for $Q^2=0.5$ (solid), $Q^2=1$
	(short-dashed), 2 (long-dashed) and 5~GeV$^2$ (dotted).
	The widths for $N \geq 1$ has been arbitrarily set at
	$\Gamma_N=100$~MeV\, with the elastic width set to
	$\Gamma_{N=0}=30$~MeV.  (From Ref.~\protect\cite{IJMV}.)}
\end{center}
\end{figure}

The structure function ${\cal S}(u,Q^2)$ is shown in
Fig.~\ref{fig:ijmv} for several finite values of $Q^2$, where for
illustration the $\delta$-functions have been smoothed by a
Breit-Wigner shape with an arbitrary but small width, $\Gamma_N$
({\em cf.} Eq.~(\ref{eq:bw})),
\begin{equation}
\delta(E_N - E_0 - \nu)
\rightarrow {\Gamma_N \over 2 \pi} \, \,
	    {f_N \over (E_N - E_0 - \nu)^2 + (\Gamma_N/2)^2}\ ,
\end{equation}
where the factor
$
f_N = {\pi}/[\pi/2 + \arctan 2 (E_N - E_0)/\Gamma_N]
$
ensures that the integral over the $\delta$-function is identical
to that over the Breit-Wigner shape.
The resonance structure is quite evident in each of the low-$Q^2$
curves, with the amplitude of oscillation decreasing with increasing
$Q^2$.
As $Q^2$ increases, each of the resonances moves out towards higher
$u$, as dictated by kinematics.
The right-most peak in each of the curves corresponds to the elastic
contribution.
At $Q^2 = 0$, this is in fact the only allowed state, and is equal
to almost half of the asymptotic value of the integral over $u$.
It remains rather prominent for $Q^2$ = 0.5~GeV$^2$, though most
of the function is by this point built up of excited states,
and it becomes negligible for $Q^2 \geq 2$~GeV$^2$.

For local duality to hold, the resonance ``spikes" would be expected
to oscillate around the scaling curve and to average to it, once $Q^2$
is large enough.
Remarkably, even the curves at lower $Q^2$ tend to oscillate around
the scaling curve.
Note that these curves are at fixed $Q^2$, but sweep over all $\nu$.
As $\nu$ is increased, more and more highly excited states are created,
making the density of states larger at smaller $u$.
In the continuum limit, where $N \to \infty$ and the density of states
becomes very large, the resonance spikes die out and the structure
function approaches its asymptotic value.
Using Stirling's formula, one can indeed show that the scaling function
takes the analytic form \cite{IJMV}
\begin{equation}
{\cal S}(u)
= {E_0 \over \sqrt{\pi} \beta }
  \exp{\left(-{(E_0-m u)^2 \over \beta^2}\right)}\ .
\label{eq:ijmv_Su}
\end{equation}
The difference between the scaling function and the curve in
Fig.~\ref{fig:ijmv} at 5~GeV$^2$ (dotted) is almost negligible.
The asymptotic scaling function therefore straddles the oscillating
resonance structure function in an apparently systematic manner.
This is quite extraordinary given the very simple nature of the model,
and points to the rather general nature of the phenomenon of duality.

\vspace*{0.5cm}
% . . . . . . . . . . . . . . . . . . . . . . . . . . . . . . . . . . . .
\paragraph{Sum of Squares vs. Square or Sums}

Simple models such as the one discussed above are clearly valuable
in providing physical insight into the dynamical origins of duality.
However, one may wonder whether some of the qualitative features of
duality and the onset of scaling here could be a consequence of the
restriction to scattering from a single quark charge.
In general, if one neglects differences between the quark flavors,
the magnitude of the structure function $F_2$ is proportional to the
sum of the squares of the (quark and antiquark) constituent charges,
$\sum_q e_q^2$.
On the other hand, the summation over resonance form factors is
implicitly driven by the coherently summed square of constituent
charges, $(\sum_q e_q)^2$, for each resonance.
The basic question arises then: {\em How does the square of the
sum become the sum of the squares?}

While the various examples above and in Sec.~\ref{sssec:conf} have
illustrated how the coherent and incoherent descriptions can be merged
at high energies, the question of the cancellation of the interference
terms $\sum_{q\not=q'} e_q e_{q'}$ has been either side-stepped or
neglected altogether in these discussions.
For instance, in the 't~Hooft model the interference term was found
not to scale following the smoothing of the $\delta$-function
discontinuity \cite{EINHORN}.
Moreover, by restricting oneself to a single quark charge, as in the
model of Isgur {\em et al.} \cite{IJMV}, the problem of interference terms
does not arise at all.
The physics of the cancellation of the interference terms, which are
related to the higher-twist matrix elements responsible for violations
of duality, is therefore not clear.

Close and Isgur \cite{CI} elucidated this problem by drawing attention
to the necessary conditions for duality to occur for the general case
of more than one quark charge.
They considered a composite state made of two equal mass scalar quarks
with charges $e_1$ and $e_2$, at positions $\vec{r}_1$ and $\vec{r}_2$,
respectively, interacting via a harmonic oscillator potential.
The ground state wave function for this system is denoted by
$\psi_0(\vec{r})$, where $\vec{r}_{1,2} \equiv \vec{R} \pm \vec{r}/2$
is defined in terms of the center of mass ($\vec R$) and relative
($\vec r$) spatial coordinates.
The amplitude for the system to absorb a photon of momentum $\vec{q}$
is proportional to
\begin{eqnarray}
& & e_1 e^{i \vec{q} \cdot \vec{r}/2}\
 +\ e_2 e^{-i \vec{q} \cdot \vec{r}/2}\ ,
\end{eqnarray}
which can be rewritten as a sum and difference of the amplitudes as
\begin{eqnarray}
& &
(e_1 + e_2)
\left(
  e^{i \vec{q} \cdot \vec{r}/2}\ +\ e^{-i \vec{q} \cdot \vec{r}/2}
\right)\
+\
(e_1 - e_2)
\left(
  e^{i \vec{q} \cdot \vec{r}/2}\ -\ e^{-i \vec{q} \cdot \vec{r}/2}
\right)\ .
\end{eqnarray}
Using the partial wave expansion
$\exp(iqz/2) = \sum_l i^l P_l(\cos \theta)\ j_l(qr/2) (2l + 1)$
to project out even and odd partial waves, the form factor is
generally given by
\begin{eqnarray}
F_{0,N(l)}(\vec{q})
&\sim& \int dr\ r^2\ \psi_l^*(r)\ \psi_0(r)\ j_l(qr/2) \nonumber\\
&    & \times
\left[ (e_1 + e_2) \delta_{l\ {\rm even}}
     + (e_1 - e_2) \delta_{l\ {\rm odd}}
\right]\ ,
\end{eqnarray}
where $N \equiv 2k + l$ with $k$ the radial quantum number, and
the wave function $\psi_l(\vec{r})$ describes the excitation of a
resonant state with angular momentum $l$.
The resulting structure function, summed over resonance excitations,
will receive even- and odd-$l$ contributions proportional to
$(e_1 \pm e_2)^2$, respectively.
For the harmonic oscillator potential the even and odd-$l$
components also correspond to even and odd $N$, {\em i.e.},
$N = 2n$ and $N = 2n+1$, respectively, with $n$ an integer.
Their contributions to the structure function can be written
\cite{CI,CZ}
\begin{eqnarray}
{\cal F}(\nu,\vec{q})
&=& \sum_{N(n)} {1 \over 4 E_0 E_N}
\left\{ (e_1^2 + e_2^2)
	(F^2_{0,2n}(\vec{q}) + F^2_{0,2n+1}(\vec{q}))\
\right.					\nonumber\\
& & \hspace*{2.5cm}
\left.
 +\ 2\ e_1 e_2\ (F^2_{0,2n}(\vec{q}) - F^2_{0,2n+1}(\vec q))
\right\}\
\delta(E_N - E_0 - \nu) \ .
\end{eqnarray}
This representation reveals the physics rather clearly.
The excitation amplitudes to resonance states contain both diagonal
($e_1^2 + e_2^2)$ and nondiagonal ($\pm 2 e_1 e_2$) terms, which are
leading- and higher-twist, respectively.
The former add constructively for any $l$, and the sum over the
complete set of states yields the scaling structure function
(``sum of squares'').
The latter, on the other hand, enter with opposite phases for even
and odd $l$, and hence interfere destructively.
This exposes the critical point that {\it at least one complete set
of even and odd parity resonances must be summed over} for duality
to hold \cite{CI}.
An explicit demonstration of how this cancellation takes place in
the SU(6) quark model and its extensions is discussed in
Sec.~\ref{sssec:qm} below.

Turning now to the more physical case of a vector photon (but still
scalar quark --- see also Ref.~\cite{JV,JV04}), the dominant structure
function at large $Q^2$ is the longitudinal response function,
\begin{equation}
R_L(\vec{q},\nu)
= \sum_{N=0}^{N_{\rm max}} {1 \over 4 E_0 E_N} |f_{0,N}(\vec{q})|^2
  (E_0 + E_N)^2 \delta(\nu + E_0 - E_N)\ ,
\label{eq:RL}
\end{equation}
where
\begin{equation}
|f_{0,N}(\vec{q})|^2\
\equiv\ (e_1^2 + e_2^2)
	\left( F^2_{0,2n}(\vec{q}) + F^2_{0,2n+1}(\vec{q}) \right)\
     +\ 2 e_1 e_2
	\left( F^2_{0,2n}(\vec{q}) - F^2_{0,2n+1}(\vec{q}) \right)\ .
\end{equation}
Once again the sum over $N$ denotes the equivalent sum over $n$ for
$N=2n$ and $N=2n+1$.
% and from kinematics $\nu_{\rm max} < |\vec{q}|$.
%
In the limit of $N$ (or $n$) $\to \infty$, the parity-even and odd
partial waves sum to the same strengths,
$\sum_{n=0}^{\infty}F^2_{0,2n}(\vec{q})
=\sum_{n=0}^{\infty}F^2_{0,2n+1}(\vec{q})$,
and the interference term proportional to $e_1 e_2$ vanishes.

The cancellation of the cross terms is explicitly realized for the
case of a harmonic oscillator potential, where the use of Stirling's
formula in the continuum limit gives rise to the scaling longitudinal
response function \cite{CZ}
\begin{eqnarray}
R_L(\vec{q},\nu)
&=& (e_1^2 + e_2^2)\
    {(\nu + 2E_0)^2 \over 4 \beta E_0 \sqrt{\pi} \nu}	\nonumber\\
& & \times
    \exp\left\{ {\nu^2 + 2 E_0 \nu \over 2 \beta^2}
	\ln\left( {Q^2+\nu^2 \over \nu^2 + 2 E_0 \nu}\right)
	  - {Q^2 + 2 E_0 \nu \over 2 \beta^2}
	\right\}\ .
\label{eq:RLscale}
\end{eqnarray}
The interference terms thus cancel exactly, leaving behind the purely
incoherent contribution proportional to the squares of the quark
charges.

Similar results have also been obtained recently by Harrington
\cite{HARRINGTON}, who performed a detailed study of the relationship
between coherent and incoherent descriptions of the structure function
within this model and the cancellation of the higher-twist interference
terms.
Summing over the orbital angular momentum for each $N$, the
contributions to the structure function from a transition to the
state $N$ were shown to be proportional to 
$e_1^2 + e_2^2 + 2 e_1 e_2 (-1)^N$, which illustrates how the
contributions from alternate energy levels tend to cancel for the
$e_1 e_2$ interference term.

\begin{figure}[ht]
\begin{center}  
\epsfig{file=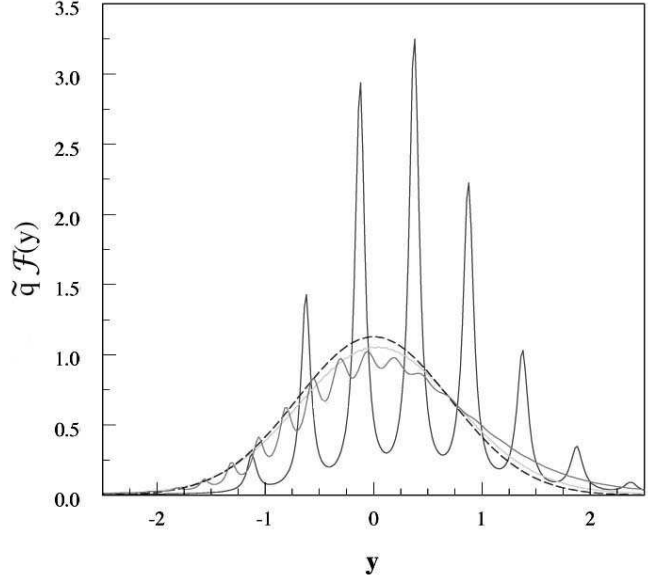,height=8cm}
\vspace*{0.5cm}
\caption{\label{fig:FigIV_harr}
	Structure function $\tilde q {\cal F}$ as a function of the
	scaling variable $y$ for dimensionless momentum transfers
	$\tilde q=4$ (large peaks), 8 (small peaks), and 32 (smooth
	curve).  The sharp energy levels have been given a width which
	increases from 0.2 to 2 as their energy increases. The dashed
	curve is the Gaussian limit of the scaled free particle.
	From Ref.~\protect\cite{HARRINGTON}.}
\end{center}
\end{figure}

For the case $e_1 = e_2 = e$, for which the charge factor in
Eq.~(\ref{eq:RLscale}) alternates between $4 e^2$ and $0$, the
contributions from the two particles cancel or add coherently
for odd- and even-parity states, respectively \cite{CI,HARRINGTON}.
The resulting (scaled) structure function $\tilde q {\cal F}(y)$
is plotted in Fig.~\ref{fig:FigIV_harr} for several values of the
dimensionless momentum $\tilde q = q/\sqrt{2m\omega}$.
The scaling variable $y$ is the dimensionless analog of the West
scaling variable in Eq.~(\ref{eq:westy}), which is related to the
component of a parton's momentum in the $\vec q$ direction before
the collision \cite{WEST}.
The resonance peaks, which have been broadened by a Breit-Wigner
form (\ref{eq:bw}), show clear oscillations about the scaling
function, as in the earlier example in Fig.~\ref{fig:ijmv}.
As $q \to \infty$, the oscillations in the curves are damped out,
and the curves approach the asymptotic scaling limit, in which the
structure function is given by a Gaussian,
$\tilde q {\cal F}(y) \to (2/\sqrt{\pi}) \exp(-y^2)$.

The approach to scaling can be further illuminated by considering
the lowest moment of the structure function.
Returning to the longitudinal response function $R_L(\vec{q},\nu)$
in Eq.~(\ref{eq:RLscale}), integration over the energy transfer $\nu$
yields the sum rule \cite{CZ}
\begin{eqnarray}
S(\vec{q})
&\equiv& \int_{-\infty}^{+\infty} d\nu\ R_L(\vec{q},\nu) 	\\
&=& e_1^2 + e_2^2 + 2 e_1 e_2\ F_{0,0}(2\vec{q})\ .
\label{eq:czSq}
\end{eqnarray}
The correction to the scaling result is thus directly proportional
to the elastic form factor, Eq.~(\ref{eq:hoff}), which clearly
illustrates how the interference term vanishes with increasing $q$.

What are the implications of these results for phenomenology?
Close \& Zhao speculate that analogous results also hold for the
$F_2$ structure function, namely, for $Q^2 \to \infty$ one has at
fixed $x$ \cite{CZ}:
\begin{eqnarray}
F_2(x)
&\to& (e_1^2 + e_2^2)
    {x^2 M^2 \over \beta \sqrt{\pi} E_0}
    \exp\left( -{M^2 \over \beta^2}
	  \left( {E_0 \over M} - x \right)^2
	\right)\ .
\label{eq:czF2}
\end{eqnarray}
The corresponding number sum rule, in analogy with Eq.~(\ref{eq:czSq}),
then becomes
\begin{eqnarray}
\int_{-\infty}^{+\infty} dx\ {F_2(x) \over x}
&=& e_1^2 + e_2^2\ ,
\end{eqnarray}
which is reminiscent of the Gottfried sum rule in the parton model
\cite{GOTT}.
Furthermore, the momentum sum rule can be written as \cite{CZ}
\begin{eqnarray}
\int_{-\infty}^{+\infty} dx\ F_2(x)
&=& (e_1^2+ e_2^2)
    \left( {E_0 \over M} + {\beta^2 \over 2 M E_0} \right)\ .
\label{eq:czmom}
\end{eqnarray}
Since the ground state energy for the harmonic oscillator potential is
$E_0 = \sqrt{3\beta^2 + m^2}$, one can identify the potential strength
$\beta$ with the Fermi momentum of the constituent,
$3\beta^2 \sim \vec p \, ^2$, and the second term in
Eq.~(\ref{eq:czmom}) can be understood as a kinetic energy
correction to the parton model result.
The physics of the parton model is recovered in the weak binding limit,
$\beta \to 0$, in which the constituents behave as if they were free.
The structure function in this case reduces to a $\delta$-function at
$x = 1/2$ (for equal mass quarks), and the momentum (energy) carried
by the constituents is then given by
\begin{eqnarray}
\int_{-\infty}^{+\infty} dx\ F_2(x)
&=& (e_1^2+ e_2^2)\ {E_0 \over M}\ ,
\end{eqnarray}
exactly as expected in the naive parton model.

In summary, we have observed the onset of scaling and the appearance of
duality in a variety of quark models in which the structure function
is explicitly obtained from sums of form factors for transitions to
excited states.
The harmonic oscillator potential is the prototypical example,
allowing the computation of the excited state spectrum to be made exact.
However, other inter-quark potentials also produce similar behavior,
which suggests that the phenomenon of quark-hadron duality may indeed
be a fundamental property of confined systems.
An important finding of these studies is the identification of the
pattern of constructive and destructive interference between resonances
by which the sum of coherent effects can be transformed into an
incoherent process, as in the parton model.
The generality of these results for arbitrary potentials, including
ones which have strong short-range repulsion \cite{IOFFE},
remains an important question for future study.
%
% Potentials which behave as $V(r) \sim 1/(r^2 \log(1/r))$ at
% short distances, for instance, give rise to logarithmic violation
% of scale invariance \cite{IOFFE}.

% -----------------------------------------------------------------------
\subsection{Local Duality: Phenomenological Applications}
\label{ssec:local}

In Section~\ref{ssec:qcd} we showed how {\em global} duality,
or moments of structure functions, can be understood within the
operator product expansion of QCD, in terms of suppression of
higher-twist contributions.
The interpretation of {\em local} duality, on the other hand,
is more elusive in QCD.
In Section~\ref{ssec:models} various dynamical models were examined
in order to gain some insight into the microscopic origin of local
duality.
To maintain clarity, and illustrate the main qualitative features of
duality, most of these models were at best gross oversimplifications
of Nature (for instance, assuming scalar quarks), with only remote
contact with experiment.
The richness of the empirical data which demonstrate duality both
in unpolarized and polarized scattering obviously calls for more
realistic theoretical descriptions, if contact with experiment is
to be achieved.

In this section we wish to explore local duality from the perspective
of its phenomenological applications, focusing in particular on the
relations between structure functions in the resonance region
(low $W$ and large $x$) and transition form factors.
We start by considering the simplest, and at the same time most
extreme, application of local duality, for the case of elastic
scattering.
Following this we discuss predictions for structure functions based
on low-lying resonances in the nonrelativistic quark model and its
extensions.

% .......................................................................
\subsubsection{Local Elastic Duality}
\label{sssec:elastic}

With accurate enough data, one can study the degree to which local
duality occurs for specific resonance regions, or even individual
resonances.
Of course, to extract information on a given resonance from inclusive
data requires understanding of nonresonant background contributions
to the structure function, as well as contributions from the tails of
neighboring resonances.
Inherently, the extraction of resonance properties is a model-dependent
procedure, and in practice one uses models, such as Breit-Wigner shapes
for resonances, to isolate the resonance and background contributions.
The one exception that does not suffer from this ambiguity is the
nucleon {\em elastic} component: below the pion production threshold
the {\em only} contribution to the cross section is from elastic
scattering.

\vspace*{0.5cm}
% . . . . . . . . . . . . . . . . . . . . . . . . . . . . . . . . . . . .
\paragraph{Drell-Yan--West Relation and Quark Counting Rules}

Exploration of the exclusive--inclusive (or form factor--structure
function) interface \cite{BLANKENBECLER,BJK,GBB,LANDSHOFF,EZAWA}
is as old as the first DIS experiments themselves.
A quantitative connection between structure functions at threshold
and elastic form factors was first made by Drell \& Yan \cite{DY} and
West \cite{WEST}, who related the high-$Q^2$ behavior of the elastic
Dirac form factor $F_1(Q^2)$,
\begin{eqnarray}
F_1(Q^2) &\sim& \left( { 1 \over Q^2 } \right)^n\ ,
		\ \ \ \ Q^2 \to \infty\ ,
\label{eq:DYWff}
\end{eqnarray}
with the threshold ($x \to 1$) behavior of the structure function
$\nu W_2(x)$,
\begin{eqnarray}
\nu W_2(x) &\sim& (1-x)^{2n-1}\ ,\ \ \ \ x \to 1\ .
\label{eq:DYWsf}
\end{eqnarray}
The power-law behavior of the form factor is simply related to the
suppression of the structure functions in the limit where one quark
carries all of the hadron's momentum.

Drell \& Yan based their derivation on earlier work on a canonical
pion-nucleon field theory in which the partons of the physical nucleon
were taken to be point-like (bare) nucleons and pions \cite{DYPI}.
The basic assumption was that in the infinite momentum frame there
exists a region in which $Q^2$ can be made larger than the transverse
components of the constituents.
Without specifying the nature of the partons, on the other hand, West
\cite{WEST} used a field-theoretic description of a nucleon in terms
of a scalar quark and a residual system of a definite mass, and derived
Eqs.~(\ref{eq:DYWff}) and (\ref{eq:DYWsf}) by requiring that the
asymptotic behavior of the nucleon--quark vertex function is damped
sufficiently at large internal momenta.

Although derived before the advent of QCD, the Drell-Yan--West
relation, as Eqs.~(\ref{eq:DYWff}) and (\ref{eq:DYWsf}) have come to
be known, can be expressed in perturbative QCD language in terms of
hard gluon exchange.
The pertinent observation is that deep inelastic scattering at
$x \sim 1$ probes a highly asymmetric configuration in the nucleon in
which one of the quarks goes far off-shell after the exchange of at
least two hard gluons in the initial state; elastic scattering,
on the other hand, requires at least two gluons in the final state
to redistribute the large $Q^2$ absorbed by the recoiling quark
\cite{LEPAGE}.
The exponent $n$ can therefore be interpreted as the minimum number
of hard gluons that need to be exchanged between quarks in the nucleon
\cite{BLANKENBECLER,GUNION}, which gives rise to the so-called
``pQCD counting rules''.
A clear prediction of the counting rules is that the $x \to 1$ limit
is dominated by scattering from quarks with the same helicity as the
nucleon \cite{FJ} (also known as ``hadron helicity conservation'').
In general, the quark distributions in a hadron $h$ are predicted to
behave as
\begin{eqnarray}
q^h(x) &\sim& (1-x)^{2n-1+2\Delta\lambda}\ ,\ \ \ \ x \to 1\ ,
\end{eqnarray}
where $\Delta\lambda = |\lambda_h - \lambda_q|$ is the difference
between the helicities of the hadron and the interacting quark
(see also Ref.~\cite{CM90}).
Scattering from quarks with helicity antialigned with respect to that
of the nucleon is therefore suppressed by a relative factor $(1-x)^2$
\cite{FJ}.
For the case of a pion, since $\lambda_\pi = 0$, the leading
behavior of the quark distribution is expected to be $(1-x)^2$
\cite{FJ}.

The relation between the power-law behavior of the form factor at
large $Q^2$ and the $x \to 1$ suppression of the structure function
in Eqs.~(\ref{eq:DYWff}) and (\ref{eq:DYWsf}) also arises at the
hadronic level from local duality, by considering the interplay between
resonances and scaling.
In the narrow resonance approximation, if the contribution of a
resonance of mass $M_R$ to the $F_2$ structure function at large $Q^2$
is given by
\begin{eqnarray}
F_2^R &=& 2 M \nu \left( G_R(Q^2) \right)^2\ \delta(W^2~-~M_R^2)\ ,
\end{eqnarray}
then a form factor behavior
\begin{eqnarray}
G_R(Q^2) &\sim& \left( { 1 \over Q^2 } \right)^n\
\label{eq:asymGR}
\end{eqnarray}
translates, for $Q^2 \gg M_R^2$, into a scaling function
\begin{eqnarray}
F_2^R &\sim& (1-x_R)^{2n-1}\ ,
\label{eq:asymF2R}
\end{eqnarray}
where $x_R = Q^2/(M_R^2 - M^2 + Q^2)$.
The asymptotic behavior of the form factor and structure function is
therefore the same as that predicted at the partonic level in the
Drell-Yan--West relation, Eqs.~(\ref{eq:DYWff}) and (\ref{eq:DYWsf}).

\vspace*{0.5cm}
% . . . . . . . . . . . . . . . . . . . . . . . . . . . . . . . . . . . .
\paragraph{Threshold Duality Relations}

The elastic contributions to the inclusive structure functions can be
expressed in terms of the elastic electric and magnetic form factors,
$G_E$ and $G_M$, by noting that for elastic scattering the helicity
amplitudes in Eqs.~(\ref{eq:F1cm})--(\ref{eq:g2cm}) reduce to
\cite{CMFF}
\begin{eqnarray}
G_+ &\to& \sqrt{ Q^2 \over 2 M^2 }\ G_M\ ,	\\
G_0 &\to& G_E\ ,				\\
G_- &\to& 0\ .
\end{eqnarray}
The elastic spin-averaged structure functions can then be written as
\begin{eqnarray}
\label{eq:elF1}
F_1^{\rm el}
&=& M \tau\
    G_M^2\ \delta\left( \nu - {Q^2 \over 2M} \right)\ ,	\\
\label{eq:elF2}
F_2^{\rm el}
&=& { 2 M \tau \over 1 + \tau }
    \left( G_E^2 + \tau G_M^2 \right)\
    \delta\left( \nu - {Q^2 \over 2M} \right)\ ,
\end{eqnarray}
where $\tau = Q^2/4M^2$.
For spin-dependent structure functions one has \cite{JI_G1,CMFF}
\begin{eqnarray}
\label{eq:elg1}
g_1^{\rm el}
&=& { M \tau \over 1 + \tau }
    G_M \left( G_E + \tau G_M \right)\
    \delta\left( \nu - {Q^2 \over 2M} \right)\ ,	\\
\label{eq:elg2}
g_2^{\rm el}
&=& { M \tau^2 \over 1 + \tau }
    G_M \left( G_E - G_M \right)\
    \delta\left( \nu - {Q^2 \over 2M} \right)\ .
\end{eqnarray}

In their original paper, Bloom \& Gilman \cite{BG1} suggested that if
one carries the idea of local duality to an extreme, and makes the
assumption that the area under the elastic peak in the measured
structure function at large $Q^2$ is the same as the area under the
scaling-limit curve, then one could relate the integral of the scaling
function below threshold to the elastic form factors.
For the $\nu W_2 (= F_2)$ structure function in Eq.~(\ref{eq:elF2}),
integrating over the Bloom-Gilman scaling variable
$\omega' = (2M\nu + M^2)/Q^2$, they find \cite{BG1}
\begin{eqnarray}
\int_1^{1 + W_t^2/Q^2} d\omega'\ \nu W_2(\omega')
&=& { 2M \over Q^2 }
    \int d\nu\ \nu W_2^{\rm el}(\nu,Q^2)		\\
&=& { G_E^2(Q^2) + \tau G_M^2(Q^2) \over 1 + \tau }\ .
\label{eq:BGel}
\end{eqnarray}
To give meaning to the integration over the $\delta$-function, the
integral in Eq.~(\ref{eq:BGel}) runs from the unphysical value
$\omega'=1$ up to an $\omega'$ corresponding to a hadron mass $W=W_t$
near the physical pion threshold.

In QCD language, De~R\'ujula {\em et al.} \cite{DGP2} showed that one
could express the threshold relation (\ref{eq:elF2}) as an integral
over the Nachtmann scaling variable $\xi$ between the pion threshold
$\xi_{\rm th}$ and $\xi=1$, which also includes the unphysical region
between the elastic nucleon pole at
$\xi_0 \equiv \xi(x=1) = 2 / (1 + \sqrt{1 + 1/\tau})$ and $\xi=1$.
Integrating the elastic structure functions over $\xi$ between the
pion threshold $\xi_{\rm th}$ and $\xi=1$, one finds for the
unpolarized and polarized \cite{CLEYMANS} structure functions
\cite{EL_QNP}:
\begin{eqnarray}
\label{eq:NmomF1}
\int_{\xi_{\rm th}}^1  d\xi\ \xi^{n-2}\ F_1(\xi,Q^2)
&=& { \xi_0^n \over 4 - 2 \xi_0 }\ G_M^2(Q^2)\ ,	\\
\label{eq:NmomF2}
\int_{\xi_{\rm th}}^1  d\xi\ \xi^{n-2}\ F_2(\xi,Q^2)
&=& { \xi_0^n \over 2 - \xi_0 }\
    { G_E^2(Q^2) + \tau G_M^2(Q^2) \over 1 + \tau }\ ,	\\
\label{eq:momG1}
\int_{\xi_{\rm th}}^1  d\xi\ \xi^{n-2}\ g_1(\xi,Q^2)
&=& { \xi_0^n \over 4 - 2 \xi_0 }\
    { G_M(Q^2) \left( G_E(Q^2) + \tau G_M(Q^2) \right)
      \over 1 + \tau }\ ,				\\
\label{eq:NmomG2}
\int_{\xi_{\rm th}}^1  d\xi\ \xi^{n-2}\ g_2(\xi,Q^2)
&=& { \xi_0^n \over 4 - 2 \xi_0 }\
    { \tau G_M(Q^2) \left( G_E(Q^2) - G_M(Q^2) \right)
    \over 1 + \tau }\ .
\end{eqnarray}
where $\xi_{\rm th} = \xi(x_{\rm th} = Q^2 / (W^2_{\rm th} - M^2 + Q^2))$,
with $W_{\rm th} = M + m_\pi$.
The local duality hypothesis is that the structure functions $F_{1,2}$
and $g_{1,2}$ under the integrals are independent of $Q^2$, and are
functions of $\xi$ only.
From Eq.~(\ref{eq:NmomF2}) De~R\'ujula {\em et al.} \cite{DGP2}
extracted the proton's $G_M^p$ form factor, assuming that the ratio
$G_E^p/G_M^p$ is sufficiently constrained, from resonance data on the
$F_2^p$ structure function at large $\xi$.

\begin{figure}[ht]
\begin{center}
\epsfig{file=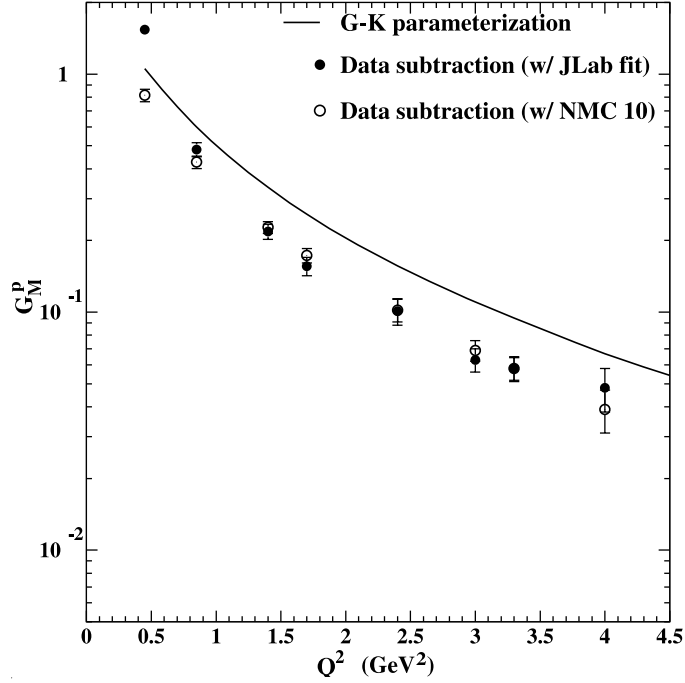,height=9cm}
\vspace*{1cm}
\caption{\label{fig:GMdual}
	Proton magnetic form factor $G_M^p$ extracted from the
	inelastic scaling curves (from NMC and JLab) using local
	duality, and compared with the Gari-Krumpelmann
	parameterization \protect\cite{GK} of the world's
	$G_M^p$ data.  (From Ref.~\protect\cite{ENT_EL}.)}
\end{center}
\end{figure}

More recently Ent {\em et al.} \cite{ENT_EL} used high-precision
Jefferson Lab data to make a quantitative test of the threshold
relations using a slightly modified extraction procedure.
Namely, the integral obtained from the resonance data, which stop at the
pion threshold $\xi_{\rm th}$ rather than at $\xi=1$, is subtracted from
the scaling integrals, and $G_M^p$ then extracted from the remaining
integrated strength.
Figure~\ref{fig:GMdual} shows the resulting proton magnetic form factor
$G_M^p$ extracted using the NMC (open circles) and Jefferson Lab
(filled circles) scaling curves for $F_2^p$.
In both cases the extracted form factor is found to be in remarkable
agreement with the parameterization of the world data on $G_M^p$
\cite{GK}.
For the case of the Jefferson Lab scaling curve, the $G_M^p$ fit is
reproduced quite well, to within $\alt 30\%$ accuracy, for $Q^2$ from
0.2~GeV$^2$ up to $\sim 4$~GeV$^2$.

Ent {\em et al.} \cite{ENT_EL} showed that one can also extract
the $G_E^p/G_M^p$ ratio from values of $R$, making use of
Eqs.~(\ref{eq:NmomF1}) and (\ref{eq:NmomF2}).
However, more precise data on $R$ at large $\xi$ are needed
\cite{ERIC} for one to be able to make quantitative predictions.

\vspace*{0.5cm}
% . . . . . . . . . . . . . . . . . . . . . . . . . . . . . . . . . . . .
\paragraph{Structure Functions in the $x \to 1$ Limit}

\begin{figure}[ht]
\begin{center}
\hspace*{0.5cm}
\epsfig{file=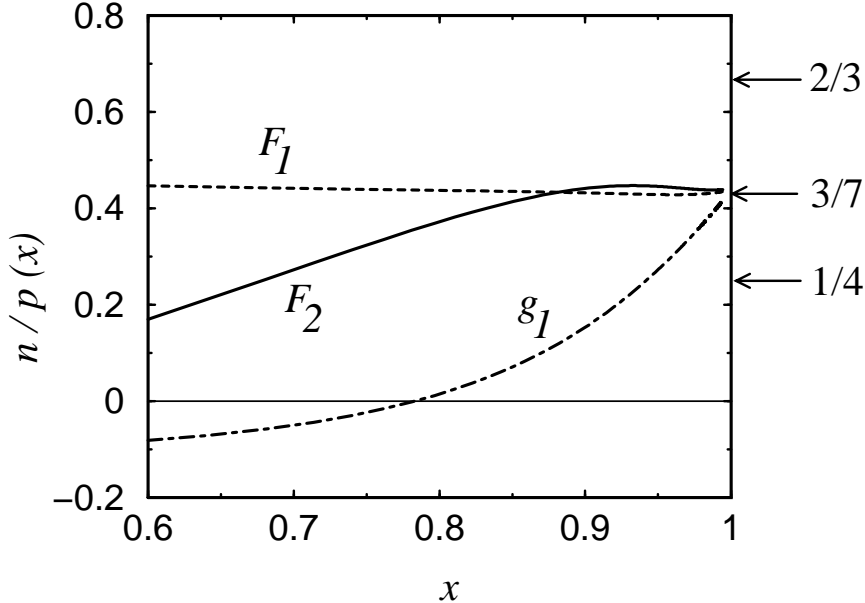,height=8cm}
\vspace*{0.5cm}
\caption{\label{fig:elNP}
	Neutron to proton ratio for $F_1$ (dashed), $F_2$ (solid)
	and $g_1$ (dot-dashed) structure functions at large $x$,
	from Ref.~\protect\cite{EL_WM}.
	Several leading-twist model predictions for $F_2$ in the
	$x \to 1$ limit are indicated by the arrows: 2/3 from SU(6),
	3/7 from SU(6) breaking via helicity conservation, and
	1/4 from SU(6) breaking through $d$ quark suppression.}
\end{center}
\end{figure}

Applying the duality argument in reverse, one can formally
differentiate the local elastic duality relations with respect
to $Q^2$ to express the scaling functions in terms of $Q^2$
derivatives of elastic form factors \cite{BG1}.
For the $\nu W_2$ structure function, for example, Bloom \& Gilman
find \cite{BG1,BG2}
\begin{eqnarray}
\nu W_2(\omega'=1+W_t^2/Q^2)
&=& { Q^2 \over 1-\omega' }
    { d \over dQ^2 }
    \left( { G_E^2 + \tau G_M^2 \over 1+\tau } \right)\ .
\label{eq:bg_largex}
\end{eqnarray}
Relations for the other structure functions can also be derived
\cite{EL_WM,EL_NSTAR,EL_ST} in terms of the elastic form factors and their
derivatives.
In particular, differentiating both sides of
Eqs.~(\ref{eq:NmomF1})--(\ref{eq:NmomG2}) with respect to $\xi$
(or $Q^2$) and changing variables from $\xi$ to $x$, one finds
\begin{eqnarray}
\label{eq:xth_F1}
F_1(x=x_{\rm th}) &=& \beta\
{ dG_M^2 \over dQ^2 }\ ,                        \\
\label{eq:xth_F2}
F_2(x=x_{\rm th}) &=& \beta
\left\{ { G_M^2 - G_E^2 \over 2 M^2 (1+\tau)^2 }
+ { 2 \over 1 + \tau }
  \left( { dG_E^2 \over dQ^2 } + \tau { dG_M^2 \over dQ^2 }
  \right)
\right\}\ ,					\\
\label{eq:xth_g1}
g_1(x=x_{\rm th}) &=& \beta\
\left\{ { G_M \left( G_M - G_E \right) \over 4 M^2 (1+\tau)^2 }
+ { 1 \over 1 + \tau }
  \left( { d(G_E G_M) \over dQ^2 } + \tau { dG_M^2 \over dQ^2 }
  \right)
\right\}\ ,					\\
\label{eq:xth_g2}
g_2(x=x_{\rm th}) &=& \beta\
\left\{
{ G_M \left( G_E - G_M \right) \over 4 M^2 (1+\tau)^2 }
+ { \tau \over 1 + \tau }
  \left( { d(G_E G_M) \over dQ^2 } - { dG_M^2 \over dQ^2 }
  \right)
\right\}\ ,
\end{eqnarray}
where the kinematic factor
$\beta = (Q^4/M^2) (\xi_0^2/x \xi^3) (2x-\xi)/(2\xi_0-4)$.
Note that the structure functions in
Eqs.~(\ref{eq:bg_largex})--(\ref{eq:xth_g2})
are evaluated at the pion production threshold, $x=x_{\rm th}$,
coming from the lower limits of integration in
Eqs.~(\ref{eq:NmomF1})--(\ref{eq:NmomG2}).
(See also Refs.~\cite{EL_CR,EL_ST} for a generalization to the case
of neutrino scattering.)
Asymptotically, each of the structure functions $F_1$, $F_2$ and $g_1$
is found to be determined by the slope of the square of the magnetic
form factor \cite{EL_WM},
\begin{eqnarray}
F_1\ ,\ \ F_2\ ,\ \ g_1\ &\sim&\ { dG_M^2 \over dQ^2 }\ ,
 \ \ \ \ Q^2 \to \infty\ ,
\end{eqnarray}
while $g_2$, which is associated with higher twists, is determined by
a combination of $G_E$ and $G_M$,
\begin{eqnarray}
g_2 &\sim& { d(G_E G_M - G_M^2) \over dQ^2 }\ , \ \ \ \ Q^2 \to \infty\ .
\end{eqnarray}
In this limit each of the structure functions can also be shown to
satisfy the Drell-Yan--West relation, Eqs.~(\ref{eq:DYWff}) and
(\ref{eq:DYWsf}).
In addition, the asymptotic behavior of $g_1$ and $F_1$ is predicted to
be the same, so that the polarization asymmetries
$A_1 \approx g_1/F_1 \to 1$ as $x \to 1$ for both the proton and neutron.
This is in marked contrast to the expectations from SU(6) symmetry,
in which the proton and neutron asymmetries are predicted to be
$A_1^p = 5/9$ and $A_1^n = 0$, respectively \cite{CLOSEBOOK}.
Recall that the symmetric SU(6) wave function for a proton polarized
in the $+z$ direction is given by
\begin{eqnarray}
| p^\uparrow \rangle
&=& {1 \over \sqrt{2}}  | u^\uparrow (ud)_0 \rangle\
 +\ {1 \over \sqrt{18}} | u^\uparrow (ud)_1 \rangle\
 -\ {1 \over 3}         | u^\downarrow (ud)_1 \rangle\ \nonumber \\
& &
 -\ {1 \over 3}         | d^\uparrow (uu)_1 \rangle\
 -\ {\sqrt{2} \over 3}  | d^\downarrow (uu)_1 \rangle\ ,
\label{eq:pwfn}
\end{eqnarray}
where the subscript 0 or 1 denotes the total spin of the two-quark
component (and similarly for the neutron, with $u \leftrightarrow d$).
Here the quark distributions for different flavors and spins are
related by the Clebsch-Gordan coefficients in Eq.~(\ref{eq:pwfn}),
with $u=2d$ and $\Delta u = -4 \Delta d$, which leads to the
familiar SU(6) quark-parton model results,
\begin{eqnarray}
R^{np} &\equiv& { F_2^n \over F_2^p }\ =\ { 2 \over 3 }\ ,\ \ \ \
A_1^p\ =\ { 5 \over 9 }\ ,\ \ \ \
A_1^n\ =\ 0\ \ \ \ \ \ \ {\rm [SU(6)]}\ .
\label{eq:su6}
\end{eqnarray}
Using parameterizations of global form factor data, the ratios of the
neutron to proton $F_1$, $F_2$ and $g_1$ structure functions are shown
in Fig.~\ref{fig:elNP} as a function of $x$, with $x$ corresponding to
$x_{\rm th}$.
Some theoretical limits for the ratios as $x \to 1$ are indicated
on the vertical axis, which range from 2/3 in the SU(6) quark model,
to 3/7 in the pQCD-inspired helicity conservation model \cite{FJ},
and 1/4 in the case where the symmetric part of the SU(6)
wave function is suppressed \cite{FeynmanBOOK,SCALAR}
(see the discussion in Sec.~\ref{sssec:qm} below).
While the $F_2$ ratio varies somewhat with $x$ at lower $x$, beyond
$x \sim 0.85$ it remains almost $x$ independent, approaching the
asymptotic value $(dG_M^{n 2}/dQ^2)/(dG_M^{p 2}/dQ^2)$.
Because the $F_1^n/F_1^p$ ratio depends only on $G_M$, it remains flat
over nearly the entire range of $x$.
The $g_1$ structure function ratio approaches the same asymptotic
limit as $F_1$, albeit more slowly, which may indicate a larger role
played by higher twists in spin-dependent structure functions than in
spin-averaged (see Sec.~\ref{sssec:ope} above).

Interestingly, the helicity conservation model prediction \cite{FJ}
of 3/7 is very close to the empirical ratio of the squares of the
neutron and proton magnetic form factors,
$\mu_n^2/\mu_p^2 \approx 4/9$.
Indeed, if one approximates the $Q^2$ dependence of the proton
and neutron form factors by dipoles, and takes $G_E^n \approx 0$,
then the structure function ratios are all determined by the
magnetic moments,
$F_2^n/F_2^p \approx F_1^n/F_1^p
	     \approx g_1^n/g_1^p \to \mu_n^2/\mu_p^2$
as $Q^2 \to \infty$.
On the other hand, for the $g_2$ structure function, which depends
on both $G_E$ and $G_M$ at large $Q^2$, the asymptotic behavior is
$g_2^n/g_2^p \to \mu_n^2 / (\mu_p (\mu_p-1)) \approx 0.73$.

Of course the reliability of the duality predictions is only as good
as the quality of the empirical data on the electromagnetic form
factors allow.
While the duality relations are expected to be progressively more
accurate with increasing $Q^2$ \cite{DGP2}, the difficulty in measuring
form factors at large $Q^2$ also increases.
Obviously more data at larger $Q^2$ would allow more accurate
predictions for the $x \to 1$ structure functions, and new experiments
at Jefferson Lab and elsewhere will provide valuable constraints.
However, the most challenging aspect of testing the validity of the
local duality hypothesis is measuring the inclusive structure functions
at high enough $x$, which will become feasible with the 12~GeV energy
upgrade at Jefferson Lab \cite{pCDR} (see also Sec.~\ref{sec:outlook}).
In particular, with data on both the $F_1$ and $F_2$ (or $g_1$ and
$F_2$) structure functions at large $x$ one will be able to extract
the $G_E$ and $G_M$ form factors separately, without having to assume
the $G_E/G_M$ ratio in current extractions \cite{DGP2,ERIC,ENT_EL}
of $G_M$ from the available $F_2$ data.

Finally, the threshold duality relations
(\ref{eq:NmomF1})--(\ref{eq:bg_largex}) have also been applied
recently \cite{MTT} in studies of the nuclear medium dependence of
nucleon structure functions at large $x$.
Recent evidence from polarized $(\vec e, e' \vec p)$ scattering
experiments on $^4$He \cite{HE4,STRAUCH} nuclei suggests that a small
change in the structure of the bound nucleon, in addition to the
standard nuclear corrections such as meson exchange currents, isobar
contributions, and final state interactions, is the most efficient way
to describe the ratio of transverse to longitudinal polarization of
the ejected protons \cite{QMC2,GUICHON,QMC1,BLU_MIL_QMC}.
Using local duality to relate the medium dependence of nucleon
electromagnetic form factors to the medium dependence of nucleon
structure functions, the recent data for a proton bound in $^4$He
\cite{HE4,STRAUCH} have been used to place strong constraints on
models of the nuclear EMC effect in which medium modification is
attributed to a deformation of the intrinsic nucleon structure
off-shell \cite{FS} (see also Ref.~\cite{MILLER_MOD}).
In particular, the results appear to rule out large bound structure
function modifications, and instead point to a small medium modification
of the intrinsic nucleon structure function, which is complemented by
standard many-body nuclear effects.
This study therefore illustrates yet another example of how
quark-hadron duality can be applied to relate phenomena which
otherwise do not appear directly related.

% .......................................................................
\subsubsection{Duality in the Quark Model}
\label{sssec:qm}

The threshold relations between structure functions near $x=1$ and
elastic form factors have met with some degree of phenomenological
success.
Their appeal is also their simplicity: there are no model-dependent
backgrounds to subtract before discussing resonant properties.
On the other hand, some of the models described in
Sec.~\ref{sssec:res} suggested that the appearance of duality was
intimately related to cancellations between states having different
angular momentum or parity quantum numbers.
At the same time, the simple nature of these models makes it difficult
to draw firm conclusions about the origins of duality in the empirical
data.
For instance, while spin degrees of freedom are not necessary to
illustrate the main qualitative features of duality, the examples
of spinless constituents involved only electric multipoles, whereas
inclusion of spin leads to both electric and magnetic multipole
contributions.
In fact, at large $Q^2$ the latter is expected to dominate.
Ultimately, therefore, one would like to study duality in models with
a closer connection to phenomenology to learn about duality in the
physical world.
For this to happen, one needs to generalize the model discussions to
the more realistic case of three valence (and possibly even sea) quarks,
instead of the simplified two-body systems considered in
Sec.~\ref{ssec:models}.

\vspace*{0.5cm}
% . . . . . . . . . . . . . . . . . . . . . . . . . . . . . . . . . . . .
\paragraph{SU(6) Symmetry}

The SU(6) spin-flavor symmetric quark model serves as a useful basis
for both visualizing the principles underpinning the phenomenon of
duality and at the same time providing a reasonably close contact with
phenomenology.
Quark models based on SU(6) spin-flavor symmetry provide benchmark
descriptions of baryon spectra, as well as transitions to excited
$N^*$ states.

In a series of classic early papers, Close, Gilman and collaborators
\cite{CG,CGK,CGPLB,COT} showed how the ratios of various deep inelastic
structure functions could be dual to sums over $N^*$ resonances in the
$l=0$ {\bf 56}-dimensional and $l=1$ {\bf 70}-dimensional
representations of SU(6).
In particular, they demonstrated that one could construct a set of
nucleon resonances, the sum of whose contributions to inclusive
structure functions replicates the results of the naive quark-parton
model.

Since the nucleon ground state wave function is totally symmetric,
the only final state resonances that can be excited have wave functions
which are either totally symmetric or of mixed symmetry, corresponding
to the positive parity ($P=(-1)^l$) {\bf 56}$^+$ and negative parity
{\bf 70}$^-$ representations, respectively \cite{CLOSEBOOK}.
% but not the totally antisymmetric {\bf 20}-dimensional representation
%
The relative weightings of the {\bf 56}$^+$ and {\bf 70}$^-$
contributions are determined by assuming that the electromagnetic
current is in a {\bf 35}-plet.
Allowing only the non-exotic singlet {\bf 1} and {\bf 35}-plet
representations in the $t$-channel, which corresponds to $q\bar{q}$
exchange, the reduced matrix elements for the {\bf 56}$^+$
and {\bf 70}$^-$ are constrained to be equal.
In the $t$-channel these appear as $\gamma\gamma \to q\bar q$,
while in the $s$-channel this effectively maps onto the leading-twist,
handbag diagram in Fig.~\ref{fig:diag}~(a), describing incoherent
coupling to the same quark.
Exotic exchanges require multi-quark exchanges, such as
$qq\bar q\bar q$ in the $t$-channel, and correspond to the ``cat's ears''
diagram in Fig.~\ref{fig:diag}~(b).
Physically, therefore, the appearance of duality in this picture
is correlated with the suppression of exotics in the $t$-channel
\cite{CGK}.

Assuming magnetic couplings and neglecting quark orbital motion, the
relative photoproduction strengths of the transitions from the ground
state, Eq.~(\ref{eq:pwfn}), to the {\bf 56}$^+$ and {\bf 70}$^-$ are
summarized in Table~\ref{tab:su6} for the $F_1$ (which is related to
$F_2$ by the Callan-Gross relation, $F_2 = 2 x F_1$, in this
approximation) and $g_1$ structure functions of the proton and neutron.
For generality, the contributions from the symmetric ($\psi_\rho$) and
antisymmetric ($\psi_\lambda$) components of the ground state nucleon
wave function,
\begin{eqnarray}
| N \rangle &=& \cos\theta_w | \psi_\rho \rangle\
             +\ \sin\theta_w | \psi_\lambda \rangle\ ,
\label{rholam}
\end{eqnarray}
have been separated, where $\theta_w$ is the mixing angle and
$\psi = \varphi \otimes \chi$ is a product of the flavor
($\varphi$) and spin ($\chi$) wave functions \cite{CLOSEBOOK}.
Defining $\rho = \cos\theta_w$ and $\lambda = \sin\theta_w$,
the SU(6) limit corresponds to $\lambda = \rho$ ($\theta_w=\pi/4$).
Remarkably, summing over the full set of states in the {\bf 56}$^+$
and {\bf 70}$^-$ multiplets, one finds in this case precisely the
same structure function ratios as in the quark-parton model,
Eq.~(\ref{eq:su6}).

\begin{table}[ht]
\begin{tabular}{||c||c|c|c|c|c||c||}
SU(6) rep.      & $^2${\bf 8}[{\bf 56}$^+$]\ \
                & $^4${\bf 10}[{\bf 56}$^+$]\ \
                & $^2${\bf 8}[{\bf 70}$^-$]\ \
                & $^4${\bf 8}[{\bf 70}$^-$]\ \
                & $^2${\bf 10}[{\bf 70}$^-$]\ \
                & total\ \                              \\ \hline\hline
$F_1^p$ & $9 \rho^2$
        & $8 \lambda^2$
        & $9 \rho^2$
        & $0$
        & $\lambda^2$
        & $18 \rho^2 + 9 \lambda^2$                     \\
$F_1^n$ & $(3 \rho + \lambda)^2/4$
        & $8 \lambda^2$
        & $(3 \rho - \lambda)^2/4$
        & $4 \lambda^2$
        & $\lambda^2$
        & $(9 \rho^2 + 27 \lambda^2)/2$                 \\ \hline
$g_1^p$ & $9 \rho^2$
        & $-4 \lambda^2$
        & $9 \rho^2$
        & $0$
        & $\lambda^2$
        & $18 \rho^2 - 3 \lambda^2$                     \\
$g_1^n$ & $(3 \rho + \lambda)^2/4$
        & $-4 \lambda^2$
        & $(3 \rho - \lambda)^2/4$
        & $-2 \lambda^2$
        & $\lambda^2$
        & $(9 \rho^2 - 9 \lambda^2)/2$                  \\
\end{tabular}
\vspace*{0.5cm}
\caption{Relative strengths of electromagnetic $N \to N^*$ transitions
        in the SU(6) quark model.  The coefficients $\lambda$ and $\rho$
        denote the relative strengths of the symmetric and antisymmetric
        contributions of the SU(6) ground state wave function.
        The SU(6) limit corresponds to $\lambda = \rho$.}
	From Ref.~\protect\cite{NOSU6}.
\label{tab:su6}
\end{table}

Although the $s$-channel sum was shown by Close {\em et al.}
\cite{CG,CGK,COT} to be dual for ratios of structure functions,
this alone did not explain the underlying reason why any individual
sum over states scaled.
The microscopic origin of duality in the SU(6) quark model was more
recently elaborated by Close \& Isgur \cite{CI}, who showed that the
cancellations between the even- and odd-parity states found to be
necessary for duality to appear, are realized through the destructive
interference in the $s$-channel resonance sum between the
{\bf 56}$^+$ and {\bf 70}$^-$ multiplets.
Provided the contributions from the {\bf 56}$^+$ and {\bf 70}$^-$
representations have equal strength, this leads exactly to the
scaling function proportional to $\sum_q e_q^2$.
In the SU(6) limit, duality will therefore not be realized unless
the {\bf 56}$^+$ and {\bf 70}$^-$ states are integrated over.
Recall that the usual assignments of the excited states in the quark
model place the nucleon and the $P_{33}(1232)$ $\Delta$ isobar in the
quark spin-1/2 octet ($^2${\bf 8}) and quark spin-3/2 decuplet
($^4${\bf 10}) representations of {\bf 56}$^+$, respectively,
while for the odd parity states the
$^2${\bf 8} representation contains the states $S_{11}(1535)$ and
$D_{13}(1520)$, the $^4${\bf 8} contains the $S_{11}(1650)$,
$D_{13}(1700)$ and $D_{15}(1675)$, while the isospin-${3 \over 2}$
states $S_{31}(1620)$ and $D_{33}(1700)$ belong to the $^2${\bf 10}
representation.

From Table~\ref{tab:su6} one sees that duality may be satisfied
for the proton (with $\lambda=\rho$) by states with
$W \alt 1.6$~GeV, since states from the $^4${\bf 8}[{\bf 70}$^-$]
and $^2${\bf 10}[{\bf 70}$^-$] representations at $W \sim 1.7$~GeV
make negligible contributions.
For neutron targets, on the other hand, one still has sizable
contributions from the $^4${\bf 8}[{\bf 70}$^-$], which necessitates
integrating up to $W \sim 1.8$~GeV.
The case of the neutron $g_1^n$ structure function is somewhat
exceptional.
Here, the SU(6) limit reveals the intriguing possibility that
duality may be localized to {\em within} each of the {\bf 56}$^+$
and {\bf 70}$^-$ representations individually: the strengths of the
$N$ and $\Delta$ transitions (with $\lambda=\rho$) in the {\bf 56}$^+$
are equal and opposite, and the octet and decuplet contributions in
the {\bf 70}$^-$ sum to zero.

Note that the region above $W \approx 1.7$~GeV also contains a
{\bf 56}$^+$ multiplet at $N=2$ in the harmonic oscillator.
In the nonrelativistic limit, to order $\vec q \, ^2 \sim 1/R^2$
the ${\bf 56}^+$ and ${\bf 70}^-$ multiplets would be sufficient
to realize closure and duality.
The analysis can be extended to higher $\vec q \, ^2$ by including
correspondingly higher multiplets, although the reliability of the
nonrelativistic harmonic oscillator may become questionable
at higher $\vec q \, ^2$ \cite{CI}.

We note again that the above results have been derived assuming
magnetic couplings, which are expected to dominate at large $Q^2$.
A realistic description of the empirical data at low $Q^2$ would
require in addition the inclusion of electric couplings, which will
give rise to a nonzero longitudinal structure function $F_L$.
Close \& Isgur showed in fact that in the SU(6) limit duality
is also realized for $F_L$ \cite{CI}.
In general, however, the interplay of magnetic and electric
interactions will make the workings of duality nontrivial.
In the $Q^2 \to 0$ limit both electric and magnetic multipoles will
contribute and the interference effects can cause strong $Q^2$
dependence \cite{CGK,CGPLB}, such as that responsible for the dramatic
change in sign of the lowest moment of $g_1^p$ in the transition
towards the Gerasimov-Drell-Hearn sum rule at $Q^2 = 0$
(see Sec.~\ref{sssec:gdh}).
Close \& Isgur suggest \cite{CI} that Bloom-Gilman duality will
fail when the electric and magnetic multipoles have comparable
strengths, although the precise $Q^2$ at which this will occur
is unknown.

\vspace*{0.5cm}
% . . . . . . . . . . . . . . . . . . . . . . . . . . . . . . . . . . . .
\paragraph{SU(6) Breaking}

While the SU(6) predictions for the structure functions hold
approximately at $x \sim 1/3$, strong deviations are expected
at larger $x$.
For instance, the neutron $F_2^n$ structure function is observed
to be much softer than the proton $F_2^p$ for $x \agt 0.5$
\cite{CLOSEBOOK,WHITLOW,NP,GOMEZ,LG,RINAT}, and although the data
are not yet conclusive, there are indications that the polarization
asymmetries show a trend towards unity as $x \to 1$ for both the
proton \cite{CLASg1} and neutron \cite{A1N}
(see {\em e.g.}, Fig.~\ref{fig:a1p_hermes}).

As discussed in Sec.~\ref{sssec:elastic}, for a given $N^*$
resonance of mass $M_R$, the resonance peak at
$x = x_R \equiv Q^2 / (M_R^2 - M^2 + Q^2)$ moves to larger $x$
with increasing $Q^2$.
If a given resonance at $x \sim 1/3$ appears at relatively low $Q^2$,
the $x \sim 1$ behavior of the resonance contribution to the structure
function will therefore be determined by the $N \to N^*$ transition
form factor at larger $Q^2$.
In the context of duality, the specific patterns of symmetry breaking
in structure function ratios as $x \to 1$ may yield information about
the $Q^2$ dependence of families of $N^*$ resonances.

At the quark level, explicit SU(6) breaking mechanisms produce
different weightings of components of the initial state wave function,
Eq.~(\ref{eq:pwfn}), which in turn induces different $x$ dependences
for the spin and flavor distributions.
At the hadronic level, on the other hand, SU(6) breaking in the
$N \to N^*$ matrix elements leads to suppression of transitions to
specific resonances in the final state, starting from an SU(6)
symmetric wave function in the initial state.
For duality to be manifest, the pattern of symmetry breaking in the
initial state must therefore match that in the final state.
It is {\em a priori} not obvious, however, whether specific mechanisms
of SU(6) breaking will be consistent with duality, and recent studies
\cite{NOSU6} have investigated the conditions under which duality can
arise in various symmetry breaking scenarios.

The most immediate breaking of the SU(6) duality could be achieved by
varying the overall strengths of the coefficients for the {\bf 56}$^+$
and {\bf 70}$^-$ multiplets as a whole.
However, since the cancellations of the $N \to N^*$ transitions for
the case of $g_1^n$ occur within each multiplet, a nonzero value of
$A_1^n$ can only be achieved if SU(6) is broken {\em within} each
multiplet rather than {\em between} the multiplets.
Some intuition is needed therefore on sensible symmetry breaking
patterns within the multiplets.

In Table~\ref{tab:su6} the SU(6) limit is obtained by assigning
equal weights for the contributions to the various $N \to N^*$
transitions from symmetric and antisymmetric components of the wave
function, $\lambda=\rho$.
On the other hand, the SU(6) symmetry can be broken if the mixing
angle $\theta_w \not= \pi/4$.
In general, for an arbitrary mixing angle $\theta_w$, summing over
all channels leads to structure function ratios given by \cite{NOSU6}:
\begin{eqnarray}
R^{np}
&=& { 1 + 2 \sin^2\theta_w \over 4 - 2 \sin^2\theta_w }\ ,\ \ \ \
A_1^p\
 =\ { 6 - 7 \sin^2\theta_w \over 6 - 3 \sin^2\theta_w }\ ,\ \ \ \
A_1^n\
 =\ { 1 - 2 \sin^2\theta_w \over 1 + 2 \sin^2\theta_w }\ .
\label{eq:rat_w}
\end{eqnarray}

If the mass difference between the nucleon and $\Delta$ is attributed
to spin dependent forces, the energy associated with the symmetric
part of the wave function will be larger than that of the
antisymmetric component.
A suppression of the symmetric $| \psi_\lambda \rangle$ configuration
at large $x$ ($\theta_w \to 0$) will then give rise to a suppressed
$d$ quark distribution relative to $u$, which in turn leads to the
famous neutron to proton ratio $R^{np} \to 1/4$
\cite{FeynmanBOOK,SCALAR,ISGUR}.
In terms of the sum over resonances in the final state, this scenario
corresponds to the suppression of the symmetric components of the
states in the {\bf 56}$^+$ and {\bf 70}$^-$ multiplets relative to
the antisymmetric, and the relative transition strengths are given in
Table~\ref{tab:su6} with $\lambda \to 0$.
In particular, since transitions to the (symmetric) $S=3/2$ or decuplet
states ($^4${\bf 8}, $^4${\bf 10} and $^2${\bf 10}) can only proceed
through the symmetric $\psi_\lambda$ component of the ground state
wave function, the $\psi_\rho$ components will only excite the nucleon
to $^2${\bf 8} states.
If the $\psi_\lambda$ wave function is suppressed, only transitions to
$^2${\bf 8} states will be allowed.

\begin{figure}[ht]
\begin{center}  
\epsfig{file=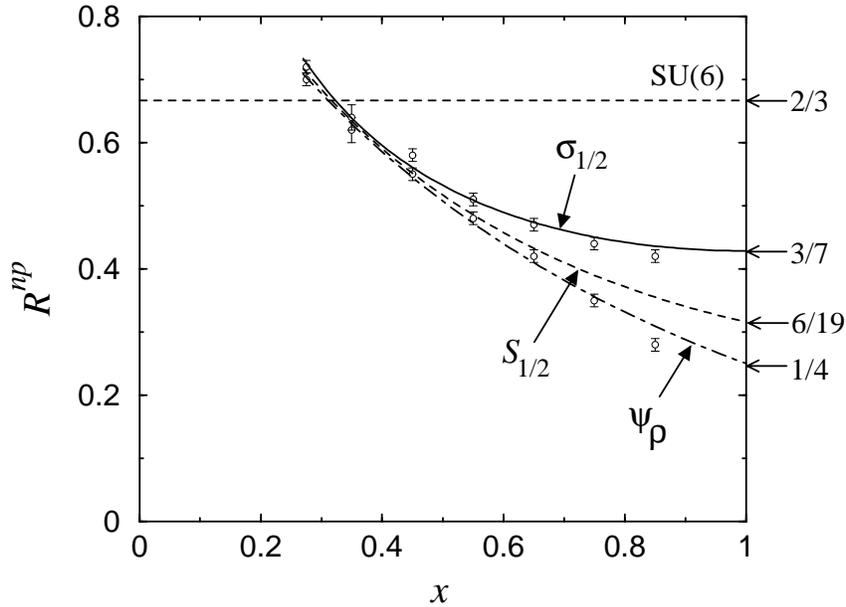,height=8cm}
\vspace*{0.5cm}
\caption{\label{fig:Rnp}
	Ratio $R^{np}$ of unpolarized neutron to proton structure
	functions from duality \protect\cite{NOSU6}, according to
	different scenarios of SU(6) breaking:
	$\psi_\rho$ dominance (dot-dashed);
	spin-1/2 ($S_{1/2}$) dominance (dashed);
	and helicity-1/2 ($\sigma_{1/2}$) dominance (solid),
	with the respective $x\to 1$ limits indicated on the ordinate.
	The data are from SLAC \protect\cite{WHITLOW,GOMEZ}, analyzed
	under different assumptions about the size of the nuclear
	effects in the deuteron \protect\cite{NP}.}
\end{center}
\end{figure}

The dependence of the structure function ratios in Eq.~(\ref{eq:rat_w})
on the mixing angle $\theta_w$ means that the SU(6) breaking scenario
with $\psi_\lambda$ suppression can be tested by simultaneously fitting
the $n/p$ ratios and the polarization asymmetries.
The $x$ dependence of $\theta_w(x)$ can be fitted to the existing data
on unpolarized $n/p$ ratios, and then used to predict the polarization
asymmetries.
Unfortunately, the absence of free neutron targets means that neutron
structure information must be inferred from deuteron structure
functions, and the current neutron $F_2^n$ data suffer from large
uncertainties associated with nuclear corrections \cite{NP}, as
illustrated in Fig.~\ref{fig:Rnp} for the neutron to proton $F_2$
ratio, $R^{np}$.

\begin{figure}[ht]
\begin{center}  
\epsfig{file=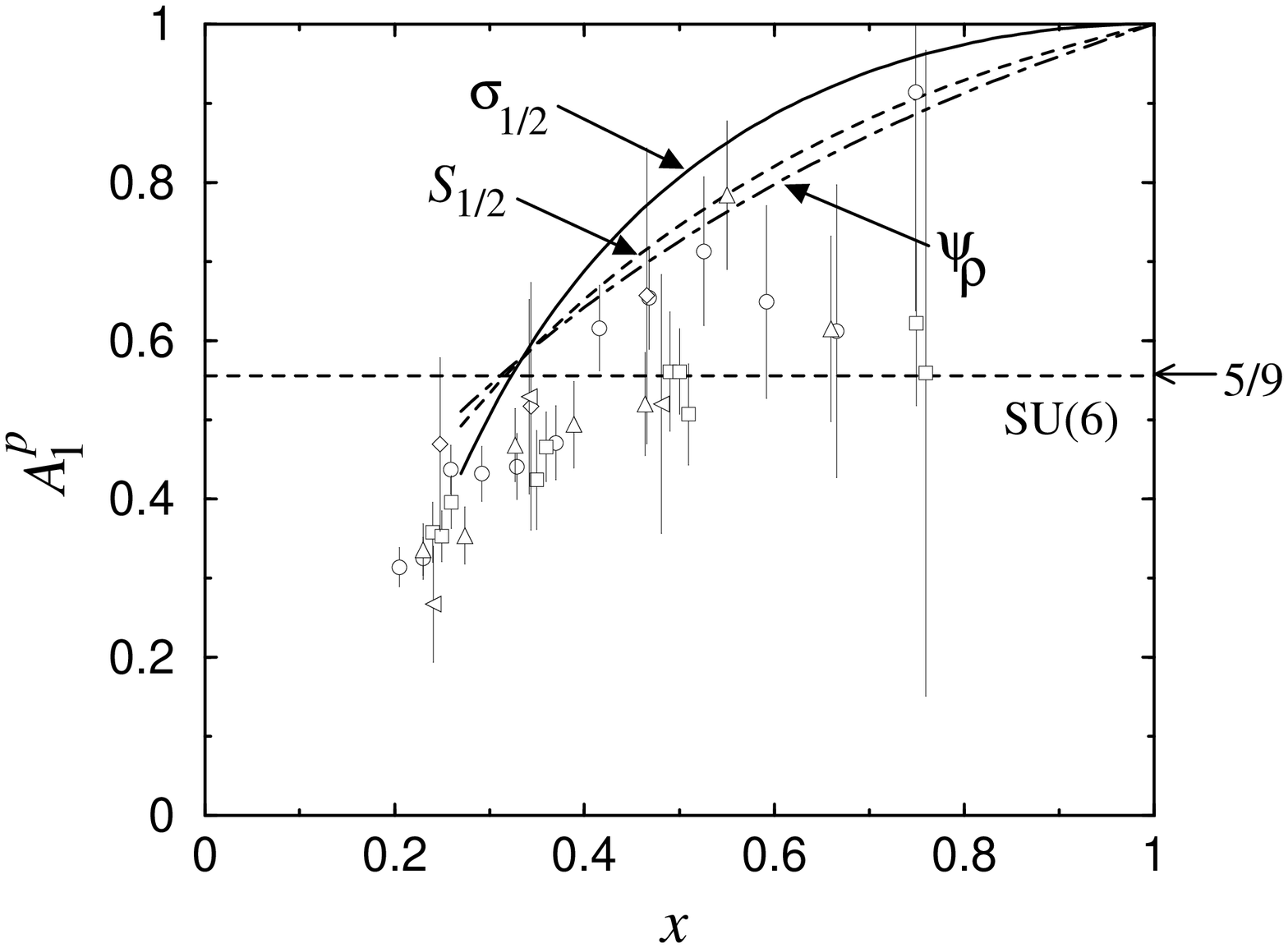,height=5.5cm}
\epsfig{file=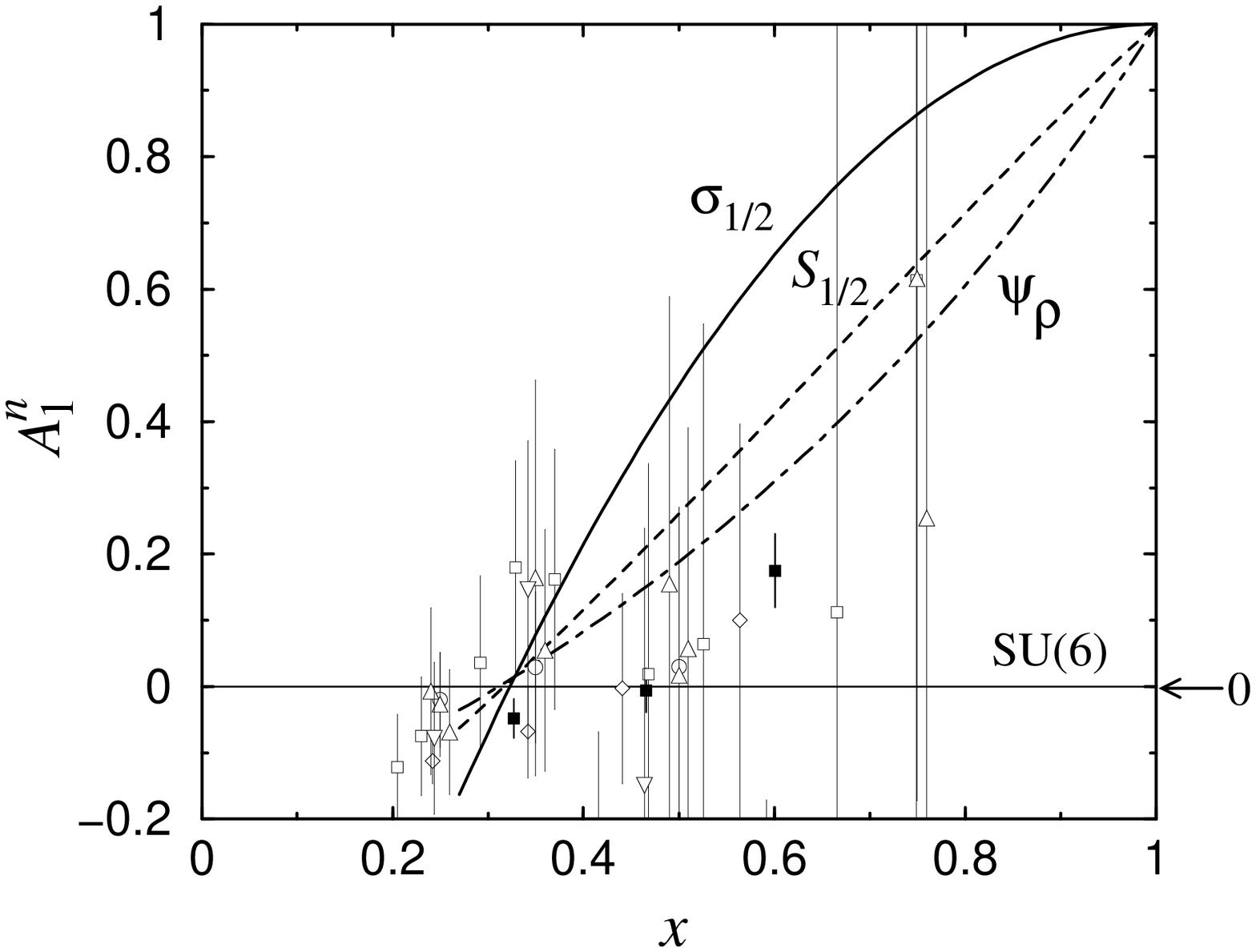,height=5.5cm}
\vspace*{0.5cm}
\caption{\label{fig:A1}
	Proton (left) and neutron (right) polarization asymmetries
	from duality \protect\cite{NOSU6}, according to different
	scenarios of SU(6) breaking, as in Fig.~\protect\ref{fig:Rnp}.
	The data are a compilation of large-$x$ results from
	experiments at SLAC \protect\cite{ABE98,E155,E142,E154},
	CERN-SMC \protect\cite{SMC},
	HERMES \protect\cite{AIR98,HERMES_n}
	and Jefferson Lab \protect\cite{A1N}.}
\end{center}
\end{figure}

A fit to the $R^{np}$ data assuming SU(6) symmetry at $x \sim 1/3$
and $\psi_\rho$ dominance at $x=1$ ($R^{np}=1/4$) is consistent with
the lower bound on the data, as indicated in Fig.~\ref{fig:Rnp}
(dot-dashed curve).
From the fitted $\theta_w(x)$, the resulting $x$ dependence of the
polarization asymmetries $A_1^p$ and $A_1^n$ is shown in
Fig.~\ref{fig:A1} (dot-dashed curves).
The predicted $x$ dependence of both $A_1^p$ and $A_1^n$ is relatively
strong; the SU(6) symmetric results which describe the data at
$x \sim 1/3$ rapidly give way to the broken SU(6) predictions as
$x \to 1$.
In both cases the polarization asymmetries approach unity as $x \to 1$
\cite{FEC74}, in contrast with the SU(6) results, especially for the
neutron.
Within the current experimental uncertainties, the $\psi_\lambda$
suppression model is consistent with the $x$ dependence of both the
$R^{np}$ ratio and the polarization asymmetries.

An interesting feature of the SU(6) quark model is that duality can
be satisfied by summing over the {\em individual} $S=1/2$ and $S=3/2$
contributions,
$S_{1/2}\equiv$ $^2${\bf 8}[{\bf 56}$^+$] + $^2${\bf 8}[{\bf 70}$^-$]
                +  $^2${\bf 8}[{\bf 70}$^-$],
and
$S_{3/2}\equiv$ $^4${\bf 10}[{\bf 56}$^+$] + $^4${\bf 8}[{\bf 70}$^-$],
separately, as well as for the total $S_{1/2} + S_{3/2}$.
If the relative contributions of the $S_{1/2}$ and $S_{3/2}$ channels
are weighted by $\cos^2\theta_s$ and $\sin^2\theta_s$, respectively,
then the unpolarized and polarized structure function ratios can be
written in terms of the mixing angle $\theta_s$ as \cite{NOSU6}
\begin{eqnarray}
R^{np}
&=& { 6 ( 1 + \sin^2\theta_s) \over 19 - 11 \sin^2\theta_s }\ ,\ \ \ \
A_1^p\
 =\ { 19 - 23 \sin^2\theta_s \over 19 - 11 \sin^2\theta_s }\ ,\ \ \ \
A_1^n\
 =\ { 1 - 2 \sin^2\theta_s \over 1 + \sin^2\theta_s }\ .
\label{eq:rat_s}
\end{eqnarray}
The presence of spin-dependent forces between quarks, such as from
single gluon exchange, can lead to different weightings of the
$S_{1/2}$ and $S_{3/2}$ components.
In particular, the expected dominance of the magnetic coupling
at high $Q^2$ leads to the suppression of $S_{3/2}$ states.
This also produces the mass splitting between the nucleon and $\Delta$,
and may be related to the anomalous suppression of the $N \to \Delta$
transition form factor relative to the elastic \cite{CM90,CM93,STOLER}.
The dominance of $S_{1/2}$ configurations ($\theta_s \to 0$) at large
$x$ leads to $R^{np} \to 6/19$, and gives unity for the polarization
asymmetries $A_1^p$ and $A_1^n$.

Fitting the mixing angle $\theta_s(x)$ to $R^{np}$ with the above
$x \to 1$ constraint, the resulting proton and neutron polarization
asymmetries are shown in Fig.~\ref{fig:A1} (dashed curves).
The predicted $x$ dependence of both $A_1^p$ and $A_1^n$ in this
scenario is similar to that in the $\psi_\lambda$ suppression model,
with a slightly faster transition to the asymptotic behavior.
The $S_{3/2}$ suppression model can be tested by studying the
electroproduction of the $l=2$ {\bf 56}$^+$ states
$P_{31}(1930)$, $P_{33}(1920)$, $F_{35}(1905)$ and $F_{37}(1950)$.
In the absence of configuration mixing, transitions to each of these
resonances should die relatively faster with $Q^2$ than for the
$^2${\bf 8} and $^2${\bf 10} states, especially for the $F_{37}(1950)$,
where mixing should be minimal.

As discussed above, duality implies that structure functions at
large $x$ are determined by transition form factors at high $Q^2$.
At large enough $Q^2$ one expects these to be constrained by
perturbative QCD, which predicts that photons predominantly couple
to quarks with the same helicity as the nucleon \cite{FJ,GNB}.
Since for massless quarks helicity is conserved, the $\sigma_{3/2}$
cross section is expected to be suppressed relative to the
$\sigma_{1/2}$ cross section.
The question then arises: Can duality between leading-twist quark
distributions and resonance transitions exist when the latter are
classified according to quark {\em helicity} rather than spin?

To answer this, consider the relative strengths of the
helicity-1/2 and helicity-3/2 contributions to the cross section
to be weighted by $\cos^2\theta_h$ and $\sin^2\theta_h$, respectively.
Using the coefficients in Table~\ref{tab:su6}, the ratios of structure
functions can then be written in terms of the mixing angle $\theta_h$
as \cite{NOSU6}
\begin{eqnarray}
R^{np}
&=& { 3 \over 7 - 5 \sin^2\theta_h }\ ,\ \
A_1^p\
 =\ { 7 - 9 \sin^2\theta_h \over 7 - 5 \sin^2\theta_h }\ ,\ \
A_1^n\
 =\ 1 - 2 \sin^2\theta_h\ .
\label{eq:rat_h}
\end{eqnarray}
In the $\theta_h \to 0$ limit the $\sigma_{3/2}$ suppression scenario
predicts that $A_1 \to 1$ for both protons and neutrons, and that the
neutron to proton ratio $R^{np} \to 3/7$.
This latter result is identical to that obtained in the classic quark
level calculation of Farrar \& Jackson \cite{FJ} on the basis of
perturbative QCD counting rules.
Again, fitting the $x$ dependence of the mixing angle $\theta_h(x)$
to the $R^{np}$ data with the corresponding $x \to 1$ constraint, the
resulting predictions for $A_1^{p,n}$ are shown in Fig.~\ref{fig:A1}
(solid curves).
Compared with the $S_{3/2}$ and $\psi_\lambda$ suppression scenarios,
the $\sigma_{1/2}$ dominance model predicts a somewhat faster approach
to the asymptotic $x \to 1$ limits.
In particular, it seems to be disfavored by the latest $A_1^n$ data
at large $x$ from Jefferson Lab \cite{A1N}, which suggest
a less rapid rise in $A_1^n$ with increasing $x$.
While it is possible that at $x \approx 1$ the structure function is
governed by helicity conservation, it appears that in the kinematical
region currently accessible perturbative QCD is not yet applicable.

Before concluding the discussion of duality in the quark model,
we should note that whereas each of the symmetry breaking scenarios
described above are consistent with duality, other scenarios are not.
For instance, suppression of the $\Delta$ or other decuplet
contributions ($^4${\bf 10} in the {\bf 56}$^+$ and $^2${\bf 10}
in the {\bf 70}$^-$) leads to inconsistent results.
Namely, the ratio of $\Delta u/u$, extracted from the $A_1^p$ and
$A_1^n$ polarization asymmetries and $R^{np}$, becomes greater than
unity, thereby violating a partonic interpretation of the structure
functions \cite{NOSU6}.
The reason for this is that removing $\Delta$ states from the
$s$-channel sum spoils the cancellation of exotic exchanges in the
$t$-channel, which cannot be interpretated as single parton
probabilities, resulting in the failure of duality in this scenario.
Inclusion of $\Delta$ states, as well as the nucleon elastic,
is vital for the realization of duality.

% .......................................................................
\subsubsection{Duality in Electron-Pion Scattering}
\label{sssec:pion}

The discussion of duality thus far has focussed on scattering from
the nucleon.
As the simplest $q\bar q$ bound state, the pion plays a unique role
in QCD: on the one hand, its anomalously small mass suggests that
it should be identified with the pseudo-Goldstone mode of dynamical
breaking of chiral symmetry in QCD; on the other, high-energy
scattering experiments reveal a rich substructure which can be
efficiently described in terms of current quarks and gluons.
The complementarity of these pictures may also reflect, in a loose
sense, a kind of duality between the effective, hadronic description
based on symmetries, and a microscopic description in terms of partons.

Shortly after the original observations of Bloom-Gilman duality for the
proton \cite{BG1,BG2}, generalizations to the case of the pion were
explored.
By extending the finite-energy sum rules \cite{DHS} devised for the
proton duality studies, Moffat and Snell \cite{MOFFAT} derived a local
duality sum rule relating the elastic pion form factor $F_\pi(Q^2)$
with the scaling structure function of the pion,
$\nu W_2^\pi \equiv F_2^\pi$,
\begin{eqnarray}
[F_\pi(Q^2)]^2
&\approx& \int_1^{\omega_{\rm max}} d\omega\ \nu W_2^\pi(\omega)\ ,
\label{eq:localpi}
\end{eqnarray}
where $\nu W_2^\pi$ here is a function of the scaling variable
$\omega \equiv 1/x$.
The upper limit of the integration
$\omega_{\rm max} = 1 + (W^2_{\rm max} - m_\pi^2)/Q^2$
was set in Ref.~\cite{MOFFAT} to $W_{\rm max} \approx 1.3$~GeV,
in order to include most of the effect of the hadron pole, and not too
much contribution from higher resonances.

\begin{figure}[htb]
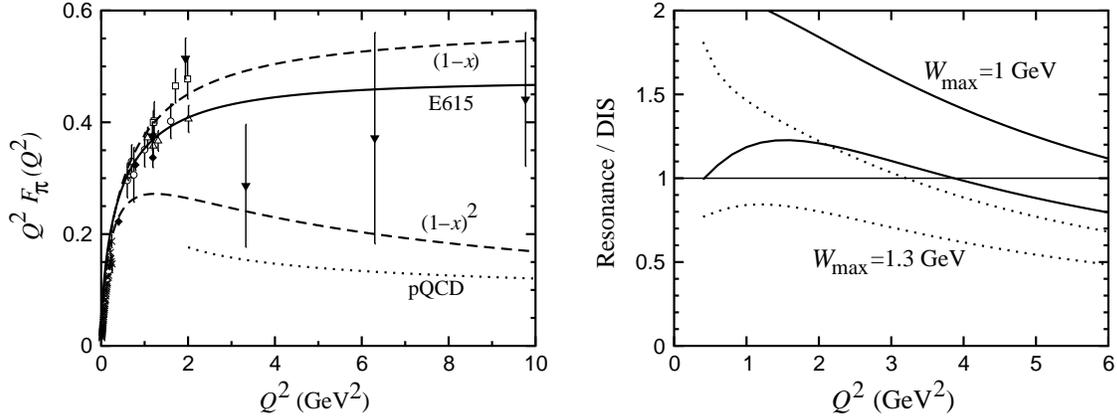

\begin{center}
% \hspace*{-0.5cm}\epsfig{file=FIGS/FigIV_F2pi.eps,height=5.5cm}
\hspace*{-0.5cm}\epsfig{file=FIGS/Fig_63a.eps,height=5.5cm}
% \hspace*{0.5cm}\epsfig{file=FIGS/FigIV_piro.eps,height=5.5cm}
\hspace*{0.5cm}\epsfig{file=FIGS/Fig_63b.eps,height=5.5cm}
\vspace*{0.5cm}
\caption{\label{fig:F2pi}
	(Left panel) Local duality prediction \protect\cite{WMPILOC}
	for the pion form factor, using phenomenological pion structure
	function input from the Fermilab E615 Drell-Yan experiment
	\protect\cite{E615} (solid), and the forms
	$F_2^\pi(x) \sim (1-x)$ and $(1-x)^2$ (dashed)
	\protect\cite{WMPI}.
	The asymptotic leading-order pQCD prediction
	\protect\cite{FJPI} (dotted) is shown for reference.
	(Right panel) Ratio of the pion resonance
	(elastic + $\pi$$\to$$\rho$ transition) contributions relative
	to the DIS continuum, for different values of $W_{\rm max}$.
	The two sets of upper and lower curves reflect the
	uncertainties in the $\pi\to\rho$ transition form factor.}
\end{center}
\end{figure}

The validity of the finite-energy sum rule relation (\ref{eq:localpi})
was tested in early analyses \cite{MOFFAT,MAHAPATRA} using Regge-based
models of the pion structure function.
More recently, data from the Drell-Yan process have allowed the
duality relation to be tested using phenomenological inputs only
\cite{WMPILOC}.
Using the fit to the $F_2^\pi(x)$ data from the E615 experiment at
Fermilab \cite{E615}, the resulting form factor $F_\pi(Q^2)$
extracted from Eq.~(\ref{eq:localpi}) is shown in Fig.~\ref{fig:F2pi}
(left panel, solid curve).
The agreement appears remarkably good, although the magnitude of
the form factor depends somewhat on the precise value chosen for
$W_{\rm max}$.
Nevertheless, the shape of the form factor is determined by the $x$
dependence of the structure function at large $x$.
In particular, while a $(1-x)$ behavior leads to a similar $Q^2$
dependence to that for the E615 fit,
% (since there $F_2^\pi \sim (1-x)^{1.2-1.3}$ \cite{E615}),
assuming a $(1-x)^2$ behavior gives a form factor which drops more
rapidly with $Q^2$.
This simply reflects the kinematic constraint $(1-1/\omega) \sim 1/Q^2$
at fixed $W$.

Although the apparent phenomenological success of the local duality
relation (\ref{eq:localpi}) is alluring, there are theoretical reasons
why its foundations may be questioned.
In fact, the workings of local duality for the pion are even more
intriguing than for the nucleon.
Because it has spin 0, elastic scattering from the pion contributes
only to the longitudinal cross section.
% (the transverse pion structure function $F_T^\pi(x=1,Q^2)=0$).
On the other hand, the spin-1/2 nature of quarks guarantees
that the deep inelastic structure function of the pion is dominated at
large $Q^2$ by the transverse cross section \cite{DGP1,DGP2,FJ}.
Taken at face value, the relation (\ref{eq:localpi}) would suggest a
nontrivial duality relation between longitudinal and transverse cross
sections, in contradiction with the parton model expectations.

While the elastic form factor of the pion is purely longitudinal, the
$\pi \to \rho$ transition on the other hand is purely transverse.
It has been suggested \cite{DGP2} that the average of the pion elastic
and $\pi\to\rho$ transition form factors may instead dual the deep
inelastic pion structure function at $x \sim 1$.
Taking a simple model \cite{WMPI} for the low-$W$ part of the pion
structure function in which the inclusive pion spectrum at
$W \alt 1$~GeV is dominated by the elastic and $\pi \to \rho$
transitions, one can estimate the degree to which such a duality
may be valid.
Generalizing Eq.~(\ref{eq:localpi}) to include the lowest-lying
longitudinal and transverse contributions to the structure function,
one can replace the left hand side of (\ref{eq:localpi}) by
$[F_\pi(Q^2)]^2 + \omega_\rho [F_{\pi\rho}(Q^2)]^2$, where
$\omega_\rho = 1 + (m_\rho^2 - m_\pi^2)/Q^2$.

The sum of the lowest two ``resonance'' contributions (elastic + $\rho$)
to the generalized finite-energy sum rule is shown in
Fig.~\ref{fig:F2pi} (right panel) as a ratio to the corresponding
leading-twist DIS structure function over a similar range of $W$.
The upper and lower sets of curves envelop different models of
$F_{\pi\rho}(Q^2)$ \cite{F_PIRO}, which can be seen as an indicator
of the current uncertainty in the calculation.
Integrating to $W_{\rm max} = 1$~GeV, the resonance/DIS ratio at
$Q^2 \sim 2$~GeV$^2$ is $\sim 50 \pm 30\%$ above unity, and is
consistent with unity for $Q^2 \sim 4$--6~GeV$^2$.
As a test of the sensitivity of the results to the value of
$W_{\rm max}$, the resonance/DIS ratio is also shown for
$W_{\rm max} = 1.3$~GeV.
In this case the agreement is better for $Q^2 \sim 1$--3~GeV$^2$,
with the ratio being $\sim 30 \pm 20\%$ below unity for
$Q^2 \sim 4$--6~GeV$^2$.

Given the simple nature of the model used for the excitation spectrum,
and the poor knowledge of the $\pi\to\rho$ transition form factor, as
well as of the pion elastic form factor beyond $Q^2 \approx 2$~GeV$^2$,
the comparison can only be viewed as qualitative.
However, the agreement between the DIS and resonance contributions
appears promising.
Clearly, data on the inclusive $\pi$ spectrum at low $W$ would be
valuable for testing the local duality hypothesis more quantitatively.
In addition, measurement of the individual transverse and longitudinal
inelastic cross sections of the pion, using LT-separation
techniques, would allow duality to be tested separately for the
longitudinal and transverse structure functions of the pion.

\begin{figure}[t]
\begin{center}
\epsfig{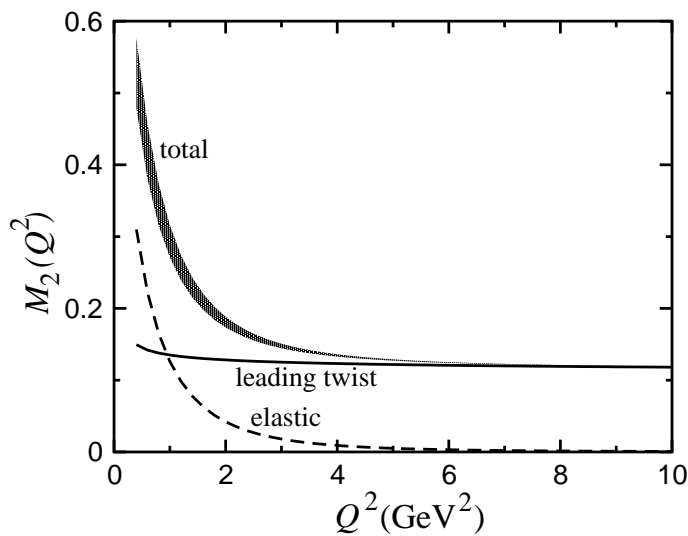}
\vspace*{0.5cm}
\caption{\label{fig:pimom}
	Lowest ($n=2$) moment of the pion structure function.
        The leading-twist (solid) and elastic (dashed)
        contributions are shown, and the shaded region represents
        the total moment using different models for the $\pi\to\rho$
        transition \protect\cite{WMPI}.}
\end{center}
\end{figure}

Going from the discussion of local duality to global duality, one can
use the available pion structure function data to perform a QCD moment
analysis, similar to that in Sec.~\ref{sssec:ope} for the proton, to
quantify the role of higher twists in $F_2^\pi$ \cite{WMPI}.
The $n=2$ moment of the pion $F_2^\pi$ structure function is shown
in Fig.~\ref{fig:pimom} as a function of $Q^2$, together with the
leading-twist and elastic contributions.
Assuming that the spectrum of $\pi \to \pi^*$ transitions is
dominated at low $W$ by the elastic and $\pi\to\rho$ transitions,
the contribution of the resonance region (which can be taken as
$W \alt 1$~GeV) to the lowest moment of $F_2^\pi$ is $\sim 50\%$
at $Q^2 \approx 2$~GeV$^2$, and only falls below 10\% for
$Q^2 \agt 5$~GeV$^2$.
The pion elastic component, while negligible for $Q^2 \agt 3$~GeV$^2$,
is comparable to the leading-twist contribution at
$Q^2 \approx 1$~GeV$^2$.
Combined, this means that the higher-twist corrections to the $n=2$
moment are $\sim 50\%$ at $Q^2 = 1$~GeV$^2$, $\sim 30\%$ at
$Q^2 = 2$~GeV$^2$, and only become insignificant beyond
$Q^2 \approx 6$~GeV$^2$.

The size of the higher-twist contribution at $Q^2 \sim 1$~GeV$^2$
is larger than that found in similar analyses of the proton $F_2$
\cite{JI_F2} and $g_1$ \cite{JI_G1} structure functions.
This can be qualitatively understood in terms of the intrinsic
transverse momentum of quarks in the hadron, $\langle k_T^2 \rangle$,
which typically sets the scale of the higher-twist effects.
Since the transverse momentum is roughly given by the inverse size
of the hadron, $\langle k_T^2 \rangle \sim 1/R^2$, the smaller
confinement radius of the pion means that the average
$\langle k_T^2 \rangle$ of quarks in the pion will be larger than
that in the nucleon.
Therefore the magnitude of higher twists in $F_2^\pi$ is expected
to be somewhat larger (${\cal O}(50\%)$) than in $F_2^p$.
%
% The E615 Collaboration indeed finds the value
% $\langle k_T^2 \rangle = 0.8 \pm 0.3$~GeV$^2$,
% within the higher twist model of Ref.~\cite{BB}.

% -----------------------------------------------------------------------
\subsection{Duality in Semi-Inclusive Reactions}
\label{ssec:semi}

In the previous sections we explored the extent to which quark-hadron
duality in inclusive processes can be understood within theoretical
models, and how duality can be utilized in phenomenological
applications relating deep inelastic structure functions to specific
exclusive channels.
To establish whether duality holds for a particular observable, one
obviously needs to know both its low-energy and high-energy behavior,
the latter which requires one to be in a region of kinematics where
perturbative QCD is applicable.
For inclusive structure functions scaling has been well established
over a large range of $Q^2$, even down to $Q^2 \alt 1$~GeV$^2$ in some
cases.
For exclusive observables, on the other hand, such as form factors,
empirical evidence suggests that considerably larger $Q^2$ values are
necessary for the onset of the expected pQCD behavior.

Exactly where perturbative scaling sets in is of course {\em a priori}
unknown -- generally speaking, the less inclusive an observable
the larger the scale at which a pQCD description is likely to hold.
One may expect therefore that duality may also set in later in
reactions which are more exclusive.
Although this may make the study of duality in less inclusive
observables more difficult experimentally at existing facilities,
it is nevertheless crucial to explore the extent and limitations of
duality in different reactions if one is to fully understand its
origins in Nature.

In this section we generalize the duality concept to the largely
unexplored domain of semi-inclusive electron scattering,
$e N \rightarrow e h X$, in which a hadron $h$ is detected
in the final state in coincidence with the scattered electron.
The virtue of semi-inclusive production lies in the ability to
identify, in a partonic basis, individual quark species in the
nucleon by tagging specific mesons in the final state, thereby
enabling both the flavor and spin of quarks and antiquarks to be
systematically determined.
Within a partonic description, the scattering and production
mechanisms become independent, and the cross section (at leading
order in $\alpha_s$) is given by a simple product of quark
distribution and quark $\rightarrow$ hadron fragmentation functions
(see also Eq.~(\ref{eq:factorization})),
\begin{eqnarray}
{ d\sigma \over dx dz }
&\sim&\
\sum_q e_q^2\ q(x)\ D_{q \to h}(z)\ ,
\label{eq:semi-parton}
\end{eqnarray}
where the fragmentation function $D_{q \to h}(z)$ gives the
probability for a quark $q$ to fragment to a hadron $h$ with a
fraction $z$ of the quark (or virtual photon) energy, $z=E_h/\nu$.
In the current fragmentation region the quark typically fragments
into mesons, which we shall focus on here.

A central question for the applicability of a partonic
interpretation of semi-inclusive DIS is whether the probability to
incoherently scatter from an individual parton ($x$ distribution),
and the subsequent probability that the parton fragments into a
particular meson ($z$ distribution), can be factorized as in
Eq.~(\ref{eq:semi-parton}).
While this is expected at high energies, it is not clear that this
is the case at low energies, such as those available at HERMES or
Jefferson Lab.
It is necessary therefore to explore the conditions under which
factorization can be applicable at energies where resonances still
play an important role.
In Sec.~\ref{sssec:pion} we reviewed the empirical status of
semi-inclusive pion production.
In this section we complement that discussion by illustrating
within a specific model how scaling and factorization can arise
from a hadronic description of semi-inclusive scattering.
Following this we consider a more local version of duality in
jet formation at high energies.

% .......................................................................
\subsubsection{Dynamical Models of Duality in Pion Production}
\label{sssec:semisu6}

\begin{figure}[ht]
\begin{center}
\epsfig{file=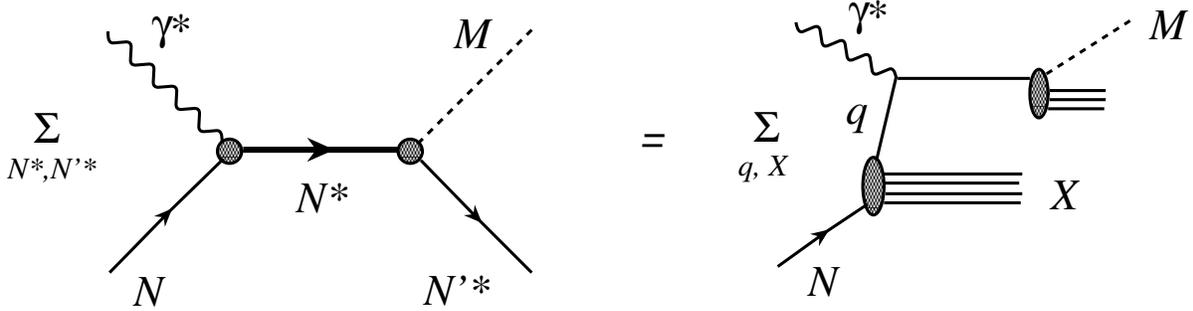,height=10cm}
\vspace*{-5cm}
\caption{\label{fig:frag}
	Duality between descriptions of semi-inclusive meson production
	in terms of nucleon resonance (left) and quark (right) degrees of
	freedom \protect\cite{CI,WM_EPIC}.}
\end{center}
\end{figure}

In terms of hadronic variables the fragmentation process can be
described through the excitation of nucleon resonances, $N^*$,
and their subsequent decays into mesons and lower-lying resonances,
which we denote by $N'^*$.
The hadronic description must be rather elaborate, however, as the
production of a fast outgoing meson in the current fragmentation
region at high energy requires nontrivial cancellations of the
angular distributions from various decay channels \cite{CI,IJMV}.
The duality between the quark and hadron descriptions of
semi-inclusive meson production is illustrated in
Fig.~\ref{fig:frag}.
Heuristically, this can be expressed as \cite{CI,WM_EPIC}
\begin{eqnarray}
\sum_{N'^*}
\left|
\sum_{N^*} F_{\gamma^* N \to N^*}(Q^2,W^2)\
	   {\cal D}_{N^* \to N'^* M}(W^2,W'^2)\
\right|^2
&=&\
\sum_q e_q^2\ q(x)\ D_{q \to M}(z)\ ,
%
% \left| \sum_{N^*,N'^*}
% F_{\gamma^* N \to N'^*}(Q^2,W^2)\
% {\cal D}_{N'^* \to N^* M}(W^2,W'^2) \right|^2
% &\sim&\ \sum_q e_q^2\ q(x)\ D_{q \to M}(z)\ , \nonumber
%
\label{eq:semi-inc}
\end{eqnarray}
where $D_{q \to M}$ is the quark $\to$ meson $M$ fragmentation function,
$F_{\gamma^* N \to N^*}$ is the $\gamma^* N \to N^*$ transition form
factor, which depends on the masses of the virtual photon and excited
nucleon ($W = M_{N^*}$), and
${\cal D}_{N^* \to N'^* M}$ is a function representing the decay
$N^* \to N'^* M$, where $W'$ is the invariant mass of the final
state $N'^*$.

The summations over hadronic states in Eq.~(\ref{eq:semi-inc}) are
considerably more involved theoretically than the corresponding sums
in inclusive scattering.
Nevertheless, there have been calculations within models, similar to
those considered in Sec.~\ref{ssec:models} for inclusive scattering,
which have attempted to carry out the resonance sums explicitly.
Close \& Isgur \cite{CI} applied the SU(6) symmetric quark model to
calculate production rates in various channels in semi-inclusive
pion photoproduction, $\gamma N \to \pi X$.
(In this model the results also generalize to virtual photoproduction.)
The pattern of constructive and destructive interference, which was
a crucial feature of the appearance of duality in inclusive structure
functions, is also repeated in the semi-inclusive case.
Defining the yields of photoproduced pions from a nucleon target as
\begin{eqnarray}
{\cal N}_N^\pi(x,z)
&=& \sum_{N'^*}
\left| \sum_{N^*}
	F_{\gamma N \to N^*}(Q^2,W^2)\
	{\cal D}_{N^* \to N'^* \pi}(W^2,W'^2)\
\right|^2\ ,
\end{eqnarray}
the breakdown of ${\cal N}_N^\pi$ into the individual states in the
SU(6) multiplets for the final $W'$ states is shown in
Table~\ref{tab:su6frag} for both proton and neutron initial states.

\begin{table}[ht]
\begin{tabular}{|c|cc|cc|}
$N'^*$ multiplet		& $\gamma p \to \pi^+ N'^*$
				& $\gamma p \to \pi^- N'^*$
				& $\gamma n \to \pi^+ N'^*$
				& $\gamma n \to \pi^- N'^*$
							\\ \hline\hline
$^2${\bf 8} [{\bf 56}$^+$]	& 100 & 0  & 0  & 25	\\
$^4${\bf 10} [{\bf 56}$^+$]	& 32  & 24 & 96 & 8	\\
$^2${\bf 8} [{\bf 70}$^-$]	& 64  & 0  & 0  & 16	\\
$^4${\bf 8} [{\bf 70}$^-$]	& 16  & 0  & 0  & 4	\\
$^4${\bf 10} [{\bf 70}$^-$]	& 4   & 3  & 12 & 1	\\ \hline
total ${\cal N}_N^\pi$		& 216 & 27 &108 & 54
\end{tabular}
\vspace*{0.5cm}
\caption{Relative strengths of SU(6) % and SU(3)$\times$SU(2)
	multiplet contributions to inclusive $\pi^{\pm}$
	photoproduction off the proton and neutron\protect\cite{CI}
	(arbitrary units).}
\label{tab:su6frag}
\end{table}

A comparison of the results of the hadronic sums with the quark level
calculation, Eq.~(\ref{eq:semi-parton}), can be made by considering
the single quark fragmentation limit, in which $z \approx 1$.
Here the scattered quark has a large probability of emerging in the
observed pion, and the hadronization process is dominated by a single
(leading) fragmentation function.
For $u$~quarks, the fragmentation into $\pi^+$ at large $z$ dominates
over that into $\pi^-$, so that $D_u^{\pi^-}/D_u^{\pi^+} \to 0$ as
$z \to 1$.
Isospin symmetry also implies that $D_d^{\pi^-} = D_u^{\pi^+}$.
This limit allows ratios of production rates to be computed directly
in terms of ratios of quark distributions.
For the case of SU(6) symmetry, where the quark distributions are
simply related by $u = 2 d$, one finds for the relative yields of
$\pi^\pm$ mesons off protons to neutrons
\begin{eqnarray}
{ {\cal N}_p^{\pi^+} \over {\cal N}_n^{\pi^+} }
&=& { {\cal N}_n^{\pi^-} \over {\cal N}_p^{\pi^-} }\
 =\ 2\ ,
\label{eq:semi-Nrat1}
\end{eqnarray}
while the ratio of $\pi^+$ to $\pi^-$ yields is
\begin{eqnarray}
{ {\cal N}_p^{\pi^+} \over {\cal N}_p^{\pi^-} }\
 =\ 8\ ,\ \ \ \ \ 
& &
{ {\cal N}_n^{\pi^+} \over {\cal N}_n^{\pi^-} }\
 =\ 2\ ,
\label{eq:semi-Nrat2}
\end{eqnarray}
for proton and neutron targets, respectively.
The total $\pi$ yield for protons versus neutrons,
$ {\cal N}_p^{\pi^+ + \pi^-} / {\cal N}_n^{\pi^+ + \pi^-} $,
is then equal to 3/2.

Comparing the parton level results with the coefficients in
Table~\ref{tab:su6frag}, one sees that these ratios coincide exactly
with those obtained from summations over coherent states in the
${\bf 56}^+$ and ${\bf 70}^-$ multiplets.
This suggests that both factorization and duality arise by summing
over all the states in the lowest-lying even- and odd-parity
multiplets.
Furthermore, the large coefficients in the first three columns of
Table~\ref{tab:su6frag} suggest that % at large $Q^2$ and $W^2$ 
an approximate duality may be obtained by including just the
${\bf 56}^+$ multiplet and the $^2{\bf 8}[{\bf 70}^-]$ states,
which phenomenologically corresponds to integrating over $W'$ up
to $\sim 1.7$~GeV.
For the ${\cal N}_p^{\pi^-}$ and ${\cal N}_n^{\pi^+}$ channels,
duality is saturated to $\approx 90\%$ already by the nucleon
elastic and $\Delta$ states alone.
One could therefore expect factorization and approximate duality
here at $W'^2 \leq 3$~GeV$^2$.

\begin{figure}[ht]
\begin{center}
\hspace*{-2cm}
\epsfig{file=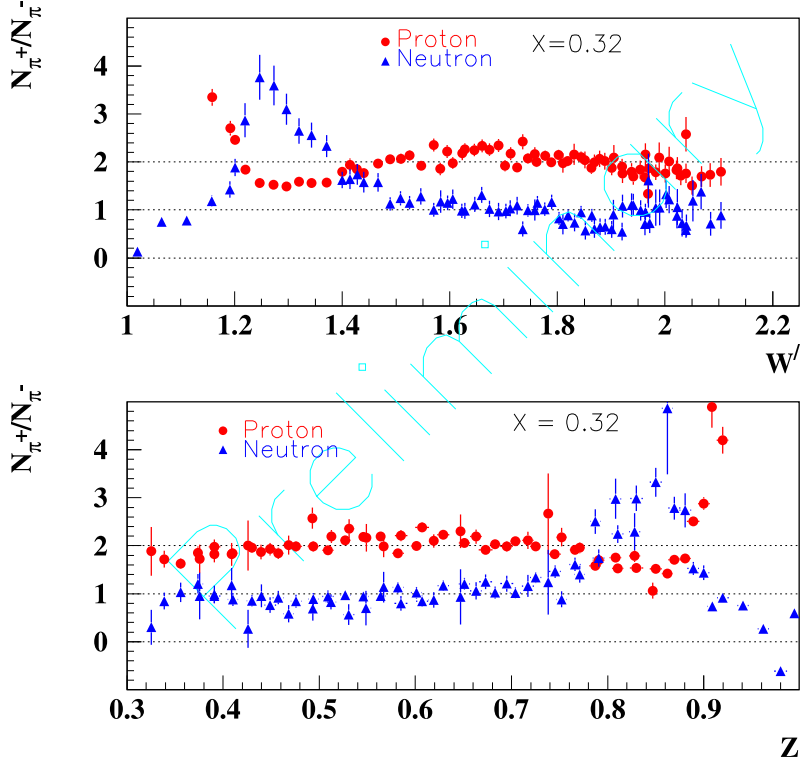,height=15cm}
\vspace*{-5cm}
\vspace*{0.5cm}
\caption{\label{fig:jlab_sidis}
	Preliminary data from Jefferson Lab experiment E00-108
	\protect\cite{E00108} for the ratio of $\pi^+$ to $\pi^-$
	semi-inclusive cross sections from proton and neutron
	targets, as a function of the final state missing mass $W'$
	(upper panel) and $z$ (lower panel).
	The value of $x$ is fixed at $x=0.32$.}
\end{center}
\end{figure}

Preliminary results on $\pi^\pm$ electroproduction from Jefferson Lab
\cite{E00108} are shown in Fig.~\ref{fig:jlab_sidis} for the ratio of
$\pi^+$ to $\pi^-$ mesons from proton and neutron targets
(the neutron data are obtained from the difference of deuteron and
proton yields).
Several interesting features are evident here.
Because $Q^2$ is relatively small ($Q^2 \sim$ few GeV$^2$),
some resonant features are clearly visible in the data,
especially at low final state hadron mass $W'$.
For the nucleon elastic contribution, since only $\pi^+$ production
is possible from the proton, and $\pi^-$ from the neutron,
the proton ratio in Fig.~\ref{fig:jlab_sidis} rises steeply
as $W' \to 1$~GeV, while the neutron ratio drops rapidly to 0.
The same feature is seen in the $z$ dependence as $z \to 1$.

In the region of the $\Delta$ resonance ($W' \approx 1.25$~GeV)
a pronounced peak is seen in the neutron ratio, but a trough
appears in the corresponding proton data.
Qualitatively, this is in agreement with the values for the
$\pi^+/\pi^-$ ratios in Table~\ref{tab:su6frag}, which are 9 times
larger for the neutron $\to \Delta$ transition than for the
proton $\to \Delta$.
Of course, we do not expect quantitative agreement with the model,
since, for instance, the results in Table~\ref{tab:su6frag} do not
include nonresonant backgrounds, which would tend to dilute the
ratios for larger $W'$ and generally bring them closer to unity.
For larger $W'$ the proton and neutron ratios are inverted again,
reflecting the stronger production rates of the $[{\bf 70}^-]$
states off the proton than off the neutron, as predicted in
Table~\ref{tab:su6frag}.

The results in Table~\ref{tab:su6frag} also suggest an explanation
for the smooth behavior of the ratio of fragmentation functions
$D^-/D^+ \equiv D_d^{\pi^+}/D_u^{\pi^+}$ for a deuterium target in
Fig.~\ref{fig:dminusdplus} of Sec.~\ref{ssec:pion}, even though
the data span the resonance region.
Since the ratio
$D^-/D^+ \approx (4 - {\cal N}^{\pi^+}/{\cal N}^{\pi^-}) /
		 (4 {\cal N}^{\pi^+}/{\cal N}^{\pi^-} - 1)$,
from the relative weights of the matrix elements in
Table~\ref{tab:su6frag} one observes that the sum of the $p$ and $n$
coefficients for $\pi^+$ production is always 4 times larger than for
$\pi^-$ production.
In the SU(6) limit, therefore, the resonance contributions to this
ratio cancel exactly, leaving behind only the smooth background,
as would be expected at high energies.
This may account for the glaring lack of resonance structure in
the resonance region fragmentation functions in
Fig.~\ref{fig:dminusdplus}.

While these results are certainly encouraging, one should caution,
however, that the coefficients in Table~\ref{tab:su6frag} apply
strictly only to the imaginary parts of the $\gamma N \to \pi N'^*$
amplitudes.
In principle one should also consider $u$-channel processes, with
the $\pi$ emitted prior to the photoabsorption.
These diagrams would give inverted ratios for $\pi^+/\pi^-$ in
Table~\ref{tab:su6frag}, and dilute the overall predictions.
On the other hand, Barbour {\em et al.} \cite{BMM} have shown that,
at least at small $Q^2$, using fixed-$t$ dispersion relations
the $s$- and $u$-channel resonances cancel to some extent for
the real part of the amplitude, so that the charge ratios in
Eqs.~(\ref{eq:semi-Nrat1}) and (\ref{eq:semi-Nrat2}) may not be
affected too strongly \cite{CI}.
Finally, while these results are restricted to the case of SU(6)
symmetry, extensions to incorporate explicit SU(6) breaking, along
the lines of those in Sec.~\ref{sssec:qm} for inclusive structure
functions, are also possible, and would be valuable in establishing
a closer connection with phenomenology.

% ........................................................................
\subsubsection{Jet Formation}
\label{sssec:jet}

At high energies a characteristic feature of semi-inclusive single
particle spectra is the production of jets in the current
fragmentation region.
In terms of resonances, the formation of a jet can be thought of as
arising from strong constructive interference in the forward region,
and destructive interference in the backward hemisphere \cite{CI}.
In principle this may be achieved by summing over different partial
waves $l$ which have specific angular distributions associated with
the respective spherical harmonics, although in practice this remains
to be demonstrated in specific dynamical models.
A first attempt in this direction within the SU(6) quark model was
described in the preceding section.
%
% In particular, from the uncertainty principle, summation over $l$
% leads one to expect an angular spread of the jet given by
% $\Delta\phi \sim 1/l_{\rm max}$ \cite{CI}.

\begin{figure}[ht]
\begin{center}
\epsfig{file=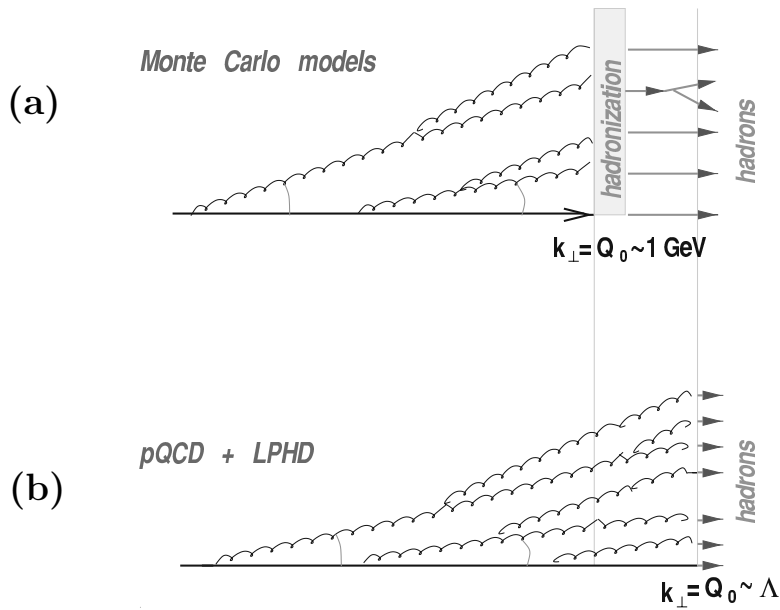,height=8cm}
\vspace*{1cm}
\caption{\label{fig:lphd}
	Schematic illustration of hadron production within
	(a) Monte Carlo fragmentation models in which hadronization
	occurs below a scale $Q_0$, and
	(b) within a pQCD framework together with
	local parton-hadron duality (LPHD).
	(Adapted from Ref.~\protect\cite{CHEKANOV}.)}
\end{center}
	\vspace*{-10cm}
	\hspace*{2cm}
	\large {\bf (a)}

	\vspace*{4.5cm}
	\hspace*{2cm}
	{\bf (b)}

	\vspace*{5cm}
	\normalsize
\end{figure}

A somewhat different application of quark-hadron duality in jet
formation in DIS was proposed by Azimov {\em et al} \cite{LPHD}.
At large $Q^2 \gg \Lambda_{\rm QCD}^2$, the conventional description
of final state formation is in terms of hard partonic scattering,
giving rise to a partonic cascade, followed by soft fragmentation
into hadrons.
Monte Carlo fragmentation models are usually used to generate the
perturbative cascade down to a scale $Q_0^2 \sim 1$~GeV$^2$,
below which nonperturbative models are invoked to describe hadron
formation --- see Fig.~\ref{fig:lphd}~(a).

The Monte Carlo models describe many properties of hadronic final
states in high energy reactions, albeit with the aid of a large number
of free parameters.
As an alternative to the Monte Carlo methods, a local correspondence
between parton and hadron distributions in hadronic jets,
termed ``Local Parton Hadron Duality'' (LPHD) was proposed \cite{LPHD}
as a way of describing the hadronic final state.
This LPHD hypothesis states that sufficiently inclusive hadronic
observables may be described entirely at the partonic level,
without {\em any} reference to hadronization.
The key assumption in LPHD is that the perturbative cascade can be
evolved down to a very low scale $Q_0 \sim \Lambda_{\rm QCD}$,
with the conversion of partons into hadrons involving only small
momentum transfers \cite{OCHS}, see Fig.~\ref{fig:lphd}~(b).
The prediction here is that hadronic spectra become proportional
to those of partons as the cut-off scale $Q_0$ is decreased towards
$\Lambda_{\rm QCD}$.

\begin{figure}[ht]
\begin{center}
\vspace*{-4cm}
\epsfig{file=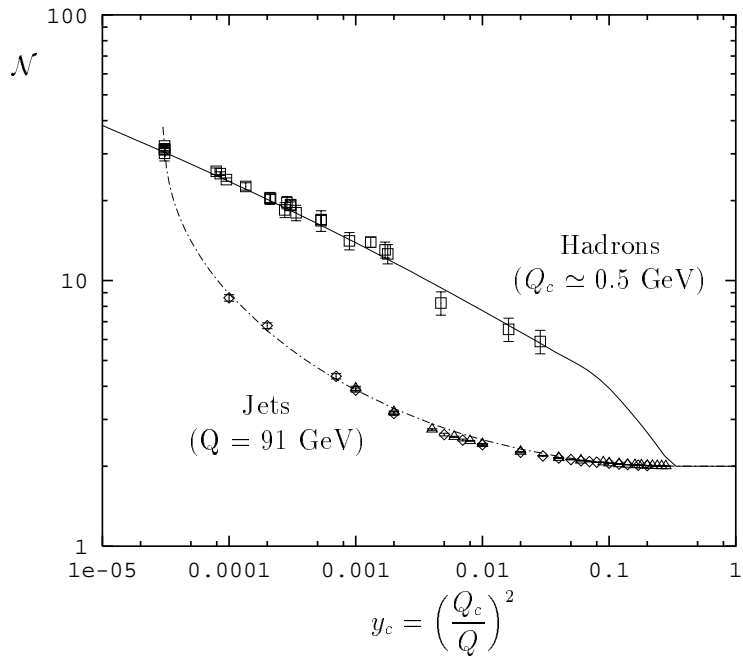,height=20cm}
\vspace*{-6cm}
\caption{\label{fig:jets}
	Average jet multiplicity $\protect{\mathcal N}$ as a
	function of the resolution parameter $y_c = (Q_c/Q)^2$, 
	at fixed $Q = 91$~GeV (lower set, ``Jets''),
	and for different energies $Q=3$--91~GeV with
	fixed $Q_c = 0.508$~GeV (upper set, ``Hadrons'')
	\protect\cite{LUPIA,Oc99}.
	The curves are described in the text.}
\end{center}
\end{figure}

\begin{figure}[ht]
\begin{center}
\epsfig{file=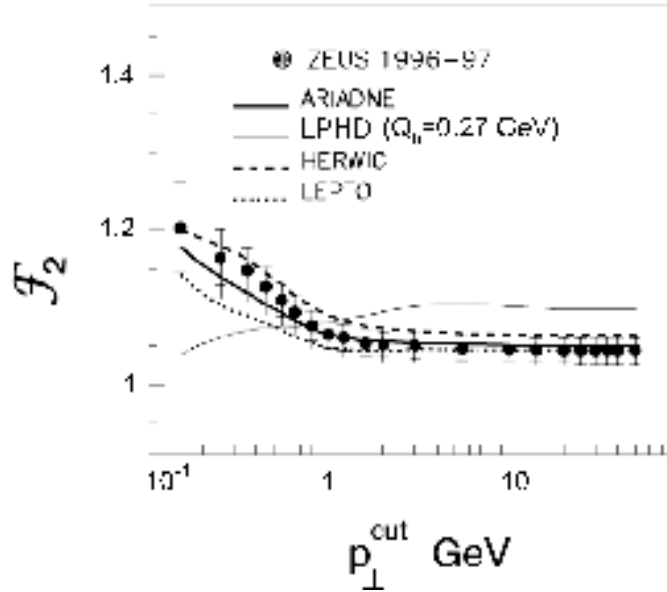,height=8cm}
\vspace*{0.5cm}
\caption{\label{fig:multip}
	Factorial moment ${\cal F}_2$ for charged particles in
	the current region as a function of $p_\perp^{\rm cut}$
	from ZEUS \protect\cite{LPHD_VIOL}, compared
	with Monte Carlo fragmentation models
	(ARIADNE, HERWIG, LEPTO),
	and the parton level calculation using LPHD.
	(Adapted from Ref.~\protect\cite{LPHD_VIOL}.)}
\end{center}
\end{figure}

Data on fragmentation in $e^+ e^-$ collisions indeed show that the
broad features of hadronic jets, such as particle multiplicities,
correlations, and inclusive spectra, calculated at the parton level
agree surprisingly well with the measured ones, as would be expected
from the LPHD hypothesis \cite{OCHS}.
Figure~\ref{fig:jets} shows results \cite{LUPIA,Oc99} on jet
multiplicities in $e^+ e^-$ collisions at LEP \cite{L3,OPAL}
as a function of the resolution parameter $y_c = (Q_c/Q)^2$,
where $Q$ here is the center of mass energy, and $Q_c$ is the
parton transverse momentum infrared cut-off scale \cite{LUPIA}.
The lower data set corresponds to jets produced at $Q = 91$~GeV,
and the curve through the data is obtained from a numerical solution
of the parton jet evolution equations \cite{JET_EV}.
The curve diverges for small cut-off $Q_c$ as $\alpha_s(k_\perp)$
becomes singular for small parton transverse momenta $k_\perp$.
The upper data set shows the average multiplicities at different
energies between $Q = 3$ and 91~GeV, calculated at fixed
$Q_c = 0.508$~GeV.
The solid curve is based on the duality picture, in which the partonic
final state corresponds to a hadronic final state at
$k_\perp \sim Q_0 \to Q_c$, with $Q_c \approx 0.5$~GeV.
The result is interpreted as indicating that in the LPHD scenario
hadrons correspond to narrow jets with resolution $Q_0 \approx 0.5$~GeV,
and that the final stage of jet evolution is reasonably well
represented by a partonic cascade even though $\alpha_s$ is large
\cite{Oc99}.

To further quantify the validity of LPHD, the ZEUS Collaboration
at HERA measured multiplicity distributions in $e^+p$ deep inelastic
scattering at very large $Q^2$ ($Q^2 > 1000$~GeV$^2$) in restricted
phase-space regions.
The particle multiplicities were studied in terms of the normalized
factorial moments, defined as
\begin{eqnarray}
{\cal F}_q(\Omega)
&=& { \langle n (n-1) \cdots (n-q+1) \rangle \over
      \langle n \rangle^q }\ ,\ \ \ q=2, 3, \ldots
\end{eqnarray}
where $q=2, 3, \ldots$ is the degree of the moment, and $n$ is the
number of particles measured inside a specified phase-space region
$\Omega$, with $\langle \cdots \rangle$ denoting the average over
all events.
The factorial moments are convenient tools to characterize the
multiplicity distributions.
As a reference point, for uncorrelated particle production within
$\Omega$, one has ${\cal F}_q=1$ for all $q$.

Correlations between particles lead to a broadening of the
multiplicity distribution and dynamical fluctuations.
Figure~\ref{fig:multip} shows a typical factorial moment,
${\cal F}_2$, measured by ZEUS \cite{LPHD_VIOL} as a function
of a transverse momentum cut, $p_\perp^{\rm cut}$.
As $p_\perp^{\rm cut}$ decreases below 1~GeV, the moment is seen
to rise, which agrees with the various Monte Carlo models of
fragmentation (ARIADNE, HERWIG, LEPTO).
On the other hand, the data disagree with perturbative calculations
(labeled ``LPHD'') in which the partonic cascade is evolved down to
$Q_0 = 0.27$~GeV, which predict a downturn in the moment with decreasing
$p_\perp^{\rm cut}$.
In fact, the pQCD prediction is
\begin{eqnarray}
{\cal F}_q(p_\perp^{\rm cut})
&\simeq& 1 + { q (q-1) \over 6 }
	 { \log(p_\perp^{\rm cut}/Q_0) \over \log(E/Q_0) }\ ,
\end{eqnarray}
where $E$ is the initial energy of the outgoing quark.
Thus the moments are predicted to approach unity for 
$p_\perp^{\rm cut} \sim Q_0$, in contrast to the rise observed
in the data.

Similar behavior is seen in the other factorial moments
(for $q=3,4,\cdots$), suggesting that additional nonperturbative
effects related to the proton remnant are necessary to explain the
data.
The results indicate therefore that the strict LPHD hypothesis, with
``one parton--one hadron'' equivalence, is violated at a
quantitative level for the hadronic multiplicities.
This suggests that the LPHD concept is applicable to more inclusive
or global (averaged) variables, and does not apply at a too exclusive
level.
Instead, a correspondence between averaged local phase-space
densities of partons and hadrons is more appropriate \cite{OCHS}.
In other words, as we have seen in other contexts,
local duality cannot be too local!

% -----------------------------------------------------------------------
\subsection{Duality in Exclusive Reactions}
\label{ssec:excl}

In the previous sections we have reviewed examples of some of the
successes and limitations of duality in inclusive and semi-inclusive
electron scattering.
The general folklore, as mentioned above, is that duality works better
for inclusive observables than for exclusive, partly because
perturbative behavior appears to set in at higher $Q^2$ for the latter,
and partly because there are fewer hadronic states over which to
average.
For exclusive processes, such as the coincidence production of a meson
$M$ and baryon $B$ in the final state, $e N \to e' M B$, duality may
be more speculative.
Nevertheless, there are correspondence arguments which relate the
exclusive cross sections at low energy to inclusive production rates
at high energy.
In this section we review the exclusive--inclusive correspondence
principle, and illustrate this with phenomenological examples in
Compton scattering and exclusive pion production.

% .......................................................................
\subsubsection{Correspondence Principle}
\label{sssec:correspond}

This exclusive--inclusive connection in hadronic physics dates back to
the early dates of deep inelastic scattering and the discussion of
scaling laws in high energy processes.
Bjorken \& Kogut \cite{BJK} articulated the correspondence relations
by demanding the continuity of the dynamics as one goes from one (known)
region of kinematics to another (which is unknown or poorly known).
The authors in fact draw an analogy with Bohr's use of the
correspondence principle in quantum mechanics, whereby the behavior
of a quantum theory is connected with the known classical limit,
which in turn leads to insights into the quantum theory itself.

\begin{figure}[ht]
\hspace*{2.5cm}
\epsfig{file=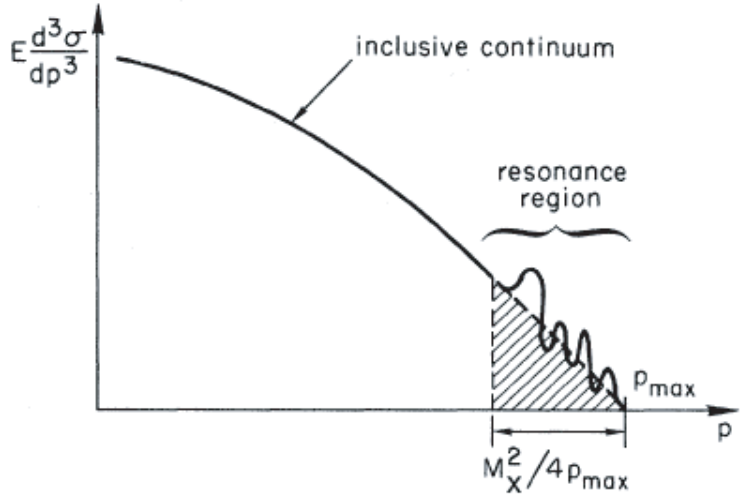,height=7cm}
\vspace*{0.5cm}
\caption{\label{fig:corresp}
	Momentum spectrum of produced hadrons in the inclusive hadron
	production reaction $\gamma^* N \to M X$.
	From Ref.~\protect\cite{BJK}.}
\end{figure}

For two-body processes, such as $\gamma^* N \to M B$, the
correspondence principle connects properties of exclusive (resonant)
final states with inclusive particle spectra, described in terms of
the differential cross section, $E d^3\sigma/dp^3$, for the
corresponding reaction $\gamma^* N \to M X$, where $E$ and $p$ are
the energy and momentum of one of the observed final state particles.
An illustration of a typical inclusive momentum spectrum for the
observed particle $M$ is shown in Fig.~\ref{fig:corresp}.
As $p$ increases, one steps from the inclusive continuum to the region
dominated by resonances.
The correspondence argument states that the magnitude of the resonance
contribution to the cross section should be comparable to the
continuum contribution extrapolated from high energy into the
resonance region,
\begin{eqnarray}
\int_{p_{\rm max} - M_X^2/4 p_{\rm max}}^{p_{\rm max}} dp\
\left. E { d^3\sigma \over dp^3 } \right|_{\rm incl}
&\sim&\ \ \sum_{\rm res}
       \left. E { d\sigma \over dp_T^2 } \right|_{\rm excl}\ ,
\label{eq:corresp}
\end{eqnarray}
where the integration region over the inclusive cross section
includes contributions up to a missing mass $M_X$.
The inclusive cross section $d^3\sigma/dp^3$ is generally a function
of the longitudinal momentum fraction $x$, the transverse momentum
$p_T$, and the invariant mass squared $s$,
\begin{eqnarray}
{ 1 \over \sigma }
{ E d^3\sigma \over dp^3 }
&=& f(x,p_T^2,sQ^2)\ .
\end{eqnarray}
At large $s$ (or equivalently large $Q^2$) this effectively reduces
to a function of only $x$ and $p_T^2$,
\begin{eqnarray}
f(x,p_T^2,sQ^2) &\to& f(x,p_T^2)\ ,\ \ \ \ s \to \infty\ .
\end{eqnarray}
Although the relation (\ref{eq:corresp}) does not represent an exact
equality, it does imply that there should be no systematic variation
of either side of the equation with external parameters.

Examples of applications of the correspondence relation
(\ref{eq:corresp}) were given by Bjorken \& Kogut \cite{BJK} for
various hadronic reactions, as well as for $e^+ e^-$ annihilation
into hadrons (see Sec.~\ref{ssec:ee} below).
For inclusive electroproduction, it was used to derive the
Drell-Yan--West relation between the asymptotic behavior of the
elastic form factor and structure function in the $x \to 1$ limit
(Sec.~\ref{sssec:elastic}).
One of the most direct application is to (real and virtual)
Compton scattering, which we discuss next.

% .......................................................................
\subsubsection{Real Compton Scattering}
\label{sssec:wacs}

\begin{figure}[h]
\hspace*{3.5cm}
\epsfig{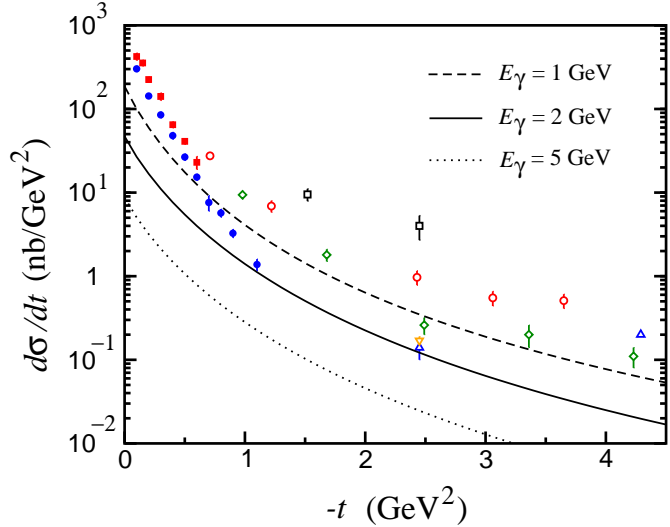}
\vspace*{0.5cm}
\caption{\label{fig:wacs}
	Cross section for wide-angle Compton scattering using
	the inclusive--exclusive correspondence relation
	(\protect\ref{eq:wacsdual}).
	The data are from SLAC \protect\cite{RCS_SLAC} for
	$E_\gamma$ between 5 and 17~GeV (filled symbols)
	and from Cornell \protect\cite{RCS_CORNELL}
	for $E_\gamma$ between 2 and 6~GeV (open symbols).}
\end{figure}

Soon after Bjorken \& Kogut suggested the exclusive-inclusive
correspondence, it was used \cite{SCOTT_WACS} to predict the behavior
of the real Compton scattering (RCS) cross section off the proton,
$\gamma p \to \gamma p$, at large angles in the center of mass frame.
At high energy the inclusive cross section for the reaction
$\gamma p \to \gamma X$ can be written (at leading order in $\alpha_s$)
in terms of quark distribution functions \cite{BJP},
\begin{eqnarray}
E { d^3\sigma \over dk^3 }
&=& { 2 \alpha^2 (s+u) (s^2+u^2) \over s^2 t^2 (-u) }
    \sum_q e_q^4\ x\ q(x)\ ,
\label{eq:wacs}
\end{eqnarray}
where $k$ is the momentum of the outgoing photon,
$s$, $t$ and $u$ are the usual Mandelstam invariants ($s+t+u=2 M^2$),
and $x = -t/2 M \nu$ is identified with the proton's longitudinal
momentum fraction carried by the quark.
(Note that here $-t$ plays the role of the large momentum scale,
in analogy with $Q^2$ in DIS.)
For large $x$ the cross section is dominated by valence quarks,
and the sum over quark charges in Eq.~(\ref{eq:wacs}) can be
replaced by
\begin{eqnarray}
\sum_q e_q^4\ x\ q(x)
&\to& { 2 (16 + d(x)/u(x)) \over 9 (4 + d(x)/u(x)) }
      F_1^p(x)\ .
\end{eqnarray}
At $x \approx 1$, one can use the local Bloom-Gilman duality relation
in Eq.~(\ref{eq:NmomF1}) to replace the $F_1^p$ structure function by
the proton magnetic form factor, $G_M(t)$, in which case the RCS
cross section at large $-t$ takes the simple form \cite{SCOTT_WACS}
\begin{eqnarray}
{ d\sigma \over dt }
&\approx& { 8 \pi \alpha^2 \over 9 }
    { s^2 + u^2 \over s^3 (-u) }
    \left( G_M^p(t) \right)^2\ ,
\label{eq:wacsdual}
\end{eqnarray}
where one has assumed $d/u \ll 1$ at $x=1$.
Using the parameterization of the proton magnetic form factor from
Ref.~\cite{MMD}, the differential cross section $d\sigma/dt$ is plotted
in Fig.~\ref{fig:wacs} versus $-t$ for several photon beam energies,
$E_\gamma = (s-M^2)/2M$.

The curves are compared with wide-angle Compton scattering data from
SLAC \cite{RCS_SLAC} at $-t \alt 1$~GeV$^2$ and from Cornell
\cite{RCS_CORNELL} for $-t \agt 1$~GeV$^2$. 
Although the calculation underestimates the magnitude of the data
somewhat, it does follow the general trend of the data, becoming a
less steep function of $t$ at large $-t$.
Upcoming data from the Jefferson Lab experiment E99-114
\cite{E99-114} will extend the kinematical range to $-t = 6$~GeV$^2$
for $E_\gamma=3$--6~GeV, which will allow more comprehensive tests
of the correspondence relation at higher~$-t$.

% .......................................................................
\subsubsection{(Deeply) Virtual Compton Scattering}
\label{sssec:dvcs}

An extension of the study of duality in Compton scattering can be
made to the case of virtual photons, and the corresponding virtual
Compton scattering (VCS) process, $e p \to e \gamma p$.
The ability to vary the virtual photon mass allows one to compare
cross sections at the same $s$ for different values of $Q^2$,
and track the behavior of resonances as one moves from low $Q^2$
to high $Q^2$.
In fact, this is precisely what led to the observation of Bloom-Gilman
duality in deep inelastic scattering, with the difference here that
one probes the real part of the virtual Compton scattering amplitude
rather than the imaginary part.

The analogy with DIS can be brought even closer by considering VCS in
the limit of large $Q^2$, known as deeply virtual Compton scattering
(DVCS).
Interest in this reaction has been fostered by the realization that at
high $Q^2$ DVCS provides access to generalized parton distributions
(GPDs), which are generalizations of parton distribution functions in
which the initial and final hadron momenta are no longer identical
\cite{GPD_FORT,GPD_JI,GPD_RAD}.
GPDs have come to prominence in recent years as a means of extracting
information on the orbital angular momentum carried by quarks and
gluons in the nucleon, and hence on the decomposition of the nucleon
spin into the various components.
Being functions of both the longitudinal and transverse parton momenta,
they also offer the prospect of mapping out a complete, three-dimensional
representation of partons in the nucleon \cite{GPD_3D}.

\begin{figure}[t]
\hspace*{4cm}
\epsfig{file=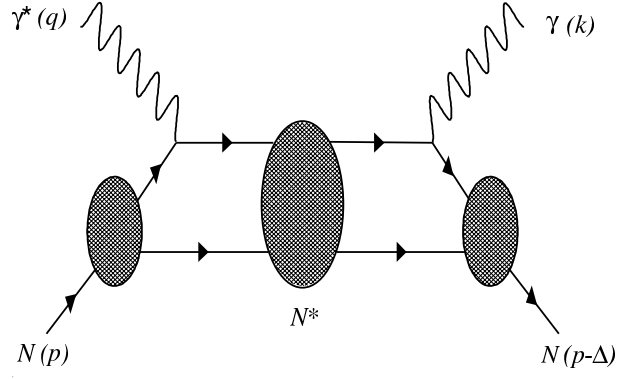,height=5cm}
\vspace*{0.5cm}
\caption{\label{fig:gpd}
	Schematic diagram for quark-hadron duality in non-forward
	Compton scattering.  (Adapted from Ref.~\protect\cite{CZ}.)}
\end{figure}

One of the important practical questions is whether the GPD formalism
is applicable at intermediate energies, such as at HERMES or at
Jefferson Lab, and it is here that one may appeal to duality for an
answer.
In particular, if one can demonstrate that duality applies also to the
case of DVCS, then a partonic interpretation of the scattering, for both
the real and imaginary parts, may be valid down to low $Q^2$.

This problem has been investigated recently by Close \& Zhao \cite{CZ}
in a generalization of the scalar constituent quark model with a
harmonic oscillator potential discussed in Sec.~\ref{sssec:res} for
the deep inelastic structure functions \cite{CI,IJMV}.
The non-forward Compton scattering process is illustrated in
Fig.~\ref{fig:gpd}, where the initial and final photon momenta 
are denoted by $q$ and $k$, respectively, and the internal ``blob''
represents coherent intermediate resonant states.
In the limit $k \to q$ the results must obviously collapse to the
forward scattering case.
For an idealized nucleon target composed of scalar quark constituents,
the generalized longitudinal response can be written in analogy with
the response for forward scattering in Eq.~(\ref{eq:RL}).
For $N$ even ($= 2n$) or $N$ odd ($= 2n+1$) excited states, one has
\cite{CZ}
\begin{eqnarray}
\widetilde R_L(\nu,\vec{q},\vec{k},\Delta^2)
&=& \sum_{N(n)} {1 \over 4 E_0 E_N}
    (E_0 \pm E_N)^2 \delta(\nu + E_0 \mp E_N)	\nonumber\\
&\times&
\left\{ \sum_{l=0(1)}^N
  \left[ (e_1 + e_2)^2
	 F_{0,2n}^{(l)}(\vec{q})
	 F_{0,2n}^{(l)}(\vec{k})\
  \right.
\right.						\nonumber\\
& & \hspace*{1cm}
\left.
  \left.
      +\ (e_1 - e_2)^2
	 F_{0,2n+1}^{(l)}(\vec{q})
	 F_{0,2n+1}^{(l)}(\vec{k})
  \right]\ \sqrt{{4\pi \over (2l+1)}} Y_{l0}(\theta)
\right\}\ ,
\label{eq:czRL}
\end{eqnarray}
where $\theta$ is the angle between the initial and final $\gamma$
momenta $\vec k$ and $\vec q$, and $\Delta = q - k$ represents the
degree to which the process is non-forward.
In contrast to the forward case, the form factors are now evaluated
at different momenta.
After performing the summation over $l$, the sum (or difference) over
all states $N$ gives \cite{CZ}
\begin{eqnarray}
\left( \sum_{N={\rm even}} \pm \sum_{N={\rm odd}} \right)
F_{0,N}(\vec{q})\ F_{N,0}(\vec{k})
&=& \exp\left( -{(\vec{q} \mp \vec{k})^2 \over 4 \beta^2} \right)\
    \equiv\ F_{0,0}(|\vec{q} \mp \vec{k}|)\ .
\end{eqnarray}
Note that in the forward limit, $\vec k = \vec q$, the sum over all
states yields unity, corresponding to completeness of states, while the
difference is equal to the elastic form factor evaluated at a momentum
$2 \vec q$.

Integrating the sum of the non-forward response over energy $\nu$,
the generalization of the sum rule in Eq.~(\ref{eq:czSq}) becomes:
\begin{eqnarray}
\widetilde S(\vec{q},\vec{k})
&\equiv& \int_{-\infty}^{+\infty} d\nu\
	 \widetilde R_L(\nu,\vec{q},\vec{k},\Delta^2)	\\
&=& (e_1^2 + e_2^2) F_{0,0}(|\vec q - \vec k|)\
 +\ 2 e_1 e_2 F_{0,0}(|\vec q + \vec k|)\ .
\label{eq:czSqk}
\end{eqnarray}
In the $Q^2 \gg |\Delta^2|$ limit the first term dominates, leading to
a partonic interpretation of the sum rule in terms of squares of quark
charges weighted by the elastic form factor \cite{CZ},
\begin{eqnarray}
\widetilde S(\vec{q},\vec{k})
&\longrightarrow& (e_1^2+e_2^2)\ F_{0,0}(\vec\Delta^2)\ .
\end{eqnarray}

The emergence of the scaling behavior from duality in this model
is due to the mass degeneracy between multiplets with the same $N$
but different $l$, which causes a destructive interference between
all but the elastic contribution.
The presence of interactions which break the $l$-degeneracy within
a given $N$ multiplet will in general spoil the exact cancellations
and give rise to violations of duality, which would lead to
oscillations about the smooth scaling law behavior at high energies
\cite{COMPTON90}.
The observation of such oscillations may therefore indicate mass
splittings {\em within} a given $N$ multiplet, in contrast to the
case of splittings {\em between} multiplets of different $N$ for
the case of inclusive structure functions discussed in
Sec.~\ref{sssec:res}.
Such patterns of oscillations have in fact been found in exclusive
pion photoproduction reactions, $\gamma p \to \pi^+ n$, which we
discuss in the next section.

% .......................................................................
\subsubsection{Exclusive Hard Pion Photoproduction}
\label{sssec:excl_pi}

While DVCS measures charge-squared weighted combinations of GPDs, in
analogy with charge-squared weighted PDFs in DIS, replacing the final
state $\gamma$ with mesons enables one to probe different combinations
of GPDs.
The process of exclusive hard meson production thus shares many
similarities with DVCS, or with wide-angle Compton scattering for real
photons.
In this section we focus our attention on the case of hard pion
production with real photons.

Implications of the inclusive--exclusive correspondence principle for
exclusive photoproduction of pions at large transverse momentum have
been investigated by several authors.
An early study of hard pion production using local duality at threshold
was made by Scott \cite{SCOTT_PI}, who used the correspondence relation
(\ref{eq:corresp}) to express the inclusive cross section near
threshold in terms of the exclusive cross section, as for Compton
scattering in Eq.~(\ref{eq:wacsdual}).

More recently, Eden {\em et al.} \cite{EHK} addressed the question of
the applicability of a leading-twist description of hard pion
production and the validity of local duality for the reaction
$\gamma p \to \pi^+ X$.
Calculating the hard scattering at the quark level in terms of the
$\gamma q \to \pi^+ q$ subprocess, and replacing the proton structure
function by the square of the magnetic form factor, Eden {\em et al.}
find in the limit of large $s$ and $t$ \cite{EHK}
\begin{eqnarray}
{d\sigma \over dt}(\gamma p \to \pi^+ n)
&\approx& 16\pi^2 \alpha\ {\alpha_s^2\ f_\pi^2 \over |t|\ s^2}
	  \left( G_M^p(-t) \right)^2\ ,
\label{eq:hoyer}
\end{eqnarray}
where one has assumed $s \ll |t|$, and that $d/u \ll 1$ in the
$x \to 1$ limit.

Experimentally, the $\gamma p \to \pi^+ n$ cross section is found to
be proportional to $1/s^2$ for $|t| \alt 2$~GeV$^2$ \cite{ANDERSON},
in agreement with Eq.~(\ref{eq:hoyer}).
However, the absolute cross section at low $s$ ($E_\gamma \alt 7$~GeV)
is underestimated by the local duality prediction by a factor $\sim 50$.
At larger $|t|$ the $\gamma p \to \pi^+ n$ cross section falls rapidly
with energy, $\sim 1/s^6$ \cite{ANDERSON}, and one expects that the
duality relation (\ref{eq:hoyer}) may be more applicable at larger $s$
and $|t|$ ($E_\gamma \sim 20$~GeV) \cite{EHK}.

The $\gamma p \to \pi^+ n$ reaction was considered by Afanasev
{\em et al.} \cite{ACW} for $s \sim |t|$, who studied duality in the
limit of fixed center of mass scattering angle, $\theta_{\rm cm}$.
Good agreement with data \cite{ANDERSON} is observed for the energy
dependence at $\theta_{\rm cm} = 90^\circ$.
However, Eden {\em et al.} \cite{EHK} point out that at fixed angle this
underestimates the measured cross section by about two orders of
magnitude, due to additional diagrams involving more than a single
quark in the nucleon which cannot be neglected in this limit.
The appropriate limit for duality, and more generally factorization,
to hold in semi-exclusive reaction is the $|t| \ll s$ limit
\cite{BDHP}.

As well as requiring an appropriate choice of kinematics, part of
the apparent failure of duality in exclusive reactions also stems
from the restriction to a single hadronic state.
Duality arises when sufficiently many intermediate hadronic states
are summed over, resulting in cancellations of non-scaling
contributions.
Certainly in Nature the cancellations are not exact, however, and
give rise to duality violations present at any finite kinematics.

\begin{figure}[ht]
\hspace*{3cm}
\hspace*{1cm}
\epsfig{file=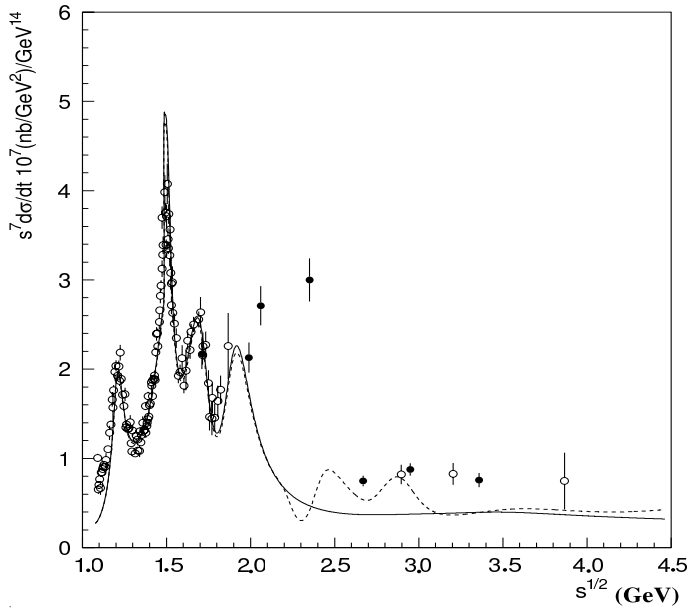,height=8cm,width=9cm}
\vspace*{0.5cm}
\caption{Energy dependence of the $\gamma p \to \pi^+ n$ cross section
	at $\theta_{\rm cm} = 90^\circ$, from Ref.~\protect\cite{ZC}.
	The solid (dashed) curve corresponds to degeneracy breaking
	for $N \leq 2$ ($N \leq 4$).
	The data are from Refs.~\protect\cite{PI_OSC} (open circles)
	and \protect\cite{JLAB_OSC} (filled circles).}
\label{fig:piosc}
\end{figure}

A novel application of duality and duality violation in exclusive
$\pi^+$ photoproduction was considered recently by Zhao \& Close
\cite{ZC}, as a possible explanation for some spectacular
oscillations seen in $\gamma p \to \pi^+ n$ cross sections
at $\theta_{\rm cm}=90^\circ$ \cite{PI_OSC,JLAB_OSC}
(see Fig.~\ref{fig:piosc}).
Using the simple pedagogical model of two scalar constituents bound
by harmonic oscillator forces from Sec.~\ref{sssec:res}, Zhao \& Close
suggest that the oscillations result from the non-degeneracy of states
with the same principal quantum number $N$ for different orbital
quantum numbers $l$.

For a degenerate spectrum, summation over resonance excitations
produces destructive interference of coherent contributions, giving
rise to scaling behavior.
In fact, since the $l$-odd terms are proportional to
$\cos\theta_{\rm cm}$, only parity-even (and hence $N$-even)
contributions will be nonzero at $\theta_{\rm cm} = 90^\circ$
\cite{ZC}.
At high energies the large number of overlapping resonances makes
the cancellations, and hence duality, appear locally.
At lower energy, however, where fewer resonances are encountered,
the different partial waves will not cancel locally if the resonances
with different $N$ are not degenerate, and one can expect deviations
from the smooth scaling behavior at $\theta_{\rm cm} = 90^\circ$
arising from interference between the non-local resonances.

In Fig.~\ref{fig:piosc} the differential cross section $s^7 d\sigma/dt$
is plotted as a function of the center of mass energy $\sqrt{s}$ for two
scenarios of degeneracy breaking: for $N \leq 2$ and for $N \leq 4$
states.
Sizable oscillations are clearly evident in both cases, which persist
to several GeV, but with decreasing amplitude for larger $\sqrt{s}$.
The agreement with the data at low $\sqrt{s}$ is quite remarkable
given the simplicity of the model.

If violations of local duality are indeed responsible for the observed
oscillations \cite{PI_OSC,JLAB_OSC}, one would expect a specific $Q^2$
dependence for these, in contrast with some of the alternative proposed
explanations in terms of charm thresholds \cite{OSC_CHARM},
or the interference between short and long-distance effects
\cite{OSC_RALSTON,OSC_JM}.

In particular, if a set of resonances is suppressed at large $Q^2$
(as discussed for example in Sec.~\ref{sssec:qm}), there should be
strong $Q^2$ dependence in the oscillations, with neither the position
nor the magnitude displaying any simple periodicity \cite{ZC}.
%
% Furthermore, if the main degeneracy breaking mechanism is due to
% partial wave dependence at $\theta_{\rm cm} = 90^\circ$, then
% destructive interference should occur within states of a given $N$,
% so that even- and odd-parity states could be isolated.
%
Furthermore, one can also expect oscillations arising from violations
of duality in other processes, such as vector meson production.
Some of these and other future tests of duality will be discussed
in Sec.~\ref{sec:outlook} below.

\clearpage

%%%%%%%%%%%%%%%%%%%%%%%%%%%%%%%%%%%%%%%%%%%%%%%%%%%%%%%%%%%%%%%%%%%%%%%%%
\section{Quark-Hadron Duality in Related Fields}
\label{sec:related}

In the previous sections we have reviewed the experimental status of
duality in structure functions and discussed its theoretical
interpretations within various models, and more formally using the
operator product expansion in QCD.
To put this discussion in a broader context, in this section we
consider examples of duality observed in areas other than electron
scattering.
We review several famous examples, ranging from the prototypical case
of duality in $e^+ e^-$ annihilation into hadrons, and the celebrated
application in semileptonic decays of heavy mesons, to a more recent
speculative example of duality in $p\bar p$ annihilation.
These examples will illustrate many features in common with
Bloom-Gilman duality in electron scattering, suggesting a common
origin of these phenomena in QCD.
To begin with, we first review one of the most extensive
{\em theoretical} applications of duality in hadronic physics,
namely that in QCD sum rules.

% -----------------------------------------------------------------------
\subsection{QCD Sum Rules}
\label{ssec:qcdsum}

The method of QCD sum rules \cite{SVZ,SVZ_APP,SVZ_QM} has enjoyed
tremendous success in the computation of a wide range of hadronic
ground state properties, as well as form factors \cite{SR_PIFF} and
(moments of) structure functions \cite{SR_SF} % in various channels
(for reviews see Refs.~\cite{RRY_REV,DEREK_REV,RADYUSH_REV}).
The basic premise behind QCD sum rules is that physical quantities are
obtained by matching results calculated in terms of quark-gluon degrees
of freedom, using asymptotic freedom, with those calculated in terms of
hadrons via dispersion relations.
The partonic side of the sum rule is often referred to as the
``theoretical'' part, while the hadronic side is referred to as the
``phenomenological'' part.
At the heart of the sum rule method lies quark-hadron duality ---
the ability to relate low-energy observables to their asymptotic
high-energy behavior.
Indeed, as Shifman, Vainshtein and Zakharov remark in their classic
paper \cite{SVZ_APP}, ``QCD sum rules can be considered as a
justification and refinement of the duality relations'' between
resonance and continuum cross sections.

Calculation of the partonic side of the sum rule relations (as in deep
inelastic scattering) relies on factorization of the short-distance
amplitudes from the long-distrance amplitudes, with the latter
parameterized in terms of quark and gluon vacuum condensates.
The hadronic (phenomenological) side, on the other hand, requires an
accurate representation of the hadronic spectrum.
In cases where the ground state is dominant, the properties of the
ground state itself can be extracted, in a way reminiscent of the
local duality discussed in Sec.~\ref{sssec:elastic}.

In the following we consider two pedagogical examples which
graphically illustrate the interplay between confinement and
asymptotic freedom implicit in duality.
While at present the accuracy of quark-hadron duality cannot be
rigorously determined in QCD, the QCD sum rule method is quite general,
and many features can be explored by considering simple models for
which exact solutions are known.
One can then try to draw lessons from the simple models to more
realistic cases, which can help us to understand the origin and
phenomenological consequences of duality in QCD.
In fact, one doesn't even need to consider quantum field theory ---
the essential elements of duality can already be seen at work in
quantum mechanics.
We shall review one such example in the next section: the quantum
mechanical harmonic oscillator.
Following this we describe the extension of the sum rule method
to field theory, by applying the operator product expansion to the
$\rho$ meson.

% .......................................................................
\subsubsection{Quantum Mechanics}
\label{sssec:SRqm}

One of the simplest examples of an exactly soluble model which
illustrates the basic elements of duality is the quantum mechanical
harmonic oscillator \cite{SVZ_QM}.
This model in fact provides an ideal laboratory to address the
question of whether asymptotic sum rules can be obtained even if they
are saturated by a single resonance.
Here we use the simplified case of a 2+1 dimensional harmonic
oscillator, as discussed by Radyushkin \cite{RADYUSH_REV}, which avoids
the unnecessary algebraic complications of the 3+1 dimensional case.
(The latter was considered by Vainshtein {\em et al.} \cite{SVZ_QM}
--- see also Refs.~\cite{SR_BB,SR_DURAND,SR_BLOK}.)
Here we shall follow closely the notations of
Refs.~\cite{SVZ_QM,RADYUSH_REV}.

In quantum mechanics the time-dependent Green's function for the
propagation of a particle in an external field from a point
$(\vec 0,0)$ to the point $(\vec x,t)$ is given by
\begin{eqnarray}
G(\vec x,t)
&=& \sum_{n=0}^\infty
    \psi_n^*(\vec 0)\ \psi_n(\vec x)\ e^{i E_n t}\ ,
\end{eqnarray}
where $\psi_n(\vec x)$ is the eigenfunction describing the particle
in the $n$-th excited state with energy $E_n$.
The time evolution of the Green's function turns out to be easier to
study in imaginary time.
Performing a Wick rotation to Euclidean space, $t \to i \tau$, and
taking $\vec x=0$, the Green's function then becomes \cite{RADYUSH_REV}
\begin{eqnarray}
G(\vec 0,\tau)
&=& \sum_{n=0}^\infty
    \left| \psi_n(\vec 0) \right|^2\ e^{-E_n \tau}\ .
\end{eqnarray}
One can show that for small Euclidean time intervals, $\tau \to 0$,
the interacting Green's function approaches the free Green's function,
as it would for the case of asymptotic freedom (see also
Ref.~\cite{FEYNHIB}).
In order to make the analogy with the OPE more apparent, it will be
more convenient to express the Green's function in terms of the
conjugate parameter $\epsilon \equiv 1/\tau$, and to define the
function
\begin{eqnarray}
M(\epsilon)
&\equiv& G(\vec 0,1/\epsilon)\
 =\ \sum_{n=0}^\infty
    \left| \psi_n(0) \right|^2\ e^{-E_n/\epsilon}\ .
\end{eqnarray}
For a 2+1 dimensional harmonic oscillator potential,
$V = {1 \over 2} m \omega^2 r^2$, the energy levels are given by
\begin{eqnarray}
E_n &=& (2n + 1)\ \omega\ ,
\label{eq:SREn}
\end{eqnarray}
where $\omega$ is the oscillator frequency, $m$ is the particle mass,
and the wave function at the origin,
$|\psi_0(\vec 0)|^2 = m \omega/\pi$, is independent of the excitation
level $n$.
Performing the sum over $n$, the function $M(\epsilon)$ can be written as
\begin{eqnarray}
M(\epsilon)
&=& {1\over 2} m \omega \sum_{n=0}^\infty e^{-E_n/\epsilon}\
 =\ { m\ \omega \over 2 \pi \sinh \omega\tau }\ .
\label{eq:QMho}
\end{eqnarray}
In the $\epsilon \to \infty$ limit, $M(\epsilon)$ collapses to its free
limit,
\begin{eqnarray}
M_0(\epsilon)
&=& { m\ \epsilon \over 2\pi }\ .
\end{eqnarray}
Note that even though each term in the series in Eq.~(\ref{eq:QMho})
depends on $\omega$, the total sum in the asymptotic limit is
independent of $\omega$.
This can be compared with the scale independence of the DIS structure
function, when summed over $Q^2$-dependent resonance form factors.
In fact, expanding $M(\epsilon)$ in Eq.~(\ref{eq:QMho}) for large
$\epsilon$ in powers of $1/\epsilon$, one has \cite{RADYUSH_REV}
\begin{eqnarray}
M(\epsilon)
&=& M_0(\epsilon)
    \left( 1 - {1 \over 6} {\omega^2 \over \epsilon^2}
	     + {7 \over 360} {\omega^4 \over \epsilon^4}
	     - {31 \over 15120} {\omega^6 \over \epsilon^6}
	     + \cdots
    \right)\ ,
\end{eqnarray}
where the $1/\epsilon^2$ corrections resemble the $1/Q^2$ corrections
in the twist expansion of Sec.~\ref{sssec:ope}.

\begin{figure}[ht]
\begin{center}
\hspace*{0.5cm}
\epsfig{file=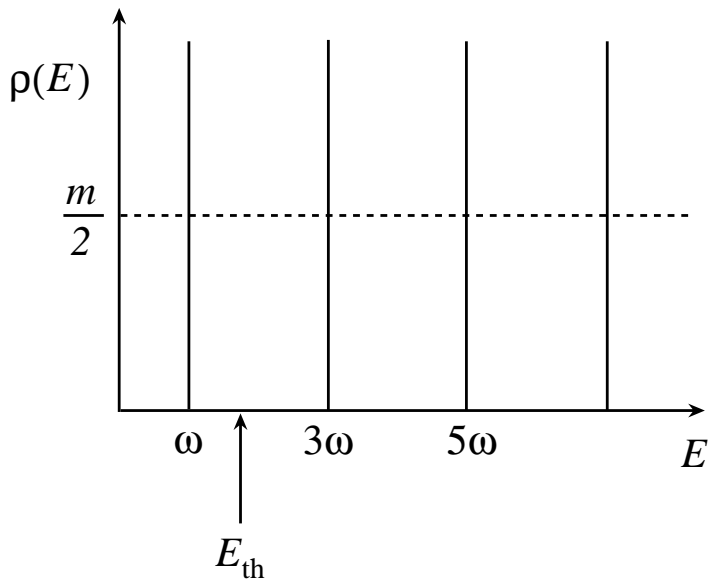,height=9cm}
\caption{\label{fig:ho}
	Spectral density $\rho(E)$ as a function of energy for the
	2+1 dimensional harmonic oscillator.
	The harmonic oscillator levels ($\delta$-functions) are
	indicated by solid vertical lines, and the free particle
	level is indicated by the dashed horizontal line at $m/2$.
	The approximate location of the continuum threshold
	$E_{\rm th}$ is indicated by the arrow.}
\end{center}
\end{figure}

To demonstrate how the sum over bound state wave functions coincides
with free states, one can use the spectral representation for the
Green's function,
\begin{eqnarray}
M(\epsilon)
&\to& {1 \over \pi} \int_0^\infty dE\ \rho(E)\ e^{-E/\epsilon}\ ,
\label{eq:SR_Mho}
\end{eqnarray}
where the harmonic oscillator spectral function,
\begin{eqnarray}
\rho(E)
&\equiv& m \omega \sum_{n=0}^\infty \delta(E - E_n)\ ,
\end{eqnarray}
is a superposition of $\delta$-functions in the energy $E$, with the
harmonic oscillator energy levels $E_n$.
The asymptotic function $M_0(\epsilon)$ can also be expressed in terms
of an analogous free spectral function,
\begin{eqnarray}
\rho_0(E)
&=& {1 \over 2} m\ \theta(E)\ ,
\end{eqnarray}
as illustrated in Fig.~\ref{fig:ho}.
Clearly the free and interacting spectral densities cannot be more
orthogonal to each other!
However, by integrating the latter between the mid-points of adjacent
$\delta$-functions, one finds an {\em exact local duality} between the
free and interacting spectral densities,
\begin{eqnarray}
\int_{2n\omega}^{2(n+1)\omega} dE\
\left( \rho(E) - \rho_0(E) \right)
&=& 0\ ,\ \ \ \forall\ \ n \geq 0\ .
\end{eqnarray}
In addition to the lowest moment, a similar duality holds also for the
first moment of $\rho$,
\begin{eqnarray}
\int_{2n\omega}^{2(n+1)\omega} dE\
E\ \left( \rho(E) - \rho_0(E) \right)
&=& 0\ .
\end{eqnarray}
Generalizing this to the exponential-weighted densities in
$M(\epsilon)$, one has
\begin{eqnarray}
\int_0^\infty dE\ e^{-E/\epsilon}\
\left( \rho(E) - \rho_0(E) \right)
&=& \sum_{n=1}^\infty c_n \left( { 1 \over \epsilon } \right)^n\ ,
\end{eqnarray}
where $c_n$ are coefficients, and once again the power corrections
$1/\epsilon$ are reminiscent of the twist expansion in QCD.
Note that since there is no ${\cal O}(1)$ term in the difference,
one also has an {\em exact global duality} in the $\epsilon \to \infty$
limit,
\begin{eqnarray}
\int_0^\infty dE\
\left( \rho(E) - \rho_0(E) \right)
&=& 0\ .
\label{eq:SRho_global}
\end{eqnarray}
Similar results have been obtained for potentials other than the
harmonic oscillator, such as the infinite spherical well, or a linear
potential --- in fact, any potential which is nonsingular at the origin
will satisfy a relation similar to Eq.~(\ref{eq:SRho_global})
\cite{SVZ_QM}.

The utility of the sum rule method lies in the possibility of
extracting properties of the ground state $\psi_0$ from the
asymptotic sum.
Both the ground state energy $E_0$ and wave function $\psi_0$ are
obtained by matching the $n \geq 1$ contributions to $M(\epsilon)$
in (\ref{eq:QMho}) with the free result above some threshold,
$E > E_{\rm th}$.
This is achieved by replacing the lower limit of integration in
Eq.~(\ref{eq:SR_Mho}) by $E_{\rm th}$, and the interacting density
$\rho$ by $\rho_0$, which leads to the relation
\begin{eqnarray}
\left| \psi_0(\vec 0) \right|^2\ e^{-E_0/\epsilon}
&\approx& { m \epsilon \over 2\pi }
	  \left( 1 - e^{-E_{\rm th}/\epsilon}
		   - {1 \over 6} {\omega^2 \over \epsilon^2}
		   + {7 \over 360} {\omega^4 \over \epsilon^4}
		   + \cdots
	  \right)\ .
\label{eq:SRthresh}
\end{eqnarray}
Differentiating both sides of (\ref{eq:SRthresh}) with respect to
$1/\epsilon$ then enables one to solve for $E_0$ in terms of
$E_{\rm th}$ and $\epsilon$, and compare with the exact expression
(\ref{eq:SREn}).
In the $\epsilon \to \infty$ limit (which corresponds to asymptotic
freedom), one obtains exact duality for the ground state alone,
\begin{eqnarray}
\left| \psi_0(\vec 0) \right|^2
&\to& { m E_{\rm th} \over 2 \pi }\ ,
\end{eqnarray}
where the corresponding ``duality interval'' in this limit is given
by $E_{\rm th} \to 2 \omega$.
At finite $\epsilon$ the accuracy of the sum rule estimate depends on
the order at which one truncates the series.
Truncating at order $\omega^3$, Radyushkin \cite{RADYUSH_REV} finds 
the minimum dependence on $\epsilon$ for $E_0 = 0.95\ \omega$, which
corresponds to an energy threshold $E_{\rm th} = 1.75\ \omega$.
For these values, the wave function
$|\psi_0(\vec 0)|^2 \approx 0.9 m \omega/\pi$.
Up to this order, one therefore finds that the parameters describing
the ground state can be described with $\sim 10\%$ accuracy.
The main source of error is associated with the somewhat crude
treatment of the spectral density for the $n \geq 1$ states.
In contrast, while the lowest state is narrow in Nature, the higher
excited states are usually rather broad, so that approximating these
by free quark states may lead to even better convergence.
The convergence of the series may also be improved by performing a
Borel summation, as discussed in the next section.

% .......................................................................
\subsubsection{Duality for the $\rho$ Meson}
\label{sssec:SRrho}

In field theory the Green's function $G(\vec x, t)$ generalizes to
a current-current correlator, $\Pi(q^2)$, defined in the momentum
representation as the vacuum expectation value of the time-ordered
product of currents $J_\mu = \bar \psi(x) \gamma_\mu \psi(x)$,
\begin{eqnarray}
(q_\mu q_\nu - q^2 g_{\mu\nu})\ \Pi(q^2)
&=& i \int d^4x\ e^{i q \cdot x}\
    \langle 0 | T( J_\mu(x) J_\nu(0) ) | 0 \rangle\ ,
\end{eqnarray}
where $\psi(x)$ is the quark field.
To illustrate the practical application of the QCD sum rule method
we shall determine the properties of the ground state in the spin-1,
isospin-1 channel, corresponding to the $\rho$ meson.

In analogy with the twist expansion in DIS, at large $q^2$ the
correlator $\Pi(q^2)$ can be expanded using the OPE in terms of
expectation values of local operators $\widehat {\cal O}$ multiplied
by hard Wilson coefficients $C_n(q^2)$,
\begin{eqnarray}
\Pi(q^2)
&=& \sum_n\ C_n(q^2)\ \langle 0 | \widehat{\cal O} | 0 \rangle\ .
\end{eqnarray}
For space-like momenta $Q^2 \equiv -q^2 > 0$ the correlator
satisfies a standard dispersion relation,
\begin{eqnarray}
\Pi(Q^2)
&=& \Pi(0)
 - { Q^2 \over 12 \pi^2 }
   \int_0^\infty ds\ { R(s) \over s\ (s+Q^2) }\ ,
\end{eqnarray}
where
\begin{eqnarray}
R(s) &=& { \sigma_h(e^+e^- \to {\rm hadrons}) \over 
	   \sigma(e^+e^- \to \mu^+\mu^-) }\
\end{eqnarray}
is the ratio of $e^+ e^-$ annihilation cross sections into hadrons
(in the $I=1$ channel) to that into muons, at a value $s$ of the
total center of mass energy squared of the $e^+ e^-$ pair
(see also Sec.~\ref{ssec:ee} below).
The elementary muon cross section is given by
\begin{eqnarray}
\sigma(e^+ e^- \to \mu^+ \mu^-)
&=& { 4 \pi\, \alpha^2 \over 3\, s }\ ,
\label{eq:muons}
\end{eqnarray}
where $\alpha$ is the electromagnetic fine structure constant.

As mentioned in the preceding section, the convergence properties of
the sum rule can be improved by making a Borel transformation of both
the OPE (partonic) and dispersion (hadronic) sides of the sum rule,
as defined by the operation
\begin{eqnarray}
\widehat B\ f(Q^2) &\to& \widetilde f(M^2)\ ,
\end{eqnarray}
where
\begin{eqnarray}
\widehat B
&=& \lim_{Q^2, n \to \infty \atop M^2\ {\rm fixed}}
    { 1 \over (n-1)! }\ (Q^2)^n
    \left( -{d \over dQ^2} \right)^n\ .
\end{eqnarray}
Here $M^2 \equiv Q^2/n$ is the square of the Borel mass, which sets the
scale at which the long- and short-distance expansions are matched.
Application of the Borel transform to the correlator $\Pi(Q^2)$ leads
to the sum rule \cite{SVZ}
\begin{eqnarray}
\int_0^\infty ds\ e^{-s/M^2}\ R(s)
&=& {3 \over 2} M^2
    \left\{ 1 + { \alpha_s(M) \over \pi }
	      - { 2 \pi^2 f_\pi^2 m_\pi^2 \over M^4 }
	      + { \pi \over 3 M^4 }
		\langle \alpha_s G \cdot G \rangle
	      - { 448 \pi^3 \over 81 M^6 }\
		\alpha_s \langle \bar q q \rangle^2
    \right\}\ ,		\nonumber\\
& &
\label{eq:SRrs}
\end{eqnarray}
where the coefficients of the $1/M$ power corrections are given
in terms of quark and gluon vacuum condensates,
$\langle \bar q q \rangle \equiv \langle 0 | \bar q q | 0 \rangle$
and
$\langle \alpha_s G \cdot G \rangle
\equiv \langle 0 | \alpha_s G_{\mu\nu}^a G^{a, \mu\nu} | 0 \rangle$,
with
$f_\pi^2\ m_\pi^2
= -2\ \langle m_u \bar u u + m_d \bar d d \rangle$.
Note the absence of ${\cal O}(1/M^2)$ corrections on the right hand side
of Eq.~(\ref{eq:SRrs}).
The ${\cal O}(1)$ term in Eq.~(\ref{eq:SRrs}) corresponds to the
free quark result for $R(s)$,
\begin{eqnarray}
R_0(s)
&=& {3 \over 2} \left( 1 + {\alpha_s(s) \over \pi} \right)\ ,
\label{eq:R0}
\end{eqnarray}
evaluated to order $\alpha_s$.
The famous factor of 3/2 in $R_0(s)$ arises from the number of
quark colors (3), and the square of the isovector quark charge,
$(e_q^{I=1})^2=((e_u-e_d)/2)^2$, multiplied by 2
(for $u$ and $d$ quarks).
Just as in the DIS case, the $1/M$ power corrections in the sum rule
(\ref{eq:SRrs}) parameterize the effects of confinement, and control
the behavior of the resonance contributions.
Taking the $M^2 \to \infty$ limit, the exponent $e^{-s/M^2} \to 1$,
and the power corrections vanish, leaving an exact duality between
the hadronic contributions and the free result,
\begin{eqnarray}
\int_0^\infty ds\ \left( R(s) - R_0(s) \right)
&\to& 0\ .
\label{eq:SRasym}
\end{eqnarray}
Using phenomenological values for the quark and gluon condensates,
$\langle m_u \bar u u + m_d \bar d d \rangle
\approx$ --(0.114~GeV)$^4$,
$|\langle \bar q q \rangle|
\approx$ (0.25~GeV)$^3$,
and
$\langle \alpha_s G \cdot G \rangle
\approx$ (0.44~GeV)$^4$,
the sum rule (\ref{eq:SRrs}) gives
\begin{eqnarray}
\int_0^\infty ds\ e^{-s/M^2}\ R(s)
&\approx& {3 \over 2} M^2
    \left\{ 1 + { \alpha_s(M) \over \pi }
              + 0.1 \left( {0.6\ {\rm GeV}^2 \over M^2} \right)^2
	      - 0.14 \left( {0.6\ {\rm GeV}^2 \over M^2} \right)^3
    \right\}\ .
\end{eqnarray}
Choosing the Borel mass to be equal to the $\rho$ mass,
$M^2 = m_\rho^2 \approx 0.6$~GeV$^2$, the power corrections appear to
be relatively small compared with the free quark term, even though
for such a value of $M^2$ the (physical) cross section integral is
dominated by a single ($\rho$) resonance!

In the narrow resonance approximation the $\rho$ contribution
is given by
\begin{eqnarray}
R^{(\rho)}(s)
&=& { 12 \pi^2 m_\rho^2 \over g_\rho^2 }\ \delta(s-m_\rho^2)\ ,
\end{eqnarray}
where the coupling constant $g_\rho$ is defined in terms of the matrix
element of the vector current,
$\langle 0 | J_\mu | \rho \rangle = \epsilon_\mu (m_\rho^2/g_\rho)$.
One can attempt therefore to extract the $\rho$ properties by
neglecting both the power corrections to the asymptotic result
$R_0(s)$, and the higher-mass resonances above the $\rho$ pole,
\begin{eqnarray}
\int_0^\infty ds\ e^{-s/m_\rho^2}\ R^{(\rho)}(s)
&\approx& { 12 \pi^2 m_\rho^2 \over e\ g_\rho^2 }\ .
\label{eq:SRrho}
\end{eqnarray}
This then leads to a remarkable prediction for the $\rho$ coupling
constant in terms of the fundamental constants $e$ and $\pi$
\cite{SVZ_APP},
\begin{eqnarray}
{ g_\rho^2 \over 4 \pi }
&\approx& { 2 \pi \over e }\
 \approx\ 2.3\ ,
\end{eqnarray}
which is very close to the phenomenological value $2.36 \pm 0.18$.
Therefore asymptotic freedom severely constrains the properties
of a single resonance!
Contrast this with the extraction of the nucleon elastic form factors
from DIS structure functions at $x \sim 1$ using local duality, in
Sec.~\ref{sssec:elastic}.

\begin{figure}[ht]
\begin{center}
\hspace*{0.5cm}
\epsfig{file=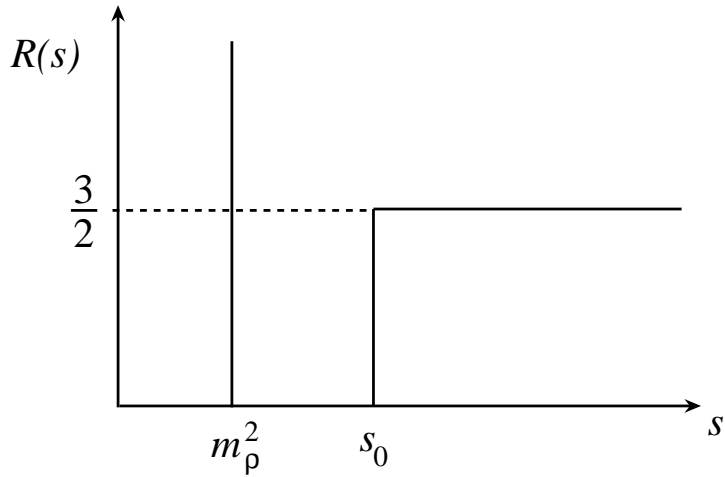,height=9cm}
\vspace*{-1cm}
\caption{\label{fig:specdens}
	Simple model of the spectral density for the ratio $R(s)$ of
	$e^+ e^-$ annihilation cross sections in the $I=1$ channel
	for hadrons to muons.  The pole contribution is at
	$s = m_\rho^2$ and the continuum begins at $s=s_0$.
	The free quark result, $R(s)=3/2$, at $s \to \infty$ is
	indicated by the dashed extension.}
\end{center}
\end{figure}

An improvement on the simple model with the single resonance can be made
by adopting the ``pole + continuum'' ansatz for the spectral density,
illustrated in Fig.~\ref{fig:specdens}, in which the hadronic ratio
above the continuum threshold $s > s_0$ is assumed to be reliably
evaluated in terms of the free quark ratio,
\begin{eqnarray}
R(s) &=& R^{(\rho)}(s)\
      +\ R_0(s)\ \theta(s-s_0)\ .
\end{eqnarray}
This in fact amounts to a statement of duality,
\begin{eqnarray}
\int_0^{s_0} ds\ R(s) &=& \int_0^{s_0} ds\ R_0(s)\ ,
\label{eq:SRglobal}
\end{eqnarray}
of which Eq.~(\ref{eq:SRrho}) is a particular example.
Adopting a similar strategy as for the harmonic oscillator study in
Sec.~\ref{sssec:SRqm}, the coupling constant, $\rho$ mass, and
threshold $s_0$ can be extracted from the sum rule by identifying the
region where the results are most stable with respect to variation
of $s_0$ and $M^2$.
In this manner one obtains $m_\rho^2 \approx 0.6$~GeV$^2$,
$g_\rho^2/4\pi \approx 2.4$, with $s_0 \approx 1.5$~GeV$^2$,
which is within the anticipated $\sim 10\%$ accuracy of the
sum rule method \cite{SVZ_APP,RADYUSH_REV}.

Despite the simplicity of the model for the spectral density adopted,
this example illustrates the power of QCD sum rules and the
effectiveness of the quark-hadron duality assumption underpinning
this method.
In the next section we discuss more practical applications of duality
in $e^+ e^-$ annihilation by considering more realistic models.

% -----------------------------------------------------------------------
\subsection{Electron-Positron Annihilation}
\label{ssec:ee}

\begin{figure}[ht]
\begin{center}
\epsfig{file=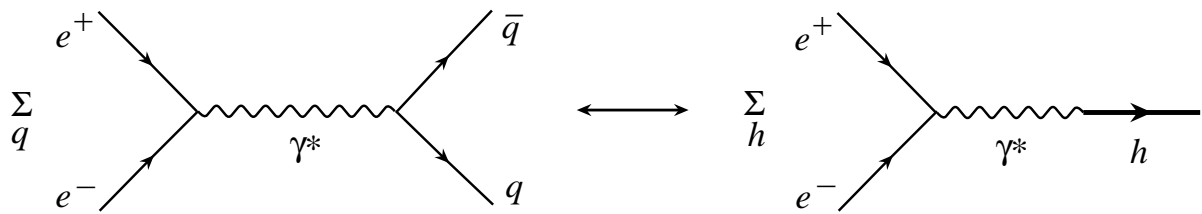,height=8cm}
\vspace*{-4cm}
\caption{\label{fig:ee}
	Quark-hadron duality in $e^+ e^-$ annihilation into hadrons:
	the sum over free $q\bar q$ pairs (left) is dual to the
	average of hadrons $h$ (right).}
\end{center}
\end{figure}

One of the classic manifestations of quark-hadron duality is in
inclusive $e^+ e^-$ annihilation into hadrons.
The annihilation reaction $e^+ e^- \to X$ proceeds through a
virtual photon coupling to a $q\bar q$ pair, which
subsequently hadronizes into physical hadrons $X$,
$e^+ e^- \to q\bar q \to X$.
At low energies the $q\bar q$ pair forms a series of bound states;
at higher energies the $q\bar q$ states appear as broad resonances
which merge into a smooth continuum.
The continuum cross section is well described by the production of a
``free'' $q\bar q$ pair followed by fragmentation into the observed
hadrons.
Duality in $e^+ e^-$ annihilation relates appropriate averages of the
highly structured physical hadronic cross section, $\sigma_h$, to the
smooth cross section for quark pair production, $\sigma_{q\bar q}$,
which can be calculated perturbatively,
\begin{eqnarray}
\langle \sigma_{h} \rangle
&\approx& \langle \sigma_{q\bar q} \rangle\ ,
%  +\ {\rm corrections}\ .
\label{eq:eedual}
\end{eqnarray}
where the brackets $\langle \cdots \rangle$ denote averaging.
The duality relation (\ref{eq:eedual}), illustrated schematically
in Fig.~\ref{fig:ee}, has been used extensively in many applications,
such as the extraction of quark masses from data, prediction of
leptonic widths, and fundamental tests of QCD.

In Sec.~\ref{sssec:eevmd} we consider an illustration of duality in
$e^+ e^-$ annihilation in hadronic, pre-QCD language, in terms of the
vector meson dominance model.
More formal discussions of duality, in terms of quark degrees of
freedom, are presented in Sec.~\ref{sssec:eepot} for potential models,
both nonrelativistically and relativistically, and examples of duality
violating corrections are given.
The appearance of duality in the exactly soluble case of QCD in 1+1
dimensions with a large number of colors is described in
Sec.~\ref{sssec:eeqcd2}.
Before proceeding to the theoretical descriptions of $e^+ e^-$ duality,
however, we firstly consider the important issue of smearing, and how
to relate the quark level calculation with that at the hadronic level.
This will illuminate the resonance averaging which was inherent in the
observation of duality in inclusive electron--nucleon scattering.

% .......................................................................
\subsubsection{Smearing Methods}
\label{sssec:eesmear}

While the production of $q\bar q$ pairs in $e^+ e^-$ annihilation can
be calculated in QCD using perturbation theory, a direct comparison
with the measured hadronic cross sections is of course more
problematic.
Such a comparison can be made at large $s$, where the conversion of
$q\bar q$ pairs into hadrons produces a smooth dependence on $s$.
However, just as in inclusive electron scattering, at low $s$ the
cross section is dominated by resonances and multihadron thresholds,
giving a rich structure whose description is far beyond the scope
of perturbative QCD.

As was found for Bloom-Gilman duality in inclusive DIS, one can
nevertheless try to relate the calculated $q\bar q$ cross section to
the observed hadronic cross section at low $s$ by suitably averaging
or smearing the hadronic cross section.
Some examples of smearing techniques were previously encountered in
Sec.~\ref{sec:thy}, where averages of resonances were found to closely
resemble scaling structure functions.
Here we consider several specific methods of smearing which, although
applied to $e^+ e^-$ annihilation, can be generalized to other
processes, including deep inelastic structure functions.

One method of smoothing the $e^+ e^-$ ratio $R(s)$ considered by
Adler \cite{ADLER} and De~R\'ujula \& Georgi \cite{DG} involved
extrapolating the experimental data from the timelike to the
spacelike regions via dispersion relations.
Comparisons with perturbative QCD predictions could then be made
for the extrapolated quantity
\begin{eqnarray}
D(Q^2)
&\equiv& Q^2 \int_{4 m_\pi^2}^\infty ds\
	 { R(s) \over (s + Q^2)^2 }\
 =\ {3\pi \over \alpha}\ Q^2
    \left.{d\Pi(s) \over ds}\right|_{s=-Q^2}\ ,\ \ \ \ \ \ Q^2 < 0\ ,
\end{eqnarray}
where $\Pi(s)$ is the vacuum polarization amplitude.
Clearly the integration over $s$ has the effect of smearing any
structures in $R$, which results in a smoothed quantity $D(Q^2)$.
The disadvantage of this technique is that one must make assumptions
about the behavior of $R$ at high energies, outside the measured
region, in order to make use of the dispersion relations.

Another technique for smearing the ratio $R$ directly in the timelike
region was proposed by Poggio, Quinn \& Weinberg \cite{PQW}, borrowing
ideas from the smoothing of neutron cross sections in nuclear reactions
\cite{NUCLEARSMEAR}.
Defining the smeared ratio
\begin{eqnarray}
\bar R(s,\Delta)
&=& { \Delta \over \pi } \int_{4 m_\pi^2}^\infty ds'
    { R(s') \over (s'-s)^2 + \Delta^2 }\ ,
\end{eqnarray}
where $\Delta$ is a phenomenological parameter, the integral averages
out both the quark-gluon thresholds in the theoretical cross sections
and the hadronic thresholds and resonances in perturbation theory.
Poggio {\em et al.} \cite{PQW} argue that as long as $\Delta$ is
sufficiently large, $\bar R(s,\Delta)$ can be calculated with some
number $N_\Delta$ of terms in perturbation theory.
On the other hand, keeping $\Delta$ as small as possible ensures
that maximal information can be extracted from the data.
Since $N_\Delta$ decreases with decreasing $\Delta$, making the
averaging too fine grained may lead to $N_\Delta$ being as small as 1.
To order $\alpha_s$, and including only contributions from quarks,
$R(s)$ is given in perturbative QCD by
\begin{eqnarray}
R_0(s)
&=& {3 \over 2} \sum_q e_q^2\ v_q\ (3-v_q^2)\
    \left( 1 + {4 \over 3}\ \alpha_s\ f(v_q) \right)\ ,
% +\ {1 \over 2} \sum_l v_l\ (3-_l^2)\ ,
\label{eq:Rth}
\end{eqnarray}
where $v_q = \sqrt{1 - 4 m_q^2/s}$ and
\begin{eqnarray}
f(v_q)
&=& {\pi \over 2 v_q}\
 -\ {3 + v_q \over 4} \left( {\pi \over 2} - {3 \over 4\pi} \right)\ .
\end{eqnarray}
In the limit $s \gg m_q^2$ one has $v_q \to 1$, and the ratio
\begin{eqnarray}
R_0(s)
&\to& 3 \sum_q\ e_q^2\ \left( 1 + {\alpha_s(s) \over \pi} \right)\ .
\label{eq:R0ee}
\end{eqnarray}
In this limit the right hand side of Eq.~(\ref{eq:R0ee}) reduces to
the expression in Eq.~(\ref{eq:R0}) for the vector-isovector channel.

\begin{figure}[ht]
\begin{center}
\epsfig{file=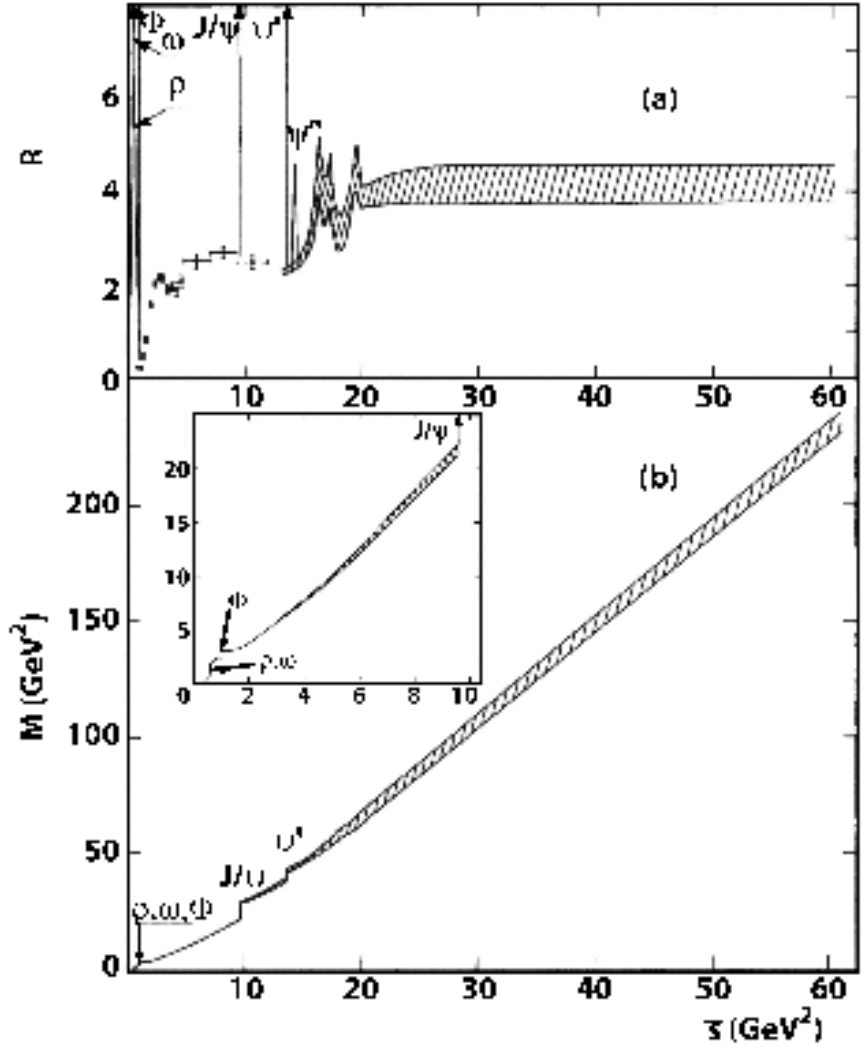,height=14cm}
\vspace*{0.5cm}
\caption{\label{fig:greco}
	(a) Ratio $R$ versus $s$ (in GeV$^2$).
	The shaded bands represent experimental uncertainties at
	large $s$.
	(b) Zeroth moment $M(\bar s)$ as a function of $\bar s$.
	The inset highlights the small-$\bar s$ ($< 10$~GeV$^2$)
	region.  (Adapted from Ref.~\protect\cite{GRECO_SMEAR}.)}
\end{center}
\end{figure}

A different smearing method, using the lowest moment of $R$,
was proposed by Shankar \cite{SHANKAR} and Greco {\em et al.}
\cite{GRECO_SMEAR}, in which the experimental ratio
was smeared by integrating over $s$ up to some maximum value
$\bar s$ \cite{GRECO_SMEAR},
\begin{eqnarray}
M(\bar s) &=& \int_{4 m_\pi^2}^{\bar s} ds\ R(s)\ .
\end{eqnarray}
While the ratio $R$ itself displays prominent resonance
structures at low $s$, as illustrated in Fig.~\ref{fig:greco}~(a),
these structures have almost disappeared in the moment $M(\bar s)$
shown in Fig.~\ref{fig:greco}~(b).
Apart from shoulders in $M(\bar s)$ corresponding to thresholds of
the prominent resonances, the $\bar s$ dependence is very smooth.

The inset in Fig.~\ref{fig:greco}~(b) shows in more detail $M(\bar s)$
in the low-$\bar s$ region, where the bumps at $\bar s \sim 1$~GeV$^2$
correspond to the $\rho$, $\omega$ and $\phi$ meson thresholds.
Extrapolating the smooth curve at large $\bar s$ down to zero,
one sees that the extrapolated curve would roughly bisect the
structures associated with the low-mass resonances.
The implication of this is that the value of $R$ extrapolated
from large $s$ approximately coincides with the value averaged
over the $\rho$, $\omega$ and $\phi$ resonances.
This scenario exactly parallels the low-$s$ (or low-$W$) structures
in inclusive electron--nucleon structure functions, which are
averaged by the high-$s$ scaling function extrapolated into the
resonance region --- see Sec.~\ref{sec:bgstatus}.
The approximate equality of the structure functions integrated over
resonances with those integrated over the smooth deep inelastic
continuum is also reminiscent of the leading-twist dominance of
moments of structure functions, even when these are dominated by
resonance contributions at low $Q^2$.

\newpage

% ........................................................................
\subsubsection{Vector Meson Dominance}
\label{sssec:eevmd}

Duality in $e^+ e^-$ annihilation can be studied more quantitatively
within dynamical models.
Early, pre-QCD attempts to link the behavior of the $e^+ e^- \to X$
cross section at low energies with that at high energies were made
using the phenomenological vector meson dominance model and its
generalizations.
Here at low $s$ the photon interacts through the standard ($\rho$,
$\omega$, $\phi$) vector mesons, while at high energies it couples
to a continuum of hadronic vector states with a linear mass spectrum.
The $1/s$ behavior of the total hadronic cross section at large $s$
arises from an infinite series of vector meson peaks, which add
together to build up a smooth scaling continuum in much the same way
as the scaling curve in DIS is obtained from a sum over an infinite
series of $s$-channel resonances (see also Sec.~\ref{sssec:res}).

Quantitatively, the total cross section $\sigma_h$ corresponding to
the coupling of the photon to vector mesons can be written as
\cite{Q2DUALITY}
\begin{eqnarray}
\sigma_h(s)
&=& \sigma_{\mu^+ \mu^-}(s)\
\sum_n { 12 \pi \over f_n^2 }
       { m_n^3\ \Gamma_n \over (s-m_n^2)^2 + m_n^2 \Gamma_n^2 }\ ,
\label{eq:eevmd}
\end{eqnarray}
where $f_n$ gives the strength of the coupling to a vector meson of
mass $m_n$ and width $\Gamma_n$, and $\sigma_{\mu^+ \mu^-}$ is the
$\mu^+ \mu^-$ production cross section, Eq.~(\ref{eq:muons}).
The sum in Eq.~(\ref{eq:eevmd}) must include an infinite number
of vector meson states if $\sigma_h(s) \sim 1/s$ on average.
Furthermore, the $1/s$ behavior imposes a constraint between the
density of meson states $\rho_n$ per unit mass squared interval
and the coupling $f_n$,
\begin{eqnarray}
\rho_n &\propto& { f_n^2 \over m_n^2 }\ ,
\end{eqnarray}
for all states $n$.
For a linear mass spectrum,
\begin{eqnarray}
m_n^2 &=& m_\rho^2\ (1 + a\ n)\ ,
\end{eqnarray}
where $a=2$ corresponds to a pure Veneziano-like mass spectrum, one
finds asymptotically the remarkable relation \cite{Q2DUALITY,GRECO}
\begin{eqnarray}
R(s) &=& \lim_{s\to\infty}
	{ \sigma_{h}(s) \over
	  \sigma_{\mu^+ \mu^-}(s) }\
   =\ { 8 \pi^2 \over f_\rho^2 }\ .
\end{eqnarray}
Namely, at asymptotically large $s$ the ratio is determined entirely
by the coupling of a photon to the ground state $\rho$.
For $f_\rho \sim 6$ \cite{Q2DUALITY}, numerically the ratio is $\sim 2$
({\em cf.} the asymptotic QCD sum rule result in Eq.~(\ref{eq:R0})).

Sakurai \cite{Q2DUALITY} further developed the relation between the
hadronic cross section and its asymptotic limit $\sigma_0(s)$ by
formulating a ``finite energy sum rule'' version of the duality
relation (also known as ``$Q^2$ duality''),
\begin{eqnarray}
\int_{4 m_\pi^2}^{\bar s} ds\ s\ \sigma_h(s)
&\approx& \int_{s_0}^{\bar s} ds\ s\ \sigma_0(s)\ ,
\label{eq:eefesr}
\end{eqnarray}
where $\sigma_0$ is given by the imaginary part of the vacuum
polarization amplitude $\Pi_0(s)$,
\begin{eqnarray}
\sigma_0(s)
&=& { 4 \pi \alpha \over s }\ \Im m\ \Pi_0(s)\ ,
\label{eq:vacpol}
\end{eqnarray}
and $s_0$ denotes the onset of the absorptive cut of $\Pi_0(s)$ in
the $s$-plane.
This relation provides a vivid analogy with dual models of the strong
interactions: the vector meson contributions on the left hand side of
Eq.~(\ref{eq:eefesr}) at low energies average to the asymptotic
cross section extrapolated down to low $s$.
One can further speculate that this duality holds locally and that
the finite energy sum rule is satisfied even if the maximum value
of $s$, $\bar s$, is chosen not far above the $\phi$ meson mass.

While phenomenologically the ``$Q^2$ duality'' appears to be
reasonably well satisfied, the formulation of the model entirely
in terms of hadronic degrees of freedom leaves the way open for a
deeper understanding of the duality phenomenon at the quark level.
We turn to this in the following.

% ........................................................................
\subsubsection{Potential Models}
\label{sssec:eepot}

The duality between the $e^+ e^-$ hadronic cross section at low energy
and its high-energy behavior can also be described microscopically in
terms of the underlying quark degrees of freedom.
Here we consider several models in which this duality can be made
explicit, firstly by considering the nonrelativistic limit, which
should be valid for heavy quarks, and then generalizing to the
relativistic case.

For nonrelativistic, free quarks the $e^+ e^- \to q\bar q$
cross section (for one quark flavor) is given by \cite{DURAND_NR}
\begin{eqnarray}
\sigma^{\rm nr}_{q\bar q}
&=& { 6 \pi \alpha^2 \over s }\
    e_q^2\ v^{\rm nr}\ |\psi^{\rm nr}_E(0)|^2\ ,
\end{eqnarray}
where $\psi^{\rm nr}_E(0)$ is the free $q\bar q$ wave function for
energy $E = \sqrt{s} - 2 m_q$, and $v^{\rm nr} = \sqrt{E/m_q}$ is the
nonrelativistic velocity.
For non-interacting particles, with conventional plane-wave
normalization, one has $|\psi^{\rm nr}_E(0)|^2 = 1$.

The nonrelativistic cross section for producing $q\bar q$ bound states
(in some confining potential), on the other hand, is given by
\cite{DURAND_NR,BHADURI}
\begin{eqnarray}
\sigma^{\rm nr}_{\rm bound}
&=& { 24 \pi^2 \alpha^2 \over m_q^2\ s }\ e_q^2\
    \sum_n |\psi^{\rm nr}_n(0)|^2\ \delta(E - E_n)\ ,
\label{eq:bound_nr}
\end{eqnarray}
where $n$ is the radial quantum number of the bound state with
excitation energy $E_n$, and $\psi^{\rm nr}_n(0)$ is the bound state
wave function at the origin.
For non-singular potentials the wave function $\psi^{\rm nr}_n(0)$
is related to the density of states, $\rho_n^{\rm nr} \equiv dn/dE_n$,
according to \cite{KRAMMER,QUIGROS,BELL}
\begin{eqnarray}
|\psi^{\rm nr}_n(0)|^2
&=& { m_q^2 \over 4 \pi^2 }\
    { v_n^{\rm nr} \over \rho_n^{\rm nr} }\ ,
\label{eq:psi_nr}
\end{eqnarray}
where $v_n^{\rm nr} = \sqrt{ (E_n-V(0))/m_q }$ is the velocity of
a free quark with energy $(E_n-V(0))/2$.
For duality to exist the averaged free quark cross section should be
equal to the bound state cross section smoothed over an appropriate
energy interval ({\em cf.} Eq.~(\ref{eq:eedual})),
\begin{eqnarray}
\langle \sigma^{\rm nr}_{\rm bound} \rangle
&\approx& \langle \sigma^{\rm nr}_{q\bar q} \rangle\ .
\label{eq:eenr}
\end{eqnarray}
The energy averaging over the $\delta$-function in
Eq.~(\ref{eq:bound_nr}) can be implemented for instance by
replacing $\delta(E-E_n)$ by smooth functions ({\em e.g.}, Gaussians)
with a finite width $\Delta$ \cite{JENNINGS}.

Early work by Krammer \& Leal-Ferreira \cite{KRAMMER} and Quigg \&
Rosner \cite{QUIGROS} showed that the duality relation (\ref{eq:eenr})
indeed emerges from a nonrelativistic ($v^{\rm nr} \ll 1$) potential
model in the Wentzel-Kramers-Brillouin (WKB) approximation.
Bell \& Pasupathy \cite{BELL} generalized the results to higher
partial waves, using the Thomas-Fermi approximation for the density of
a Fermi gas with one particle per level in a non-singular potential.
Later Durand \& Durand \cite{DURAND_NR} showed that the
energy-averaged cross sections in Eq.~(\ref{eq:eenr}) can be related
by a Fourier transform to the short-time behavior of the quark
propagator.
For a given confining potential, the short-time propagator is then
related to the free propagator, with calculable corrections.

Bhaduri \& Pasupathy \cite{BHADURI} used a different smoothing
procedure, in which the energy-averaged bound state cross section
is related to the {\em unsmeared} free quark cross section.
For a harmonic oscillator potential, $V(r) = m \omega^2 r^2/2$, where
$m = m_q/2$ is the reduced mass of the $q\bar q$ system, the averaged
bound state cross section is expanded in powers of $\hbar$ to yield
\cite{BHADURI}
\begin{eqnarray}
\langle \sigma^{\rm nr}_{\rm bound} \rangle
&=& \sigma^{\rm nr}_{q\bar q}
    \left\{ 1 + {1 \over 16} \left({\hbar \omega \over E}\right)^2
	      + {\cal O}(\hbar^4)
    \right\}\ .
\label{eq:bound_nr_ho}
\end{eqnarray}
To lowest order the averaged bound state cross section is manifestly
equal to the free quark cross section $\sigma^{\rm nr}_{q\bar q}$.
More generally, for any non-singular potential $V(r)$
% such as a linear potential $V(r) \sim r$, 
one can write \cite{BHADURI}
\begin{eqnarray}
\langle \sigma^{\rm nr}_{\rm bound} \rangle
&=& \sigma^{\rm nr}_{q\bar q}
    \left\{ 1 + {\hbar^2 \over 8 m_q} {V^{''}(0) \over E^2}
	      + {5 \hbar^2 \over 32 m_q} {|V'(0)|^2 \over E^3}
	      + {\cal O}(\hbar^4)
    \right\}\ ,
\end{eqnarray}
where the correction terms are given by derivatives of $V(r)$ at
the origin.
One can verify that for the harmonic oscillator potential this
expression reduces to Eq.~(\ref{eq:bound_nr_ho}).

While the nonrelativistic duality may be relevant phenomenologically
for heavy quarks, for light quarks one needs to demonstrate that
duality is also valid relativistically.
A proof of relativistic duality in $e^+ e^-$ annihilation was given
by Durand \& Durand \cite{DURAND_REL} in the framework of the
Bethe-Salpeter equation.
The relativistic free quark cross section is given by \cite{BELL}
\begin{eqnarray}
\sigma_{q\bar q}
% &=& \sigma_{\mu^+ \mu^-}\
%     e_q^2\ {3 \over 2 s} v_q (3-v_q^2)\ \theta(s-4 m_q^2)\  \nonumber\\
&=& { 2 \pi \alpha^2 \over s }\
    e_q^2\ v_q (3-v_q^2)\ \theta(s-4 m_q^2)\ ,
\end{eqnarray}
where $v_q$ is the velocity of the quark in the center of mass system
({\em cf.} Eq.~(\ref{eq:Rth})).
The corresponding relativistic bound state cross section can be
written
\begin{eqnarray}
\sigma_{\rm bound}
&=& 6\pi^2 \sum_n { \Gamma_n(e^+ e^-) \over s }\
    \delta(\sqrt{s}-M_n)\ ,
\label{eq:sigbound_rel}
\end{eqnarray}
where the width $\Gamma_n(e^+ e^-)$ is given by \cite{DURAND_REL}
\begin{eqnarray}
\Gamma_n(e^+ e^-)
&=& { 16 \pi \alpha^2 \over M_n^2 }\ e_q^2\
    |\psi_n(0)|^2\ (1-\Delta_n)\ ,
\label{eq:gam_rel}
\end{eqnarray}
with $\psi_n(0)$ the ``large'' component of the $S$-state
Bethe-Salpeter wave function at zero space-time quark separation.
The term $\Delta_n$ includes $D$-state effects and terms arising
from the ``small'' components of the wave function.
For relativistic systems the wave function is related to the
relativistic density of states $\rho_n \equiv dn/dM_n$
({\em cf.} Eq.~(\ref{eq:psi_nr})) via \cite{DURAND_REL}
\begin{eqnarray}
|\psi_n(0)|^2
&\approx& { M_n^{\prime 2} \over 16 \pi^2 }
          { v'_n \over \rho_n }\ (1-\Delta'_n)\ ,
\label{eq:psi_rel}
\end{eqnarray}
where $v'_n$ is the relativistic velocity of a free quark with energy
$M'_n/2 = (M_n-V(0))/2$, and $\Delta'_n$ is a correction for
retardation and radiative gluonic effects.
Once again the duality relation (\ref{eq:eedual}) is obtained by
substituting Eqs.~(\ref{eq:gam_rel}) and (\ref{eq:psi_rel}) into
the bound state cross section (\ref{eq:sigbound_rel}), averaging
the result over an appropriate energy range, and replacing the sum
over $n$ by an integral.
Relativistic duality can thus be used, for example, to estimate the
radiative corrections to the leptonic widths $\Gamma_n(e^+ e^-)$
for bound $q\bar q$ systems by using the known results for free
$q\bar q$ systems.

% .......................................................................
\subsubsection{$e^+ e^-$ Annihilation in the 't Hooft Model}
\label{sssec:eeqcd2}

The duality relations in the potential models discussed above raise
the question of whether and how duality in $e^+ e^-$ annihilation
can be shown to arise in QCD.
A step towards answering this question was made by Einhorn
\cite{EINHORN_EE} and Bradley {\em et al.} \cite{SHAW} who considered
$e^+ e^-$ annihilation in QCD in 1+1 dimensions in the
$N_c \to \infty$ limit (the 't~Hooft model).

As discussed in Sec.~\ref{sssec:conf}, the 't Hooft model
\cite{THOOFT} is a fully soluble theory, which has the features of
confinement and asymptotic freedom.
The mass spectrum consists of an infinite sequence of narrow bound
states which become equally spaced in $m_n^2$ ($n=0,1,2,\ldots$) at
large $n$, reminiscent of linear Regge trajectories.
In terms of the vacuum polarization amplitude $\Pi(s)$, which is
related to the cross section as in Eq.~(\ref{eq:vacpol}), the bound
state contribution to the imaginary part of $\Pi(s)$ is given by
\cite{SHAW}
\begin{eqnarray}
\Im m\ \Pi(s)
&=& \sum_{n=0}^{\infty} g_n^2\ \delta(s - m_n^2)\ ,
\end{eqnarray}
where the couplings $g_n$ are zero for $n$ odd (due to parity).
Because of the completeness relation for the $q\bar q$ wave functions,
the couplings satisfy $\sum_n g_n^2 = 1$.
(Values for the couplings and masses can be obtained \cite{SHAW}
by solving the 't Hooft equation \cite{THOOFT} numerically.)
In the limit $s \to \infty$ the vacuum polarization amplitude becomes
\begin{eqnarray}
\Pi(s)
&\to& -{1 \over \pi s}
      \int_{s_0}^\infty ds\ \Im m\ \Pi(s)\
   =\ -{1 \over \pi s}\ .
\end{eqnarray}
It is instructive to also consider the case of large but finite $N_c$.
Here one may approximate the sum of $\delta$-functions by a sum of
Breit-Wigner resonances \cite{EINHORN_EE},
\begin{eqnarray}
\Pi(s)
&\sim& \sum_n g_n^2
       { m_n\ \Gamma_n \over (s-m_n^2)^2 + m_n^2 \Gamma_n^2 }\ ,
\end{eqnarray}
where $\Gamma_n$ is the width of meson $n$.
As $N_c \to \infty$, $\Gamma_n \to {\cal O}(1/N_c)$, so that on the
resonance peak the cross section is ${\cal O}(N_c^2)$, whereas between
the resonances ($s \not= m_n^2$) it is ${\cal O}(1)$.

Asymptotically, the absorptive part of the vacuum polarization
amplitude calculated in terms of free quarks is \cite{SHAW}
\begin{eqnarray}
\Im m\ \Pi_0(s)
&=& 2 m_q^2\ (s - 4 m_q^2)^{1/2} s^{-3/2}\ \theta(s-4 m_q^2)\ ,
\end{eqnarray}
which leads to
\begin{eqnarray}
\Pi_0(s)
&\to& -{1 \over \pi s}
      \int_{4 m_q^2}^\infty ds\ \Im m\ \Pi_0(s)\
   =\ -{1 \over \pi s}\ .
\end{eqnarray}
Thus the global duality relation,
\begin{eqnarray}
\int_{s_0}^\infty ds\ \Im m\ \Pi(s)
&=& \int_{4 m_q^2}^\infty ds\ \Im m\ \Pi_0(s)\ ,
\end{eqnarray}
between the absorptive part corresponding to free $q\bar q$ pairs,
and that associated with narrow resonance poles and confined quarks
is explicitly verified.
Bradley {\em et al.} \cite{SHAW} further consider the question of whether
this equality also holds for more local averages.
Comparing the couplings $g_n$ extracted from duality with those
calculated explicitly by solving the 't Hooft equation numerically,
one indeed finds good agreement, which becomes exact in the limit
$n\to\infty$.

Once again the close analogy with the appearance of duality in
structure functions in the large-$N_c$ limit, as discussed in
Sec.~\ref{sssec:conf}, suggests that the phenomenon of duality
between bound state resonances and the free quark continuum is
a general feature of strongly interacting systems.
An even more vivid realization of this is seen in the case of weak
decays of heavy mesons, which we turn to in the next section.

% -----------------------------------------------------------------------
\subsection{Heavy Meson Decays}
\label{ssec:hq}

Weak decays of heavy mesons have provided a fertile testing ground for
studying the origin of quark-hadron duality in strong interactions.
Here systematic expansions, such as those based on heavy quark
effective theory (HQET) \cite{IW,EICHTEN,CGG,VSOPE}, can be used to
expand decay rates or widths in inverse powers of the heavy quark mass,
$1/m_Q$.
In the heavy quark limit, $m_Q \to \infty$, duality has indeed been
shown to be exact, even down to zero recoil energy.
Even though the quarks are heavy and move with small momentum,
there is typically a large energy release and a correspondingly
large number of final states that contribute to the total width.
As we have seen in the case of Bloom-Gilman duality in
Section~\ref{sec:thy}, this is one of the necessary conditions
needed to ensure the emergence of duality.

One of the vital practical applications of duality in heavy quark
decays is the determination of the Cabibbo-Kobayashi-Maskawa (CKM)
matrix elements $V_{cb}$ and $V_{ub}$.
A major source of uncertainty in their extraction is the deviations
from duality expected at higher orders in the $1/m_Q$ expansion.
A better understanding of duality would also have clear implications
for the identification of physics beyond the Standard Model.

By examining the mechanisms behind duality in heavy meson decays, we
shall try to gain insight into the mechanisms underlying Bloom-Gilman
duality.
In this section we consider both semileptonic and nonleptonic decays
of heavy hadrons.
More extensive accounts of heavy meson decays can be found in the
recent reviews in Refs.~\cite{SHIFMAN_REV,SHIFMAN03,BIGI_REV,LIGETI}.
Before discussing the phenomenology of duality and its violations
in semileptonic and nonleptonic heavy meson decays, it will be
useful to consider a simple pedagogical example which illustrates
the essential physics of duality in heavy quark systems.
This will allow us to compare and contrast this with Bloom-Gilman
duality.

% ....................................................................... 
\subsubsection{Duality in Heavy Quark Systems: a Pedagogical Example}
\label{sssec:hqped}

Our discussion follows closely that of Isgur {\em et al.} \cite{IJMV},
who considered a simple toy model in which a heavy quark $Q$ bound
to a light antiquark $\bar q$ decays to a heavy quark $Q'$ after
emitting a scalar particle $\phi$,
$(Q \bar q)_0 \rightarrow (Q' \bar q)_n + \phi$.
The subscript $n$ denotes the possible excitations of the final state
heavy meson that are allowed kinematically.
At the free quark level the decay $Q \to Q' + \phi$ produces the
heavy $Q'$ quark with recoil velocity $\vec v$, with the $\phi$
emitted with kinetic energy $T_{\rm free}$.
At the hadronic level, in the narrow resonance approximation, the
$\phi$ will emerge with any of the sharp kinetic energies allowed
by the strong interaction spectra of these two mesonic systems.

If $M_{(Q \bar q)_n}$ and $M_{(Q' \bar q)_n}$ are the masses of the
heavy-light bound states of $(Q \bar q)$ and $(Q' \bar q)$,
respectively, then in the heavy quark limit
$M_{(Q \bar q)_n} - M_{(Q' \bar q)_n} \simeq m_Q - m_{Q'}$,
and the mass difference between the heavy meson and heavy quark
can be neglected,
$M_{(Q \bar q)_n} \simeq m_Q$, and
$M_{(Q' \bar q)_n} \simeq m_{Q'}$.
In this limit the hadronic spectral lines cluster around $T_{\rm free}$,
and as $m_Q \to \infty$ they indeed coincide with $T_{\rm free}$
exactly.
Since for $m_Q, m_Q' \to \infty$ the decay proceeds as though the
quarks $Q$ and $Q'$ were free, the sum of the strengths of the
spectral lines clustering around $T_{\rm free}$ is equal to the free
quark strength, ensuring exact duality in this limit \cite{IJMV}.

One can now proceed to unravel this duality to understand how the
required ``conspiracy'' of spectral line strengths arises physically.
Because the recoiling heavy quark $Q'$ carries off a negligible
kinetic energy, but a large momentum, its recoil velocity $\vec v$
is only slightly changed by the strong interaction.
In the rest frame of the recoiling meson, this configuration requires
that the two constituents have a relative momentum $\vec q$ which
increases with $\vec v$.
For $\vec v \to 0$ only the ground state process
$(Q \bar q)_0 \to (Q' \bar q)_0 + \phi$ is allowed.
Since the masses and matrix elements for the transitions
$(Q \bar q)_0 \to (Q' \bar q)_0 + \phi$ and $Q \to Q' + \phi$ are
identical, the hadronic and quark spectral lines and strengths are
also identical and duality is valid at $\vec q \, ^2 = 0$.

For nonzero $\vec v$ (and therefore $\vec q$), the elastic form factor
decreases from unity, so its spectral line carries less strength.
However, since $\vec q$ is nonzero, excited states $(Q' \bar q)_n$ can
now be created, with a strength which exactly compensates for the loss
of elastic rate.
The excited state spectral lines also coincide with $T_{\rm free}$
and duality is once again exact.
Regardless of how large $\vec q \, ^2$ becomes, all of the excited
states produce spectral lines at $T_{\rm free}$ with strengths that
sum to that of the free quark spectral line
\cite{IJMV,BJORKEN_HQ,IWBJ}.

For finite quark masses duality violation occurs, although this is
formally suppressed by two powers of $1/m_Q$
\cite{CGG,SHIFMAN_REV,BIGI_REV} (see Sec.~\ref{sssec:hqsl} below).
In this case the spectral lines are still clustered around
$T_{\rm free}$, but no longer coincide exactly with it.
Although the spectral line strengths differ from those of the heavy
quark limit, they do so in a way which compensates for the duality
violating phase space effects from the spread of spectral lines
around $T_{\rm free}$.
In addition, because $m_Q - m_{Q'}$ is now finite, some of the high
mass resonances required for exact duality are kinematically
forbidden, which also leads to duality violation \cite{IJMV}.

From this discussion it is clear that the strong interaction dynamics
of heavy meson decays has a number of similarities to that of electron
scattering.
Essentially the same model was used in Sec.\ref{ssec:models} to
describe scattering from a heavy-light $(Q\bar q)$ bound state,
and the emergence of a scaling function from the
$(Q \bar q)_0 \to (Q' \bar q)_n$ transitions.
The crucial point is that the system must in each case respond
to a relative momentum kick $\vec q$.
An important difference, however, is that in a decay to a fixed mass
$\phi$ only a single magnitude $\vec q \, ^2$ is produced at the
quark level, while in electron scattering a large range of
$\vec q \, ^2$ and $\nu$ is accessed.
This pedagogical example should serve to remind us that even though
the physical origin of duality may be similar, exactly how it
manifests itself must of course depend on the specific process at hand.

% .......................................................................
\subsubsection{Semileptonic Weak Decays}
\label{sssec:hqsl}

Having illustrated the essential workings of duality in heavy meson
decays in a simple toy model, we now examine the more realistic case
of duality in semileptonic weak decays of heavy mesons.
Historically, the exact duality between semileptonic decay rates of
heavy mesons calculated at the quark and hadronic levels was first
pointed out by Voloshin \& Shifman \cite{VS}.
In the extreme nonrelativistic limit, one assumes that the initial
$(Q)$ and final $(Q')$ quarks are heavy, and satisfy the relation
\begin{eqnarray}
m_Q + m_{Q'} &\gg& m_Q - m_{Q'}\ \gg\ \Lambda_{\rm QCD}\ ,
\label{eq:SVlimit}
\end{eqnarray}
where $m_Q$ and $m_{Q'}$ are the respective heavy quark masses.
This is usually referred to as the ``small velocity'' or
Shifman-Voloshin (SV) limit.
In the rest frame of the heavy quark $Q$, the kinetic energy of
the quark $Q'$ produced in the reaction
\begin{eqnarray}
Q &\to& Q'\ l\ \bar\nu_l
\label{eq:Q1dk}
\end{eqnarray}
will be small compared with its mass, but large compared with
$\Lambda_{\rm QCD}$.
The semileptonic decay rate for the process (\ref{eq:Q1dk})
calculated at the quark level is given by \cite{VS}
\begin{eqnarray}
\Gamma^q
&=& { G_F^2\ \delta m^5 \over 15 \pi^2 }\ |V_{QQ'}|^2\ ,
\label{eq:hqlimit}
\end{eqnarray}
where $G_F$ is the Fermi decay constant,
$V_{QQ'}$ is the $Q \to Q'$ CKM matrix element,
and $\delta m = m_Q - m_{Q'}$.
The mass difference between the heavy quark and the corresponding
heavy-light meson $M_{(Q \bar q)}$, where $\bar q$ is a light
antiquark, is negligible in the SV limit, so that
$\delta m = M_{(Q\bar q)} - M_{(Q'\bar q)}$.

\begin{figure}[ht]
\begin{center}
\hspace*{-3cm}
\epsfig{file=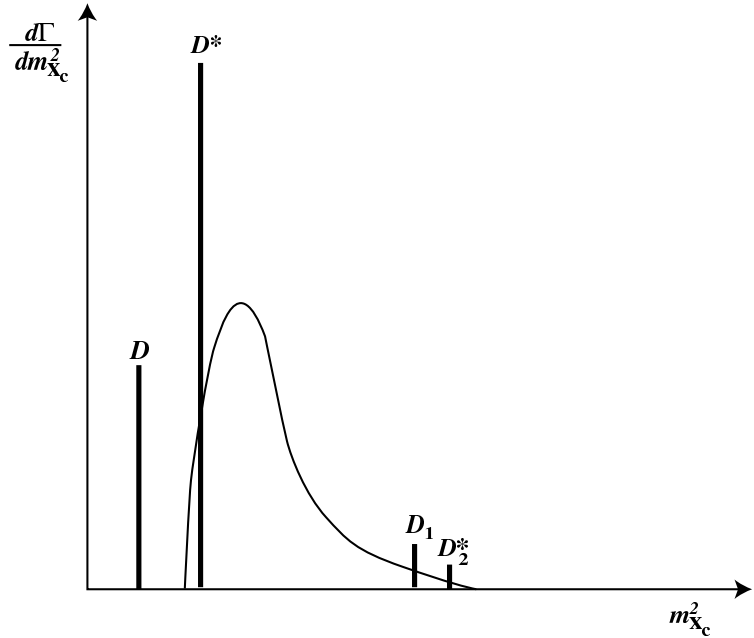,height=20cm}
\vspace*{-11cm}
\caption{\label{fig:HQisgurSL}
	A sketch of the $b \to c$ inclusive semileptonic decay
	spectrum calculated at the quark level (continuous curve)
	compared to the known $B \to D^{(*)}$ spectrum (resonance
	lines).  (From Ref.~\protect\cite{ISGUR_VIOL}.)}
\end{center}
\end{figure}

For sufficiently large quark masses a duality arises between the
partonic rate $\Gamma^q$ and the rate $\Gamma^h$ calculated at the
hadronic level involving a sum over a set of hadronic final states
containing $Q'$.
The remarkable feature of the SV limit is that the rate $\Gamma^h$
for the hadronic decay
\begin{eqnarray}
(Q \bar q) &\to& X_{Q'}\ l\ \bar\nu_l\ 
\label{eq:M1dk}
\end{eqnarray}
is saturated by just {\em two} exclusive channels, $X_{Q'} = P$
and $V$, corresponding to pseudoscalar and vector states,
respectively.
In particular, the rates for the individual $P$ and $V$ channels
in the SV limit are \cite{VS}
\begin{eqnarray}
\Gamma^P
&\to& { G_F^2\ \delta m^5 \over 60 \pi^2 }\ |V_{QQ'}|^2\ ,	\\
\Gamma^V
&\to& { G_F^2\ \delta m^5 \over 20 \pi^2 }\ |V_{QQ'}|^2\ ,
\end{eqnarray}
so that the total hadronic rate is exactly dual to the free quark rate,
\begin{eqnarray}
\Gamma^h &=& \Gamma^P + \Gamma^V\ \leftrightarrow\ \Gamma^q\ .
\label{eq:HQdual}
\end{eqnarray}
The physical situation where this duality is realized most precisely
is in the semileptonic decay of $B$ to $D$ and $D^*$ mesons, where the
measured hadronic rates are used to extract the $V_{cb}$ matrix element.
In Fig.~\ref{fig:HQisgurSL} we show a sketch of the
$B \to X_c\ l\ \bar\nu_l$ decay width as a function of the squared mass
$M_{X_c}^2$ of the final state charmed meson $X_c$ \cite{ISGUR_VIOL}.
The known $B \to D^{(*)}$ spectrum is illustrated by the narrow
resonance lines, while the inclusive quark rate $b \to c\ l\ \bar\nu_l$
is shown by the continuous curve.
Duality is realized by integrating over the mass spectrum $M_{X_c}$.

% Smearing...

Of course, in general the hadronic level and quark level rates cannot
be identical even at very high energies due to the structure of the
singularities in the multiparticle thresholds and quark-gluon
production thresholds.
However, in computing the semileptonic widths one integrates over the
leptonic variables, which amounts to a smearing of the quark level
width.
In analogy with electron scattering, the equality between the smeared
quark and hadronic widths is referred to as global duality, whereas
local duality refers to equality between the unsmeared widths.
For the example of saturation of the $B \to X_c\ l\ \bar\nu_l$ rate
by two hadronic final states, $D$ and $D^*$, a local duality clearly
cannot be defined in terms of the mass of the final state $M_{X_c}$:
duality in this case sets in at threshold since even as $\delta m$
approaches zero as $m_b \to \infty$, the heavy recoiling $c$ quark
has an energy much greater than $\Lambda_{\rm QCD}$.
In the SV limit it must therefore hadronize with unity probability
as $D$ and $D^*$ \cite{ISGUR_DUAL}.
As in electron scattering, the issue is not whether duality holds in
semileptonic heavy meson decay, but rather how accurately it holds.

% 1/m_Q corrections...

While the relationship between the quark and hadronic results in the
heavy quark limit is clear, a debate has existed over the leading
corrections in the $1/m_Q$ expansion, and the degree to which duality
is violated at finite $m_Q$.
In particular, there has been some controversy about whether the
leading corrections to the heavy quark limit enter at ${\cal O}(1/m_Q)$
or ${\cal O}(1/m_Q^2)$ --- see {\em e.g.},
Refs.~\cite{BIGI_REV,ISGUR_VIOL,ISGUR_DUAL,BUV,BOYD}.

Using the operator product expansion method developed by Voloshin
and Shifman \cite{VSOPE}, Bigi {\em et al.} \cite{BUV} expanded the
inclusive decay width for a heavy meson in terms of local operators
containing $Q\bar Q$ and gluon fields.
The imaginary part of the forward $Q \to Q$ amplitude was expressed
as a series of local operators of increasing dimension, with
coefficients proportional to powers of $1/m_Q$.
Bigi {\em et al.} \cite{BUV} find that there are no terms in this
expansion which are linear in $m_Q$: the leading nonperturbative
corrections to the widths arise only at order $1/m_Q^2$.
The reason for this is that the only operators containing $Q\bar Q$ that
can induce $1/m_Q$ terms are of dimension four, and these operators
either vanish (in the case of total derivatives) or can be reduced via
the equations of motion to the original quasi-free operator $Q\bar Q$.

On the other hand, it was suggested \cite{ISGUR_VIOL} that hadronic
thresholds can lead to violations of duality at ${\cal O}(1/m_Q)$
which do not appear explicitly in the OPE, and which could
significantly compromise the reliability with which $V_{cb}$ can
be extracted.
In the infinite mass limit, duality in this picture arises through a
cancellation between the fall-off of the ground state contribution
and the corresponding rise of the excitations.
At finite mass, however, there is some mismatch near zero recoil,
which could be of order $1/m_Q$.
Isgur \cite{ISGUR_VIOL} and Le~Yaouanc {\em et al.} \cite{LEYAOUANC} have
studied these possible violation of duality using nonrelativistic
quark models.

These results have been challenged, however, by Bigi \& Uraltsev
\cite{BIGI_REV}, who argue that the analyses in
Refs.~\cite{ISGUR_VIOL,LEYAOUANC} are based on a different OPE
scheme to that used in Refs.~\cite{BUV,BSUV}, and that these
differences lead precisely to the terms which are interpreted in
\cite{ISGUR_VIOL,LEYAOUANC} as violating duality at order $1/m_Q$.
Given that the extraction of fundamental Standard Model parameters
--- the CKM matrix elements $V_{cb}$ and $V_{ub}$ --- relies critically
on duality, it is clearly vital to understand the degree to which
duality holds for $B$ decays, and the size of the duality violations.
We shall see in the next section that a similar understanding of
duality and its violations is important in nonleptonic decays of heavy
mesons.

% ........................................................................ 
\subsubsection{Nonleptonic (Hadronic) Weak Decays}
\label{sssec:hqnl}

The discussion of duality in nonleptonic ({\em i.e.}, hadronic) weak
decays of heavy mesons follows closely that of semileptonic decays,
albeit with some important differences.
Whereas for semileptonic decays the heavy meson decays into one heavy
meson in the final state, the presence of two hadronic currents in
nonleptonic decays requires a factorization of the final state
hadrons.
In general such factorization has been demonstrated only in special
cases, such as for QCD with $N_c \to \infty$.
Moreover, since for nonleptonic decays there are no lepton momenta
to be integrated over, a more local version of duality needs to exist
in order to relate hadronic observables with those calculated from
the OPE \cite{CDN}.
Consequently duality violations in nonleptonic decays have also
been the subject of controversy
\cite{BUV,CDN,AMPR,GRIN_LEBED,GRIN01,GRIN02}.

Working in a special limit which combines the heavy quark and
large-$N_c$ limits, Shifman \cite{SHIFMAN_NL} considered to what
extent factorization of weak matrix elements can hold in the
presence of finite $1/N_c$ corrections.
The process considered involves a heavy quark $Q$ decaying into
two heavy quarks $Q'$ and $Q''$ and a light antiquark $\bar q'$,
\begin{eqnarray}
Q &\to& Q' + Q'' + \bar q'\ ,
\end{eqnarray}
with the kinematics defined such that
\begin{eqnarray}
m_{Q'}
&=& m_{Q''} \equiv M\ ,\ \ \ \ \ m_Q\ =\ 2 M + \Delta\ ,\ \ \ \ \
M\ \gg\ \Delta\ \gg\ \Lambda_{\rm QCD}\ .
\label{eq:SVgen}
\end{eqnarray}
This generalizes the SV limit in that the heavy quarks produced are
extremely slow and at the same time very energetic.
In this generalized limit Shifman \cite{SHIFMAN_NL} showed that a quark
level description of the process $Q \to Q' + Q'' + \bar q'$ duals the
hadron level description of the transition of the initial meson
$(Q\bar q)$ into two heavy final state mesons,
\begin{eqnarray}
(Q\bar q) &\to& (Q'\bar q) + (Q''\bar q')\ ,
\label{eq:NL1}
\end{eqnarray}
or
\begin{eqnarray}
(Q\bar q) &\to& (Q'\bar q') + (Q''\bar q)\ .
\label{eq:NL2}
\end{eqnarray}
In this case the two-meson final state saturates the partonic decay,
with each additional final state meson suppressed by a power of $1/N_c$.
To leading order in the $1/N_c$ expansion, one finds that the total
weak decay amplitude factorizes for the process in Eq.~(\ref{eq:NL1})
as \cite{SHIFMAN_NL}
\begin{eqnarray}
& & \left\langle\ (Q' \bar q)\ (Q'' \bar q')\
  \left|\ \bar Q'\ \Gamma_\mu\ Q\
	  \bar Q''\ \Gamma^\mu\ q'\
  \right|
(Q \bar q) \right\rangle			\nonumber\\
&=&
\left\langle (Q' \bar q)
  \left|\ \bar Q'\ \Gamma_\mu\ Q\
  \right|
(Q \bar q) \right\rangle\ 
\left\langle (Q'' \bar q')
  \left| \bar Q''\ \Gamma^\mu\ q'\
  \right| 0 \right\rangle\ ,
\end{eqnarray}
where $\Gamma_\mu = \gamma_\mu (1 - \gamma_5)$,
and similarly for the process in Eq.~(\ref{eq:NL2}).
The total quark decay width at leading order in $1/M$ then becomes
\cite{SHIFMAN_NL}
\begin{eqnarray}
{ d\Gamma^q \over dE_{q'} }
&=& 2 N_c
    \left( \kappa_1^2 + \kappa_2^2
	 + {2 \kappa_1^2 \kappa_2^2 \over N_c }
    \right) 
    { M^2\ E_{q'}^2 \over \pi^3 }
    \sqrt{ {\Delta - E_{q'} \over M} }\ ,
\label{eq:nlq}
\end{eqnarray}
where $\kappa_1$ and $\kappa_2$ give the respective strengths of the
transitions (\ref{eq:NL1}) and (\ref{eq:NL2}), respectively.

Assuming that the ground state of the initial meson $(Q \bar q)$ is
a pseudoscalar (as for $B$ or $D$ mesons), the hadronic width is
calculated by summing over exclusive transitions from the ground
state to pseudoscalar and vector $(Q' \bar q)$ or $(Q' \bar q')$
states, and over transitions from the vacuum to scalar and vector
$(Q'' \bar q')$ or $(Q'' \bar q)$ states.
If the mass of the excited state is $M + \delta$, with
$\delta \gg \Lambda_{\rm QCD}$, the total (integrated) hadronic
rate is given in the $N_c \to \infty$ limit by \cite{SHIFMAN_NL}
\begin{eqnarray}
\Gamma^h
&=& 2 N_c \left( \kappa_1^2 + \kappa_2^2 \right)
    { M^2 \over \pi^3 }
    \int_0^\Delta d\delta\
	\delta^2\ \sqrt{ {\Delta - \delta \over M} }\ ,
\end{eqnarray}
which coincides exactly with the quark rate (\ref{eq:nlq}) if one
identifies $\delta$ with $E_{q'}$.
In this limit one therefore not only observes duality, but the
duality is also local on the Dalitz plot.

Of course the limit (\ref{eq:SVgen}) is somewhat far from reality ---
for the decay $B \to D \pi$ it would correspond to $m_c = m_d = M$,
while $m_u=0$!
However, it does provide a useful illustration of the workings of
duality in hadronic decays of heavy mesons.

% -----------------------------------------------------------------------
\subsection{Proton-Antiproton Annihilation}
\label{ssec:ppbar}

To complete our survey of quark-hadron duality in reactions other
than electron scattering, we consider a novel application of
duality discussed recently in the context of proton--antiproton
annihilation into photons, $p\bar p \to \gamma\gamma$ \cite{PP_CZ}.
The similarity of this process with Compton scattering, viewed in
the crossed channel, suggests that under certain kinematic conditions
it may be described in terms of generalized parton distributions
(see Sec.~\ref{sssec:dvcs}) through the dominance of the handbag
diagram \cite{PPBAR,PP_WEISS,PP_DFJK,PIRE}.
The appearance of duality in this reaction may therefore have
elements in common with Bloom-Gilman duality in DIS.

In contrast to forward Compton scattering, where the ``diquark"
system is a spectator, in the $p\bar p \to \gamma \gamma$ process
the diquarks effectively annihilate into the vacuum, without emitting
additional particles.
In the limit $s \to \infty$ this constrains the diquarks to have zero
momentum, and the annihilating $q\bar{q}$ pair must therefore carry
all of the momentum of the hadrons, $x \to 1$
\cite{PPBAR,PP_WEISS,PP_DFJK}.
However, as Close \& Zhao \cite{PP_CZ} point out, for $s \to \infty$
there are coherent, higher-twist contributions associated with the
``cat's ears'' topologies (see Fig.~\ref{fig:diag}(b)) which are of
the same order of magnitude as the handbag diagram.

Phenomenologically the descriptions of Compton scattering and other
processes in terms of leading-twist contributions have been relatively
successful, even at intermediate values of $s$.
Close \& Zhao suggest \cite{PP_CZ} how the handbag dominance of
$p\bar p$ annihilation could arise from quark-hadron duality when
a suitable average over coherent contributions is made.

As discussed in Sec.~\ref{sssec:res} for inclusive electron scattering,
the excitation of positive and negative parity intermediate state
resonances gives rise to constructive interference for the incoherent
contributions (proportional to $e_q^2$), but destructive for the
coherent ($e_q e_{q'}$) terms.
Similarly in the crossed channel, $p \bar p \to \gamma \gamma$
(or equivalently $\gamma \gamma \to p\bar p$), one finds that terms
proportional to $\sum_{q \ne q'} e_q e_{q'}$ are suppressed because of
destructive interference between even and odd parity excitations in
the intermediate state.

Using a generalized form for the structure function from the model
study of generalized parton distributions described in
Sec.~\ref{sssec:dvcs}, Close \& Zhao \cite{PP_CZ} suggest a factorized
ansatz for the crossed channel structure function in the limit where
the momentum fractions of the annihilating partons in the $p$ and
$\bar p$ are same, $\xi = x_p - x_{\bar p} \to 0$,
\begin{eqnarray}
F_2(x,\xi \to 0,s)
&=& \sum_q e_q^2\ x q(x)\ F_{\rm el}(s)\ ,
\end{eqnarray}
where $F_{\rm el}(s)$ is the elastic form factor, and
\begin{eqnarray}
{1 \over 2} \int_{\xi=-1+|x|}^{1-|x|} {d\xi \over x} F_2(x,\xi,s=0)
&=& \sum_i e_i^2 [\theta(x) q(x) -\theta(-x) \overline{q}(-x)]\ .
\end{eqnarray}
A further consequence of this simple model is that the annihilation
cross sections for $p\bar{p}$ and $n\bar{n}$ are determined by the
constituent quark charges,
$\sigma(p\bar{p})/\sigma(n\bar{n})
= (2e_u^2 + e_d^2)/(2e_d^2 + e_u^2)
= 3/2$.

While more elaborate models would be needed for quantitative
comparisons with data, the scheme described here gives a plausible
mechanism to support the dominance of the leading-twist process
in a region where its justification is otherwise questionable.
Furthermore, it underscores the versatility of quark-hadron duality
in finding application in a wide range of phenomena.

% -----------------------------------------------------------------------
\subsection{Reprise}
\label{ssec:reprise}

The examples highlighted in this section indeed give strong support
to the thesis that quark-hadron duality is not an isolated phenomenon
but is a general feature of the strongly interacting landscape,
of which Bloom-Gilman duality is one, albeit particularly striking,
manifestation.
The common features of hadronic sums leading to observables
characterized by independence of scale can be seen in many different
physical phenomena, such as the seminal $e^+ e^-$ annihilation into
hadrons, as well as in numerous theoretical applications, most
overtly in QCD sum rules.
The scale independence is most conveniently accounted for by the
presence of free, point-like constituents of hadrons, which is
naturally accommodated through the existence of asymptotic freedom
in QCD.

The crucial element in this ``global'' duality is the availability
of a complete set of hadronic states, which is realized more
effectively with increasing energy.
The existence of duality in QCD is thus an inevitable consequence
of confinement, which guarantees the orthogonality of quark-gluon
and hadronic states and ensures no double counting, and asymptotic
freedom, which allows perturbative descriptions at the quark level.
Of course the details of the physical realization of duality must
depend on specific applications, so that the energy at which duality
can be said to hold at a given level will in general not be universal.

More intriguing perhaps, and more challenging from a theoretical
perspective, is the appearance of ``local'' duality, in which a
quark-hadron correspondence exists even when a small subset of
hadronic states is summed over.
As we have seen in the example of semi-leptonic weak decays of heavy
mesons, as few as two final states can be sufficient to saturate the
quark level result.
In some cases local duality relations can be derived between a single
hadronic state and the high-energy continuum, as illustrated in the
vector meson dominance picture of $e^+ e^-$ annihilation.
With clear parallels to the threshold relations between elastic
form factors and leading-twist structure functions at $x \sim 1$,
occurrences of local duality in other applications are suggestive,
although understanding the dynamics responsible for local duality
in QCD remains an important future pursuit.

Clearly, the concept of duality is an indispensable one in many areas
of hadronic physics, and new applications continue to be found
\cite{PPBAR,HYBRID,BIGI_OSC,EIDEMULLER,LIUTI03,KALLONIATIS,HOFMANN}.
While the study of the origin of duality provides us with important
clues to the inner workings of QCD, it is equally vital to understand
violations of duality.
As illustrated by the example of heavy meson decays, knowledge of the
magnitude of higher-twist ($1/m_Q^2$) corrections within the OPE is
essential for the extraction of CKM matrix elements.
Similarly, control over higher twists has enormous practical benefits
in electron scattering, such as in understanding the limits of
applicability of leading-twist parton distribution function analyses,
as well as in unraveling the long-range quark-gluon correlations in
hadrons.
In the next section we delve further into the practical relevance
of duality, and outline plans for its future experimental study.

\clearpage

%%%%%%%%%%%%%%%%%%%%%%%%%%%%%%%%%%%%%%%%%%%%%%%%%%%%%%%%%%%%%%%%%%%%%%%%%
\section{Outlook}
\label{sec:outlook}

In this section we take a somewhat longer-term perspective,
and examine the prospects for experimental duality studies over
the next decade.
We discuss improvements which are expected in measurements of
inclusive structure functions, both in electron and neutrino
scattering, as well as in the relatively new realm of meson
electroproduction.
We begin, however, with a short discussion of the practical
relevance of improving our understanding of the duality phenomenon.

% -----------------------------------------------------------------------
\subsection{Why is Duality Relevant?}
\label{ssec:relevance}

Most theoretical studies of duality to date have concentrated on
establishing how coherent resonance transitions excited in
electron--nucleon scattering conspire to obtain the scaling behavior
as expected from the underlying electron--quark scattering mechanism.
Describing this transition in terms of narrow resonances built up
from valence quarks can be motivated partly by the large-$N_c$
limit of QCD, in which only resonances ``survive'' \cite{THOOFT},
as well as the experimental indication that duality seems to prevail
in the small-$Q^2$ region dominated by valence quarks \cite{F2JL2}.
A number of recent model studies have also established conditions
(such as cancellations between states of different parity) under
which a summation over nucleon resonances can lead to the results
expected from the parton model \cite{CI,NOSU6,CZ}.

It is clear, nonetheless, that to understand the duality phenomenon
in detail, one also needs to consider the role of the nonresonant
background.
Indeed, the ``two-component'' duality picture, postulated long ago
in the context of hadron--hadron scattering, invokes duality between
resonances and nondiffractive (valence) contributions on the one hand,
and between the nonresonant background and diffractive (sea) effects
on the other.
In the electromagnetic case, the requirement of this interplay is
most strikingly illustrated by the establishment of a precise
($< 10\%$ accuracy) local duality in the $F_2$ structure function
{\em solely} in the $\Delta$ resonance region.
Here local duality still appears at a $Q^2$ scale of 1~GeV$^2$,
even though there is only a single resonance contribution, with only
very small contributions from tails of higher-mass resonances.

An obvious phenomenological path to investigate the interplay of
resonant and nonresonant contributions in establishing quark-hadron
duality would be within empirical models such as the Unitary Isobar
MAID model \cite{MAID}, which provide phenomenological descriptions
of inclusive and exclusive electron scattering reactions in the
nucleon resonance region.
Beyond a verification of duality within phenomenological models,
{\em ab initio} calculations including both resonant and nonresonant
contributions, especially in the the $\Delta$ region, are needed to
understand duality at a more fundamental level.

Since duality appears to be a general phenomenon in QCD, with examples
ranging from $e^+e^-$ annihilation and deep inelastic scattering, to
inclusive decays of heavy quarks, dilepton production in heavy ion
reactions \cite{Ra02}, and hadronic $\tau$ decays \cite{SHIFMAN_REV},
one has to wonder whether duality may be a general property of
quantum field theories with inherent weak and strong coupling limits.
This is especially relevant in view of the recent revival of interest
in the relation between QCD and string theory
\cite{ST01,PO02a,PO02b,FR03}.
String theory had its beginnings in a theory of strong interactions,
but went out of fashion following the birth of QCD.
Nevertheless, the notion that these theories may be dual descriptions
has persisted, and the interplay between gauge field theories and
string theory is widely recognized.
Indeed, it has been rigorously shown that strings describe some
large-$N_c$ limits of QCD \cite{THOOFT}.
With the realization by Maldacena of a duality between descriptions
of higher-dimensional superstring theory and supersymmetric SU(N)
gauge field theories in four space-time dimensions \cite{Ma98},
exploration of the possible connections between QCD and string
theory \cite{PO02a,PO02b} will be of great interest.

If we take quark-hadron duality to be a general property of QCD,
it is still intriguing why Nature has redistributed its global
strength in specific local regions, and what the ultimate origin
of the duality violations is.
This is nowhere better illustrated than in the transition from
large $Q^2$ to $Q^2 = 0$.
Obviously, strong local duality violations from a simple parton
picture are found in the coherent, elastic channel in spin-averaged
electron scattering, or the elastic and $N-\Delta$ channels in
spin-dependent scattering.
On the other hand, the higher excitation regions have already far
more available channels contributing, and consequently mimic local
duality more closely.
Even more dramatically, for inclusive electron scattering the region
$W^2 > 4$~GeV$^2$ already has a sufficient number of electroproduction
channels that experimentally one cannot distinguish between this
region and the asymptotic high-energy limit of electron--free quark
scattering.

If duality is understood quantitatively, or if regions where duality
holds to good precision are well established, either experimentally
or theoretically, then one can imagine widespread practical
applications of duality.
The region of very high $x$, for instance, which has not been
explored at all experimentally due to the requirement of high-energy
beams with sufficiently high luminosity, will become accessible.
The $x \to 1$ region is an important testing ground for
nonperturbative and perturbative mechanisms underpinning valence
quark dynamics, and is vital to map out if we hope to achieve a
complete description of nucleon structure.
A good understanding of the large-$x$ region will also have important
consequences for future high-energy searches for new physics at the
Tevatron, Large Hadron Collider, and Next Linear Collider \cite{St03}.
Data from the nucleon resonance region, where quark-hadron duality
has been established, could be used to better constrain QCD
parameterizations of parton distribution functions, from which
the hadronic backgrounds in high-energy collisions are computed.

The large-$x$ region also constitutes an appreciable amount of the
moments of polarized and unpolarized structure functions, especially
for the higher moments.
It is precisely these moments that can be calculated from first
principles in QCD on the lattice \cite{LATTICE_MOM}, in terms of
matrix elements of local operators.
Presently, due to technical limitations, the lower moments are
typically calculated at scales $Q^2 \sim 4$~GeV$^2$.
A comparison of the moments of leading-twist parton distributions
with the measured moments at a given $Q^2$ can in principle tell us
about the size of higher-twist effects at that $Q^2$.
On the other hand, since the $x$ dependence of structure functions
cannot be calculated on the lattice directly, one cannot easily use
the lattice to learn about the degree to which duality holds locally.
Indeed, the ability to calculate a leading-twist moment on the
lattice implicitly uses quark-hadron duality to average the
resonance contributions to a smooth, scaling function.

A quantitative understanding of quark-hadron duality, or more
explicitly duality violations, may have other direct applications.
The most clear-cut example is the problem of the $B$ meson
semileptonic branching ratio \cite{SHIFMAN_REV}.
Theoretical calculations obtained in a quark-gluon framework exceed
the measured value by 10--20\%.
However, the possible local duality violations in this ratio are at
present not clear.
If the duality violations could be ruled out, such observations
could lead to more precise tests of QCD, or to new physics at
higher-energy scales.
As Shifman point out, ``short of a full solution of QCD, understanding
and controlling the accuracy of quark-hadron duality is one of the most
important and challenging problems for QCD practitioners today''
\cite{SHIFMAN_REV}.

% -----------------------------------------------------------------------
\subsection{Duality in Inclusive Electron Scattering}
\label{ssec:incl_outlook}

There are several avenues for pursuing experimental duality studies
in inclusive electron scattering which will become accessible in the
next few years.
We focus on two of them here: structure functions at low $Q^2$,
and structure functions at large Bjorken-$x$.

% ....................................................................... 
\subsubsection{Low $Q^2$ Structure Functions}
\label{sssec:lowq2_outlook}

As discussed in Section~\ref{sssec:real}, experimentally duality
is seen to hold even in the low-$Q^2$ regions where perturbative
expansions would be expected to become unreliable, as one transcends
the region where perturbative high-energy techniques are applicable,
to the strongly-coupled, nonperturbative regime at $Q^2 = 0$.
We saw, for example, that the $Q^2$ dependence of the $F_2$ structure
function at intermediate $x$ and the ratio $R$ at $x = 0.1$,
both at $Q^2 \sim 0.2$~GeV$^2$, do not follow expectations from
electromagnetic current conservation.
% or generalized vector meson dominance models \cite{BKF2}.
On the other hand, there are indications of duality between
resonance and continuum cross sections even at the real photon
point (see Fig.~\ref{fig:photon}), as well as in the resemblance
of low-$Q^2$ $F_2$ structure function data and $x F_3$ data from
neutrino scattering (Fig.~\ref{fig:xf3}).
The latter result is particularly striking: it suggests that at low
$Q^2$ the $F_2$ structure function is dominated by valence quarks,
with the sea contributions playing only a minor role.

Recent HERA experiments have shown that $F_2$ at very low $x$
($x \approx 10^{-6}$), and correspondingly very large $W^2$,
can be described by perturbative evolution down to
$Q^2 \approx 1$~GeV$^2$, provided one adopts a gluon distribution
which vanishes at low $x$ (referred to as a ``valence-like'' gluon
\cite{zeus00}) and a non-vanishing but small sea distribution.
At even lower values of $Q^2$, in the same very low-$x$ region,
the dramatic collapse of the proton structure function
(Fig.~\ref{fig:F2val}) could be viewed as evidence for a smooth
transition from pQCD to the real photon point at $Q^2 = 0$
\cite{DLF2,zeus00}.
Gauge invariance requires that the $F_2$ structure function for
inelastic channels must vanish linearly with $Q^2$ as $Q^2 \to 0$
\cite{DLF2}.

Experimentally, however, at the values of $x$ where the nucleon
resonances are visible, and for similar low $Q^2$ ($\sim$ 0.2~GeV$^2$),
the $F_2$ structure function does {\em not} vanish linearly with
$Q^2$.
On the other hand, the nucleon resonances {\em do} seem to obey
some sort of duality, so the picture is currently somewhat murky.
A possible resolution may involve a separate $Q^2$ dependence
for the vanishing of the large-$x$ strength at small $Q^2$
(governed by the nucleon resonances), and for the growth of
the small-$x$ sea \cite{MOMENTS}.

An experiment to investigate the detailed behavior of the nucleon
structure functions at low $Q^2$, through the nucleon resonance
region, has recently been carried out in Hall C at Jefferson Lab
\cite{E00-002}, and is currently being analyzed.
Data from this experiment will fill the critical gap between
$Q^2 \sim 1$~GeV$^2$ and the photoproduction limit, $Q^2 = 0$.
This should enable one to determine whether the $Q^2$ dependence
observed in the nucleon resonance region is due to the suppression
of the large nucleon sea, or a reflection of the vanishing of
valence quark distributions at low $Q^2$.
Either way, it will provide valuable information on the region
of $Q^2$ where the linear $Q^2$ behavior of the $F_2$ structure
function sets in, and the extent to which duality may be relevant
in the very low-$Q^2$ regime.

% ....................................................................... 
\subsubsection{Structure Functions at Large $x$}
\label{sssec:largex_outlook}

The standard method to determine parton distribution functions
is through global fits \cite{MRST,CTEQ,GRV98} of data on structure
functions measured in deep inelastic scattering and other hard
processes.
It has been standard practice in these analyses to omit from the data
base the entire resonance region, $W^2 < W_{\rm res}^2 = 4$~GeV$^2$
(or in some cases even $< 10$~GeV$^2$).
If one could utilize this vast quantity of resonance data, one could
not only significantly improve the statistics, but also decrease
uncertainties which arise from extrapolations of parton distributions
into unmeasured regions of $x$.
An important consequence of duality is that the strict distinction
between the resonance and deep inelastic regions becomes artificial
--- both regions are intimately related, and properly averaged
resonance data can help us understand the deep inelastic region.

Recall that for any finite $Q^2$, one is always limited by
kinematics to $x < x_{\rm res} = Q^2/(W^2_{\rm res}-M^2+Q^2)$.
Extending to very large $x$ at a finite $Q^2$, one always encounters
the resonance region, $W < W_{\rm res}$.
As discussed in Sec.~\ref{ssec:local}, there are a number of
reasons why the large-$x$ region is important.
Firstly, given better constraints on the $Q^2$ dependence at
large $x$, one could derive parameterizations for parton
distributions directly from the data without resorting to
theoretical inputs for extrapolations to $x=1$.
Secondly, the region of $x \approx 1$ is an important testing
ground for mechanisms of spin-flavor symmetry breaking in valence
quark distributions of the nucleon \cite{CLOSEBOOK,NP,ISGUR}.
Thirdly, with nuclear targets it would permit a measurement of
the nuclear medium modification of the nucleon structure function
at large $x$ (nuclear EMC effect) \cite{Ge95}, where the deviation
from unity of the ratio of nuclear to nucleon structure functions
is largest, and sensitivity to different nuclear structure models
greatest.
Finally, knowledge of quark distributions at large $x$ is essential
for determining high-energy cross sections at collider energies,
such as in searches for new physics beyond the Standard Model
\cite{St03}, where structure information at large $x$ feeds down
through perturbative $Q^2$ evolution to lower $x$ at higher values
of $Q^2$.

A quantitative description of nucleon structure in terms of parton
distribution functions relies, however, on our ability to unravel
in detail the $Q^2$ dependence of the data.
In particular, it is important to obtain more precise information
on the regions of $x$ and $Q^2$ where perturbative evolution
\cite{EVOLVE} can no longer be considered the main mechanism
responsible for the $Q^2$ dependence of the data.
There are arguments \cite{Br80}, for example, which suggest that
$x W^2$, rather than $Q^2$, is the natural mass scale of the twist
expansion (since at large $x$ the struck quark becomes highly
off-shell, with virtuality $k^2 \sim k_\perp^2/(1-x)$).
Because $x W^2 \sim Q^2 (1-x)$, the difference between evolution
in $Q^2$ and in $Q^2(1-x)$ becomes most important at large $x$.
Experimentally this seems to be confirmed using nucleon resonance
region data in a local duality framework \cite{Li02}, although
a more thorough investigation using additional large-$x$ data
should be pursued.

In perturbative QCD analyses performed so far higher-twist terms
have been extracted from data by applying a cut in the kinematics
at $W^2 > 10$~GeV$^2$ \cite{Vi92,Ya99,Al01}.
In Refs.~\cite{Li02,Ni99} it was shown that only a relatively small
higher-twist contribution, consistent with the one obtained in
Refs.~\cite{Vi92,Ya99,Al01}, is necessary in order to describe the
entire set of $F_2$ structure function data.
The low-$W^2$ region dominated by nucleon resonances was analyzed
recently by Liuti {\em et al.} \cite{Li02} within a fixed-$W^2$
framework ({\em i.e.}, for each resonance region), in the spirit
of the $Q^2 (1-x)$ evolution above.
The higher-twist contributions in the nucleon resonance region were
found to be similar to those from $W^2 > 10$~GeV$^2$, with the
exception of the $\Delta$ region where the effects were larger.
Although it may seem {\it a priori} surprising that higher-twist
effects originally derived from deep inelastic data can also be
extracted exclusively from the resonance region, this follows
automatically from quark-hadron duality.

Currently a concerted experimental effort is underway to measure both
spin-averaged and spin-dependent structure functions at intermediate
$Q^2$ ($\sim 5$~GeV$^2$) and at large $x$.
The strategy of these experiments is to firstly verify duality in
local resonance regions at some scale $Q_0^2$, and then use local
duality to extend parton distributions to larger $x$ for
$Q^2 > Q_0^2$.
At present local duality has only been well quantified for the
$F_2$ structure function at $Q^2 > 1$~GeV$^2$ \cite{F2JL1,Li02},
although as we saw in Sec.~\ref{sec:bgstatus} there are qualitative
indications of duality in other spin-averaged and spin-dependent
structure functions as well.

In the near future a series of experiments in Hall C at Jefferson
Lab will push measurements of the unpolarized structure functions
(for protons to heavy nuclei) up to the largest values of $x$ and
$Q^2$ attainable with a 6~GeV beam energy
\cite{E00116,E02109,E02019,E03103}.
These will extend existing measurements of the $F_2$ structure
function of the proton and deuteron up to $Q^2 \approx 7$~GeV$^2$
\cite{E00116}.
Since duality has been well verified for each of the individual
resonance regions already at $Q^2 \agt 1$~GeV$^2$, the new data
will allow extensions of measured parton distributions with good
precision up to $x \approx 0.9$.
In addition, the issue of $Q^2$ versus $W^2$ evolution can be
revisited with higher precision.
The new measurements will also allow high-precision extractions
of the lower moments of the $F_2$ structure function moments at
these $Q^2$ values.
Obvious extensions of this program to higher $Q^2$ and $x$,
both in the polarized and unpolarized cases, are possible with
the planned 12~GeV beam energy upgrade at Jefferson Lab \cite{pCDR}.

To experimentally verify proper extraction of $F_2$ in the
experiments using deuteron targets, the ratio $R$ of longitudinal
to transverse deuterium cross sections is required.
A dedicated effort is being made using the longitudinal--transverse
(LT) separation technique to determine both the unpolarized
structure functions $F_1^d$ and $F_2^d$ \cite{E02109} over a
similar $Q^2$ range as mapped out for the proton.
This will provide a global survey of LT-separated unpolarized
structure functions on deuterium throughout the resonance region
with an order of magnitude better precision than previously
possible.

Two further experiments \cite{E02019,E03103} will vastly extend
the present $F_2^A$ structure function data using a large range of
nuclear targets, spanning from $^3$He to $^{197}$Au.
The emphasis of these experiments will be twofold: to experimentally
verify the observed scaling behavior of $F_2^A$ data in the region
of $x > 1$ and $Q^2 > 3$~GeV$^2$, and to extend measurements of the
nuclear EMC effect to larger $x$ values \cite{Ar03} and to few-body
nuclei.
Light nuclei are of special interest due to their relatively
large neutron excess, and because theoretical calculations of the
nuclear EMC effect can be based on better determined wave functions,
in contrast to heavier nuclei.
They may help to differentiate, for example, between models of the
nuclear EMC effect based on an $A$ dependence or a density dependence
of the magnitude of the effect at $x \sim 0.6$ \cite{GOMEZ}.

To ``round off'' upcoming efforts to study the duality phenomenon
in unpolarized structure functions, a dedicated experiment has been
designed to extend the vast amount of existing electron--proton
scattering data to the neutron \cite{E03012}.
Compared to the structure of the proton, much less is known about
neutron structure due to the absence of free neutron targets,
and the theoretical uncertainties associated with extracting
information from neutrons bound in nuclei.
This is especially critical at large $x$ and in the resonance region.

\begin{figure}
\begin{minipage}[ht]{4.0in}
\hspace*{0.5cm}
\epsfig{figure=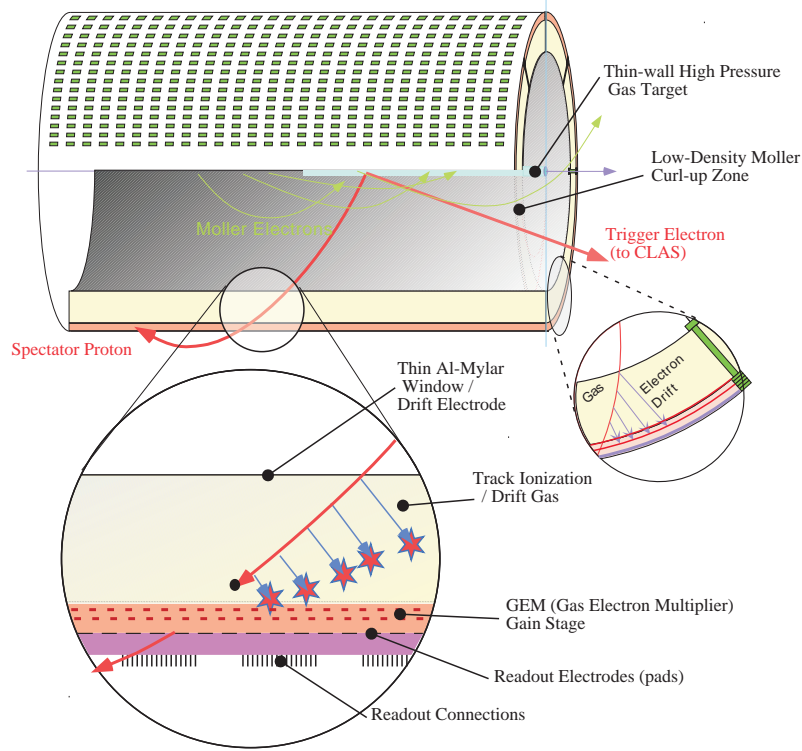, height=10cm}
\end{minipage}
\vspace*{1cm}
\begin{centering}
\caption{\label{fig:bonus}
	Sketch of the radial time projection chamber for the
	electron--(quasi-free) neutron scattering experiment as proposed
	in Ref.~\protect\cite{E03012}.  The upper portion of the figure
	shows background M\o ller electrons and a highly-ionizing
	low-energy spectator proton in a cylindrical detector inside a
	strong solenoidal magnetic field.  The lower portion shows pad
	readout of the highly-ionizing proton track.}
\end{centering}
\end{figure}

To overcome this problem, the new experiment will measure the
inclusive electron scattering cross section on an almost free
neutron using the CEBAF Large Acceptance Spectrometer (CLAS)
and a novel recoil detector with a low momentum threshold for
protons and high rate capability (see Fig.~\ref{fig:bonus}).
This detector will allow tagging of slow backward-moving spectator
protons with momentum as low as 70~MeV in coincidence with the
scattered electron in the reaction $d(e,e^\prime p)X$.
The restriction to low momentum will ensure that the electron
scattering takes place on an almost free neutron, with its initial
four-momentum inferred from the observed spectator proton.
(For an alternative method of determining inclusive neutron
cross sections using a combination of $^3$He and $^3$H targets,
see Ref.~\cite{HE3-H3}.)

The spectator tagging technique will be used to extract the
structure function $F_2^n$ over a large range of $Q^2$
(up to $\sim 5$~GeV$^2$) and $W$ (from the elastic peak to
$W=3$~GeV).
The kinematic coverage, including the elastic and resonance
regions, as well as part of the deep inelastic continuum,
will allow the first quantitative tests of quark-hadron duality
in the neutron.
As discussed in Sec.~\ref{sssec:qm}, Close and Isgur \cite{CI}
argued using a quark model that the neutron structure functions
should exhibit systematic deviations from local duality, and that
duality should occur at higher $W$ for the neutron than the proton.

For spin-dependent scattering, two experiments will access the
$g_1$ structure function in the nucleon resonance region up to
$Q^2 = 5$~GeV$^2$.
The CLAS Collaboration routinely has running periods scattering
polarized electrons from polarized NH$_3$ and ND$_3$ targets
\cite{CLASg1}.
The Hall A Collaboration has also just performed a measurement
scattering polarized electrons from a polarized $^3$He target
\cite{E01012} to test duality in the neutron $g_1^n$ structure
function, and the data are currently being analyzed.

The precision of the $g_1$ structure function data will be further
enhanced by the results of the LT-separated unpolarized structure
function data in a similar $Q^2$ region, and by two Hall~C
experiments at Jefferson Lab to fully disentangle the $g_1$ and
$g_2$ structure functions.
These latter experiments will in particular determine the $g_2$
structure function with high precision at $Q^2 = 1.3$~GeV$^2$
\cite{E01006} and $Q^2 \sim 5$~GeV$^2$ \cite{E03109},
respectively.
Data at $Q^2 = 1.3$~GeV$^2$ are presently under analysis.

The onset of duality in all spin-averaged and spin-dependent
structure functions will soon be well verified up to
$Q^2 = 5$~GeV$^2$.
The combination of LT-separated unpolarized $F_1$ and $F_2$
structure functions, in addition to separated polarized electron
scattering data determining $g_1$ and $g_2$, will allow for
unprecedented precision tests of duality in nucleon structure
functions.
It will in addition reveal the extent to which duality can be
used to access the $x \to 1$ region, and should shed considerable
light on this somewhat obscure but vital corner of phase space.

% -----------------------------------------------------------------------
\subsection{Neutrino Scattering}
\label{ssec:neutrino_outlook}

In discussing Bloom-Gilman duality in this report, we have dealt
almost exclusively with observables measured using electron
scattering.
Weak currents, on the other hand, can provide complementary
information on the quark structure of hadrons, not accessible to
electromagnetic probes.
In particular, neutrino-induced reactions can provide important
consistency checks on the validity of duality.
While deep inelastic neutrino structure functions are determined
by the same set of universal parton distribution functions as in
charged lepton scattering, the structure of resonance transitions
excited by neutrino beams is in some cases strikingly different to
that excited by virtual photons.
Although on general grounds one may expect that a Bloom-Gilman type
duality should also exist for weak structure functions \cite{CM93},
the details of how this manifests itself in neutrino scattering may
be quite different from that observed in electron scattering.

Unfortunately, current neutrino scattering data are sparse in the
resonance region \cite{Di86}, and due to the small weak cross
sections is often only available for heavy nuclei (where large
target volumes are easier to handle and are more affordable than
light nuclei) \cite{Be88}.
It has not been possible therefore to make any concrete statements
to date about the validity of duality in neutrino scattering.

The main difference between electron and neutrino scattering reactions
can be most easily understood considering specific resonance
transitions.
While a neutrino beam can convert a neutron into a proton,
it cannot convert a proton into a neutron, for example
(and {\em vice versa} for an antineutrino beam).
Similarly, there are dramatic differences between inelastic
production rates in the $\Delta$ resonance region \cite{CG,Ab72}
--- because of charge conservation, only transitions to
isospin-$3/2$ states from the proton are allowed.
The prospect of high-intensity neutrino beams at Fermilab,
as well as in Japan and Europe, offers a valuable complement
to the study of duality and resonance transitions.
For example, the recently-approved MINER$\nu$A experiment \cite{Mo02}
at Fermilab will be an exceptional tool for such measurements.
The goal of MINER$\nu$A will be to perform a high-statistics
neutrino-nucleus scattering experiment using a fine-grained detector
specifically designed to measure low-energy neutrino-nucleus
interactions accurately. 
The high-luminosity NuMI beam line at Fermilab will provide energies
spanning the range $\sim 1-15$~GeV, over both the resonance and deep
inelastic regimes, making MINER$\nu$A a potentially very important
facility to study quark-hadron duality in neutrino scattering.

A particularly interesting measurement would be of the ratio of
neutron to proton neutrino structure functions at large $x$.
Here, similar valence quark dynamics as in charged lepton scattering
are probed, but with different sensitivity to quark flavors.
At the hadronic level, quark model studies reveal quite distinct
patterns of resonance transitions to the lowest-lying positive and
negative parity multiplets of SU(6) \cite{CG,NOSU6,CGK,CGPLB}.
The contributions of the $N \to N^*$ transition matrix elements to
the $F_1$ and $g_1$ structure functions of the proton and neutron
in the SU(6) quark model are displayed in Table~\ref{tab:neusu6}.
Summation over the $N \to N^*$ transitions (for the case of equal
symmetric and antisymmetric contributions to the wave function,
$\lambda=\rho$) yields the expected SU(6) quark-parton model
results, providing an explicit confirmation of duality.
On the other hand, some modes of spin-flavor symmetry breaking
($\lambda\not=\rho$) yield neutrino structure function ratios which
at the parton level are in obvious conflict with those obtained
from electroproduction, as shown in Table~\ref{tab:su6} in
Sec.~\ref{sssec:qm}.
Neutrino structure function data can therefore provide valuable
checks on the appearance of duality and its consistency between
electromagnetic and weak probes.

\begin{table}[htb]
\begin{tabular}{||c||c|c|c|c|c||c||}
SU(6) rep.      & $^2${\bf 8}[{\bf 56}$^+$]\ \
                & $^4${\bf 10}[{\bf 56}$^+$]\ \
                & $^2${\bf 8}[{\bf 70}$^-$]\ \
                & $^4${\bf 8}[{\bf 70}$^-$]\ \
                & $^2${\bf 10}[{\bf 70}$^-$]\ \
                & total\ \                              \\ \hline\hline
$F_1^{\nu p}$ & $0$
        & $24 \lambda^2$
        & $0$
        & $0$
        & $3 \lambda^2$
        & $27 \lambda^2$                     \\
$F_1^{\nu n}$ & $(9 \rho + \lambda)^2/4$
        & $8 \lambda^2$
        & $(9 \rho - \lambda)^2/4$
        & $4 \lambda^2$
        & $\lambda^2$
        & $(81 \rho^2 + 27 \lambda^2)/2$                 \\ \hline
$g_1^{\nu p}$ & $0$
        & $-12 \lambda^2$
        & $0$
        & $0$
        & $3 \lambda^2$
        & $- 9 \lambda^2$                     \\
$g_1^{\nu n}$ & $(9 \rho + \lambda)^2/4$
        & $-4 \lambda^2$
        & $(9 \rho - \lambda)^2/4$
        & $-2 \lambda^2$
        & $\lambda^2$
        & $(81 \rho^2 - 9 \lambda^2)/2$                  \\
\end{tabular}
\vspace*{0.5cm}
\caption{Relative strengths of neutrino-induced $N \to N^*$ transitions
        in the SU(6) quark model \protect\cite{NOSU6}.  The coefficients
	$\lambda$ and $\rho$ denote the relative strengths of the
	symmetric and antisymmetric contributions, respectively, of the
	ground state wave function.  The SU(6) limit corresponds to
	$\lambda = \rho$.}
\label{tab:neusu6}
\end{table}

Similarly, it would be of particular interest to verify the onset of
duality in the $x F_3$ structure function in deep inelastic neutrino
scattering.
The $x F_3$ structure function describes the response to the
vector--axial vector interference, and is as such associated with
the parity-violating hadronic current.
Consequently in the quark-parton model the $x F_3$ structure function
measures the difference of quark and antiquark distributions, and is
insensitive to sea quarks.
As described in Sec.~\ref{sec:bgstatus}, the resemblance of the
average $F_2$ electroproduction structure function at low $Q^2$ in
the nucleon resonance region to the measured deep inelastic $x F_3$
structure function \cite{F2JL2} suggests a sensitivity to resonant
contributions only.
If the interplay between resonances and nonresonant backgrounds is an
important contributor to the onset of duality, one could anticipate
this onset to occur at larger $Q^2$ scales in the $x F_3$ structure
function.

Lastly, it is worth mentioning that a large effort to consistently
model electron, muon, and neutrino scattering is currently being
undertaken \cite{Bo02}.
Data from atmospheric neutrino experiments \cite{Fu00} and neutrinos
from the Sun \cite{SNO,KAMLAND} have been interpreted as evidence for
neutrino oscillations.
These neutrino data dictate the need for next generation,
accelerator-based, oscillation experiments using few-GeV neutrino
energies.
Good modeling of neutrino cross sections at low energies is needed
for this upcoming generation of more precise neutrino oscillation
experiments.
This is particularly true for neutrinos in the region around 1~GeV,
where, for example, single-pion production comprises about 30\% of
the total charged-current cross section. A solid understanding of
the transition between the deep inelastic
% (where good neutrino data are available and 
% perturbative-based models describe the data well)
and resonance production regimes will be crucial to this effort.  
Because of experimental resolution and Fermi motion (for nuclear
targets) a description of the average cross section in the resonance
region is expected to be sufficient, and hence duality can also be
used as a tool to model this transition.

% -----------------------------------------------------------------------
\subsection{Duality in Meson Electroproduction}
\label{ssec:pion_outlook}

While considerable insight into quark-hadron duality has already been
gained from inclusive electron scattering studies of the $F_1$, $F_2$,
and $g_1$ structure functions, duality in the case of semi-inclusive
meson photo- and electroproduction is yet to be as thoroughly tested
experimentally.
Here, duality would manifest itself in an observed scaling in the
meson plus resonance final state \cite{ACW}.

As discussed in Secs.~\ref{ssec:pion} and \ref{ssec:semi}, at high
energies one expects factorization between the virtual photon--quark
interaction, and the subsequent quark $\to$ hadron fragmentation.
In this case the $e N \to e^\prime h X$ reaction cross section,
at leading order in $\alpha_s$, is simply given by a product of the
parton distribution function and a quark fragmentation function to
a specific hadron $h$.
We will initially restrict ourselves to the hadron being an energetic
meson ($\pi$ or $K$) detected in the final state in coincidence with
the scattered electron.
The detected meson is assumed to carry most, but not all, of the
energy transfer, such that other mesons may also be produced.
We will come back to heavier mesons and baryons at the end of this
section.

By selecting only mesons carrying most of the energy transfer, one can
more cleanly separate the target and current fragmentation regions.
However, at low energies the struck quark still converts into the
meson in the vicinity of the scattering process, and it is not obvious
that here one can make the simplifying assumption of factorization.
Nonetheless, if duality holds one may see behavior consistent with the
simple high-energy factorization picture, and recent data does tend to
support factorization at lower energies than previously assumed.

The implication for semi-inclusive scattering is then that the overall
scale of scattering in the low-$W^\prime$ region must mirror that at
high $W^\prime$, where $W^\prime$ is the mass of the unobserved
hadrons.
This surprising property may come about {\it if} the various decay
channels from resonances with varying $W^\prime$ interfere in such
a way as to produce factorization.
Obviously, this would require a nontrivial interference between the
decay channels, although there are empirical indications for such
behavior from hadronic $\tau$ decays \cite{SHIFMAN_REV}, as well as
theoretically in quark model studies \cite{CI}
(see Sec.~\ref{ssec:semi}).
Schematically, the resonance region would appear here as the
exclusive limit of a high-energy fragmentation function
$D_{q \rightarrow h}(z,Q^2)$, similar to the momentum spectrum
of produced hadrons in the inclusive hadron production reaction
$\gamma^*N \rightarrow MX$ in Fig.~\ref{fig:corresp} \cite{BJK}.

These considerations strongly suggest that a beam energy of order
10~GeV will provide the right kinematical region to quantitatively
study the appearance of duality, and the associated onset of
factorization, in meson electroproduction.
In the framework of duality, separating current and target
fragmentation regions and restricting oneself to the region
$W^\prime > 2$~GeV (beyond the resonance region of the residual
system after removal of the meson) are sufficient conditions to
mimic the high-energy limit.
An upcoming Hall C experiment at JLab \cite{E00108} will investigate
this in detail, addressing two main questions:
(i) Do the $\gamma^*N \to \pi^\pm X$ cross sections factorize at
low energies and reproduce the fragmentation functions determined
from high-energy scattering?
(ii) Do nucleon resonances average around these high-energy
fragmentation functions and exhibit duality?
Duality may still be found in these processes, regardless whether
factorization (in $x$ and $z$) does or does not hold \cite{CI}.

If factorization {\em is} found to hold, it can open up new lines
of investigation into quark fragmentation and QCD at $\sim 10$~GeV
energies.
Jefferson Lab with a 12~GeV electron beam \cite{pCDR} would be an
ideal facility to study meson production in the current fragmentation
region at moderate $Q^2$, allowing the onset of scaling to be tracked
in the pre-asymptotic regime.
This would allow for unprecedented studies of the spin and flavor
dependence of duality, which can be most readily accessed through
semi-inclusive electron scattering.
In addition, one could also study the role of duality in
exclusive reactions, such as deeply virtual Compton scattering
or (hard) pion photoproduction (Secs.~\ref{sssec:dvcs}
and~\ref{sssec:excl_pi}), which may answer the important practical
question of whether the recently developed formalism of generalized
parton distributions is applicable at intermediate energies.

At higher energies ($\sim$ 100--200~GeV) experiments at CERN
\cite{Sl88} have demonstrated the existence of a clear separation
between current and target fragmentation regions.
Although these data have been accumulated at high $W'$, there does
not seem an {\em a priori} reason why such separation would not
persist into the low-$W'$ region, for sufficiently high energies.
Extending such data into the unexplored low-$W'$ region would enable
a detailed investigation of duality in the current fragmentation
region with various meson and baryon tags, in addition to a search
for duality in target fragmentation.
With new data on semi-inclusive scattering in and beyond the
resonance region, one can use tags of various mesons to test whether
sensitivity to sea quarks can be enhanced with $K^-$ or $\phi$ mesons,
where resonances are not easily produced in the residual system,
and what the vector mesons are dual to.
An understanding of duality for baryon tags and target fragments
would be the next challenge for electron scattering experiments.

\begin{figure}[t]
\begin{minipage}{3.1in}
\epsfig{figure=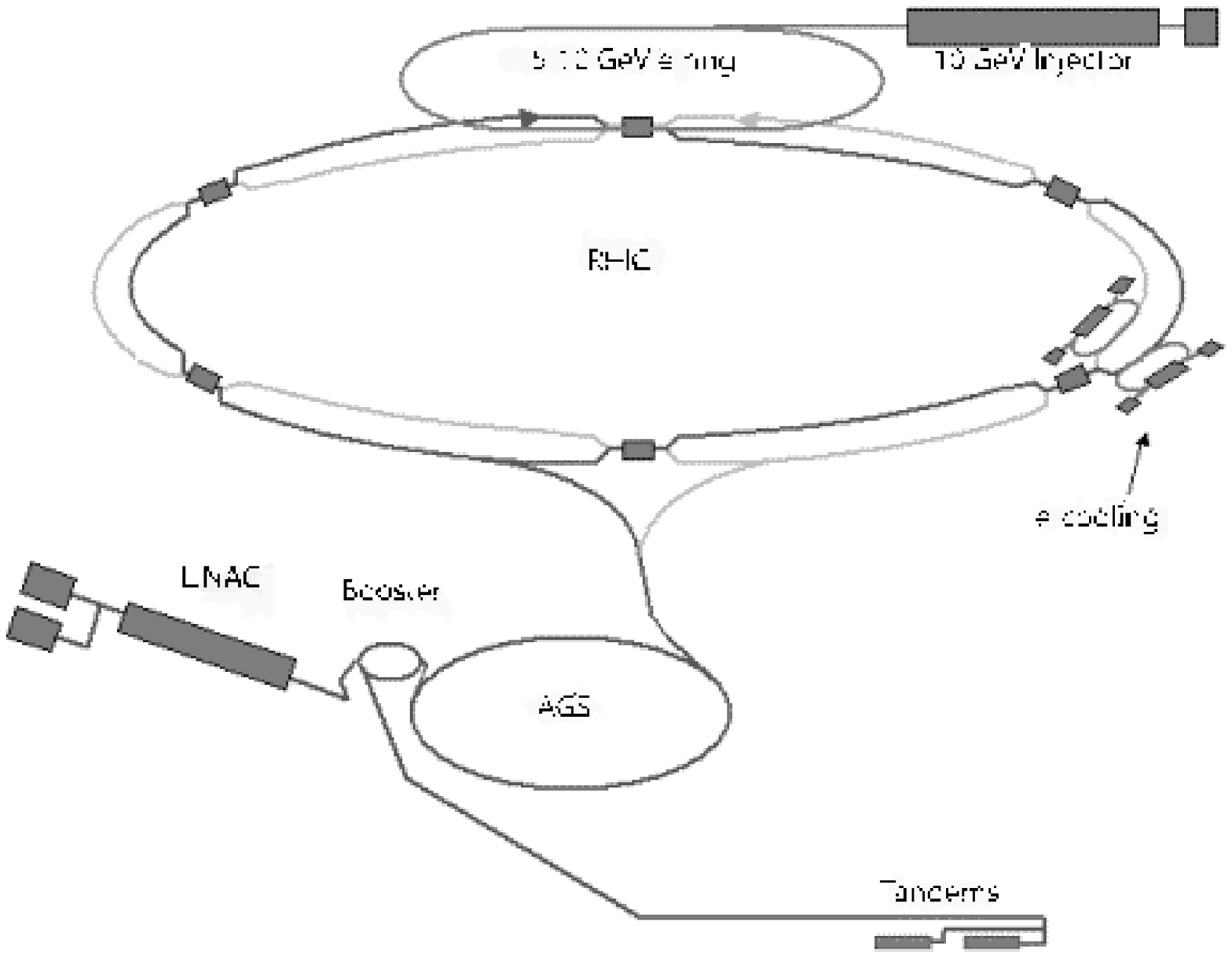, height=5.5cm} % width=7cm}
\end{minipage}
\hspace*{-0.7cm}
\begin{minipage}{3.1in}
\epsfig{figure=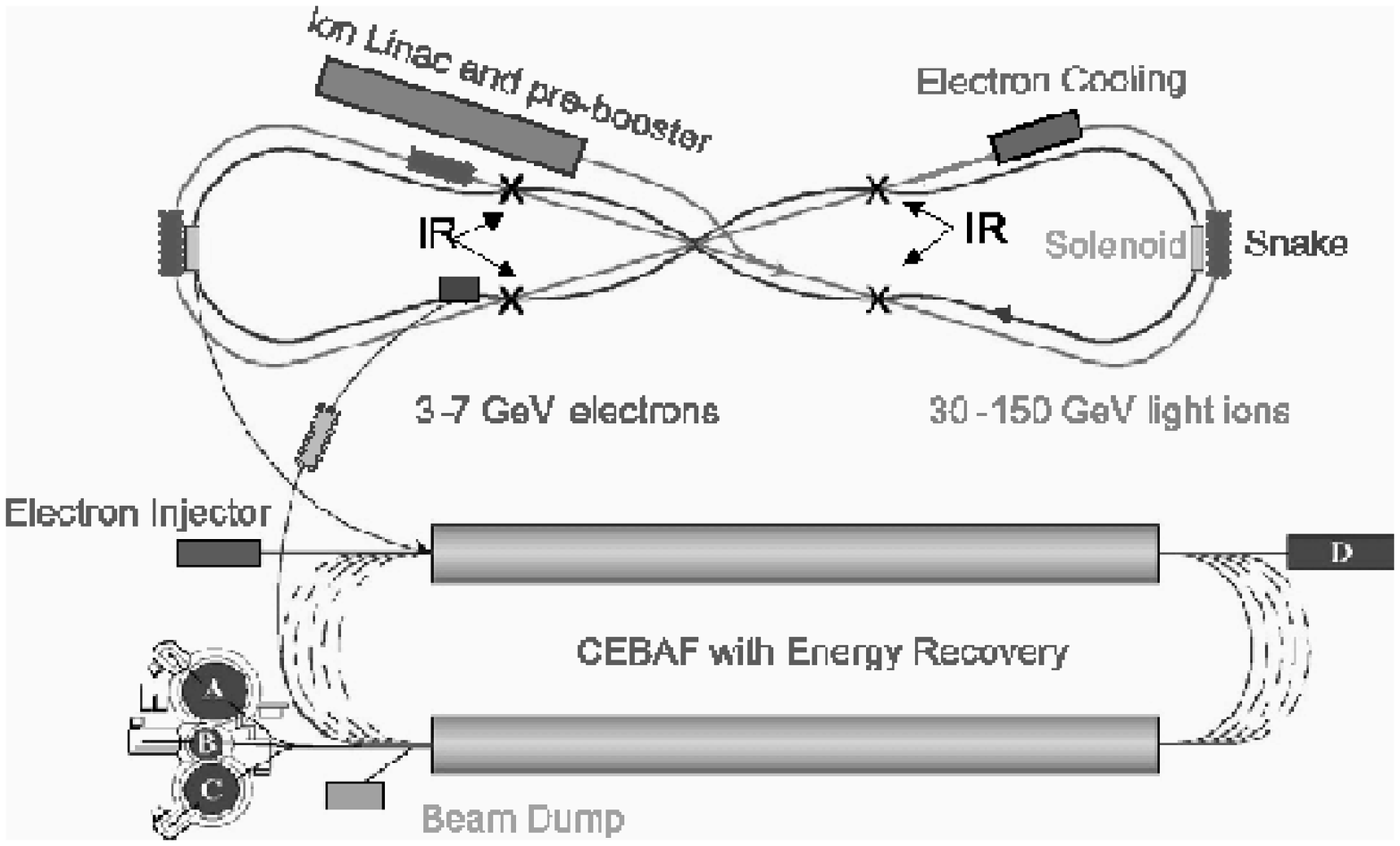, height=5.5cm} % width=7cm}
\end{minipage}
\vspace*{0.7cm}
\begin{centering}
\caption{\label{fig:colliders}
	Schematic layout of the electron-Relativistic Heavy Ion
	Collider (eRHIC) under investigation at Brookhaven National
	Laboratory \protect\cite{NSAC,EPIC} (left), and the
	Electron-Light Ion Collider (ELIC) at Jefferson Lab
	\protect\cite{Mermpc} (right).}
\end{centering}
\end{figure}

Higher electron energies would also enable one to investigate duality
in the heavy quark sector \cite{VS,ISGUR_VIOL,ISGUR_DUAL}, and that
between hadrons and jets \cite{Oc99} (Sec.~\ref{sssec:jet}).
The higher center of mass energy projected at the electron--hadron
colliders presently under discussion at both Brookhaven National
Laboratory and Jefferson Lab \cite{NSAC,EPIC}
(see Fig.~\ref{fig:colliders}) will allow for a superior tool to
perform such studies.
Despite a lower luminosity than available for fixed-target experiments,
such a collider would use its far higher center of mass energy to
enable measurements transcending the region where perturbative QCD
calculations are more readily applicable, and factorization of current
and target fragmentation regions less problematic.
The development of such facilities offers an exciting opportunity to
push our understanding of hadron structure far beyond its present
limits.

\clearpage

%%%%%%%%%%%%%%%%%%%%%%%%%%%%%%%%%%%%%%%%%%%%%%%%%%%%%%%%%%%%%%%%%%%%%%%%%%
\section{Conclusion}
\label{sec:conclusion}

The historical origins of quark-hadron duality can be traced back
to the 1960s, and the discovery of $s$- and $t$-channel duality in
hadronic reactions.
This duality reflected the remarkable relationship between low-energy
hadronic cross sections and their high-energy behavior, which, in the
context of finite energy sum rules, allowed Regge parameters
(describing high-energy scattering) to be inferred from the
(low-energy) properties of resonances.

It was natural, therefore, that the early observations of a duality
between resonance production and the high-energy continuum in inclusive
electron--nucleon deep inelastic scattering would be interpreted within
a similar framework.
Bloom \& Gilman found that by averaging the proton $F_2$ structure
function data over an appropriate energy range the resulting structure
function closely resembled the scaling function which described the
high-energy scattering of electrons from point-like partons.

With the emergence of QCD came the realization that at a fundamental
level the quark-hadron duality phenomenon reflects the transition
between the physics of nonperturbative and perturbative QCD.
In particular, the development of the operator product expansion in
high-energy physics allowed a very simple interpretation of duality
in terms of leading-twist parton distribution functions, with
violations of duality attributed to higher-twist effects associated
with nonperturbative multi-parton correlations.

Simply from unitarity, the access to a complete set of states means
that descriptions of observables in terms of either quark and gluon
or hadronic degrees of freedom must be equivalent, which itself is a
statement about the existence of confinement in QCD.
In practice, however, the computation of observables in different
kinematic regions is often made considerably easier with a different
choice of basis; in terms of hadrons at low energies, or in terms of
quarks and gluons at high energies.
The existence of regions where {\em both} truncated quark-gluon
{\em and} hadronic bases can provide accurate descriptions is one of
the remarkable consequences of duality.
The same forces of confinement also mean, however, that there must
exist a limit to how small these regions can be, and that duality
must eventually break down at a very local level.

Electron scattering provides a wonderful stage for investigating
the dynamical origin of quark-hadron duality.
The perturbative scaling of the deep inelastic structure functions
occurs here in terms of the parton light-cone momentum fraction $x$,
which can be accessed at different values of $Q^2$ and $W^2$, both
within and outside the resonance region.
Hence, both the resonance spectra and the scaling function describing
the high-energy cross section can be mapped by varying the mass $Q^2$
of the virtual photon.

Following the pioneering deep inelastic scattering experiments at
SLAC over three decades ago, the availability of (continuous wave)
high-luminosity polarized beams, together with polarized targets,
has allowed one to revisit Bloom-Gilman duality with unprecedented
precision, and disentangle its spin, flavor, and nuclear dependence,
both in local and global regions.
The results have been striking: quark-hadron duality occurs at much lower
$Q^2$ and in far less limited regimes than could have been expected.

For spin-averaged structure functions, the resonance region spectra
agree to better than 10\% precision with the perturbative scaling
function, down to $Q^2$ as low as 0.5~GeV$^2$.
This is true for both the transverse and longitudinal structure
functions, which is all the more remarkable given that the
longitudinal structure function is associated with higher-twist
contributions in QCD.
Moreover, the quark-hadron duality phenomenon is found to occur in
fairly local regions of $W^2$, working quite well even in the region
where only the $\Delta$ resonance resides.
In nuclei, the well established $\xi$-scaling behavior is found to be
just a reflection of the nucleus averaging the nucleon electromagnetic
response over a finite energy range, by virtue of the nucleon Fermi
motion.

For spin-dependent structure functions, the onset occurs at higher
$Q^2$, but indications are that also here the transition to
perturbative scaling behavior is mostly complete by
$Q^2 = 2$~GeV$^2$, even for local regions of $W^2$.
The slower onset of duality reflects the greater role played by the
$\Delta$ resonance (as the ground state of the spin-3/2 states)
here than in spin-independent scattering.
Nonetheless, the region beyond $W^2 = 2$~GeV$^2$ already closely
follows the perturbative scaling behavior for $Q^2 \agt 0.8$~GeV$^2$.
The existing evidence indicates that duality works at lower $Q^2$
for the neutron than the proton, but additional data are needed to
quantify this more precisely.

The reported experimental developments in the study of quark-hadron
duality have coincided with considerable progress made over the last
few years in our theoretical understanding of this phenomenon.
Perturbative QCD calculations are now available with high precision,
making detailed studies of the $Q^2$ dependence of structure functions
possible.
Moment analyses of structure function data show surprisingly small
amounts of higher-twist contributions to the low moments, suggesting
that single quark processes dominate the scattering mechanism even
down to $Q^2 \sim 1$~GeV$^2$ for some observables.
It is not yet understood, however, from first principles in QCD why
a leading-twist description should provide a good approximation to
structure functions at such low $Q^2$.

Indeed, while the operator product expansion allows one to identify
and organize the duality violations in terms of the matrix elements
of higher-twist operators, it cannot by itself explain
{\em why} certain higher-twist matrix elements are small, or cancel.
Physical insight into the origins of early (or ``precocious'') scaling
requires nonperturbative methods, such as lattice QCD, or QCD-inspired
models, in order to understand the dynamics responsible for the
transition to scaling.
While lattice simulations of leading-twist matrix elements are
approaching a relatively mature stage, with direct comparisons
with experiment soon feasible, calculations of higher-twist matrix
elements are still in their infancy.

In light of this, many of the recent developments on the theoretical
front have been in the context of models, with varying degrees of
sophistication, which have allowed a number of important features of
the quark-hadron transition in structure functions to be elucidated.
An important realization has been that resonances themselves constitute
an integral part of scaling structure function, and that the
traditional resonance--scaling distinction is somewhat arbitrary.
Phenomenologically, while the traditional resonance region
($W \alt 2$~GeV) contributes a significant part of the total structure
function at low $Q^2$ ($\sim 70\%$ for the $n=2$ moment of $F_2^p$ at
$Q^2=1$~GeV$^2$), the higher-twist contributions at the same $Q^2$
are considerably smaller ($\sim 10\%$ of the total $F_2^p$ moment).

Theoretically, this dichotomy can be dramatically illustrated in
the large-$N_c$ limit of QCD, where the hadronic spectrum consists of
infinitely many narrow resonances, which are protected from strong
decay by the suppression of sea quark loops.
Since the quark level calculation yields a smooth, scaling function,
some form of resonance averaging is needed to yield the required
quark-hadron duality, even in the limit as $Q^2 \to \infty$.
This can be explicitly demonstrated in the case of 1+1 dimensions
(the 't~Hooft model), or in models which provide generalizations
to 3+1 dimensions.
Other model studies have clarified how scaling can arise in the
presence of strong confining interactions responsible for the
formation of bound state resonances.
For the case of a harmonic oscillator potential, the energy spectrum
can be calculated exactly, and the inclusive structure function
obtained from a direct sum over the squares of transition form factors.

At first sight the equivalence of a coherent sum over exclusive
$N \to N^*$ transitions and an incoherent sum over individual quark
contributions to the inclusive structure function seems impossible;
the former (in the case of electric couplings) is proportional to the
square of a sum of quark charges, while the latter is given by the sum
over the squares of quark charges.
A resolution of this dilemma comes with the observation that
interference effects, such as those between even- and odd-parity
excited states, can result in cancellations of the duality-violating
cross terms (which can be identified with higher-twist, multi-parton
effects), exposing the leading-twist, diagonal contributions which
interfere constructively.
Critical to this observation is the fact that a certain minimum subset
of states must be summed over in order for duality to be saturated.
Such patterns of constructive and destructive interference can be
realized in phenomenological models, such as the nonrelativistic quark
model and its various extensions, for both
unpolarized and polarized structure functions.

Although these studies have shed considerable light upon the dynamical
origins of quark-hadron duality, there are still important questions
which need to be addressed before we come to a quantitative
understanding of Bloom-Gilman duality in the structure function data.
The observation of duality in spin-averaged structure functions in the
region of the $\Delta$ resonance, for instance, suggests nontrivial
interference effects between resonant and nonresonant (background)
physics.
Early descriptions of the resonance and background contributions
employed the so-called two-component model of duality, in which the
resonances are dual to valence quarks (associated with the exchange
of Reggeons at high energy), while the background is dual to the
$q\bar q$ sea (associated with Pomeron exchange).
In more modern language, this would call for a QCD-based derivation
in which the properties of the nonresonant background can be
calculated within the same framework as those of the resonances
on top of which they sit.

It is also clear that the quark-hadron duality phenomenon is not
restricted to inclusive electron--hadron scattering alone.
If, as we believe, it is a general property of QCD, then it should
manifest itself in other processes and in different observables.
There are, in fact, predictions for quark-hadron duality in
semi-inclusive and exclusive electroproduction reactions.
The available evidence is scant, but it does suggest that at energy
scales of a few GeV such reactions may proceed by closely mimicking
a high-energy picture of free electron-quark scattering.
This will be an exciting area of research for the next decade,
within reach of the energy and luminosity of 1--100~GeV electron
scattering facilities.

Further afield, important lessons about Bloom-Gilman duality can be
learned from duality in areas outside of electron-hadron scattering.
The prototypical reaction in which duality has been studied is
$e^+ e^-$ annihilation into hadrons.
Semileptonic and nonleptonic weak decays of heavy mesons have provided
extremely rich ground on which duality has been tested and quantified.
In fact, duality between heavy quarks and heavy mesons is vital here
for the extraction of fundamental Standard Model parameters such as
the CKM matrix elements.
More recent applications of quark-hadron duality include
deeply virtual Compton scattering, and $p\bar p$ annihilation.
Duality also underpins the entire successful phenomenology of the
QCD sum rule method of computing hadronic observables.

It is truly remarkable that in a region where we have only a few
resonances, all consisting of strongly interacting quarks and gluons,
the physics still ends up resembling a perturbative quark-gluon theory.
Quark-hadron duality is the underlying cause of the smooth transition
``on average'' from hadrons to quarks witnessed in Nature, allowing
simple partonic descriptions of observables down to relatively
low-energy scales.
Unraveling the dynamics and origin of quark-hadron duality may well
hold the key to understanding the details of the quark-hadron
transition in QCD.

\clearpage

%%%%%%%%%%%%%%%%%%%%%%%%%%%%%%%%%%%%%%%%%%%%%%%%%%%%%%%%%%%%%%%%%%%%%%
\vspace*{1cm}

{\bf ACKNOWLEDGEMENTS} \\

This work was supported in part by the U.S. Department of Energy (DOE)
contract \mbox{DE-AC05-84ER40150}, under which the Southeastern
Universities Research Association (SURA) operates the Thomas Jefferson
National Accelerator Facility (Jefferson Lab).

\vspace*{1cm}

%%%%%%%%%%%%%%%%%%%%%%%%%%%%%%%%%%%%%%%%%%%%%%%%%%%%%%%%%%%%%%%%%%%%%%%%%%
{\bf APPENDIX:}\ \ {\bf SCALING VARIABLES}
% \label{app:var}

\vspace*{1cm}

Here we collate and summarize for reference the various scaling
variables which are commonly used in the literature, or which have been
used in this report.

\begin{enumerate}

\item
The classic scaling variable derived by Bjorken is defined in terms
of the ratio of the squared momentum and energy transferred to the
nucleon,
\begin{eqnarray}
x &=& { Q^2 \over 2 M \nu }\ ,
\end{eqnarray}
where $M$ is the mass of the nucleon.
This is the correct scaling variable in the Bjorken limit, in which
both $Q^2$ and $\nu \to \infty$.
In this limit the variable $x$ corresponds to the ``plus''-component of
the light-cone momentum fraction of the nucleon carried by the struck
parton.
In the early literature one often encounters the inverse of the
Bjorken-$x$ variable, namely
\begin{eqnarray}
\omega &=& { 1 \over x }\ .
\end{eqnarray}

\item
Different variables have been suggested in order to improve the
scaling behavior at finite momentum transfer, in the pre-asymptotic
region.
Phenomenologically, Bloom \& Gilman introduced the variable
\cite{BG1}
\begin{eqnarray}
\omega^\prime &=& { 2 M \nu + M^2 \over Q^2 }\
               =\ 1 + { W^2 \over Q^2 }\
               =\ \omega + { M^2 \over Q^2 }\ ,
\end{eqnarray}
which they found gave better scaling in the $F_2$ structure function
in the resonance region.

\item
An improved scaling variable which was subsequently derived from the
kinematics of deep inelastic scattering at finite $Q^2$ is the Nachtmann
variable \cite{EONacht},
\begin{eqnarray}
\xi &=& { 2 x \over 1 + \sqrt{1 + Q^2/\nu^2} }\
     =\ { 2 x \over 1 + \sqrt{1 + 4 M^2 x^2/Q^2} }\ ,
\end{eqnarray}
which takes into account target mass corrections, $M^2/Q^2$.
Expanding $\xi$ in powers of $1/Q^2$ at high $Q^2$ gives
\begin{eqnarray}
{ 1 \over \xi }\ \approx\ \omega\ +\ { M^2 \over \omega\, Q^2 }\ ,
\end{eqnarray}
which makes apparent the origin of the Bloom \& Gilman variable
$\omega'$ above.

\item
While the Nachtmann $\xi$ variable is an improvement on Bjorken-$x$
at finite $Q^2$, it is, however, not the most general scaling variable.
It implicitly assumes massless, on-shell quarks with zero transverse
momentum.
A generalization of the Nachtmann variable to include finite quark
masses $m$ was made by Barbieri {\em et al.} \cite{BARBIERI},
\begin{eqnarray}
\xi_{\rm Barb}
&=& { Q^2 + \sqrt{ (Q^2 + 4 m^2 Q^2 }
\over 2 M \nu \left( 1 + \sqrt{1 + Q^2/\nu^2} \right) }\ .
\end{eqnarray}
Here the initial and final quarks are taken to have the same mass,
although the generalization to unequal masses is straightforward
\cite{BARBIERI}.

\item
Including in addition the effects of quark transverse momentum, the
light-cone momentum fraction of the nucleon carried by a parton can
be written in its most general form as
\begin{eqnarray}
\gamma
&=& { x \over 1 + \sqrt{ 1 + Q^2/\nu^2 } }
   \left( 1 - { k^2 - m^2 \over Q^2 } \right)
   \left( 1 + \sqrt{ 1 + { 4 (k^2 + k_T^2)/Q^2 \over
                           ( 1 - (k^2-m^2)/Q^2 )^2 } }
   \right)\ ,
\end{eqnarray}
where $k^2$ is the quark virtuality, and $k_T$ the quark
transverse momentum.
In the limit $k^2/Q^2 \sim k_T^2/Q^2 \sim m^2/Q^2 \ll 1$,
this variable reduces to the Nachtmann scaling variable,
$\gamma \to \xi$.

\item
Variables used in various model studies in this report include
the nonrelativistic scaling variable employed by Greenberg
\cite{GREENBERG}
\begin{equation}
x_{\rm nr} = { \vec q \, ^2 \over 2 M \nu }\ ,
\end{equation}
which uses $\vec q \, ^2$ rather than the four-momentum transfer
squared, $Q^2$.

\item
Taking into account the effects of the spectator ``diquark'' system,
once a quark has been removed from the nucleon, Gurvitz
\cite{GURVITZ_BSE} derived the light-cone momentum fraction carried
by a struck quark as
\begin{equation}
\tilde x
= { x + \sqrt{1 + 4 M^2 x^2/Q^2} - \sqrt{(1-x)^2 + 4 m_s^2 x^2/Q^2}
   \over 1 + \sqrt{1 + 4 M^2 x^2/Q^2} }\ ,
\end{equation}
where $m_s$ is the mass of the spectator system.
At large $Q^2$, $\tilde x$ and $\xi$ are related by \cite{GURVITZ_BSE}
\begin{equation}
\tilde x
\simeq \xi + {M^2 x^2 \over Q^2} - {m_s^2 x^2 \over (1-x) Q^2}\ ,
\end{equation}
and $\tilde x \to \xi \to x$ in the limit $Q^2\to\infty$.

\item
The scaling variable $u$ used in the heavy-light quark model of
Isgur {\em et al.} \cite{IJMV} is given by
\begin{eqnarray}
u &=& {1 \over 2 m}
      \left( \sqrt{\nu^2 + Q^2} - \nu \right)
      \left( 1 + \sqrt{1 + {4 m^2 \over Q^2}} \right)\ ,
\end{eqnarray}
which takes into account both target mass and quark mass effects
\cite{BARBIERI} ({\em cf.} the variable $\tilde x$ above).
In the Bjorken limit the variable $u$ becomes the scaled Bjorken
variable,
\begin{eqnarray}
u \to { M \over m }\ x\ .
\end{eqnarray}

\item
Finally, the West scaling variable $y$ is defined in terms of the
momentum $q$ and energy $\nu$ of the photon,
\begin{equation}
y = -{q \over 2} + {m \nu \over q}\ .
\end{equation}

\end{enumerate}

\newpage
%%%%%%%%%%%%%%%%%%%%%%%%%%%%%%%%%%%%%%%%%%%%%%%%%%%%%%%%%%%%%%%%%%%%%%%%%%

\end{document}